\RequirePackage{fix-cm}
\documentclass[a4paper]{article}
\usepackage{xcolor}

\usepackage{graphicx}
\usepackage[section]{placeins}
\makeatletter
\AtBeginDocument{%
  \expandafter\renewcommand\expandafter\subsection\expandafter{%
    \expandafter\@fb@secFB\subsection
  }%
}
\makeatother

\usepackage[utf8]{inputenc}
\usepackage{subcaption}
\usepackage{booktabs}
\usepackage{natbib}

\usepackage{algorithm}
\usepackage{algpseudocode}

\usepackage{microtype}
\usepackage[hidelinks]{hyperref}

\usepackage [english]{babel}
\usepackage [autostyle, english = american]{csquotes}
\MakeOuterQuote{"}

\usepackage{amsmath}

\usepackage{multirow}
\usepackage[normalem]{ulem}
\useunder{\uline}{\ul}{}

\usepackage{cleveref}

\begin{document}

\title{The impact of social influence in Australian real-estate: market forecasting with a spatial agent-based model}

\author{
Benjamin Patrick Evans         \and
        Kirill Glavatskiy   \and
        Michael S. Harr\'e \and
        Mikhail Prokopenko
}

\date{Centre for Complex Systems, The University of Sydney, Sydney, NSW, Australia \footnotetext{The authors are thankful to Paul Ormerod, Adri\'an Carro and Markus Brede for many helpful discussions of the baseline model. The authors acknowledge the HPC service at The University of Sydney for providing HPC resources that have contributed to the research results reported within this paper. The authors would also like to acknowledge the Securities Industry Research Centre of Asia-Pacific (SIRCA) and CoreLogic, Inc. (Sydney, Australia) for their data on Greater Sydney housing transactions.} }


\maketitle

\begin{abstract} % 150 to 250 words
Housing markets are inherently spatial, yet many existing models fail to capture this spatial dimension. Here we introduce a new graph-based approach for incorporating a spatial component in a large-scale urban housing agent-based model (ABM). %with several hundred thousand agents (households).
The model explicitly captures several social and economic factors that influence the agents' decision-making behaviour (such as fear of missing out, their trend following aptitude, and the strength of their submarket outreach), and interprets these factors in spatial terms. The proposed model is calibrated and validated with the housing market data for the Greater Sydney region. The ABM simulation results not only include predictions for the overall market, but also produce area-specific forecasting at the level of local government areas within Sydney as arising from individual buy and sell decisions. In addition, the simulation results elucidate agent preferences in submarkets, highlighting differences in agent behaviour, for example, between first-time home buyers and investors, and between both local and overseas investors.

\end{abstract}

\section{Introduction}

%Economic markets establish a price of goods and services within the marketplace, and developing economic models to understand and forecast these economic markets has long been a goal of economists (\cite{friedman2013fortune}). For example, Lawrence Klein won the Nobel Prize in 1980 for his work on macroeconomic forecasting (\cite{evans1967wharton, klein1955econometric}), likewise,  Trygve Haavelmo won in 1989 for his contributions to probabilistic econometric forecasting (such as in \cite{haavelmo1944probability}). 

%Economic forecasting is non-trivial due to the range of endogenous and exogenous factors affecting predictions, which have traditionally led to a variety of simplifying assumptions being made in many economic models (\cite{stiglitz2002information}), although new computational models may help overcome such assumptions (\cite{calder2018computational, lehtinen2007computing}) thanks to the increasing availability of computational power. Here, we specifically address our focus on these relatively new computational economic models (namely Agent-Based Models) and their applications to housing market modelling, due to a range of interesting and unique characteristics with great economic interest.

Within economic markets, housing markets are unique for a variety of reasons. The combination of durability, heterogeneity and spatial fixity amplifies the role of the dwellings' perceived value and the buyers and sellers' expectations~\citep{alhashimi2004there}. The extremely high cost of entry and exit into the market (with moving fees, agent fees, etc.) further complicates the decision-making of participating households~\citep{huang2009house}. 
There are long time delays in the market response as houses can not be erected instantaneously to accommodate an increase in demand~\citep{bahadir2014housing}. 
The fact that real-estate can be seen as both an investment asset and a consumption good \citep{piazzesi2007housing} (and even a status good, \cite{wei2012status}) magnifies the impact of social influence on both decision-making and resultant market dynamics and structure.

Consequently, housing markets are notoriously difficult to model as the ensuing market dynamics generates volatility, with non-linear responses, and "boom-bust" cycles \citep{sinai2012house, miles2008boom, burnside2016understanding},  making traditional time-series analysis insufficient. Non-linear dynamics of housing markets are ubiquitous, being observed throughout the world, from Tokyo \citep{shimizu2010housing} to Los Angeles \citep{cheng2014wall}. 

Traditional economic modelling methods, such as dynamic stochastic general equilibrium models (DSGE), typically use representative aggregated agents while making strong assumptions about the behaviour of the markets (rational and perfect competition). Such representative agents may be limiting for economic models \citep{gallegati1999beyond}. Furthermore, these assumptions (and many other traditional economic assumptions) are known to be inadequate in housing markets, motivating a well-recognised need for change in housing market modelling \citep{mcmaster1999economics}. In addressing this need, a specific type of models, called agent-based models (ABM) has been applied. ABMs aim to capture markets from the "bottom-up" \citep{tesfatsion2002agent}, i.e., by focusing on the decision-making of individual agents in the market, possibly influenced by non-economic factors. In this sense, ABMs are capable of modelling macroeconomies from micro (i.e., agent-specific) behaviour \citep{tesfatsion2008modeling} and analysing the economic decision-making in counter-factual settings. While ABMs have shown promise in housing market modelling \citep{geanakoplos2012getting} (and wider economic modelling, \cite{poledna2019economic}), current ABMs themselves are not exempt from some limitations. Many of the existing housing ABMs tend to introduce at least one of the following constraints: the spatial structure of markets is neglected (meaning submarkets are not considered), perfect information is still assumed, and/or the impact of social influence on decision-making of individual agents is underestimated.

A fine-resolution model of spatiotemporal patterns within such markets is desirable: it would give an understanding of how market dynamics shape within local areas, explaining how the pricing structure directly affects the agents' mobility over time (i.e., by forcing households out of certain regions due to gentrification and higher cost of living). Furthermore, within such housing markets (and in fact, many economic markets, \cite{conlisk1996bounded}), it is also known that agents do not act perfectly rational \citep{wang2018agent}, instead following bounded rationality \citep{simon1955behavioral, simon1957models}).  Firstly, humans are often influenced by social pressure (e.g., herd mentality), with the decisions being made purely based on social pressure rather than a perfectly rational choice. Secondly, it is difficult to process all the relevant information in the market (i.e., it is impractical for an agent to be able to view every dwelling listing within the housing market). Thus, it is unreasonable to assume that agents act in a perfectly rational manner, yet this is what many current housing market models assume (despite ABMs not intrinsically requiring these assumptions to be made).

To address these limitations, we introduce a spatial agent-based model, in which the constraints imposed by various search and mobility costs create effective spatial submarkets. These submarkets are modelled graph-theoretically, with a graph-based component used in both representing imperfect information and modulating social influences. The spatial ABM is then capable of capturing the "boom-bust" cycles observed in Australia (in particular, Greater Sydney) over the last 15 years as arising from individual agent decisions. That is, we show complex non-linear market dynamics as arising through individual buy and sell decisions in the market, with introduced factors which affect and explain the decision-making behaviour. The model succeeds in forecasting nonlinear pricing and mobility trends within specific submarkets and local areas. In exploring the pricing dynamics, we focus on the influence of imperfect spatial information and the role of social influence on agent decision-making.  In doing so, we identify the salient parameters which drive the overall dynamics and decision-making, and pinpoint the parameter thresholds, beyond which the resultant dynamics exhibit strong nonlinear responses. These thresholds allow us to distinguish between different configurations of the market (e.g., markets with supply or demand dominating) and differences in agent behaviour.  In addition, we identify and trace specific interactions of parameters, in particular the interplay of social influences, such as the fear of missing out and the trend following aptitude, in presence of imperfect spatial information.

The remainder of the paper is organised as follows. In \cref{secBackground} we provide an overview of agent-based models of housing markets. In \cref{secModel} we outline the baseline model, while in \cref{secExtension} we outline the proposed spatial extensions and new parameters. In \cref{secBayesian} we analyse the sensitivity and parameters of the model, before presenting the results and discussion in \cref{secResults}. In \cref{secConclusions} we provide conclusions and highlight future work.

\section{Background}\label{secBackground}

\subsection{Agent-Based Models of Housing Markets}

One of the pioneering works for agent-based modelling (ABM) of the housing markets was by \cite{geanakoplos2012getting} (and extended further in \cite{axtell2014agent, goldstein2017rethinking}), where the Washington DC market was modelled from 1997--2009 in an attempt to understand the housing boom and crash. Macroeconomic experiments were then conducted to see how changing underlying factors, such as interest rates or leverage rates, would affect this pricing trend.

\cite{baptista2016macroprudential} model the UK housing market to see the effects that various macroprudential policies have on price cycles and price volatility. \cite{gilbert2009agent} also looks at the English housing market, varying exogenous parameters and policies, and tracking the effect these have on median house prices.
Likewise, \cite{carstensen2015agent} explore the Danish housing market and macroprudential regulations, such as income and mortgage rate shocks.

\cite{ge2017endogenous} analyse how housing market bubbles can form (and bust) purely endogenously without external shocks, due to leniency and speculation of agents. \cite{kouwenberg2015endogenous} also show an ABM can "endogenously produce boom-and-bust cycles even in the absence of fundamental news".

A recent ABM of the Australian housing market proposed by \cite{glavatskiy2020explaining} explained the volatility of prices over three distinct historic periods, characterised by either steady trends or trend reversals and price corrections. This model highlighted the role of the agents' trend-following aptitude in accurately generating distinct price dynamics, as detailed in \cref{secModel}.
In this paper, we further develop this model by introducing several features directly capturing social influences and bounded rationality in agent decision-making, as elaborated in \cref{secExtension}.

Traditionally, the modelling goal is to explain the housing market pricing (rather than predict its trajectory), and trace how possible macroeconomic policy changes may have affected the dynamics. Here, instead, we focus on predicting the pricing dynamics beyond the period covered by the current datasets (i.e., presenting out-of-sample forecasting) as arising from individual buy and sell decision-making behaviour in the market (as opposed to a black-box technical analysis machine learning-based approach), as well as illuminating agent preferences within specific housing submarkets. This motivation is aligned with the growing suggestions that agent-based models should be predictive \citep{polhill2018social} (which has admittedly been met with some resistance, \cite{edmonds2018using}).  Agent-based models have recently been shown to outperform traditional economic models, such as vector autoregressive models and DSGE models for out-of-sample forecasting of macro-variables (GDP, inflation, interest rates etc.) \citep{poledna2019economic}. For example, it was demonstrated that an ABM can outperform standard benchmarks for out-of-sample forecasting in the US housing market~\citep{kouwenberg2014forecasting}, while successful out-of-sample forecasting was carried out by \cite{geanakoplos2012getting} as well. 

\subsubsection{Spatial Models}

Spatial distribution of houses and dependencies between market trends on spatial patterns have been recognised as important and desirable features \citep{goldstein2017rethinking}. For example, \cite{baptista2016macroprudential} describe the spatial component as one that is 
"highly-desirable", yet "this approach greatly increases the complexity of the models and hence most spatial ABMs in the field listed below make use of a highly simplified
representation of the environment, often in the shape of small grids".

Spatial agent-based models have also shown to be useful in a variety of other areas such as epidemic modelling \citep{chang2020modelling, cliff2018investigating}, cooperative behaviour \citep{power2009spatial}, and symbiotic processes \citep{raimbault2020spatial}. Despite the promise shown by housing ABMs, there are currently only relatively few spatial housing market models with the capacity to accurately forecast nonlinear price dynamics. 

The seminal works in spatial housing ABMs  are by \citep{ge2013creates} and \citep{ustvedt2016agent}. Both use a matrix-based approach, with the region being arranged on a 2-dimensional grid. 
In \cite{ge2013creates}, each cell (row/column) in the grid is assigned a neighbourhood quality (endogenous) and a nature quality (exogenous).  The neighbourhood quality is a measure of attractiveness which aims to capture concepts such as safety, and is dependent on agents that live in that region (which can change in the model, thus endogenous). In contrast, nature quality is based on outside factors not changed by the model, such as distance to a beach or weather (thus exogenous).  Data used in this work is abstract, that is, it is not calibrated to a particular city, but rather used to trace how these factors affect the trends. 

\cite{ustvedt2016agent} also use a 2-dimensional grid for a NetLogo model. However, an important additional spatial step is made: district borders
are incorporated using GIS data (somewhat similar to what we propose in our model with the graph-based approach, however, there are important differences which we outline below), and the model is calibrated based on Oslo, Norway.

Another work is \citep{pangallo2019residential}, which models theoretical (i.e., not calibrated to any specific region) income segregation and inequality, using a spatial agent-based model, and the effect this may have on house prices. Again, this approach uses a 2-dimensional grid for the spatial component. This model assumes a monocentric city, and measures the "attractiveness" of a location, based on the distance to the (generic) city centre. The main contribution of this work is a mathematically tractable spatial model for capturing income segregation. The effects are related to the house prices, where unequal income is shown to lower the house price globally.

\paragraph{Unexplored Extensions}

The spatial work outlined above provide strong models which achieve their purpose of policy understanding and effect of market shocks in \citep{ge2013creates, ustvedt2016agent}. However, the spatial component is often considered a secondary point of the models, which means there are some key additional insights which have yet been unexplored. Particularly, the presence of area-specific submarkets have not been explicitly modelled, and analysis into the spatial preferences of agents within the market (and submarkets) have not been explored.

For example, in studies of \cite{ge2013creates} and \cite{ustvedt2016agent} the spatial contribution affects the initial price and the supply of dwellings for a given grid cell (location and population). However, at every simulated step the effects of the spatial component do not extend to varying search costs or probability of listing for agents. Furthermore, the spatial component is not utilised to initialise agent characteristics (such as income, wealth, etc.), based on the areas in which they reside. These area-specific agent characteristics and behaviours become particularly important for capturing submarkets across various areas, especially when real world data is available.

Another assumption of existing spatial models is that of a monocentric city, meaning a distance metric such as "distance to centre" is used for measuring attractiveness, which becomes problematic for polycentric cities or with agents who have no desire to live within the "centre" when analysing agent preferences. Furthermore, computing these distances in a 2d grid can often be misinformative, as moving across a region border (i.e., into a new zone) often incurs a far larger cost than moving a cell inward into the same zone. To address this, we explicitly capture this feature in the proposed graph-based spatial extension. Distances are measured as the shortest path through the graph, with nodes representing various regions (contained within boundaries).
The graph-based approach is particularly useful, as no monocentric assumption is made.

Futhermore, as detailed area-specific analysis remains largely unexplored, the distributions of prices within individual areas is often not considered, instead, using a representative mean or median rather than sampling from the actual underlying distributions for each area, which may fail to capture certain area trends. This is often caused by the lack of underlying data. In our work, we use several contemporary datasets, such as SIRCA-CoreLogic and the Australian Census datasets, constructing the relevant pricing probability density functions for each area.

In summary, in contrast to the grid-based approach commonly used, in this work, we propose an extensible graph-based approach which is described in \cref{secExtension}. Such an approach allows us to further exploit the spatial component by introducing submarkets with graph-based search costs and initialisation of agent characteristics based on areas, giving insights into area-specific submarkets which have not yet been explored. In addition, the graph-based approach does not assume a monocentric region, allowing for polycentric cities (which Greater Sydney is developing towards \cite{greater2018greater, crosato2020}) to be modelled more effectively.

\section{The Baseline Model}\label{secModel}

This work extends the work of \cite{glavatskiy2020explaining}, which we will refer to as the "Baseline" method.  %The outline here is intended to serve as enough of a background to understand the model and proposed extensions entirely, however, for a full description of all model characteristics, we refer you to the original paper \citep{glavatskiy2020explaining}. Throughout the paper, we refer to the method outlined in this section as the "Baseline" method.
In this section, we describe basic features of the baseline model, which the present work carries over.

\subsection{Agents}
There are three key agent types in the model: dwellings, households, and the bank.

A dwelling is a "physical" property, e.g. a  house, apartment or condo. Each dwelling has an intrinsic quality, which reflects its hedonic value (e.g. large house or existence of a pool). The quality is fixed during the simulation. The quality of a dwelling is used as a reference for determining its listing price and dwelling payments. All dwellings have an owner. Dwellings can be rented or sold.  A rental contract is a binding agreement between the owner and renting household. Dwellings may be vacant at any period (which means that they are not rented out).

A household represents a person or group of people (i.e., a family), which reside within Greater Sydney. Additionally, the model contains overseas agents, which can participate in the market but do not reside in the Greater Sydney region. Households have heterogeneous monthly incomes and liquid cash levels. Households can own several dwellings, but can only reside in one (overseas agents do not reside in any dwelling). When purchasing a dwelling, households always choose the most expensive dwelling they can afford. If they can afford to buy a dwelling, they always attempt to do so, putting a market bid (see below). Households that own more than one dwelling attempt to rent the additional dwellings out, and, if successful, receive rental payments as a contribution to their liquid cash. Households that do not own a dwelling rent one. Households pay tax based on their income and ownership. 

The bank combines the functions of a commercial bank and the regulatory body, controlling various financial characteristics, such as income tax rates, mortgage rates, overseas approval rates, mortgage approvals, and mortgage amounts (how much can be lent to a particular household).

\subsection{Behavioural rules}

The agents' behaviour is governed by the price they are willing to sell their dwelling for, the listing price, and the price they can afford to buy a new dwelling, the bid price. The basis for the pricing equations comes from the pioneering housing market ABM of \cite{axtell2014agent}.

The household bid price, i.e., their desired expenditure, is modulated by the household's monthly income $I[t]$ according to \cref{eqBidPrice}. Following \cite{axtell2014agent}, the desired expenditure formulation is the result of an analysis into income dynamics, motivated by the $\frac{1}{3}$ of income on housing expenses heuristic, with modifications to capture a wider range of heterogeneity and expenditures.

\begin{equation}\label{eqBidPrice}
    B[t] = H \frac{U_b[t]  \phi_b I[t]^{\phi_I}}{\phi_{M}[t] + \phi_H - h * \Delta_{HPI}[t]}
\end{equation}

$H$ is a uniformly random value between $1 \pm b_h/2$ for each bid, where $b_h = 0.1$ is the listing heterogeneity parameter. This allows variation in bid prices around a central value (i.e., uniform prices within a range around the centre, in this case, $\pm 5\% $). $U_b[t]$ is the urgency of a household to buy a dwelling, which is equal to $1$ if the household has recently not sold any dwelling, and is larger than $1$ by a term proportional to the number of months since the last sale otherwise. Furthermore, $\phi_I$ and $\phi_b$ are the income modulating parameters, which are calibrated from the mortgage-income regression for that period of interest. The resulting expenditures are sublinear with income, meaning high-income agents spend a lower percentage of their income on housing than their lower-income counterparts. In addition, $\phi_{M}[t]$ is the mortgage rate at time $t$, while $\phi_H$ is the annual household maintenance costs. Finally, $h$ is the trend-following aptitude and $\Delta_{HPI}[t]$ is the change in house price index (HPI) over the previous year.

The bid price of the overseas investors is determined as an average, given the total volume and quantity of the approved overseas investments by the Foreign Investment Review Board. Several aspects of the overseas investments are detailed in \cref{secOverseasApprovals}.  

The dwelling list price $P[t]$ at time $t$ is modulated by the quality of the dwelling $Q$ according to \cref{eqListPrice}. Again, the basis for the listing price equation comes from the estimation results of \cite{axtell2014agent}, where the formulation arose from a comprehensive set of real estate transaction data.

\begin{equation}\label{eqListPrice}
    P[t] = H \frac{b_\ell   \overline{Q}_h S[t] ^ {b_s} (1 + D_h[t]) ^{b_d}}{U_\ell[t]}
\end{equation}

$H$ is again a uniformly random parameter that behaves the same as in \cref{eqBidPrice}. $b_\ell = 1.75$ is the listing price factor, showing the extent to which the seller tends to increase the listing price. Furthermore, $\overline{Q}_h$ is the average sale price of the 10 dwellings with the most similar quality to the dwelling $h$ for sale. In addition, $S[t]$ is the market average of the sold-to-list price ratio, $b_s = 0.22$ is the sold to list exponent parameter, $D_h[t]$ is the number of months the dwelling has been on the market, and $b_d = -0.01$ is the number of months exponent parameter. Finally, $U_\ell[t]$ is the urgency to sell the dwelling, which is equal to $1$ if the household is not in financial stress, and increases proportionally to the number of months in financial stress otherwise. 

Banks approve households desired expenditure based on the bank's lending criteria. The bank uses the households liquidity and monthly income for determining an appropriate amount to lend and offers the corresponding loan to the household. If the loan amount is greater than $p_{d} \times B[t]$ (with $p_{d}=0.6$)  then the household accepts the loan, otherwise, the household skips this round of the market.

The parameters from the baseline method are presented in \cref{tblBaseParams}, and sensitivity analysis around the default values of these parameters is given in \cref{appendixBaseParams}, and more in depth explanation is given in the study of \cite{glavatskiy2020explaining}.

\subsection{Market algorithm}

An equilibration period is run at the start of the simulation to build up a "history" of the market, allowing for sales etc. to take place before the true simulation begins. Household and housing characteristics (such as income or construction) are smoothly extrapolated in the equilibration period, to match the corresponding values at the beginning of the actual simulation.

The model runs in sequential steps, with each step representing one month of actual time. That is, one sequential step (or tick) in the simulation, corresponds to one month in the actual housing market. During every step, the model makes several market updates:

\begin{enumerate}
    \item The city demographics are updated (new dwellings and households created to match the actual projected numbers for the Greater Sydney region).
    
    \item Each household receives income and pays its living costs: non-housing expenses, maintenance fees and taxes (if owning a dwelling), rent (if renting). The balance is added to or subtracted from the household's liquid cash.
    
    \item Expiring rental contracts are renewed.

    \item Dwellings are placed on sale.

    \item Households put their bids for buying.

    \item The buyers and sellers are matched (described below)
    
    \item The households receive mortgages and mortgage contracts from the bank, and the balance sheets of both buyers and sellers are updated. 
\end{enumerate}

To match buyers and sellers, bids and listings are sorted in descending order. Each listed dwellings is then attempted to match with the highest bid. If the bid price is higher than the list price, then the deal is made with probability $p_{m}=0.8$. Otherwise, the listing is considered unattended and the next one attempts to match. The pseudo-code for the process is given in \cref{appendixMarketClearing}.

\section{Model Extension}\label{secExtension}

 In this section, we develop the spatial component of an ABM housing market, effectively introducing submarkets into the model. Specifically, we investigate how social aspects influence selling a dwelling, as well as account for the agents' preferences to buy in a nearby neighbourhood when purchasing a dwelling. In doing so, we investigate how pricing dynamics are driven by individual agent decisions. The spatial component also affects how households are initialised (i.e., what neighbourhood they should belong to), and how the prices of nearby dwellings may affect the listing price.

\subsection{Spatial Component}

\begin{figure}[ht]
    \centering
    \begin{subfigure}[b]{0.45\textwidth}
    \includegraphics[width=\textwidth]{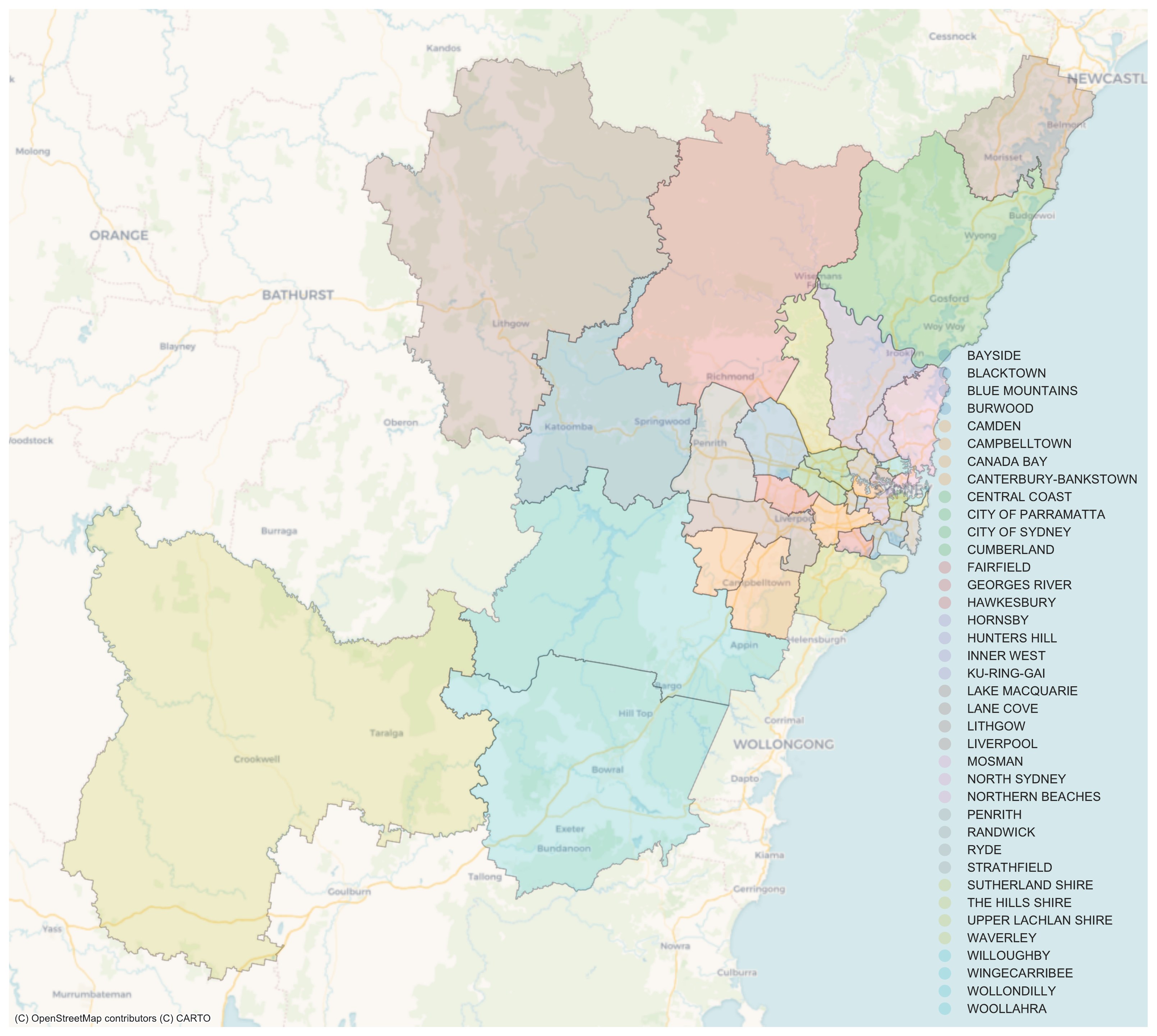}
    \caption{LGA Boundary Map}
    \label{figMap}
    \end{subfigure}
    \hfill
    \begin{subfigure}[b]{0.5\textwidth}
    \includegraphics[width=\textwidth]{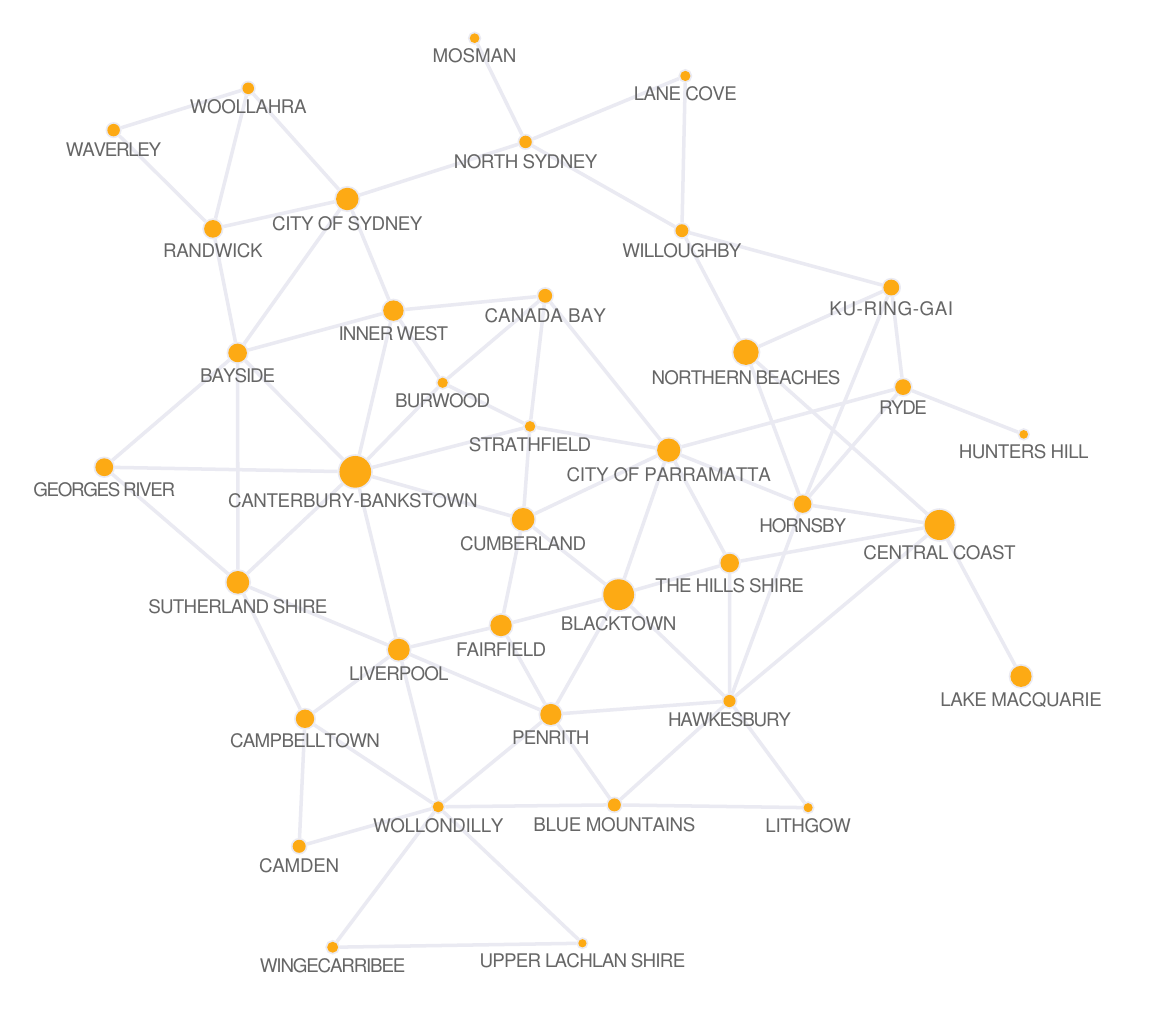}
    \caption{LGA Network}
    \label{figGraph}
    \end{subfigure}
    
    \caption{Greater Sydney LGAs. On the left, we see the raw GIS data. On the right, the processed graph (with nodes scaled based on population size).}
    
\end{figure}

Greater Sydney is composed of 38 Local Government Areas (LGAs), each of which contains several suburbs (and postcode areas). The data provides sales at a postcode level and the LGA level. However, the postcode data may be too granular as the number of listings in a given time period for small areas could be low or even zero. For this reason, we analyse the data at the LGA level, but the proposed approach is general and can be used at any level of granularity (i.e., over countries, states, cities, government areas, postcodes, suburbs, or even individual streets) assuming the data is available. The LGAs are visualised in \cref{figMap}.

\subsubsection{Graph-Based Topology} \label{graphs_maps}

To incorporate the LGA areas into the ABM, the data must be converted to an appropriate data structure. This is achieved by converting the map (from \cref{figMap}) into an undirected graph $G$, with equal edge weights of 1 (i.e., unweighted, the weighted extensions are discussed below) shown in \cref{figGraph}. In doing so, the topology of the spatial relationships of the suburbs are preserved but not the exact locations, i.e., the $x$, $y$ coordinates of latitude and longitude are not needed.

Each of the $N$ LGAs (shaded polygon) in \cref{figMap} is converted into a vertex $v_i$, $i \in [1, \ldots, N]$. The cardinality of the set of all vertices $V$ is $|V| = 38$ corresponding to the 38 LGAs. Two LGA areas $v_i$ and $v_j$ are adjacent to one another if they share a border (darker lines in \cref{figMap}) and these borders are converted to the set of edges $E$ all of weight one that form an adjacency matrix $G$ for the LGAs. Formally there is a 2-D spatial region for Sydney (the map of Sydney): $\mathcal{M}$, composed of the $N$ non-overlapping LGAs that form a complete cover of Sydney and each LGA shares at least one border with another LGA. LGA($x$) associates a graph vertex $x$ with an LGA, and Adj($l_i$,$l_j$), defined for two LGAs $l_i$ and $l_j$, is a function measuring the length of their common border in $\mathcal{M}$:
\begin{eqnarray}
    V & = & \{v_{i} \mid i \in [1, \ldots, N], \,\, \mathrm{LGA}(v_i) \in \mathcal{M}\} \label{eqVertices}, \\
    E & = & \{e_{i,j}=1 \mid v_i, v_j \in V, \ i \neq j, \  \mathrm{Adj}(\mathrm{LGA}(v_i), \mathrm{LGA}(v_j)) > 0\}. \label{eqEdges}
\end{eqnarray}
This definition implies $G(E,V)$ is a connected undirected graph, there are no disconnected subgraphs. An important edge is also added that represents the Sydney harbour bridge connecting Northern Sydney with the City of Sydney.

To calculate the distance between vertices $v_i$ and $v_j$, the minimum path length (i.e., the path with the lowest number of edges) between the two vertices is used as edges are equally weighted: $\delta(v_i, v_j)$ denotes this shortest path. Because the edges have unit weighting the shortest paths are found using a simple breadth-first search. However, future extensions could consider edge weightings based on metrics such as real distance between centroids, travel time between centroids, or even adding additional edges for public transport links. In cases of weighted edges, Dijkstra's algorithm could be used to compute $\delta(v_i, v_j)$ instead.

\subsubsection{Spatial Submarkets}

Dwellings are allocated initial prices based on the distribution of recent sales within their LGA, and also populated according to the census data for dwellings in each LGA. A full description of the process is given in \cref{appendixSpatial}. Households are also distributed into LGAs based on census data, with renters then moving to LGAs in which they can afford, as described in \cref{appendixRentalMarket}.

The original dwelling list pricing equation from \cref{eqListPrice} is also now updated to be based on each dwelling's LGA. Rather than $\overline{Q}_h$ being the average of the 10 most similar quality dwellings in the model, it is the average of the 10 most similar within the LGA. Likewise, $S[t]$ is the average sold-to-list price ratio for the dwelling's LGA (not overall).
In this sense \textit{spatial} submarkets (LGAs) \citep{watkins2001definition} capable of exhibiting their own dynamics are introduced. Recent research \citep{bangura2020housing} has shown the importance of submarkets in the Greater Sydney market, so capturing such microstructure is a key contribution of the proposed approach, as trends can be localised to specific submarkets (a feature not prominent in existing ABMs of housing markets).

\subsection{Spatial outreach}\label{secSpatialKnowledge}

In an actual housing market, a typical buyer does not review every listing in the entire city due to the high search costs and desire to live in certain areas. Rather, the buyer targets particular spatial sub-markets, relating to a given area. In particular, listings immediately around the buyer's location are likely to be viewed with a higher probability than listings which are further away.

%When considering an agent's knowledge in the ABM, it is not realistic to assume perfect information due to the high transaction costs in a housing market. That is, when a buyer is looking for a new house, they can not view every single listing in the Greater Sydney region. Likewise, from a sellers perspective, it is not possible to advertise to all buyers in the market. 
%Instead, it is much more likely the probability of seeing a listing is based on the listings proximity to the buyer's current location.

\begin{figure}[ht]
    \centering
    \begin{subfigure}[b]{0.5\textwidth}
        \centering
        \includegraphics[width=\textwidth]{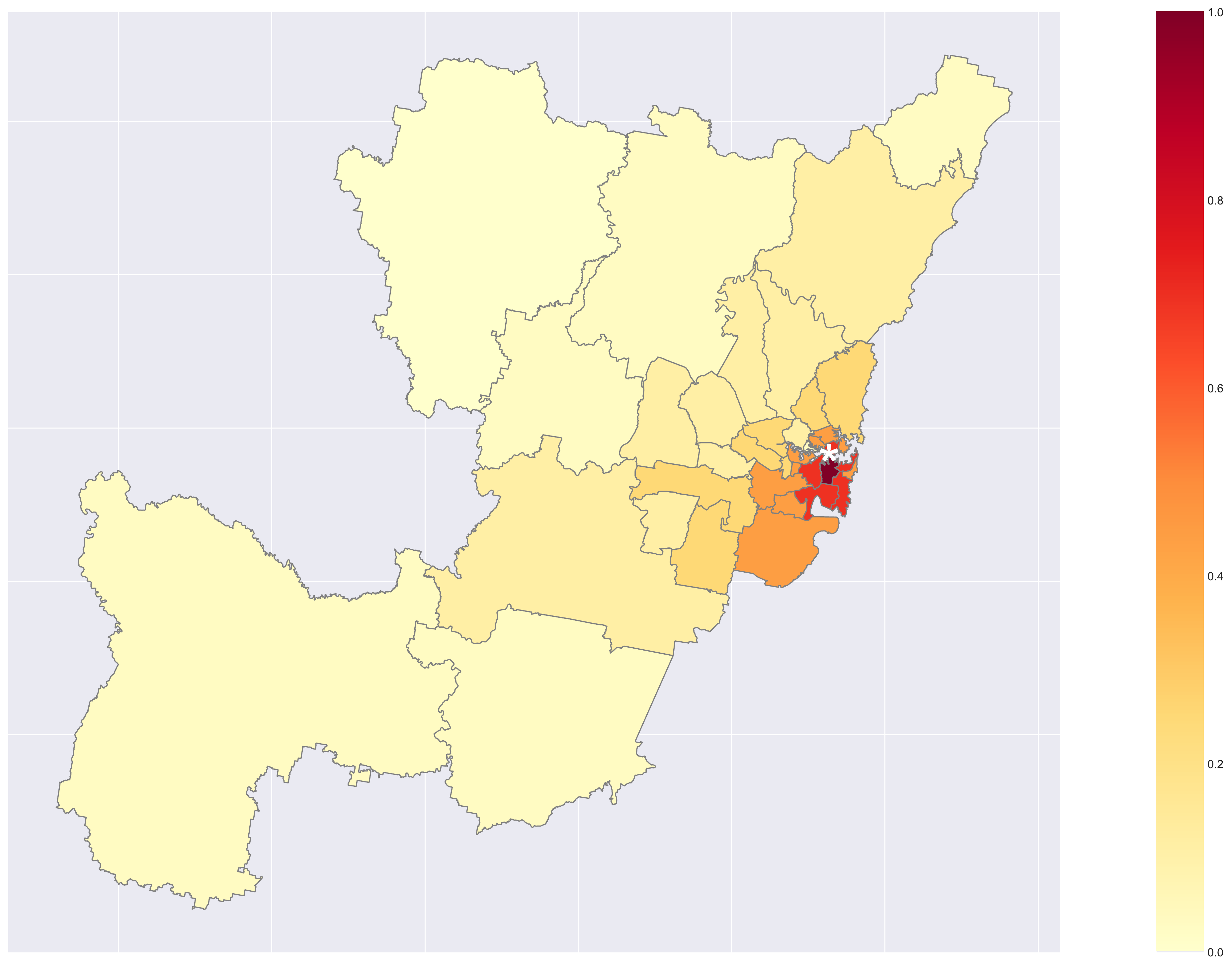}
        \caption{Overview}
    \end{subfigure}
    \hfill
    \begin{subfigure}[b]{0.475\textwidth}  
        \centering 
        \includegraphics[width=\textwidth]{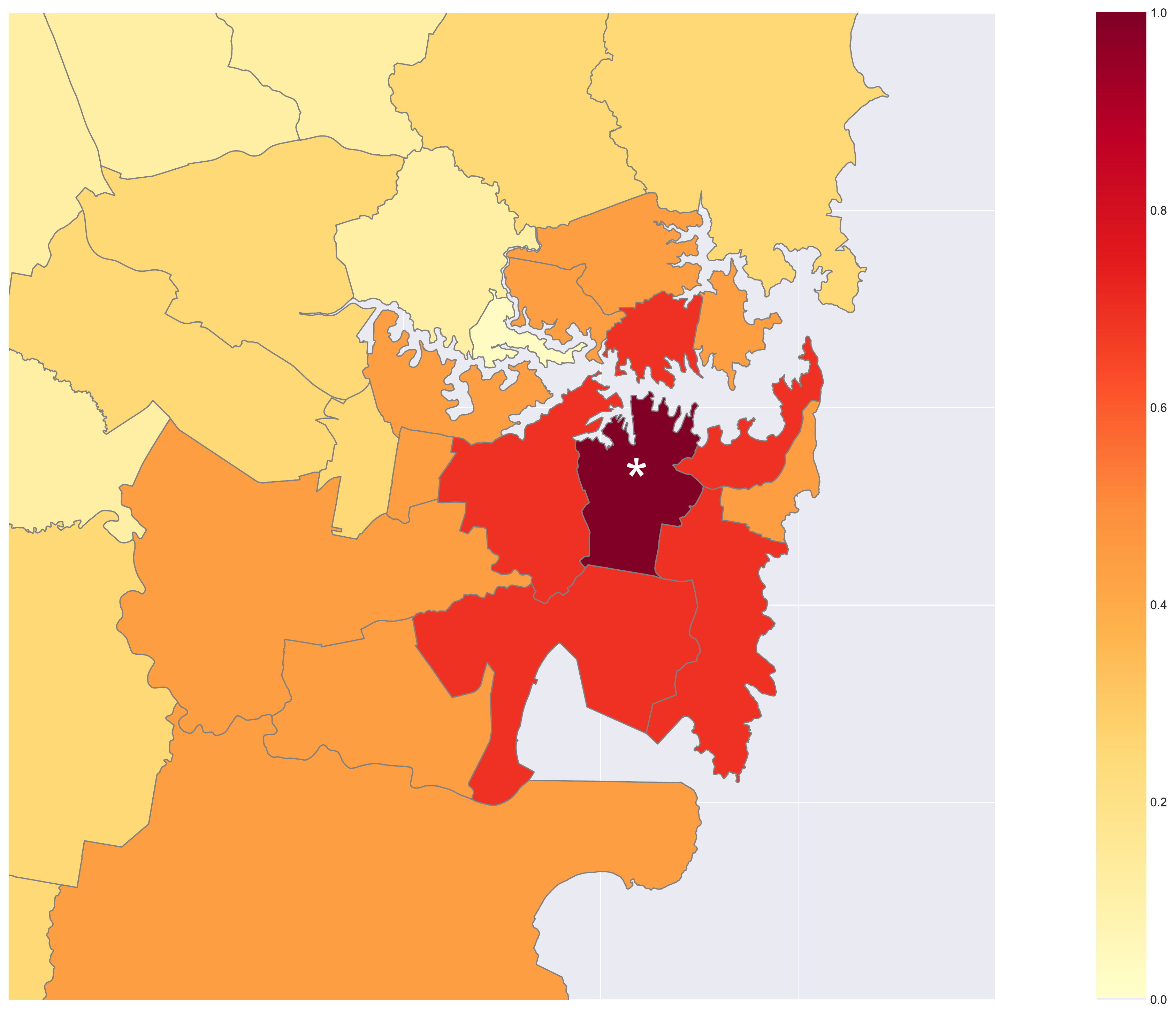}
        \caption{Zoomed}
    \end{subfigure}
    
    \caption{Probability of viewing a listing based on buyers location (in this case the City of Sydney). Dark red indicates high probability, lite yellow indicates low probability.}
    \label{figProbabilities}

\end{figure}

Therefore, it is unrealistic to assume perfect knowledge in an ABM of the housing market. To model this imperfect spatial information,  we introduce an outreach term $O$, which determines the likelihood for a buyer located at $v_i$ to view a listing located at $v_j$, as described by \cref{eqOutreach}. According to this expression, the likelihood of viewing the listing decreases with the distance between the buyer and the listed dwelling. The outreach factor is illustrated in \cref{figProbabilities} for a buyer located in the LGA "City of Sydney".

%The negation ($1-$) of the probability is because the lowest distance should have the highest probability. Other alternative outreach terms could be used,  however, we have chosen this method due to the simplicity. One candidate alternative would be a softmax outreach with a varying base parameter, however, we have chosen \cref{eqOutreach} as this does not introduce an additional free parameter (e.g. the base in softmax). Future work could compare a range of outreach terms, or consider heterogeneous outreach (varying levels of mobility) between agents.

% with the probability of selection shown by the colour of the area. 

\begin{equation}\label{eqOutreach}
   O(v_i, v_j) = 1 - 
   \frac{\delta(v_i, v_j)}
        {\max\limits_{k \in V} \delta(v_i, v_k)}
\end{equation}

While the spatial outreach makes sense for first-time home buyers, for investors, the outreach becomes uniform, as they do not necessarily desire rental properties near where they reside. So for investors, we use $O(v_i, v_j) = 1, \forall {i,j}$.

To control the strength of the outreach, we introduce a new parameter $\alpha \in (0,1)$, so the probability of viewing a listing $P_{view}(v_i, v_j)$ is given in \cref{eqProbabilty}. 
\begin{equation}\label{eqProbabilty}
   P_{view}(v_i, v_j) = \alpha O(v_i, v_j)^2
\end{equation}
where $\alpha$ modulates the spatial information on dwelling listings: the higher $\alpha$, the more listings are viewed by a potential buyer. That is, $\alpha$ adjusts the likelihood of viewing a listing, based on the distance to that listing. The effect $\alpha$ has on resulting decisions is further discussed in \cref{appendixUtility}.

\subsection{Spatial FOMO}

In the baseline model, dwellings have a fixed probability of being listed, $p_{b} = 0.01$. Here, we consider a spatial probability to list a dwelling located in a certain LGA $l_i$, which depends on the number of recent sales in $l_i$. For this, we introduce the "fear of missing out" (FOMO) parameter, denoted by $\beta$, which modulates the probability of listing a dwelling situated in $l_i$ by the number of recent sales in $l_i$ altering the agents' decision-making behaviour. In particular, if a high number of dwellings in $l_i$ have been sold, then the owners of dwellings in $l_i$ will be more likely to list their dwelling on the market. This will account for the possibility that if a certain LGA becomes a popular location for selling a dwelling, the owners of dwellings in this LGA would not want to miss an opportunity to sell their dwelling.

The spatial listing probability $p_{list}(l_i)$ is expressed by considering the difference in the fraction of dwellings currently in $l_i$'s submarket ($f_{l_i} = listings_{l_i} / dwellings_{l_i}$) with respect to the Greater Sydney average. A higher $f_{l_i}$ means dwellings in $l_i$ have been less likely to sell in the previous months (since all begin with a fixed probability of selling $p_{b}$) compared to the regions average. This spatial listing probability\footnote{$p_{list}$ can technically be  $<0$ or $>1$, so $p_{list}$ is capped to be between 0 and 1, in order to be a true probability, although this is exceptionally rare and does not appear to occur in \cref{figBayesianUnivariate}.}  is given by \cref{eqSpatialList}. 
\begin{equation}\label{eqSpatialList}
    p_{list}(l_i) 
    %= p_{b} + \beta (p_{b} * \frac{f_{l_i}} {\sum\limits_{a\in V} f_{a} / |V|}) - p_{b}
    = p_{b} +  p_{b} \beta \left[\frac{f_{l_i}} {\sum\limits_{a\in V} f_{a} / |V|} - 1 \right]
\end{equation}
%\begin{equation}
%    = p_{b} +  p_{b} \beta \left[\frac{f_{u}} {\sum\limits_{a\in V} f_{a} / |V|} - 1 \right]
%\end{equation}

%
%of listing a house (situated in location $u$) based on $\beta$. From \cref{eqSpatialList}, we can see 

The magnitude of $\beta$ controls the strength of $l_i$'s spatial submarket contribution to the listing probability. Rewriting \cref{eqSpatialList} by denoting the term in the square brackets as $x$, i.e., $$p_{b} + p_{b} \beta x$$ we see that if both $\beta$ and $x$ are negative, then the listing probability will be higher than the baseline level $p_b$. Therefore, if the number of dwellings for sale in a particular LGA is less than the average in the whole city, this means that this particular LGA's submarket has been clearing fast, and homeowners in this LGA will be more likely to list their dwelling. In contrast, if $x$ is positive, then dwellings in the current LGA are not clearing as fast as in the other LGAs, so homeowners in this LGA will be less likely to list. Conversely, a positive $\beta$ results in an opposite effect. If an LGA has comparatively few listings, the homeowner from this LGA will be less likely to list a dwelling for sale, whereas if this LGA has many listings, the probability to list a dwelling there increases. In this way, $\beta$ has a direct effect on the supply of dwelling listings, and alters the decision-making behaviour of sellers.

\section{Optimisation and Sensitivity Analysis}\label{secBayesian}

The selection of appropriate parameters is an important step in agent-based modelling, and in most existing work, parameters are selected over the entire period of interest. Here, we instead adopt a machine learning approach whereby we split the time series into a training and a testing portion, this ensures the model also performs well for the unseen (i.e., the testing) portion of data, and avoids biasing the selection of parameters by considering the entire time period.

We use Bayesian hyperparameter optimisation \citep{snoek2012practical} to find appropriate combinations of parameters. The training set is used for parameter selection, whereas the test set is only used to evaluate the goodness of fit of the model after the optimisation process is completed (in \cref{secResults}). We stress that the test set is never seen by the optimisation process. This is an important distinction from previous work \citep{glavatskiy2020explaining}, which constructs the models by considering all time points, and as such can not be considered true predictions, unlike here where the model can be seen as a true predictor of future pricing trends. Optimisation details are given in \cref{appendixBayesian}.

\subsection{Optimisation Results}\label{secOptResults}

We run several optimisation processes in order to quantify the contribution of each component on agent decision-making, and the effect these decisions have on resulting market dynamics. Each method follows the same optimisation process.

\begin{figure}[htb]

\centering
\begin{subfigure}{.3\textwidth}
  \includegraphics[width=\linewidth]{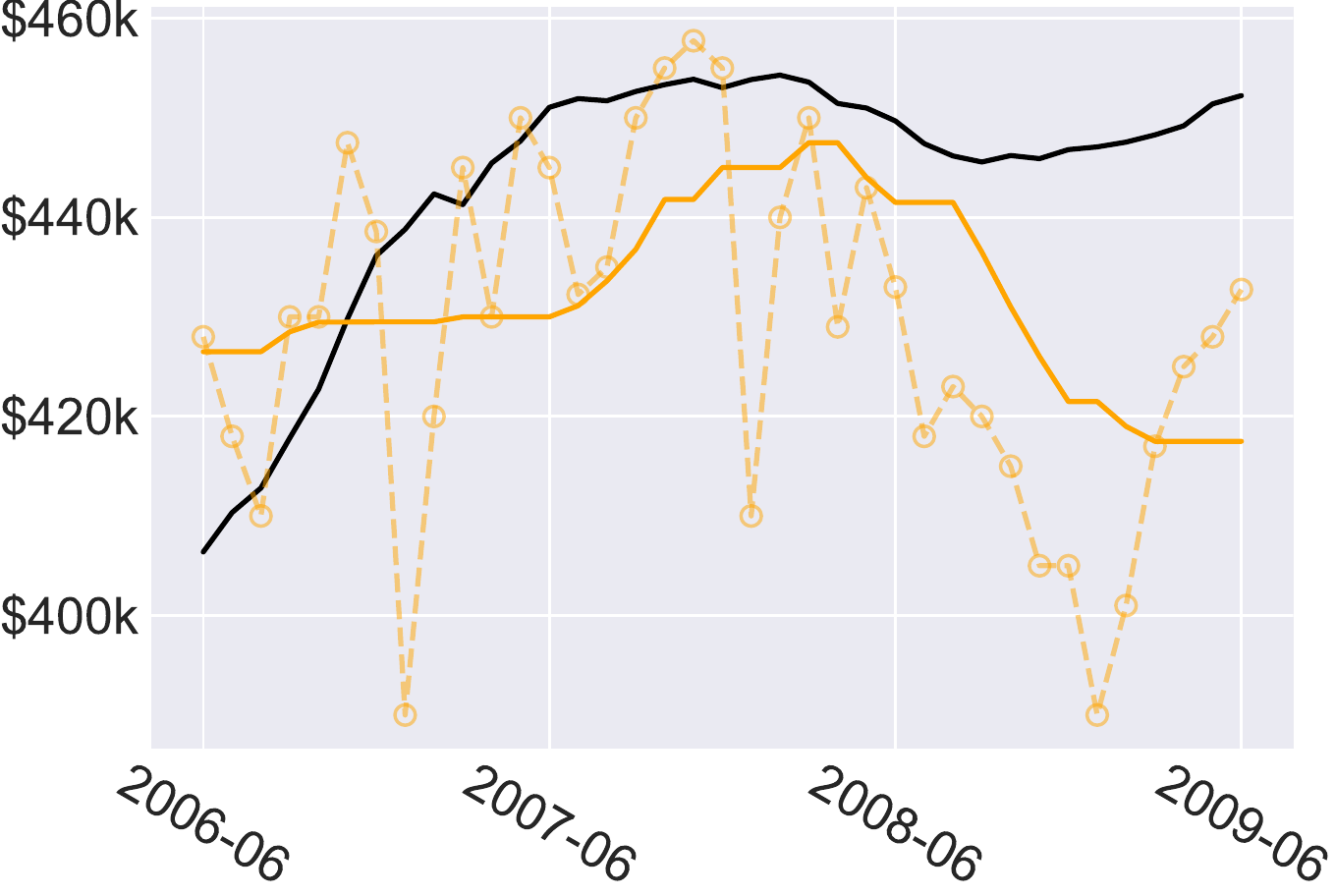}
  \caption{2006--2010 ($\ell=45726$) }
\end{subfigure}\hfil
\begin{subfigure}{.3\textwidth}
  \includegraphics[width=\linewidth]{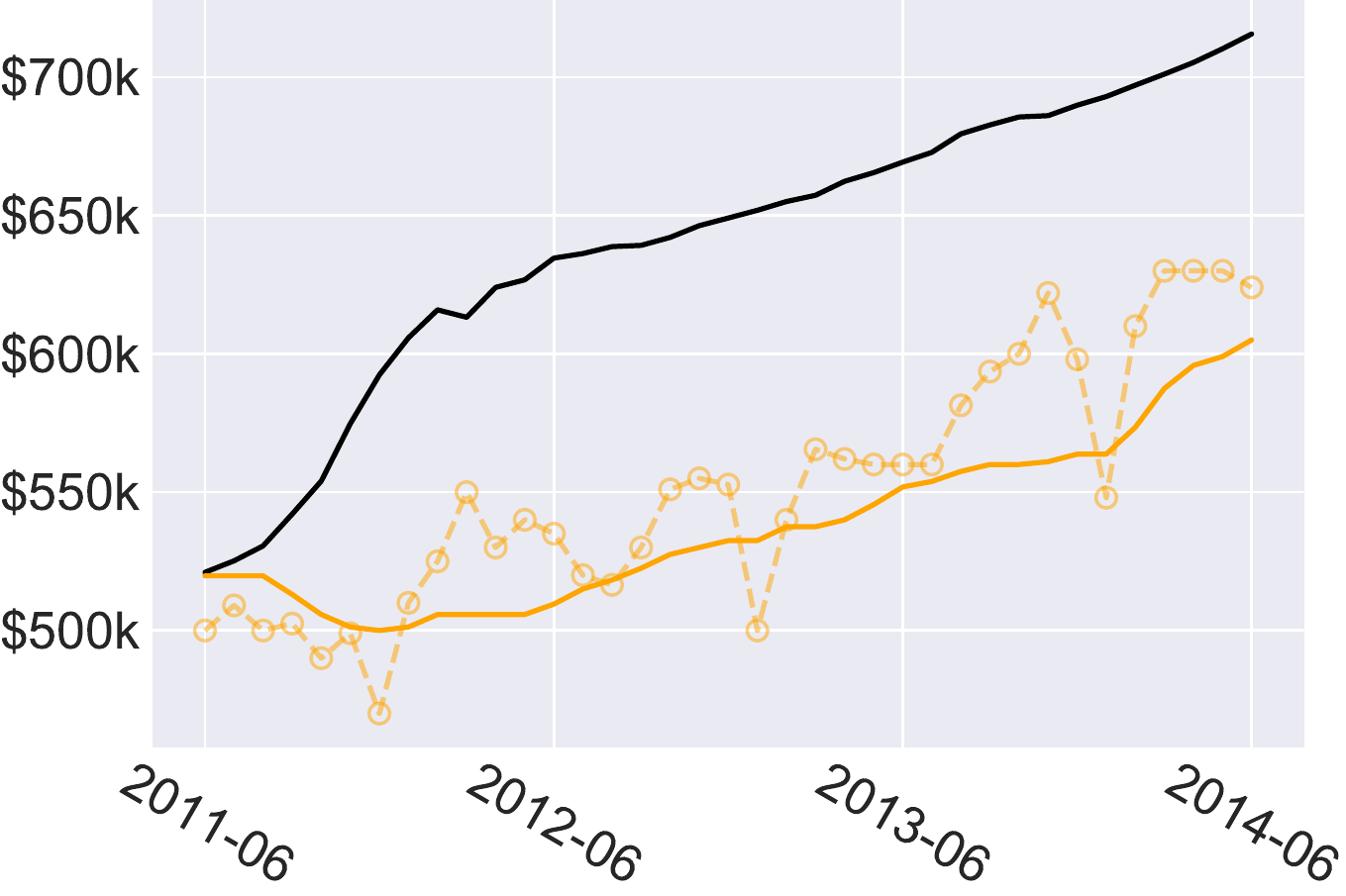}
  \caption{2011--2015 ($\ell=187549$)}
\end{subfigure}\hfil
\begin{subfigure}{.3\textwidth}
  \includegraphics[width=\linewidth]{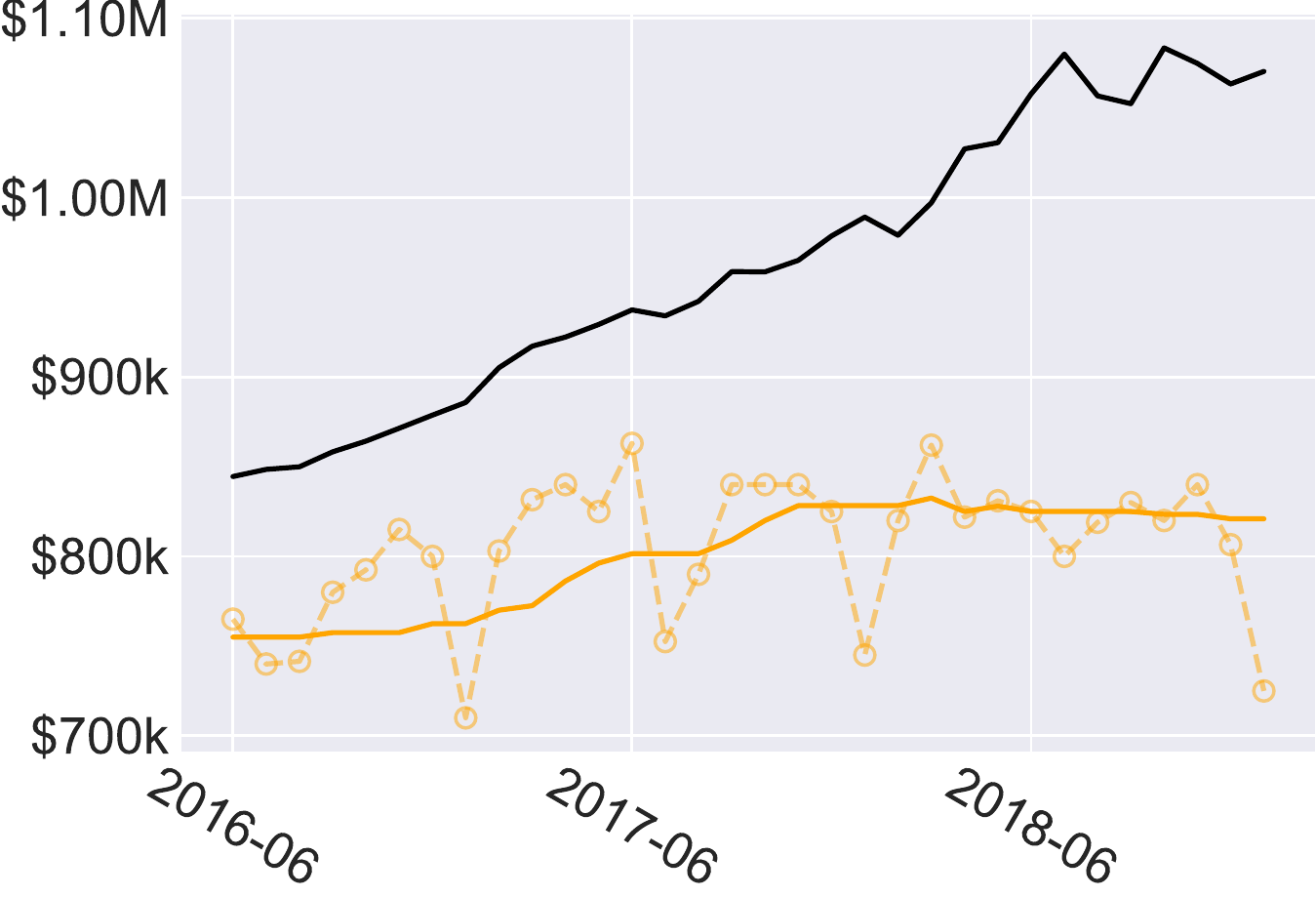}
  \caption{2016--2019 ($\ell=459482$)}
\end{subfigure}
\caption*{Baseline. No spatial component. Optimising $h$ only.}

\medskip

\begin{subfigure}{.3\textwidth}
  \includegraphics[width=\linewidth]{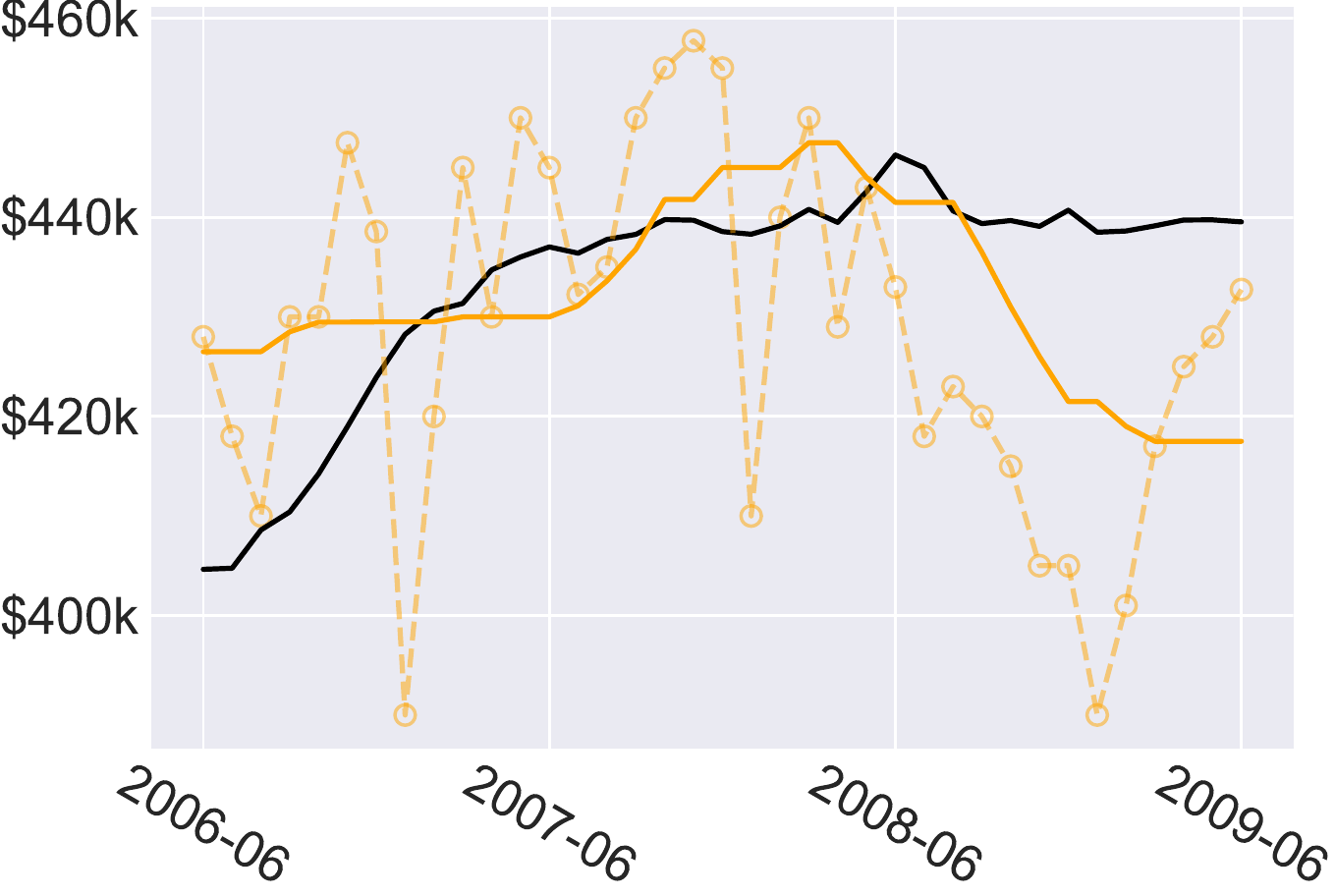}
  \caption{2006--2010 ($\ell=41613$)}
\end{subfigure}\hfil
\begin{subfigure}{.3\textwidth}
  \includegraphics[width=\linewidth]{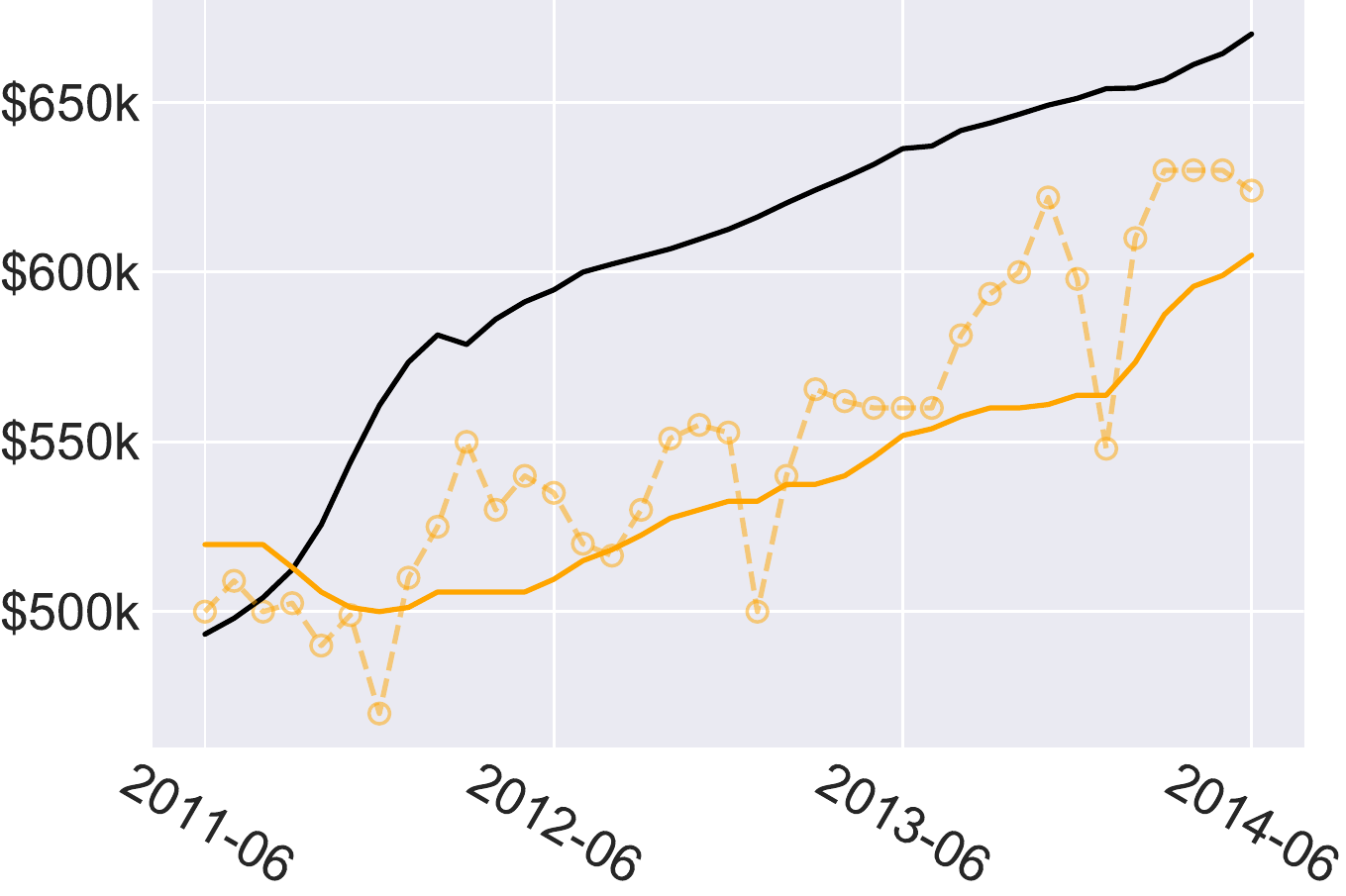}
  \caption{2011--2015 ($\ell=100745$)}
\end{subfigure}\hfil
\begin{subfigure}{.3\textwidth}
  \includegraphics[width=\linewidth]{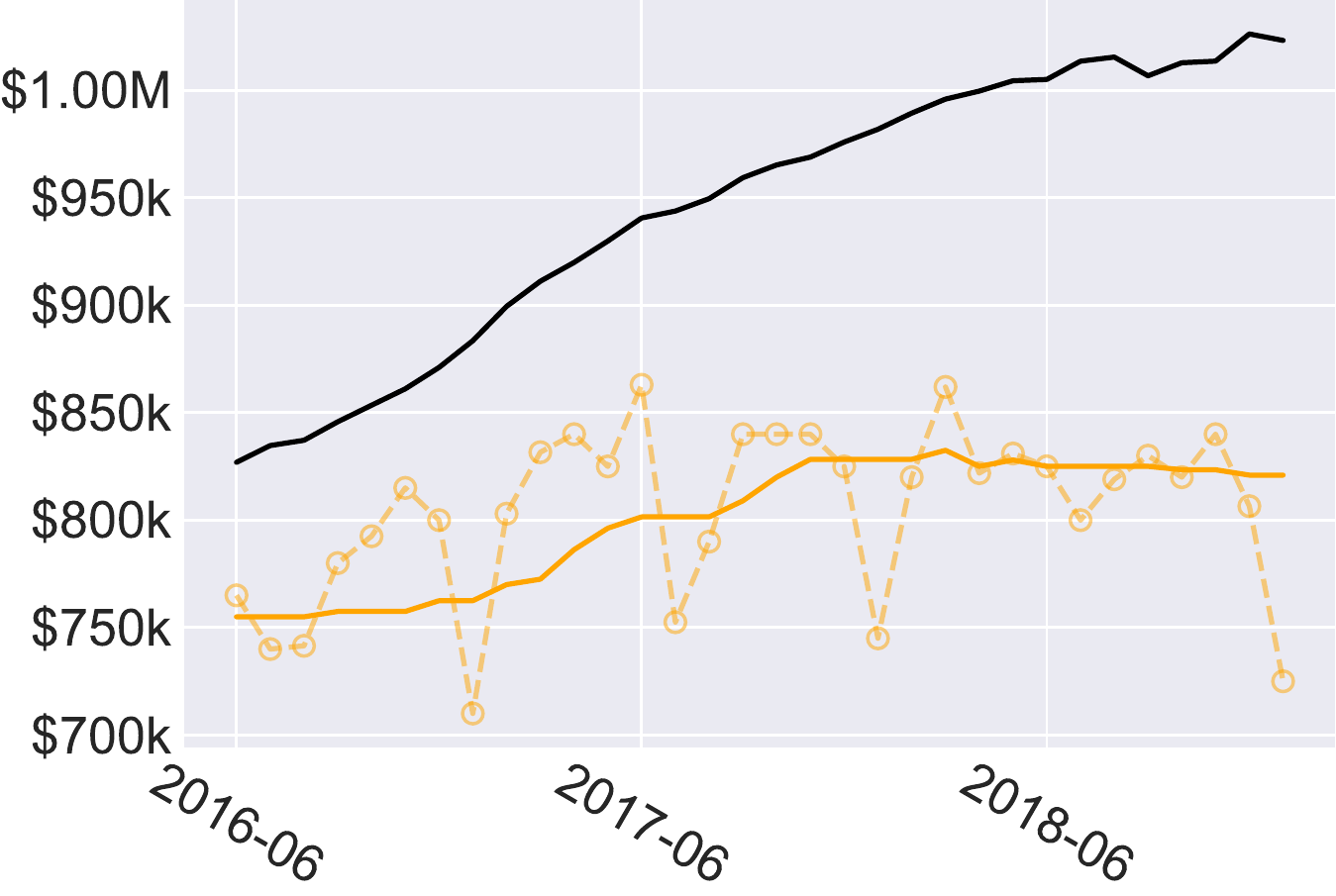}
  \caption{2016--2019 ($\ell=392208$)}
\end{subfigure}
\caption*{Spatial. Optimising $h$ only.}

\medskip

\begin{subfigure}{.3\textwidth}
  \includegraphics[width=\linewidth]{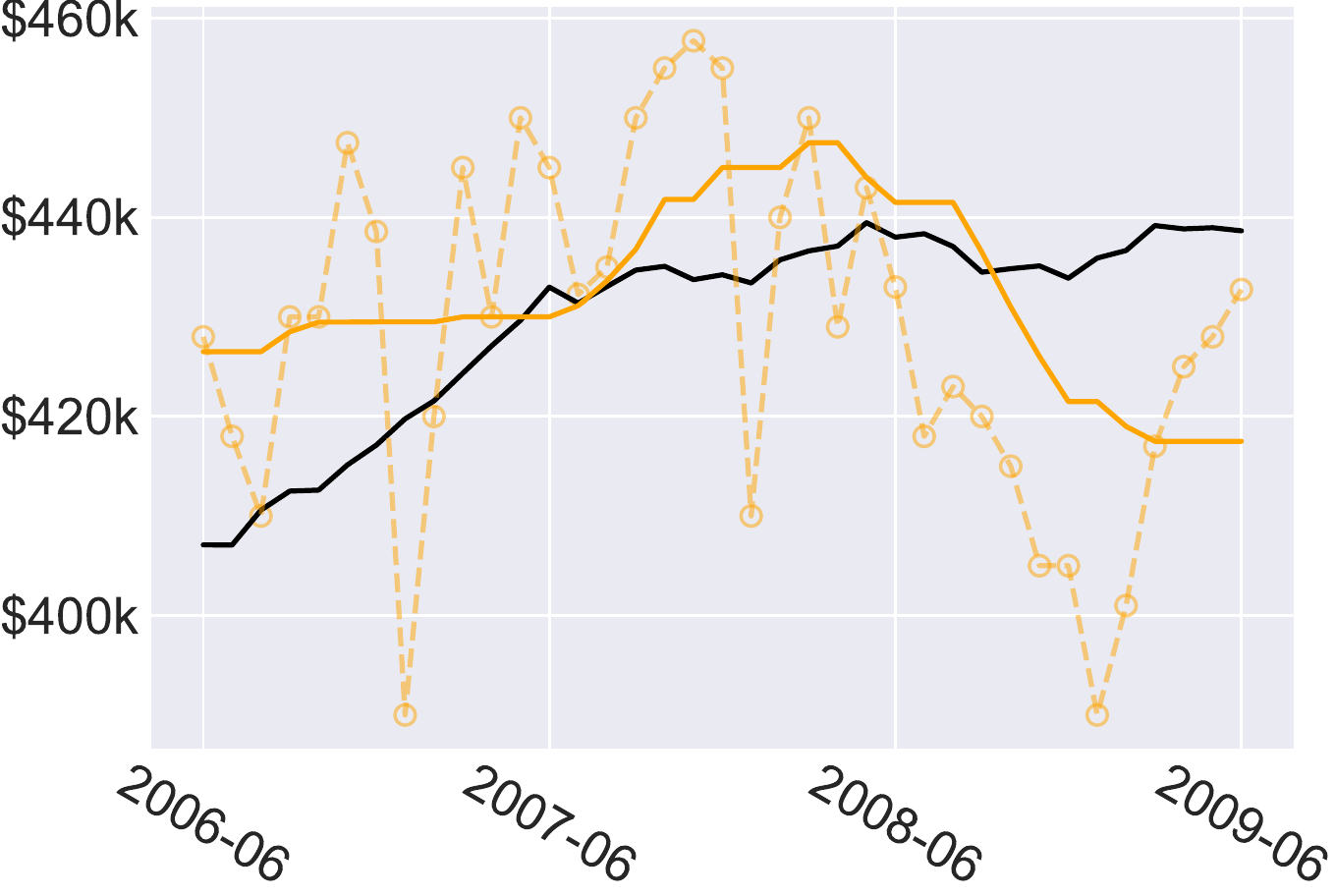}
  \caption{2006--2010 ($\ell=37523$)}
\end{subfigure}\hfil
\begin{subfigure}{.3\textwidth}
  \includegraphics[width=\linewidth]{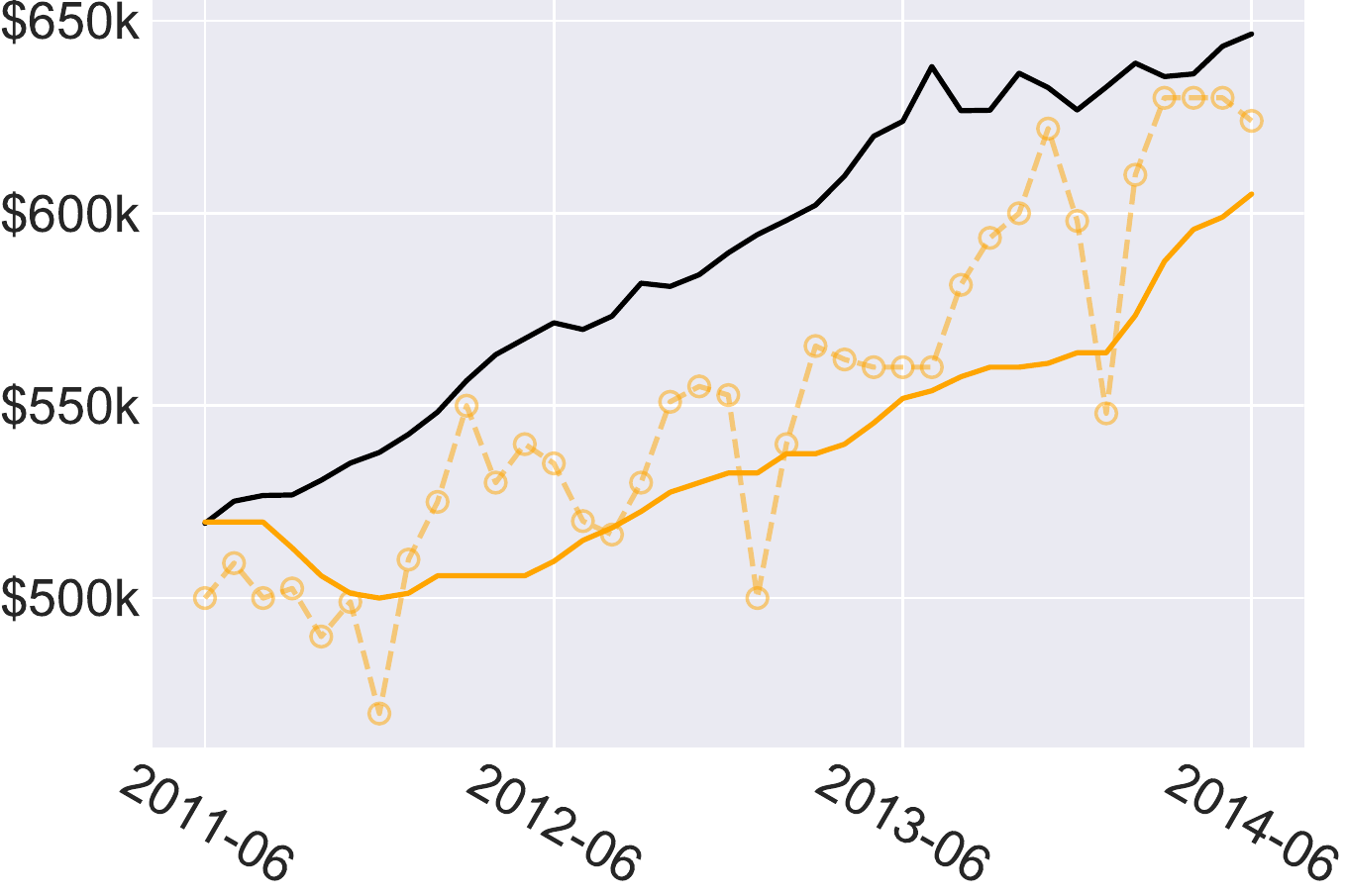}
  \caption{2011--2015 ($\ell=81553$)}
\end{subfigure}\hfil
\begin{subfigure}{.3\textwidth}
  \includegraphics[width=\linewidth]{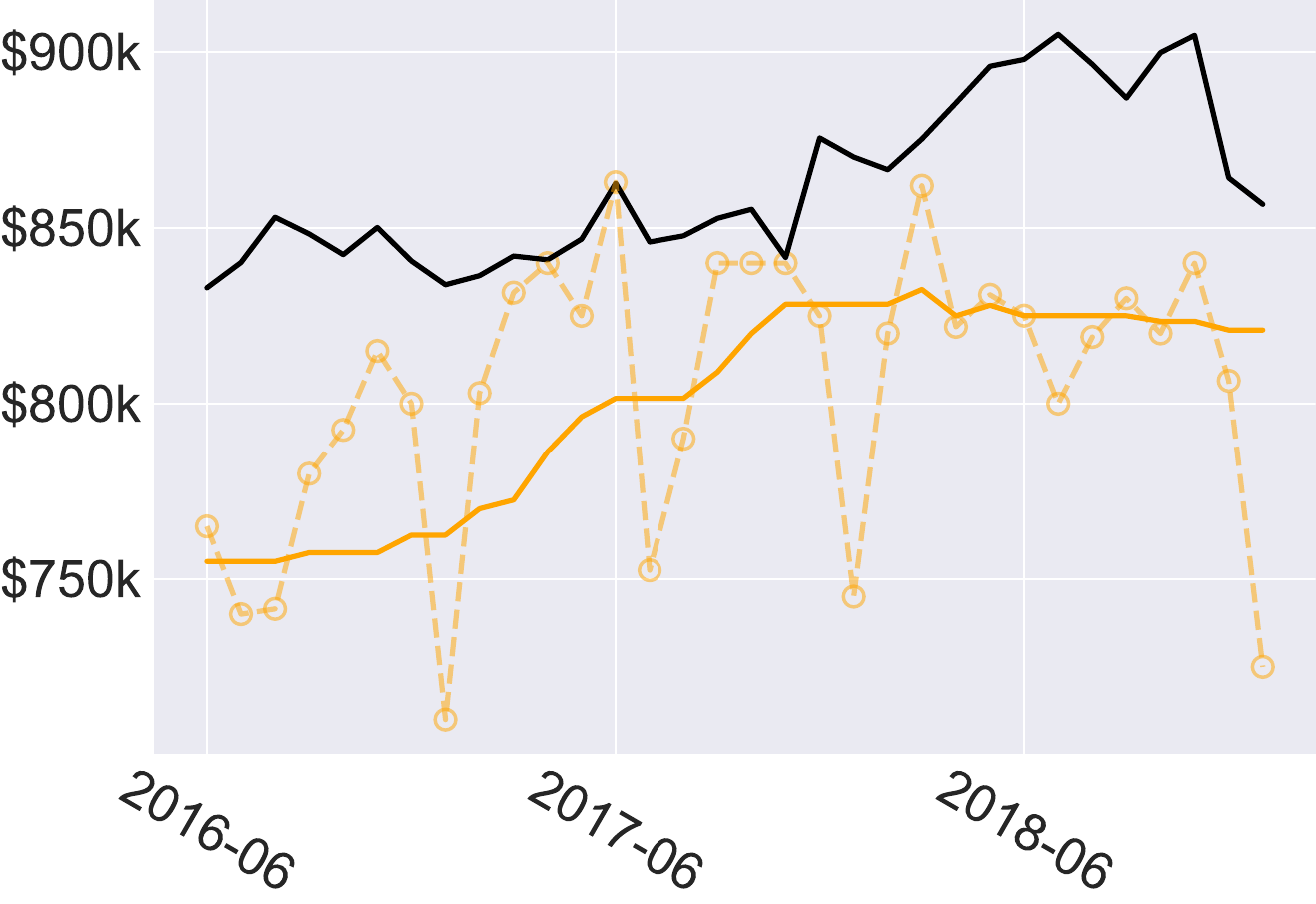}
  \caption{2016--2019 ($\ell=236476$)}
  \label{fig:6}
\end{subfigure}
\caption*{Spatial. Optimising $h$ and $\alpha$.}

\medskip

\begin{subfigure}{.3\textwidth}
  \includegraphics[width=\linewidth]{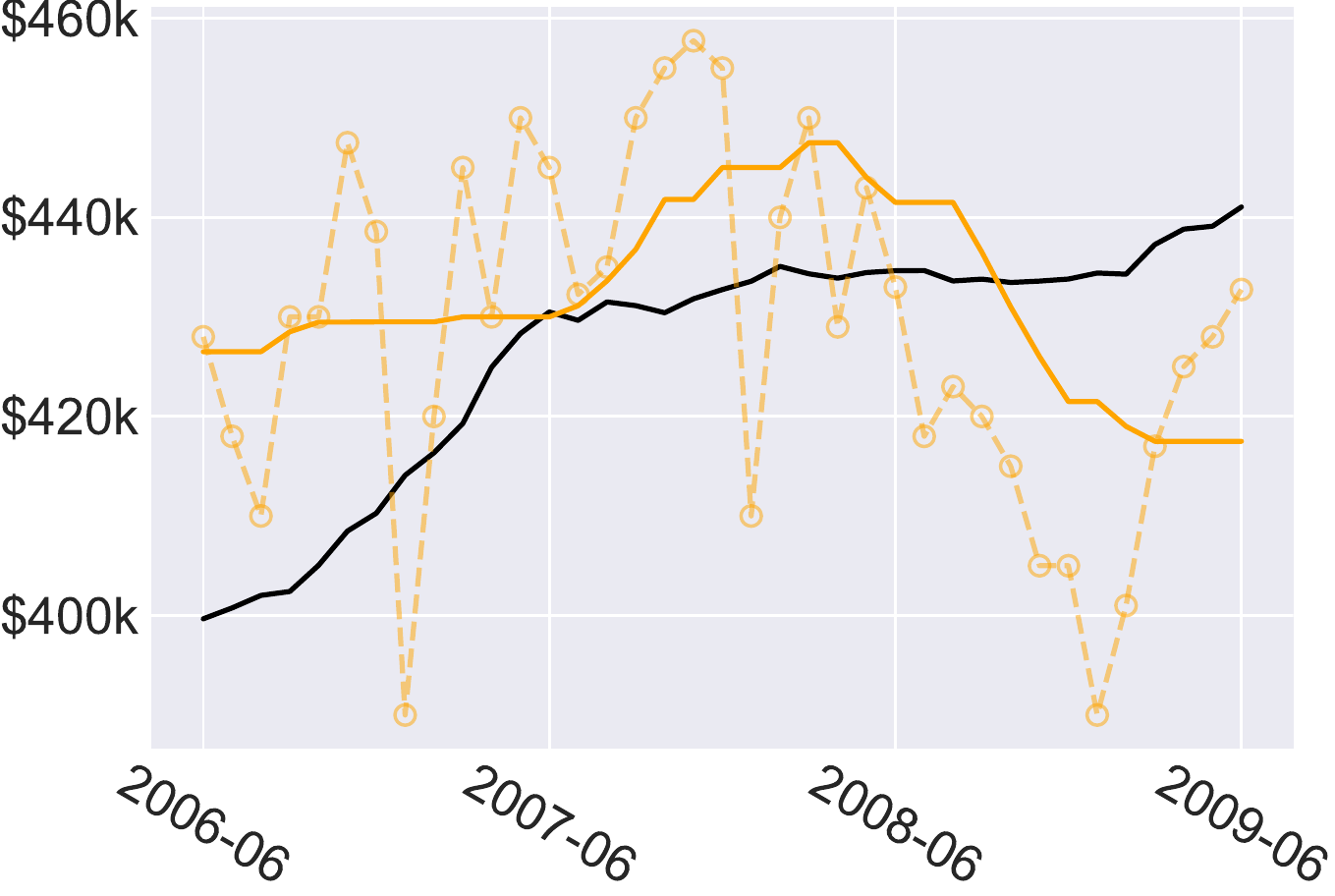}
  \caption{2006--2010 ($\ell=43459$)}
\end{subfigure}\hfil
\begin{subfigure}{.3\textwidth}
  \includegraphics[width=\linewidth]{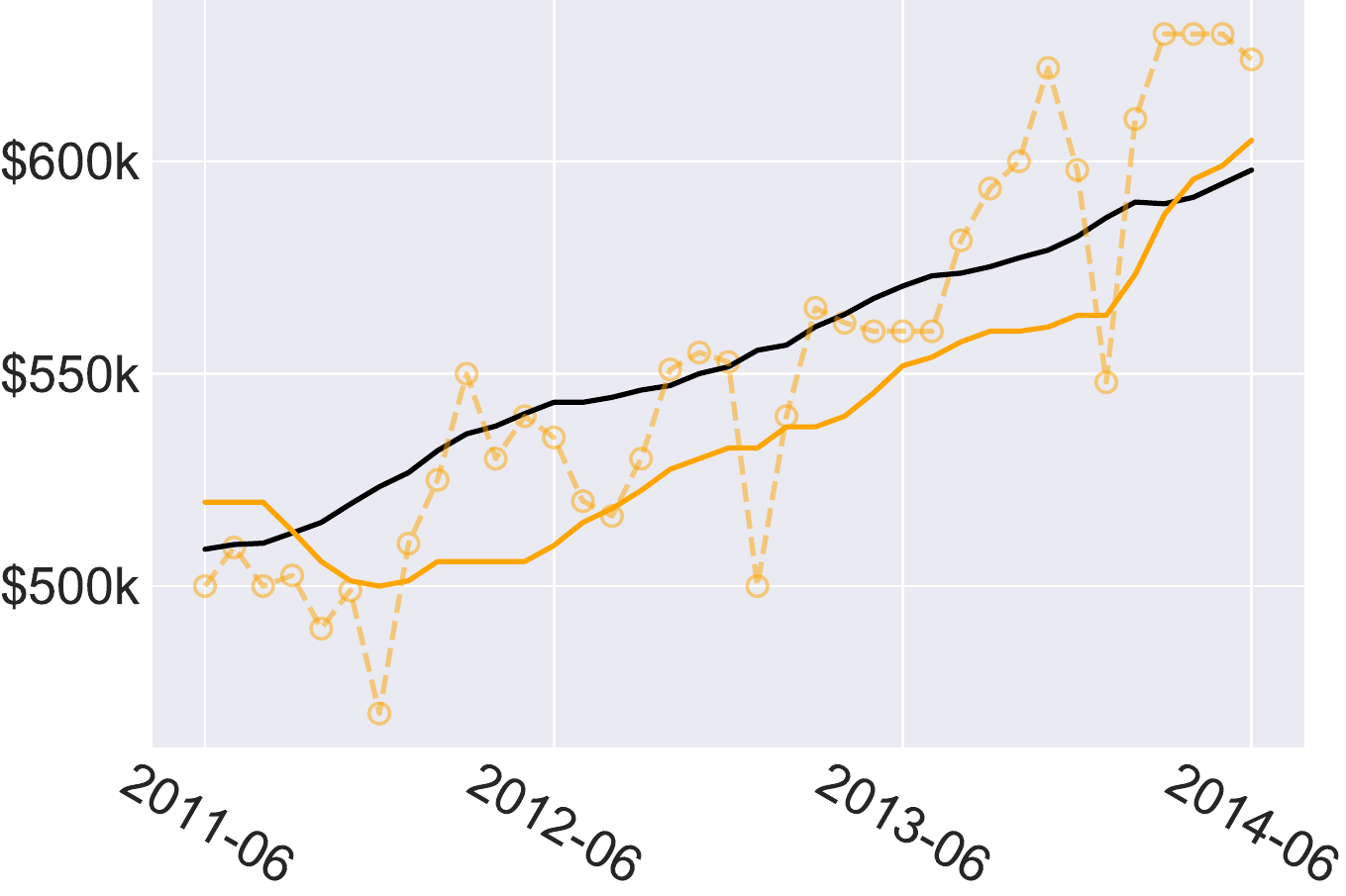}
  \caption{2011--2015 ($\ell=20321$)}
\end{subfigure}\hfil
\begin{subfigure}{.3\textwidth}
  \includegraphics[width=\linewidth]{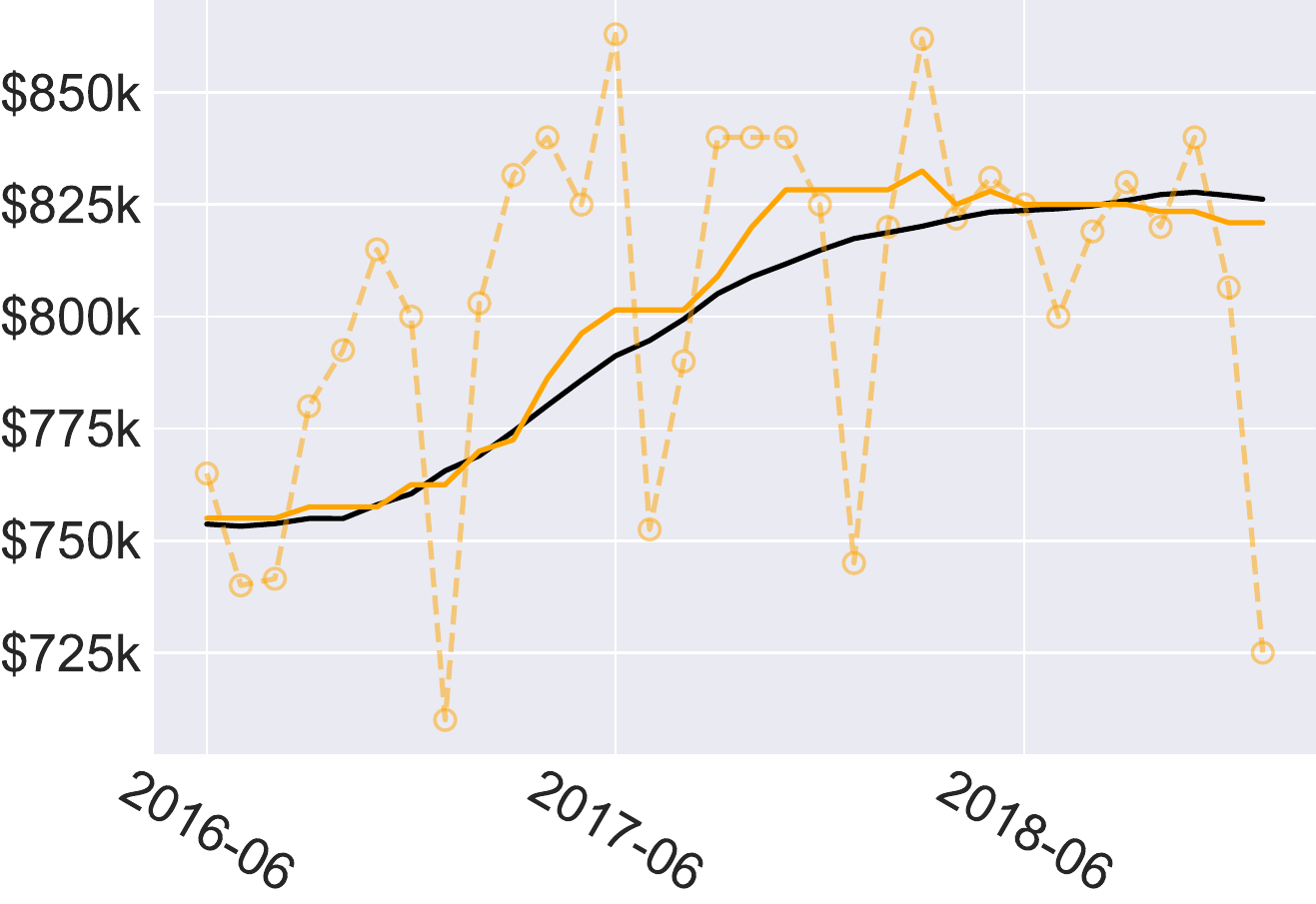}
  \caption{2016--2019 ($\ell=23674$)}
\end{subfigure}

\caption*{Spatial. Optimising $h$ and $\beta$.}

\medskip

\begin{subfigure}{.3\textwidth}
  \includegraphics[width=\linewidth]{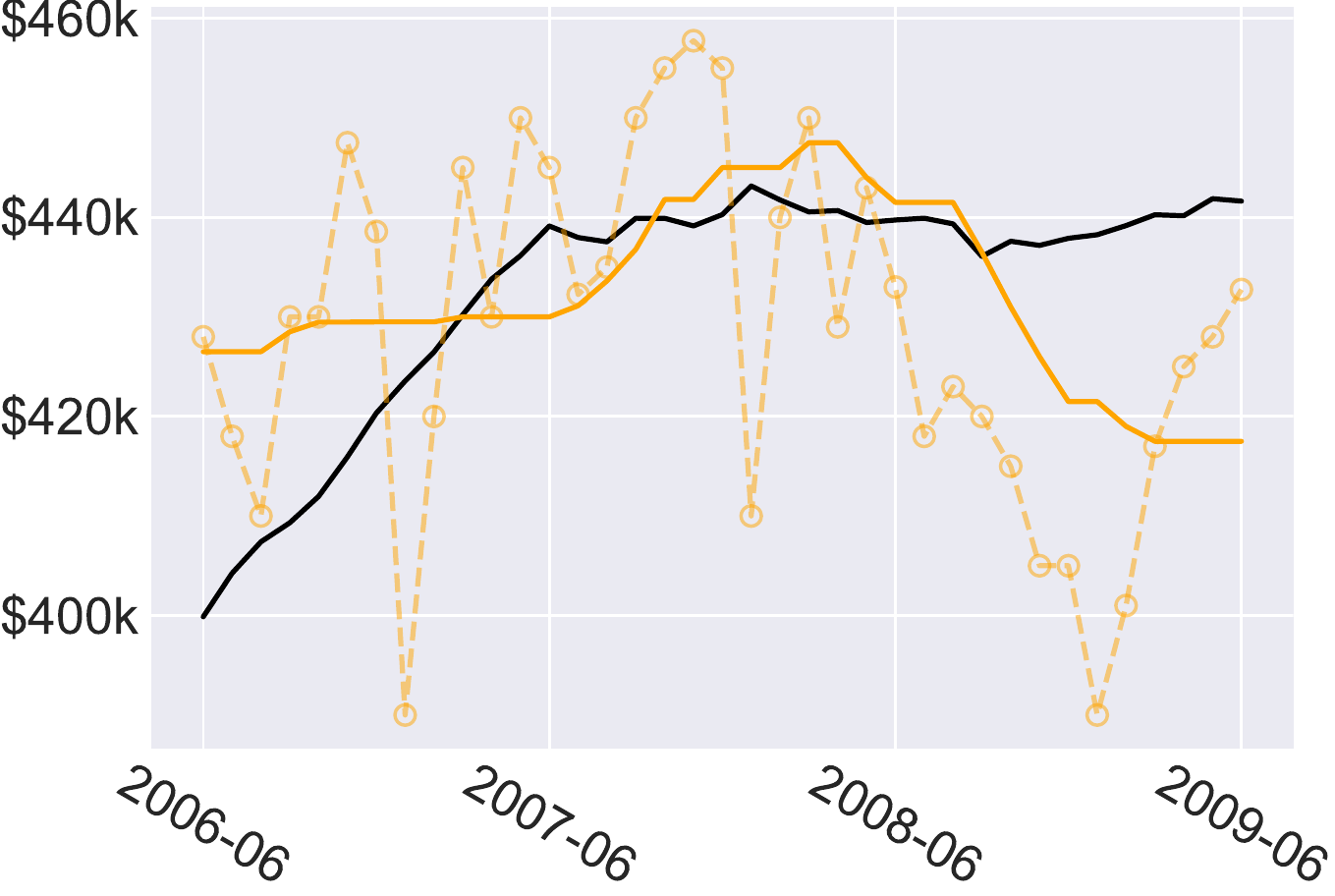}
  \caption{2006--2010 ($\ell=39786$)}
\end{subfigure}\hfil
\begin{subfigure}{.3\textwidth}
  \includegraphics[width=\linewidth]{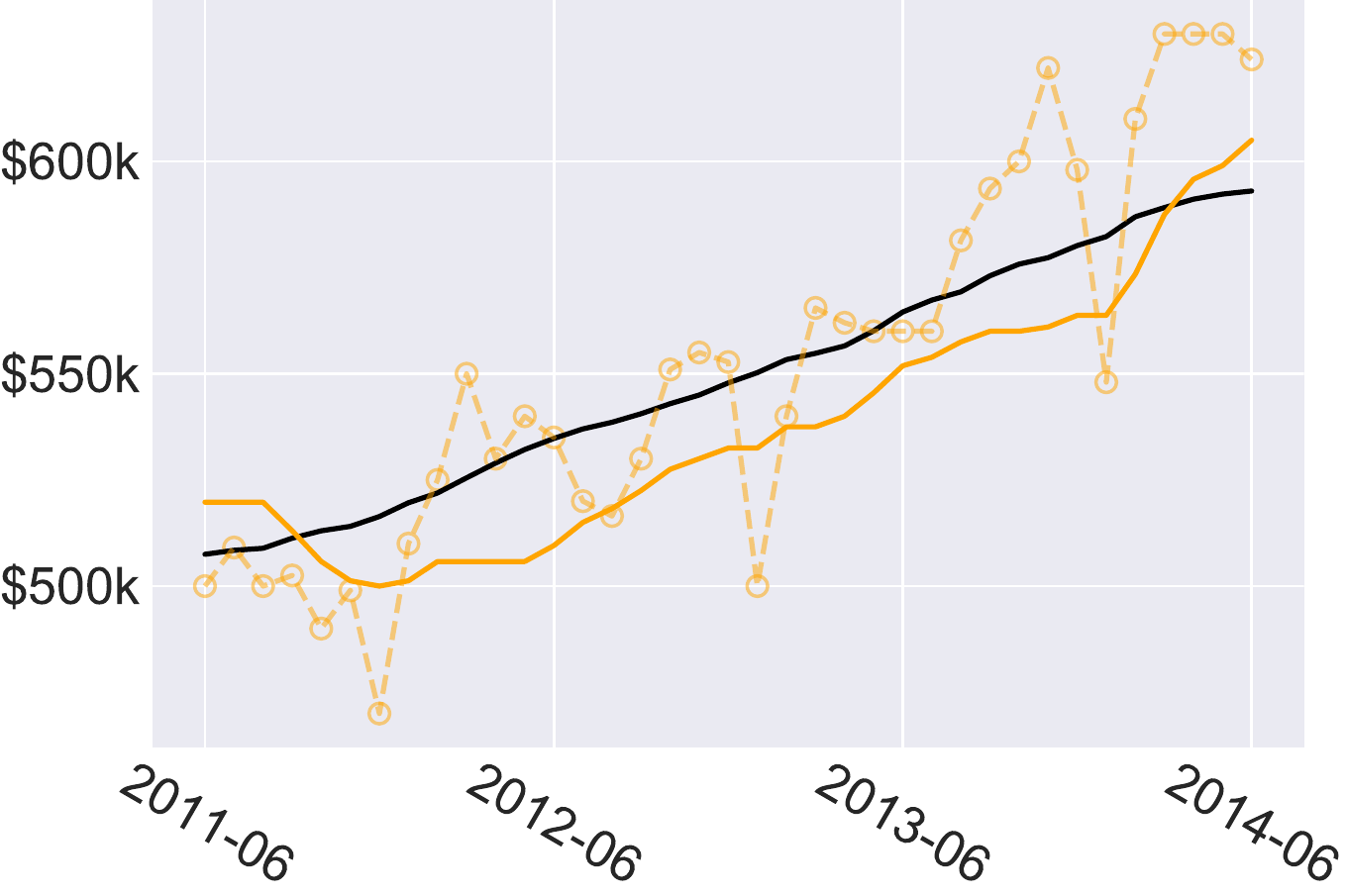}
  \caption{2011--2015 ($\ell=20106$)}
\end{subfigure}\hfil
\begin{subfigure}{.3\textwidth}
  \includegraphics[width=\linewidth]{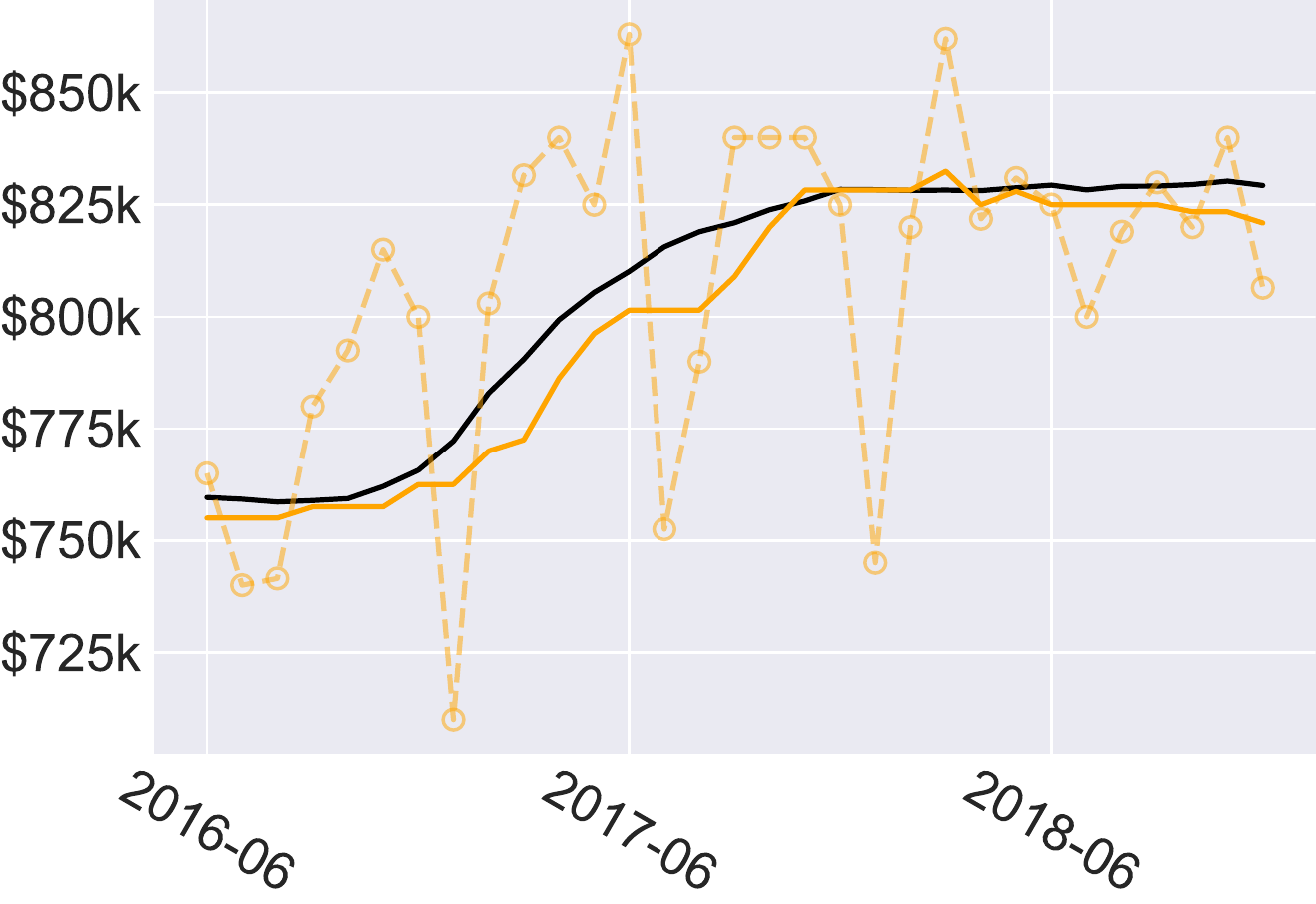}
  \caption{2016--2019 ($\ell=22492$)}
\end{subfigure}
\caption*{Spatial. Optimising $h$, $\beta$ and $\alpha$.}

\medskip
\begin{subfigure}{.3\textwidth}
  \includegraphics[width=\linewidth]{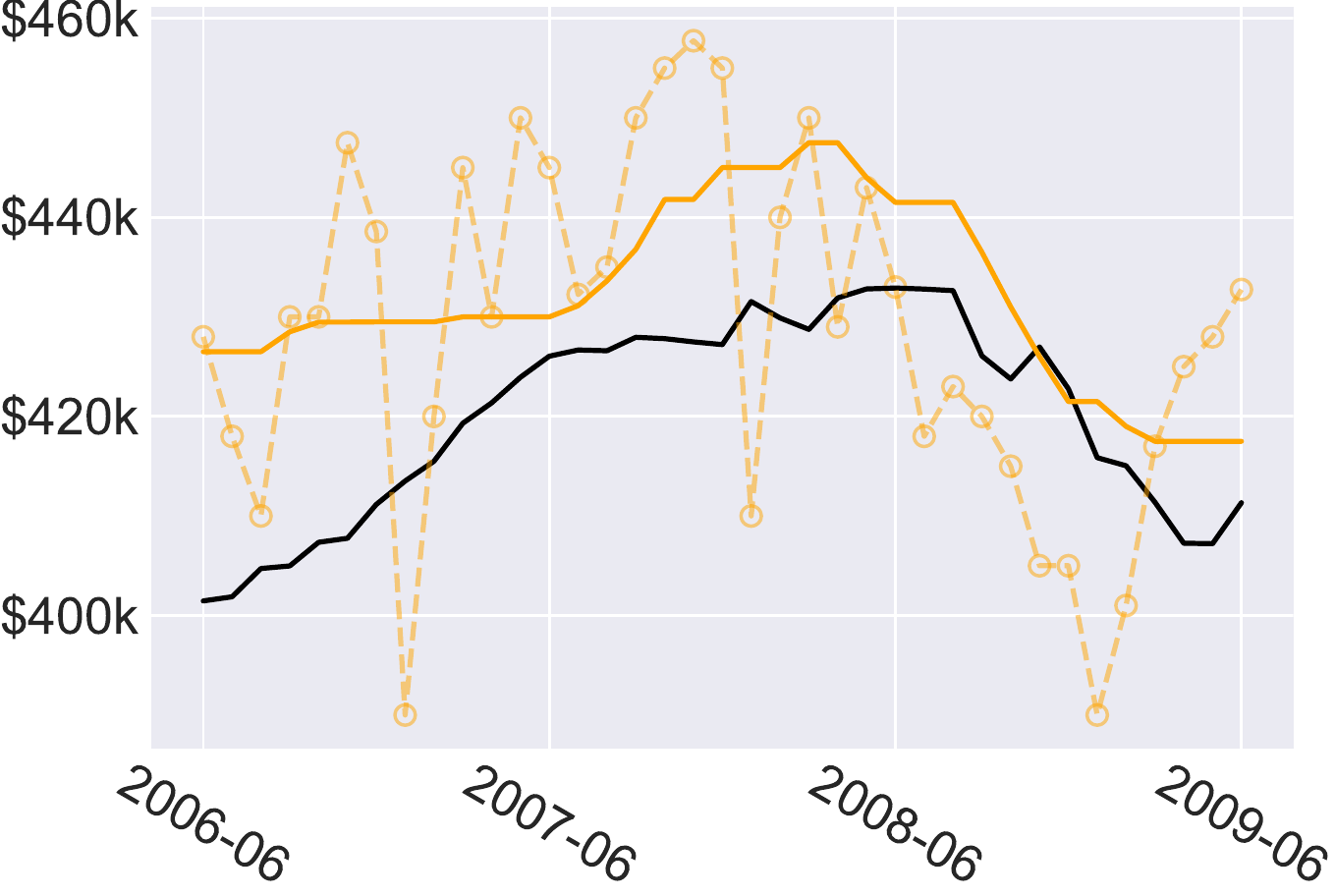}
  \caption{2006--2010 ($\ell=64334$)}\label{figGlobalConstraintLoss}
\end{subfigure}\hfil
\begin{subfigure}{.3\textwidth}
  \includegraphics[width=\linewidth]{images/bayesian/2011-Areas.pdf}
  \caption{2011--2015 ($\ell=20106$)}
\end{subfigure}\hfil
\begin{subfigure}{.3\textwidth}
  \includegraphics[width=\linewidth]{images/bayesian/2016-Areas.pdf}
  \caption{2016--2019 ($\ell=22492$)}
\end{subfigure}
\caption*{Spatial. Optimising $h$, $\beta$ and $\alpha$ with global constraint in 2006--2010.}

\caption{Optimisation of the goodness of fit for dwelling prices across all models (for the training period). The orange lines are from the SIRCA-CoreLogic data (the solid line represents the rolling median, and the dotted line represents the month-to-month median). The black line shows the best fitted path from the model.}
\label{figModelOptimisation}

\end{figure}

A "baseline" method is run, where there is only a single area (Greater Sydney), and only $h$ is optimised for (with perfect knowledge\footnote{Perfect knowledge in this paper is assumed to mean $\alpha=1, O(v_i,v_j)=1$, i.e., ability to view every listing across all of Greater Sydney, i.e., $\mathcal{M}$ in~\ref{graphs_maps}.} and no $\beta$ optimisation). A spatial version of the baseline, with the 38 Greater Sydney LGAs as areas. Again, only $h$ is optimised for (with perfect knowledge and no $\beta$). We then run pairwise combinations, so $h$ and $\alpha$, and $h$ and $\beta$. We never run without optimising $h$, as this was the key decision-making parameter in the original model. Finally, we run the proposed extensions in their entirety -- that is, we optimise over all three parameters ($h$, $\beta$, and $\alpha$). We then apply a global constraint (on the result of the training optimisation) that 2006--2010 must exhibit a peak, with details outlined in \cref{appendixConstraints}. The results are presented visually in \cref{figModelOptimisation}. We also provide additional analysis into the network architecture itself in \cref{secSensitivyArchitecture}.

Looking at the resulting plots, we can see that with the introduction of each new component, the resulting values of the loss function $\ell$ (see eq. \ref{eqLoss}) over the training period is reduced at each step, with the proposed extensions achieving the minimal $\ell$, indicating that the predictions arising from the resulting agent decisions follow more closely to those in the actual market. From this point forward, we focus on the proposed extension in its entirety, due to the improved performance in all three periods.

\subsection{Resulting Parameters}\label{bayesianExploration}

\begin{figure}[ht]
    \centering
    \begin{subfigure}{.8\textwidth}
    \centering
    \includegraphics[width=\textwidth]{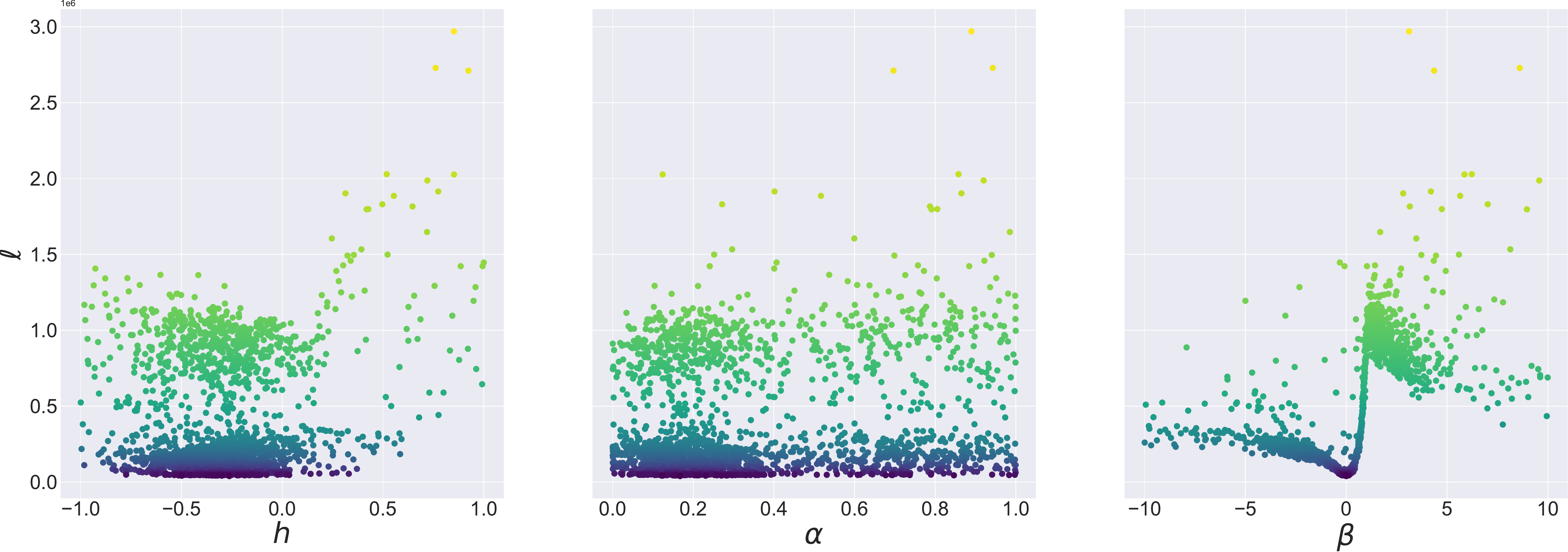}
    \caption{2006--2010}
    \end{subfigure}
    \hfill
    
    \begin{subfigure}{.8\textwidth}
    \centering
    \includegraphics[width=\textwidth]{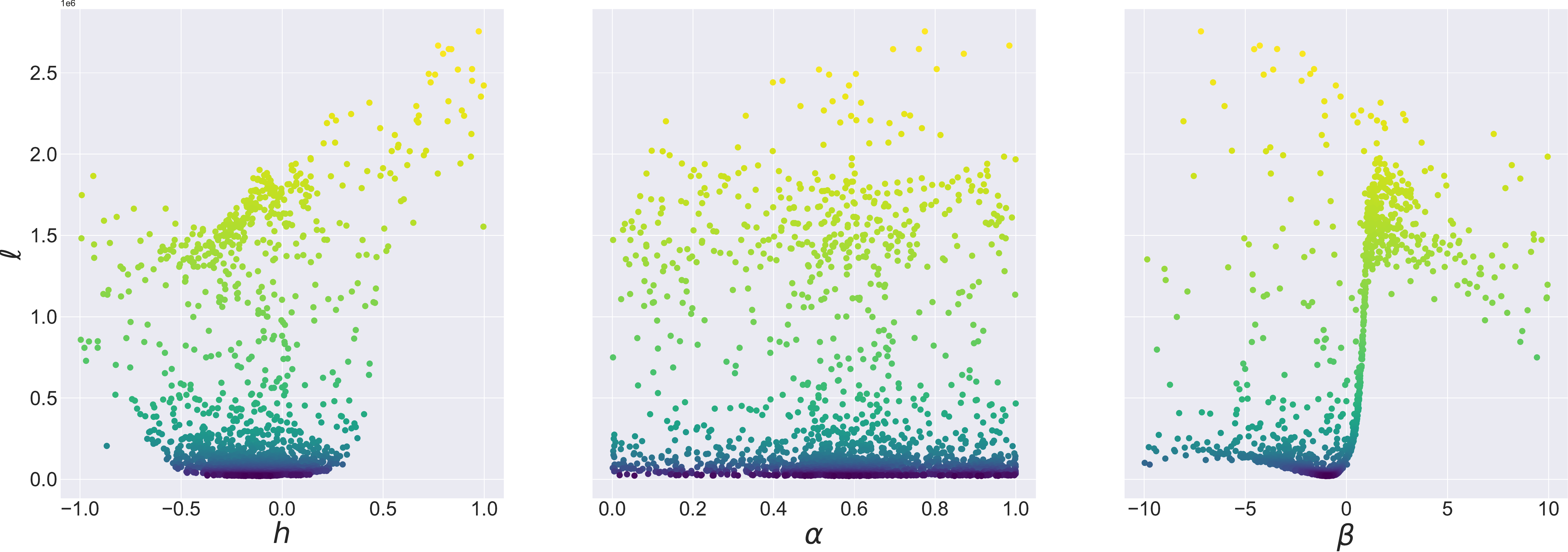}
    \caption{2011--2015}
    \end{subfigure}
    \hfill
    
    \begin{subfigure}{.8\textwidth}
    \centering
    \includegraphics[width=\textwidth]{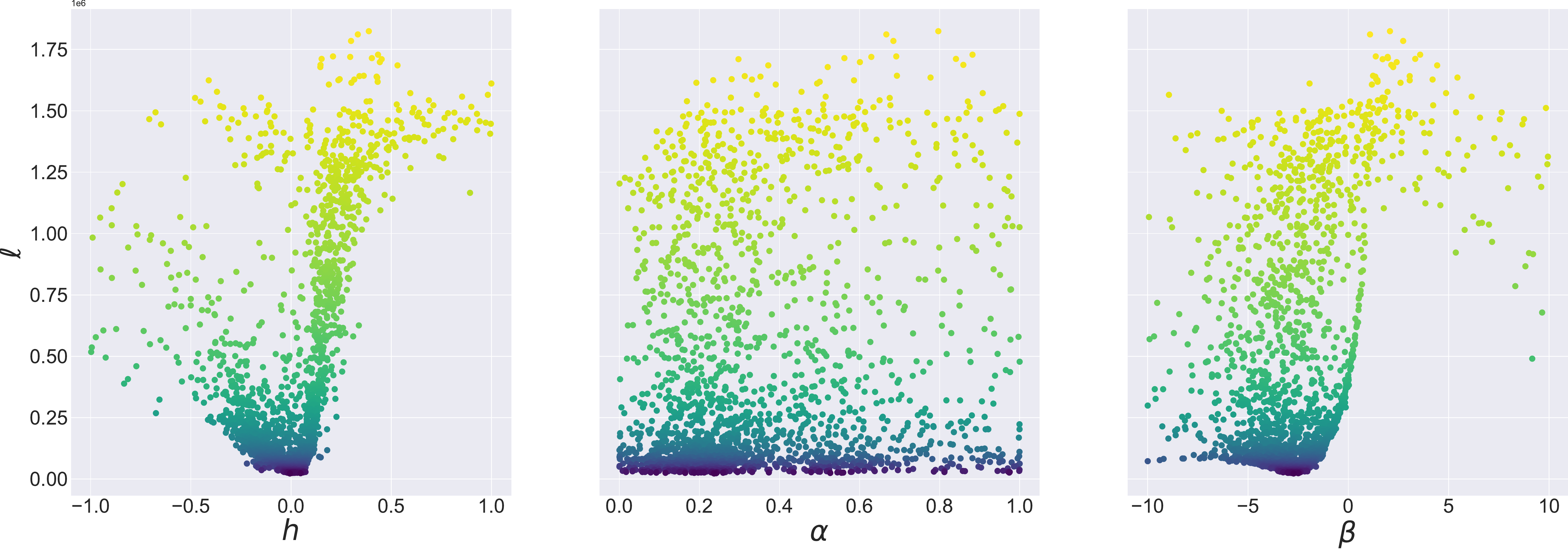}
    \caption{2016--2019}
    \end{subfigure}
    \hfill
    
    \caption{Univariate Parameter Analysis. The $x$-axis represents the parameter value, the $y$-axis represents the loss (with logarithmic colours for consistency across the various loss plots). The other parameters are averaged over to provide the 1-dimensional view.}
    \label{figBayesianUnivariate}
    
\end{figure}

Exploring the entire parameter search space would be computationally prohibitive. Bayesian optimisation intelligently explores this search space, balancing exploration and exploitation with the use of an acquisition function allowing more emphasis to be placed on well-performing or unexplored regions of the space \citep{shahriari2015taking}. The resulting exploration is visualised in \cref{figBayesianExploration}. From this, we can see the regions of interest, with dark sections indicating areas with the lowest loss $\ell$ (and more sample points being present in such areas). We can see the search space is fairly well explored in all cases, with obvious regions of well-performing parameter combinations (pairwise combinations are visualised in \cref{figBayesianInteractions}). While the 3D plot gives a high-level overview, it is difficult to visualise the contribution of each component. To facilitate this, we present a flattened 1-dimensional view of each parameter, where the results are averaged over the other 2 parameters to view the loss for 1 parameter at a time. This is shown in \cref{figBayesianUnivariate}. From these plots, it can be seen how each parameter behaves in isolation (noting that such plots do not capture the parameter interactions). The selected ABM parameters (the ones which had the lowest $\ell$) are presented in \cref{tblResultingParamaters}. A sensitivity analysis for the parameters is performed in \cref{appendixSensitivity}.

\begin{table}[!ht]
\centering
\begin{tabular}{@{}llll@{}}
\toprule
                   & \textbf{2006--2010} & \textbf{2011--2015} & \textbf{2016--2019} \\ \midrule
\textbf{$h$}       &  -0.80             &   -0.11             &   -0.005            \\
\textbf{$\beta$}    &  0.08             &   -1.03            &   -2.73             \\
\textbf{$\alpha$} &   0.28           &   0.59            & 0.24              \\ \bottomrule
\end{tabular}
\caption{The selected ABM parameters from the training period with Bayesian optimisation, rounded to 2 decimal places.}
\label{tblResultingParamaters}
\end{table}

The parameter search space is uniform across the ranges given in \cref{tblParams}, so an uninformative parameter would be sampled uniformly as well (since there would be no Bayesian preference the sampling would be approximately uniformly random), instead, we see clear uni-modal peaks in almost all cases in \cref{figBayesianInteractions}, with the $h$ and $\beta$ parameters being normally distributed around the optimal value found, and $\alpha$ with a clear peak but non-normally distributed. This indicates each parameter seems to have a useful range, which is further verified in the sensitivity analysis in \cref{appendixSensitivity}.

\paragraph{$h$:}

Recall that the HPI aptitude $h$ directly influences the agents' bid price (given in \cref{eqBidPrice}), and serves as the key trend-following parameter in the original model. We can see the value of $h$ controls the contribution of the HPI over the previous year and as such it affects the bid price based on the markets state. This relationship depends on both the current {\it HPI}, and the yearly difference in $\Delta_{HPI}$. 

The absolute value of $h$ controls the magnitude of the contribution, and we can see 2006--2010 had the highest contribution indicating the largest market effect on bidding (in the original model, this also had a large magnitude but the opposite sign). The value for 2011--2015, $h=-0.11$, is very close to that chosen in the original model $h=-0.10$, which reflects the relatively consistent price dynamics, with agents mostly ignoring the market trend.

Interestingly, in 2016--2019, the chosen $h$ is near zero. This means the denominator of \cref{eqBidPrice} simplifies to: $$ \phi_{M}[t] + \phi_H - h * \Delta_{HPI}[t] \,\, \simeq \,\, \phi_{M}[t] + \phi_H$$ meaning the price is dependant on the mortgage rate and homeownership rate, as $\phi_H$ is set based on the current value of HPI, this means the historical values are not being used and instead a much more "forgetful" market based only on the previous month's  HPI is used for setting the price, with agents paying less attention to historical trends.

The HPI aptitude $h$ also has clear optimal ranges for each period, with 2006--2010 in the range $[-0.75\dots 0]$, 2011--2015 in the range $[-0.5 \dots 0.25]$ (which are of similar widths), and 2016--2019 in the much narrower range $[-0.2 \dots 0.05]$. We can see a sharp transition occurring in 2016--2019 around 0.1, in which case the loss begins increasing drastically for any higher values.

\paragraph{$\beta$: }

This parameter (the righthand column in \cref{figBayesianUnivariate}) exhibits a very sharp transition around 0 for all years. This is because $\beta$ has a direct relation to the supply and demand in the model, which drastically changes the dynamics based on the availability of properties.  We see increasing contributions of $\beta$ throughout the years, with 2016--2019 indicating the highest levels of $\beta$. This is perhaps reflective of the market, where people are increasingly following trends when it comes to the decision to sell a dwelling (perhaps an indicator of a "bursting" bubble, with a large cascading sell-off). This indicates sharper peaks and dips are likely to occur in the future, with decision-making being made increasingly under the pressure of social influence.

\paragraph{$\alpha$:} This parameter does not present as clear of a transition, or defined range of optimal vales as $h$ or $\beta$. This indicates the value of $\alpha$ has less of an impact on the results when compared to $h$ and $\beta$ (as is confirmed in the sensitivity analysis in \cref{appendixSensitivity}).  The likely reason for this is since the buyer always attempts to purchase the most expensive viewed dwelling, the varying levels of $\alpha$ do not have as large of an effect on the outcome as the other parameters, since the most expensive dwellings still have a higher preference. Discussion on the effect $\alpha$ has on agent utilities and decisions is given in \cref{appendixUtility}. 

We can see a general preference towards various $\alpha$ ranges for each period, based on higher density of samples in these areas. The preference is never for "perfect" information, i.e., $\alpha=1$, reinforcing that such agents often act in a boundedly rational manner (with $\alpha < 1$). In the 2006--2010 period there is a general preference towards sampling lower levels of $\alpha$ (within $\approx [0.05, \dots 0.3]$, with the optimal value being $0.28$), resulting in buyers acting with less information. In 2011--2015, during the economic recovery in Australia, buyers may have been more cautious and considering a wider range of available dwellings when purchasing, reflected in higher values of $\alpha$ ($\approx [0.5, \dots 0.7]$, with the optimised value being $0.59$). In 2016--2019 the cautiousness of buyers appears to revert again (within $\approx [0.05, \dots 0.3]$, with optimised value of $0.24$), showing buyers making less "informed" choices perhaps due to the rapidly increasing dwelling prices and buyer's desires to partake, at the expense of making "optimal" choices with larger $\alpha$ values. 

\section{Results}\label{secResults}

\begin{figure}[ht]
    \centering
    \includegraphics[width=\textwidth]{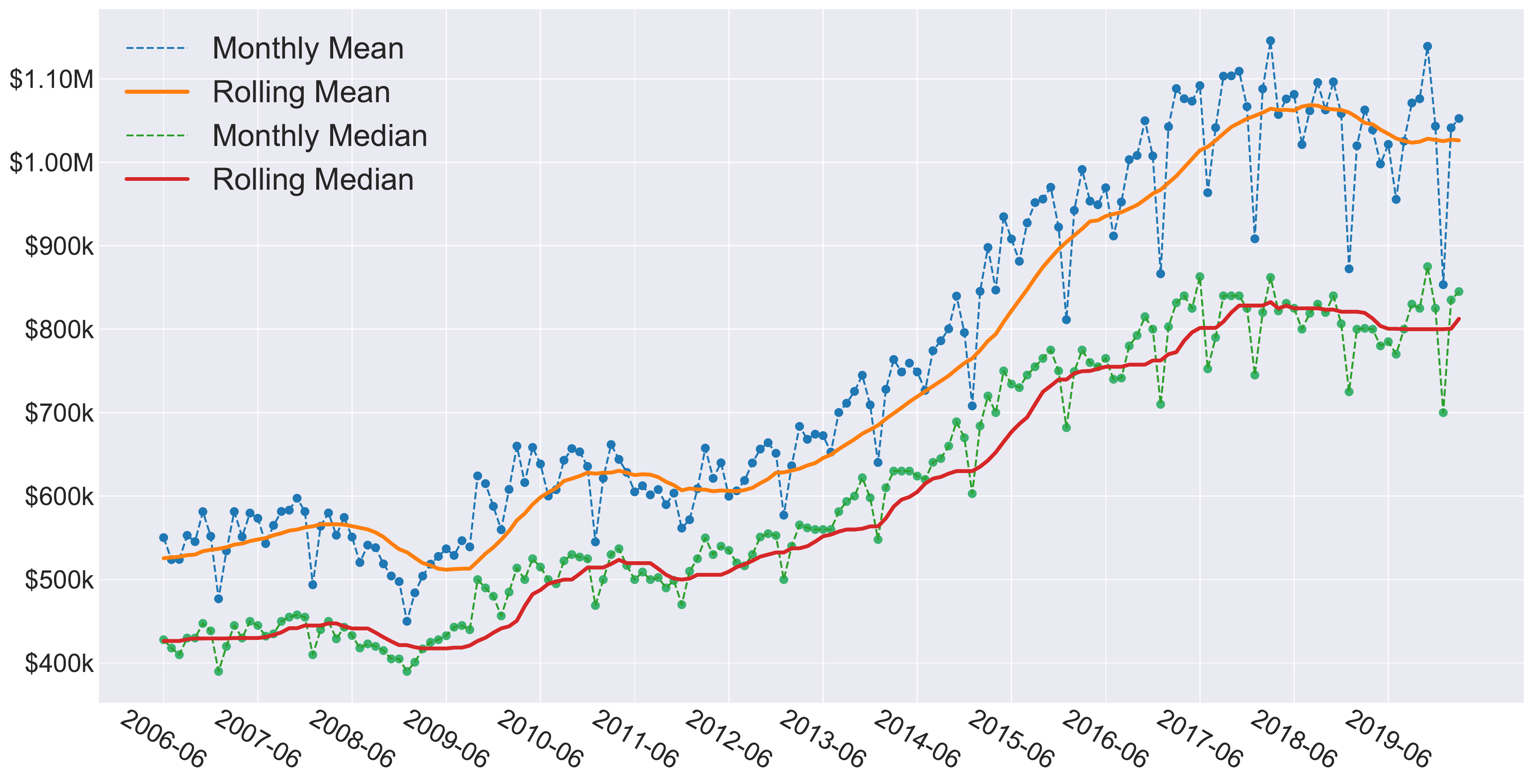}
    \caption{The actual trends of Greater Sydney house prices from June 2006 to December 2019. Source:  SIRCA-CoreLogic. The orange line represents the rolling mean, and the red line the rolling median. The raw monthly data points are also visualised (dotted lines). We use the median price trend as a more robust measure in all cases in this paper.}
    \label{figRealTrend}
\end{figure}

In this section, we present results of ABM simulations in terms of (i) price forecasting over three historic periods, aligned with the Australian Census years (2006, 2011 and 2016), and (ii) resultant agent preferences in terms of household mobility patterns. We also identify market trends across the three periods.
The considered periods 2006--2010 and 2011-2015 include 48 months, while the contemporary period, 2016--2019, covers 42 months (our SIRCA-CoreLogic dataset includes the market data until 31 December 2019). Each time period compromises a separate set of simulations (with varying optimised parameters). Discussion on the chosen time periods is given in \cref{appendixTimePeriods}.
For each period, we run 100 Monte Carlo simulations, using the model parameters optimised for the corresponding training set, as described in \cref{secBayesian}, and then obtain predictions for the remaining (testing) part of the data. For the first two periods, the first $\frac{3}{4}$ of the time-series is the training part (e.g., 36 months from 1 July 2006 to 30 June 2009), and the remaining $\frac{1}{4}$ is the testing portion not used by any optimisation process (e.g., 12 months from 1 July 2009 to 30 June 2010). For the last period, the training part includes 30 months (from 1 July 2016 to 31 December 2018), with the remaining 12 months of 2019 used for testing.  

\subsection{Price Forecasting}

\begin{figure}[ht]
    \centering 
\begin{subfigure}{0.33\textwidth}
  \includegraphics[width=\linewidth]{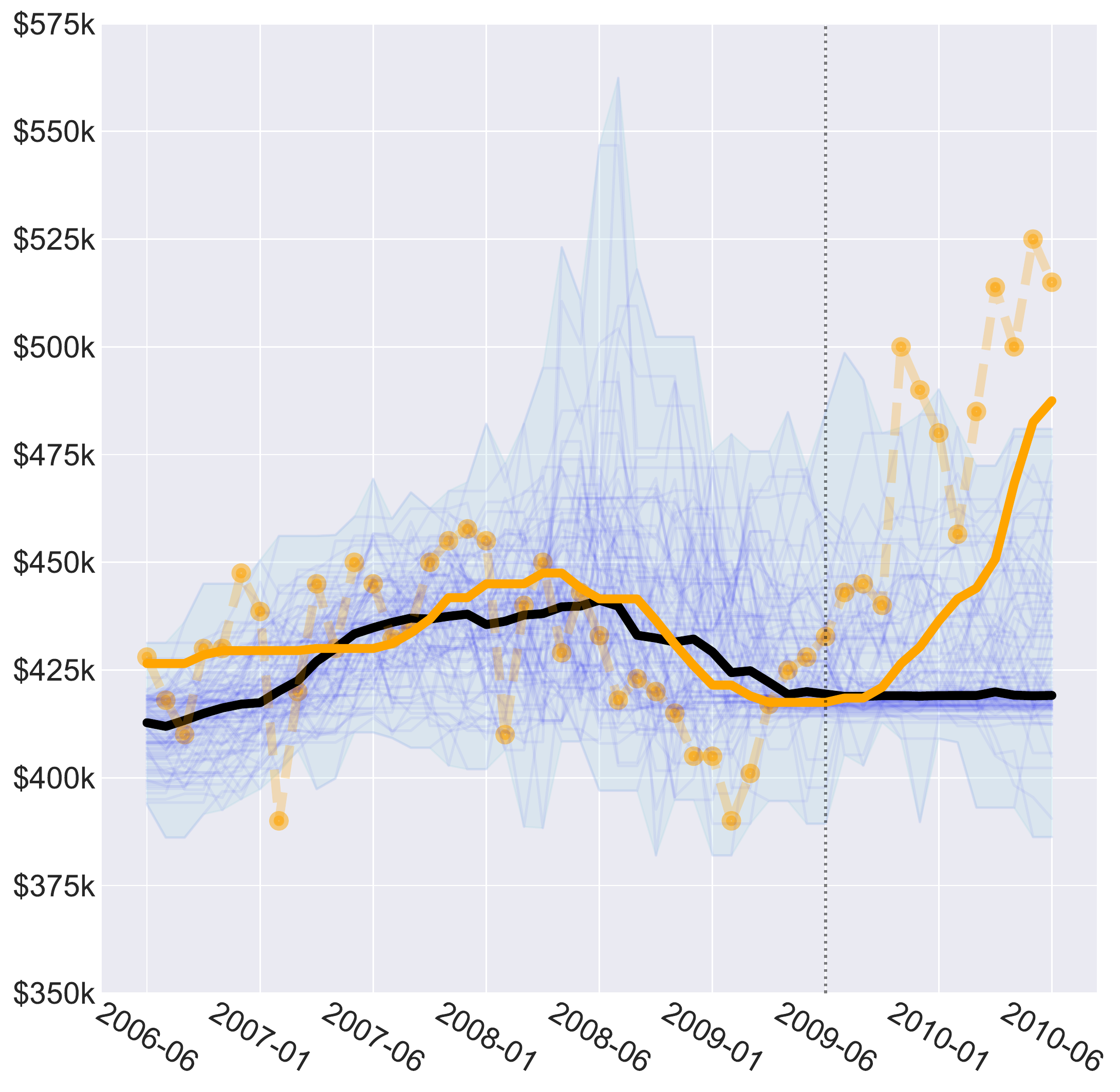}
  \caption{Median: 2006--2010}
\end{subfigure}\hfil 
\begin{subfigure}{0.33\textwidth}
  \includegraphics[width=\linewidth]{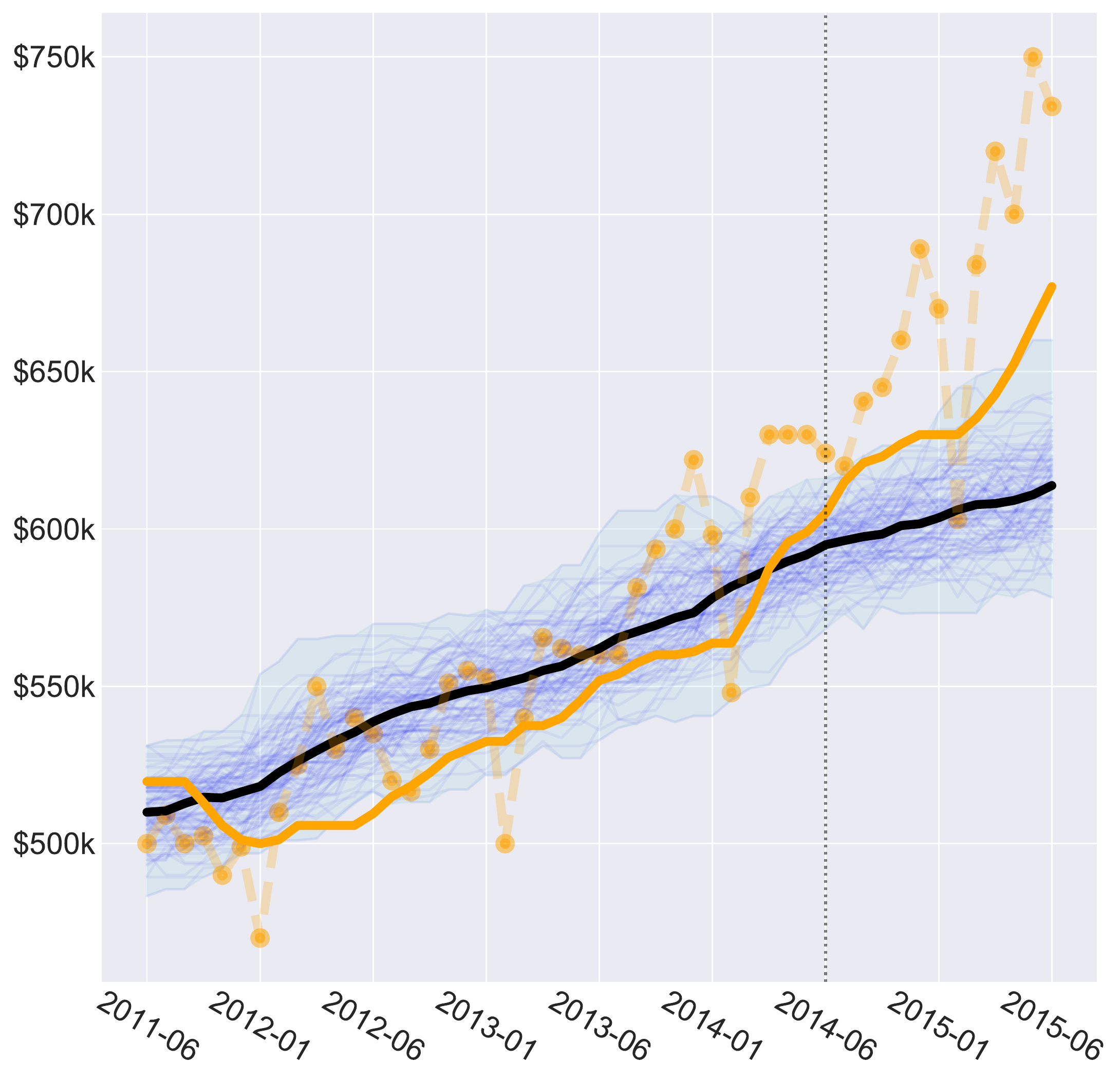}
  \caption{Median: 2011--2015}
\end{subfigure}\hfil 
\begin{subfigure}{0.33\textwidth}
  \includegraphics[width=\linewidth]{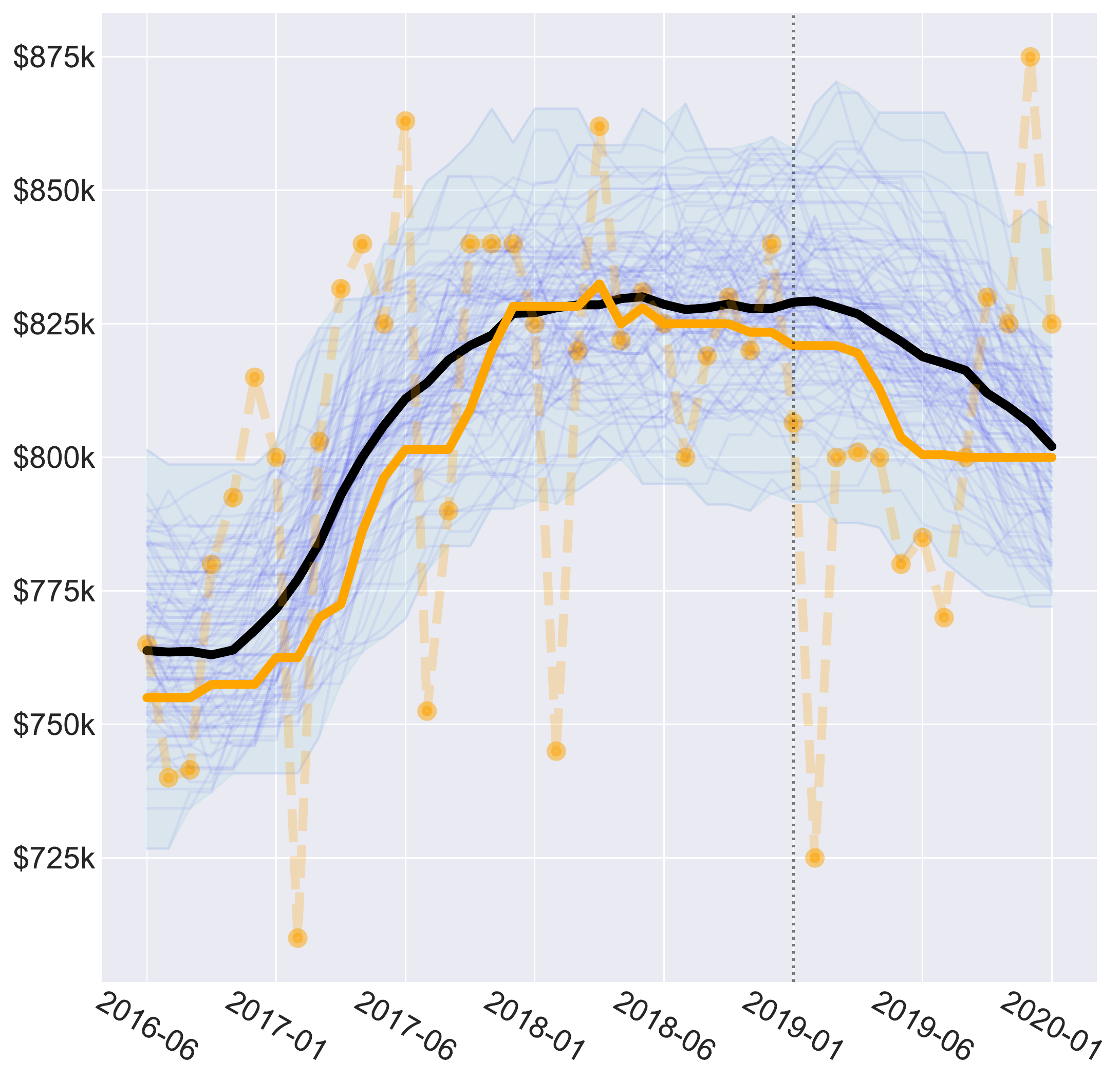}
  \caption{Median: 2016--2019}
\end{subfigure}

\medskip
\begin{subfigure}{0.33\textwidth}
  \includegraphics[width=\linewidth]{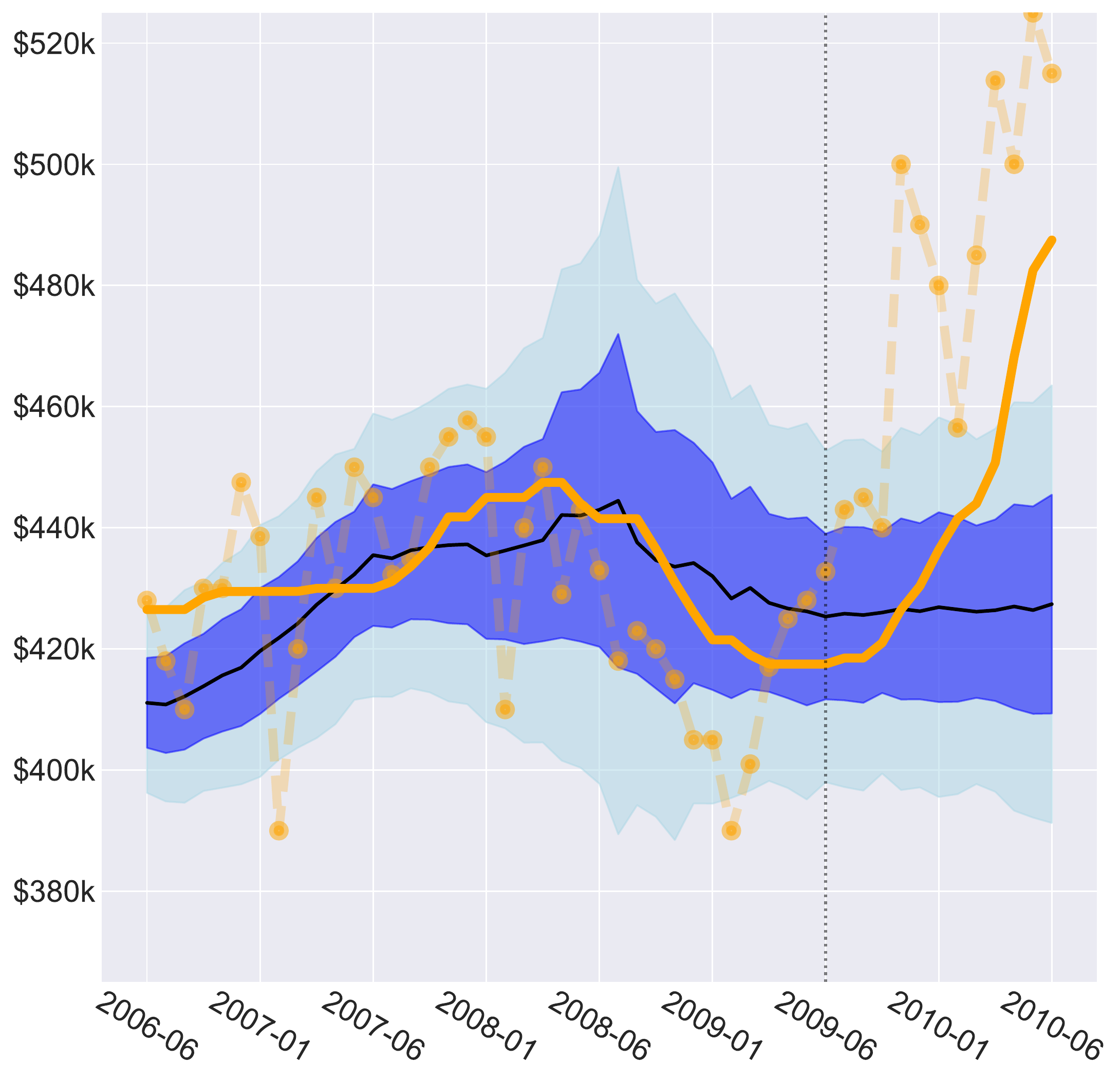}
  \caption{Mean: 2006--2010}
\end{subfigure}\hfil 
\begin{subfigure}{0.33\textwidth}
  \includegraphics[width=\linewidth]{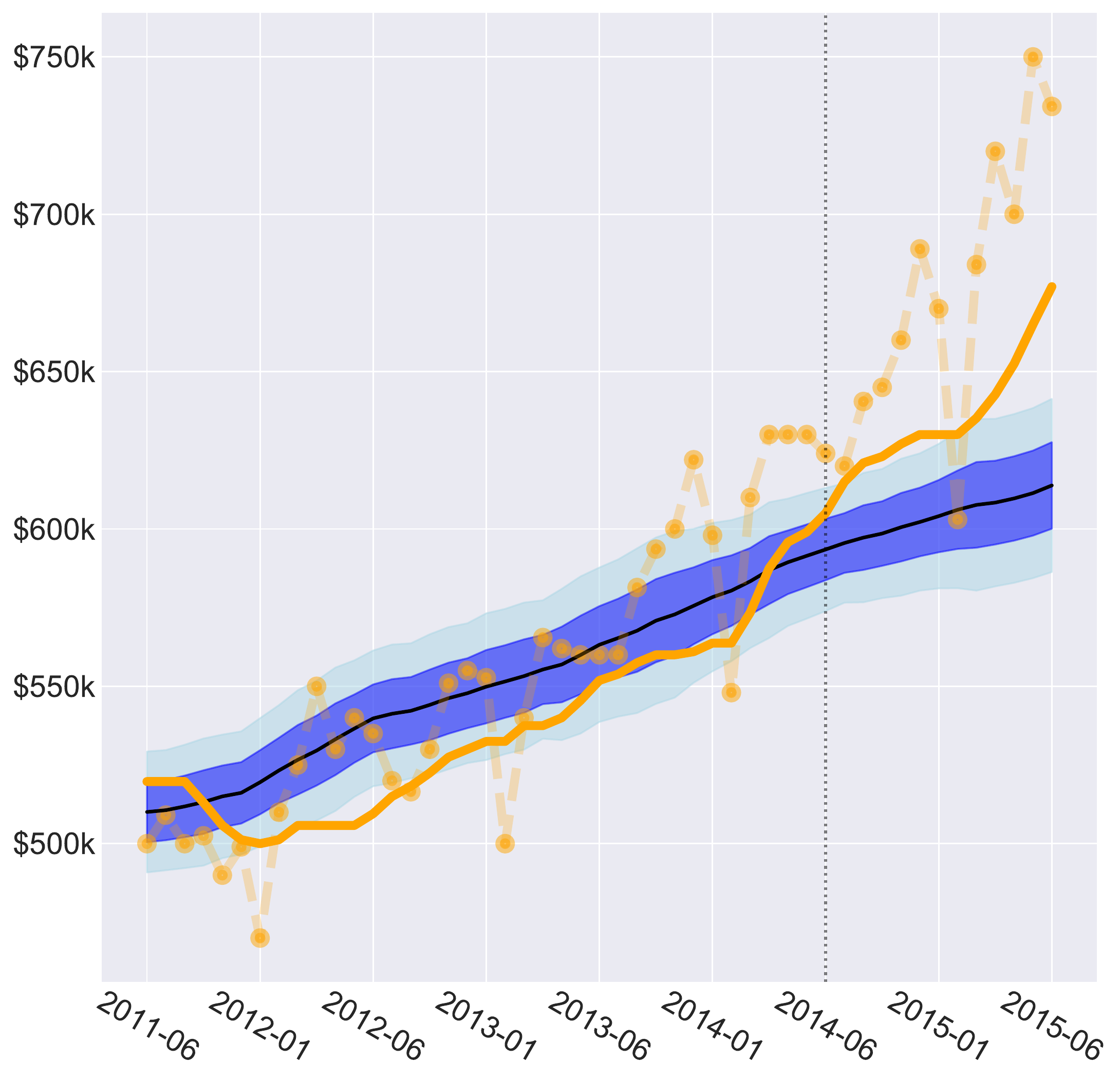}
  \caption{Mean: 2011--2015}
\end{subfigure}\hfil 
\begin{subfigure}{0.33\textwidth}
  \includegraphics[width=\linewidth]{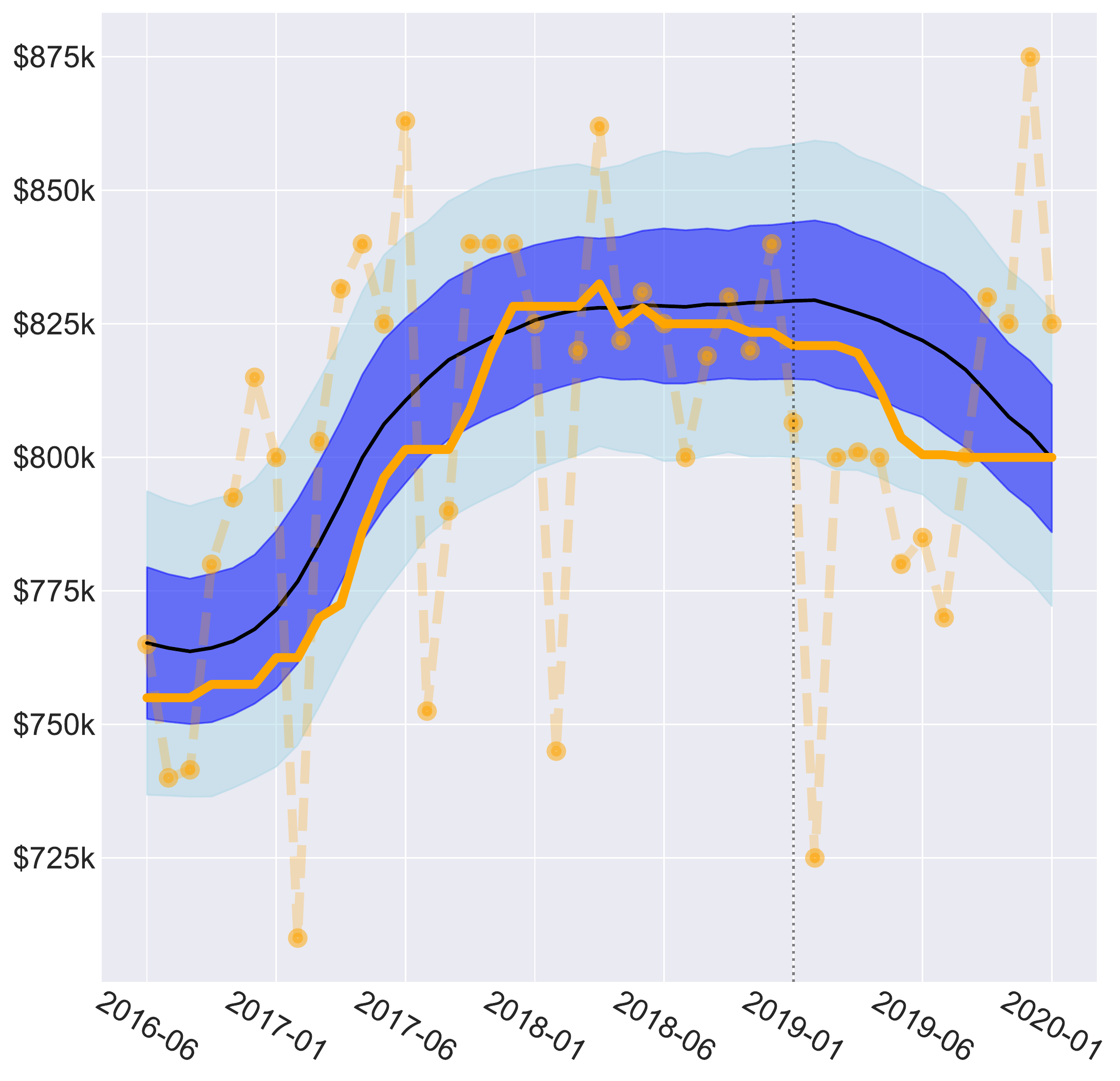}
  \caption{Mean: 2016--2019}
\end{subfigure}

\caption{ABM simulation predictions. The top row shows the median of the simulations (black lines), every individual run (blue lines), and the minimum and maximum of every run with the light blue fill. The bottom row shows the mean of the simulations (black line), and $\pm$ 1 and 2 standard deviations (dark blue and light blue respectively). The vertical dashed lines separate training and testing parts of the time series. The orange lines show the actual trend obtained from SIRCA-CoreLogic data.}
\label{figPredictions}
\end{figure}

We visualise the forecasting results in \cref{figPredictions}. 
It is evident that the model can successfully capture the key trends across the entire time-series as emerging from individual buy and sell decisions, correctly identifying the peak and dip in 2006--2010, the steady growth during 2011--2015, and the growth and slow decline in 2016--2019. 

The time period beginning in 2006 was a  period of substantial uncertainty, triggered by the Global Financial Crisis (GFC). Tracing predictions for the last quarter of the period (i.e., the testing part of the dataset), shown in \cref{figPredictions}.a and \cref{figPredictions}.d, we observe that the average market rebound pattern is not fully followed. However, when considering the range of the simulations runs, i.e., the possibilistic regions of the simulation (as defined by \cite{edmonds2018using}, and visualised by the blue boundaries in \cref{figPredictions}), we see that these contain the first segment of the fast rebound, and capture the market recovery to a good degree. In other words, the model is able to show the possibility of such a rebound. The discrepancy indicates that there was a catalyst underpinning a significant price appreciation during 2010 that was not solely driven by social influence and endogenous regulatory factors such as interest rates. It is well accepted that the market was reignited in 2009--2010 by exogenous factors, most notably by post-GFC government stimulus initiatives, such as the  First Home Owners Boost~\citep{randolph2013first}.

For the period beginning in 2011, the predicted time series correctly follows the actual trend, while slightly underestimating the slope of the growth towards the end of the period, as shown in \cref{figPredictions}.b and \cref{figPredictions}.e. It can be argued that the social influence factors, estimated during the optimisation phase, continue to affect the market dynamics during the last quarter of the period. In other words, the influence of these factors on decision-making, coupled with endogenous factors (e.g., interest rates),  results in a steady growth of the market, predicted until the end of the period.

In the last considered period, starting in 2016, the model captures both the testing and training portions very well, correctly predicting the dip from 2019 onward, as shown in \cref{figPredictions}.c and \cref{figPredictions}.f. The notable decline occurs only in the testing period,  and yet the model is able to accurately predict both the peak and the correction.  This indicates that the underlying reasons for the market reversal have developed during the first part of the period, and have been adequately captured by the parameter optimisation. 

Overall, in all three time periods, the possibilistic output of the model contains most of the actual market dynamics. Specifically, during 2006--2010, 96\% of the actual monthly prices are within the possibilistic range boundaries of the simulation (i.e., between the minimum and maximum output for each time period, the top row in \cref{figPredictions}), for 2011--2015: 84\%, and 2016--2019: 100\%. When using the mean $\pm$ 1 (2) standard deviations (bottom row in \cref{figPredictions}) as possibilistic boundaries, we observed the following probabilities of falling within the boundaries, during 2006--2010: 78\% (92\%), 2011--2015: 27\% (63\%), 2016--2019: 86\% (100\%).

\subsection{Area-specific Price Forecasting}

The model was not directly optimised for spatial submarkets. However, in this section, we evaluate the predictive capacity of the model in terms of area-specific forecasting. During initialisation, some area-specific information for LGAs is drawn from available distributions, for example, the recent sale price of dwellings in that LGA. Likewise, households are also distributed into LGAs based on the actual population sizes of the LGAs.  However, no additional optimisation is applied across different LGAs with respect to actual area-specific trends.  In other words, the spatial component is used only for initialising relevant distributions, leaving the market dynamics to develop through agent-to-agent interactions.

\begin{figure}[ht]
    \centering
    
    \begin{subfigure}{.9\textwidth}
      \includegraphics[width=\textwidth]{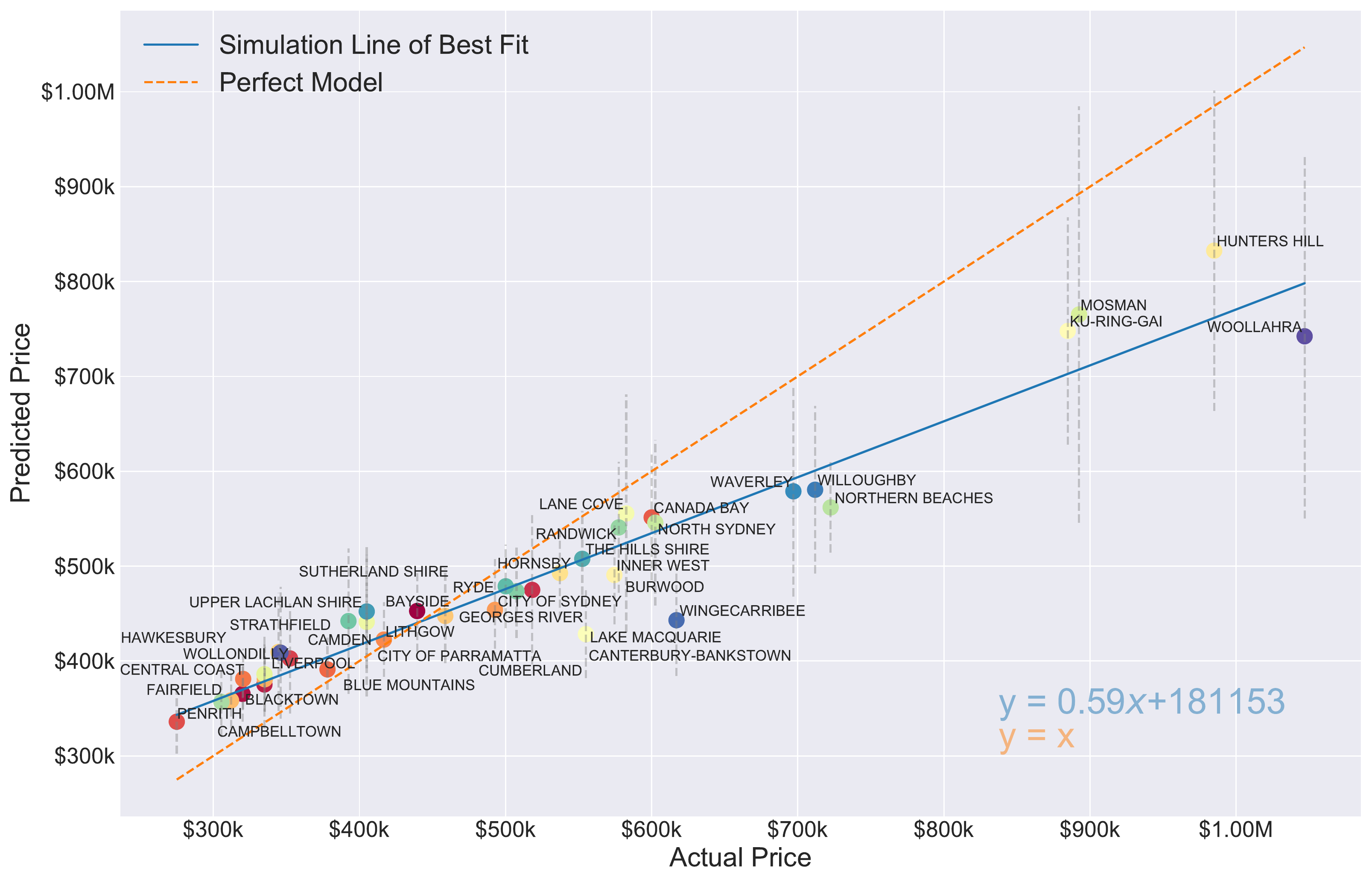}
        \caption{June 2010 predictions (end of 2006--2010 test period). $R^2=0.92$.}
        \label{figAreas2006}
    \end{subfigure}
    \hfill
    
    \begin{subfigure}{.9\textwidth}
    \includegraphics[width=\textwidth]{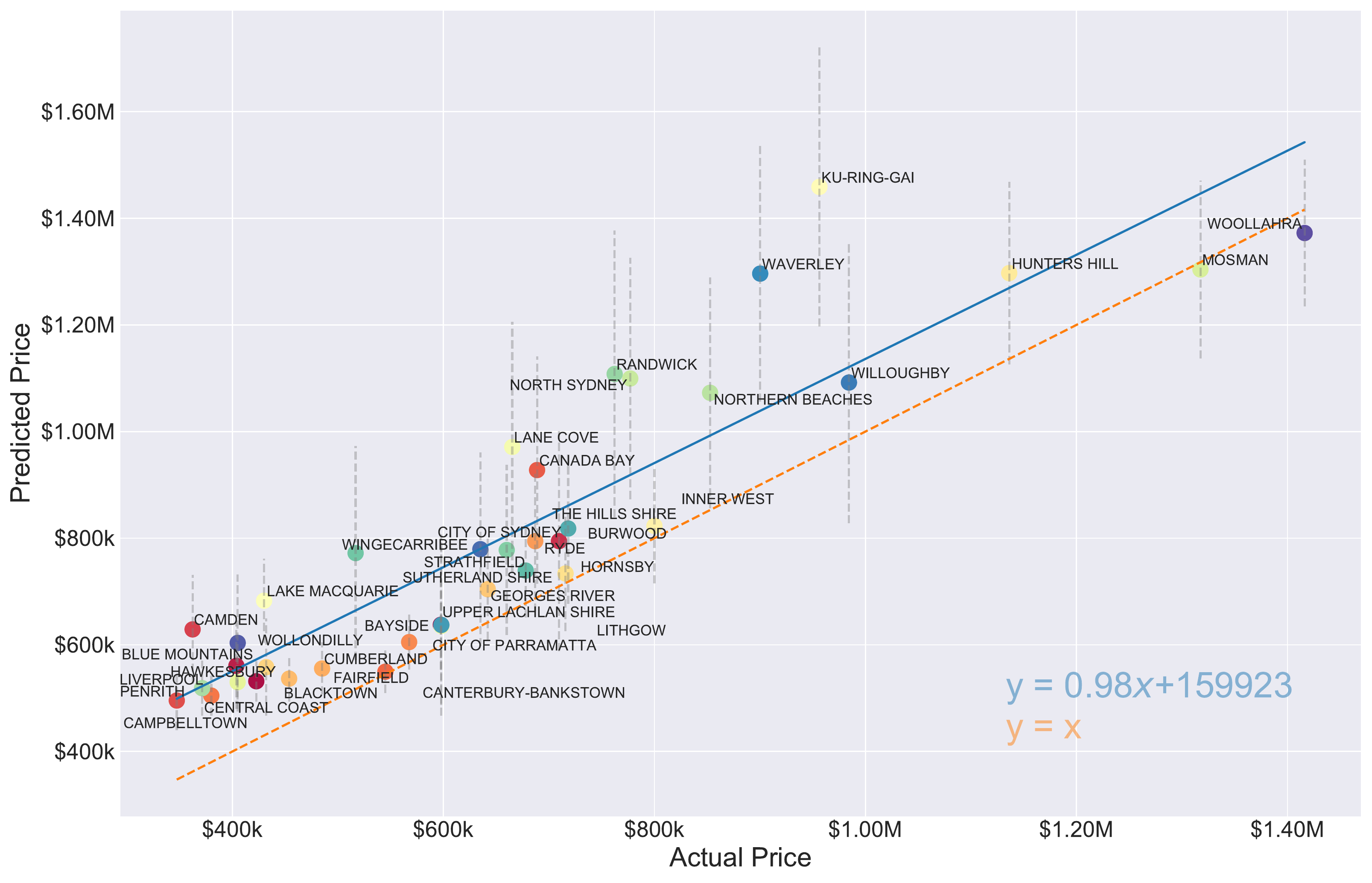}
    \caption{June 2015 predictions (end of 2011--2015 test period). $R^2=0.82$.}
    \label{figAreas2011}
    \end{subfigure}
    \hfill
    
    \begin{subfigure}{.9\textwidth}
    \includegraphics[width=\textwidth]{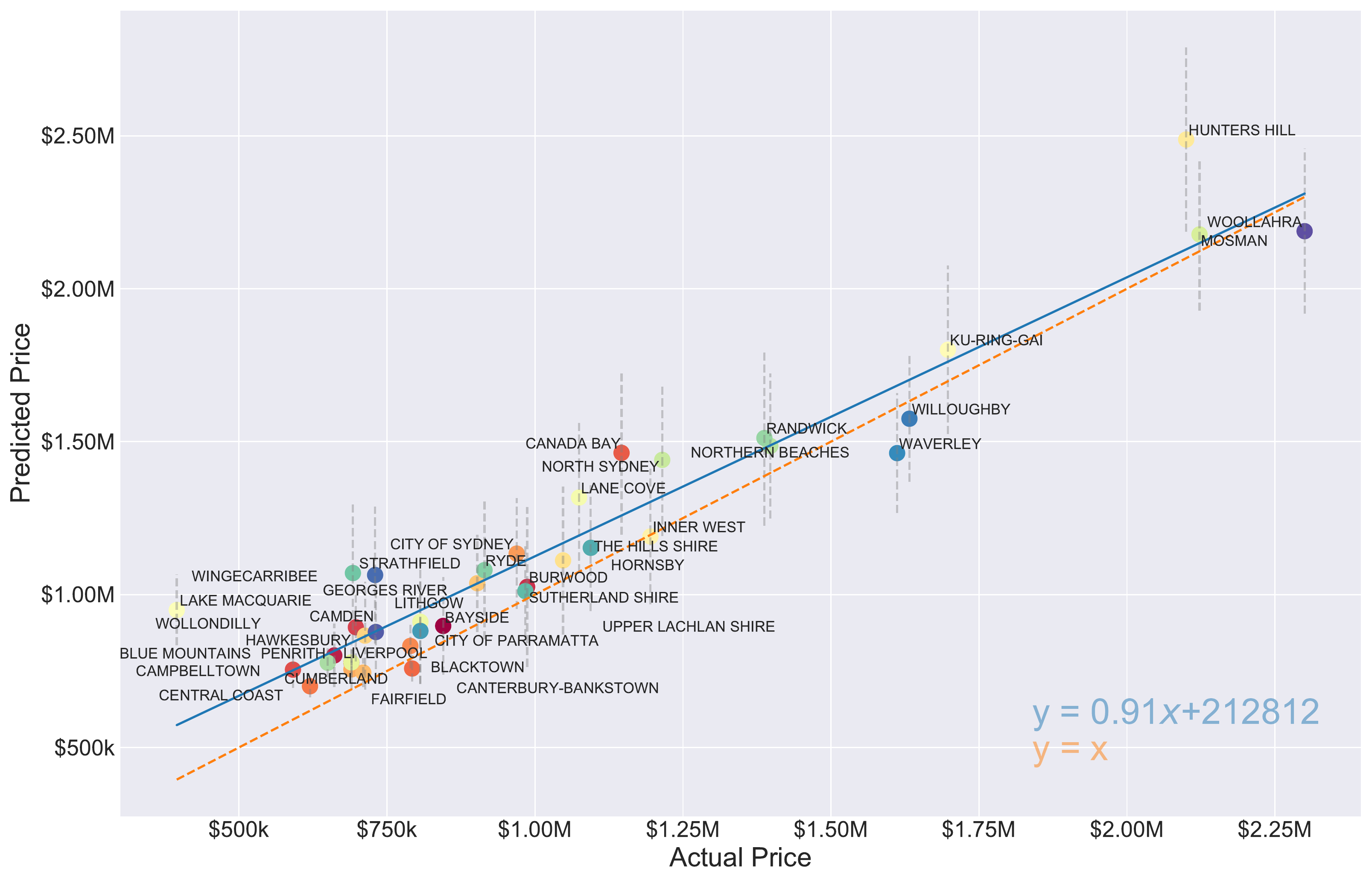}
    \caption{December 2019 predictions (end of 2016--2019 test period). $R^2=0.91$.}
    \label{figAreas2016}
    \end{subfigure}
    \hfill
    
    \caption{Predicted LGA pricing at the \textit{end} of the testing period. The actual prices (SIRCA-CoreLogic) are shown on the x-axis, with the predicted prices on the y-axis. Error bars represent the standard deviation of the prediction across simulation runs. The orange line shows the perfect model ($y=x$), the blue line shows the least-squares line of best fit for the predictions (equation given on plot).}
    \label{figAreaPricing}
    
\end{figure}

In \cref{figAreaPricing} we visualise the predicted area-specific pricing at the end of the testing period. We can see that these predictions closely follow actual data in general, despite not being directly optimised for, with all predictions characterised by high $R^2$ values. 

For the period 2006--2010 (with the end of the testing period mapping to June 2010), the simulations slightly overestimate the final price of the cheaper LGAs, but underestimate the resulting price of the most expensive LGAs, as shown in \cref{figAreaPricing}.a. However, the perfect model (orange line) tends to be within the error margin (standard deviation) of the predictions of the simulation.

For 2011--2015 (with the end of the testing period being June 2015), the slopes of both actual and predicted regressions are almost identical ($m=0.98$), as shown in \cref{figAreaPricing}.b. However, the additive constant of the regression (i.e., y-axis intercept) for the predicted line is greater than 0 (of the perfect model): as a result, we are predicting slightly higher values across the LGAs on average. Considering the LGAs that were most overpriced with respect to the spatial trend (such as Kuringai, Waverly, Northern Beaches and North Sydney), we can compare their resultant predicted prices in June 2015, \cref{figAreaPricing}.b, with the actual prices depicted in \cref{figAreaPricing}.c. The 2016--2019 plots show that this growth did eventually happen, and so the simulations produced for 2011--2015 merely predicted this appreciation for an earlier time than the actual scenario. 

The predictions for the period 2016--2019 (the end of the testing period: December 2019) produce a regression strongly aligned with the actual fit, particularly for the higher-priced LGAs. Again, there is some overestimation in the cheaper LGAs, but this seems to highlight the increasing popularity of these suburbs. Analogously to the previous period, this may be indicative of some future price growth for these areas, not yet reflected in the current actual pricing.

Overall, we observe that the resulting area-specific price predictions at the end of each testing period fit closely to the actual resulting prices. This indicates that the model successfully captured the spatial submarkets, despite having been optimised for the overall market dynamics of the Greater Sydney region as a whole.

\subsection{Agent Preferences and Mobility Patterns}\label{secHouseholdMobility}

Analysing the households' movements produced by the simulation is another key insight the spatial agent-based model can provide. In this section, we consider various agent movement patterns (which we refer to as household mobility), aiming to identify the salient trends in agent preferences. There are several key areas we focus on: first-time home buyers, investors, and new households (i.e., migrations or households splitting). Again, no direct optimisation was applied to the movements, and so the identified preferences are intrinsic results of the model, and not attributed to some actual data. However, we show that such mobility preferences are supported by evidence, thus arguing that the model is able to produce sensible local patterns for which it was not explicitly optimised for, based only on the global calibration data and proposed spatial structure.

The mechanisms shaping the process of settlement formation and generating intra-urban mobility specifically, include transitions driven by critical social dynamics, transformations of labour markets,  changes in transport networks, as well as other infrastructural developments~\citep{Kim2005,Simini2012,Barthelemy2013Planning,Louf2013,Arcaute2016,Barthelemy2016book, crosato2018critical,Barbosa2018,Piovani2018,slavko2019dynamic,Barthelemy2019}. Types of homeownership, in particular, are known to affect mobility patterns~\citep{crosato2020}. In this work, we focus solely on the movements resulting from the housing market dynamics, which in turn incorporate the imperfect spatial information and other subjective factors such as the FOMO and trend following aptitude. We do not model any structural changes across the regions, i.e., the LGAs boundaries, transport and other infrastructure topologies, etc. remain fixed.

\begin{figure}[ht]
    \centering
    \begin{subfigure}[b]{0.32\textwidth}
        \centering
        \includegraphics[width=\textwidth]{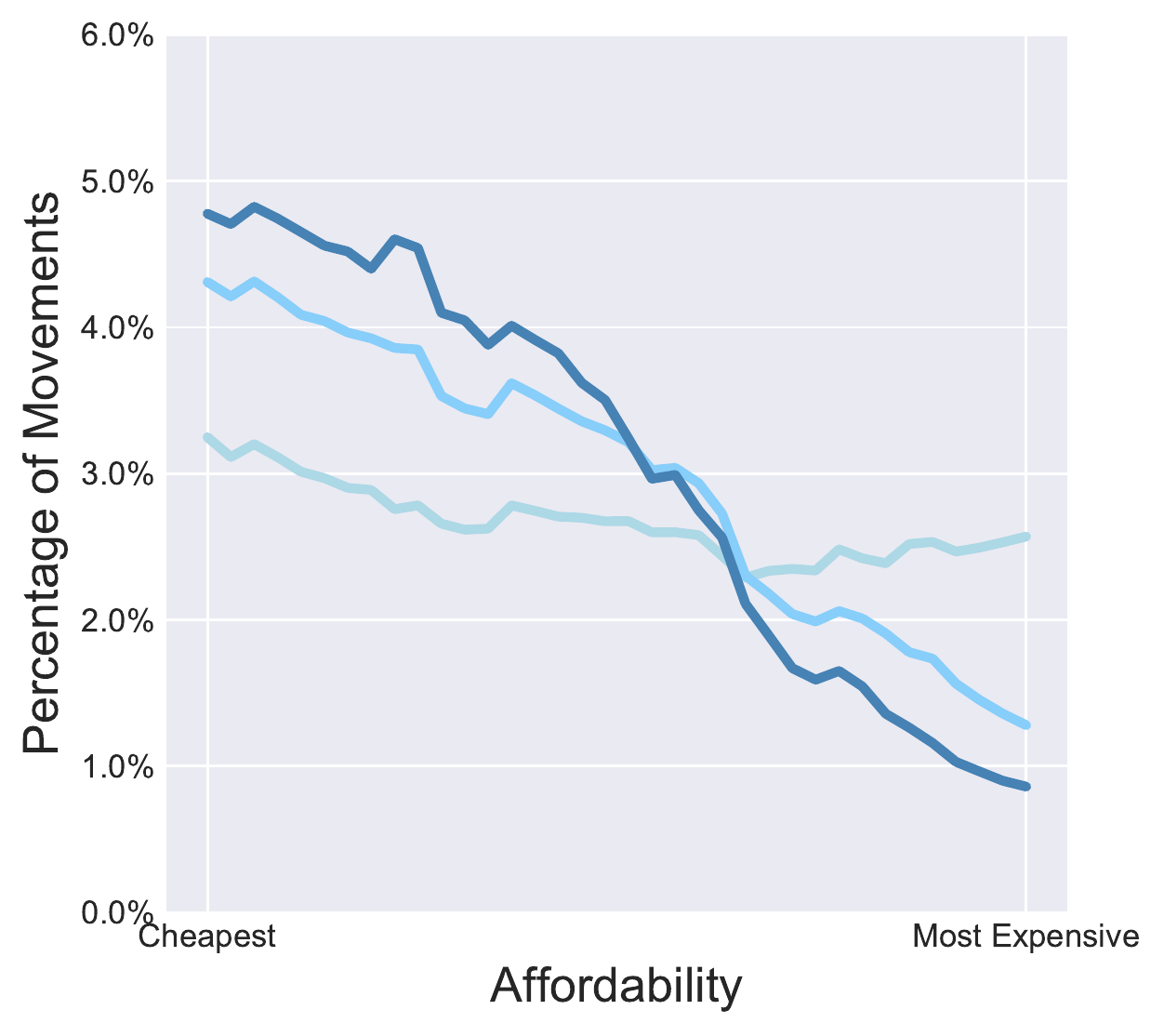}
        \caption{First-time home buyers}\label{figFirstHomeownerComp}
    \end{subfigure}
    \hfill
    \begin{subfigure}[b]{0.3\textwidth}
        \centering
       \includegraphics[width=\textwidth]{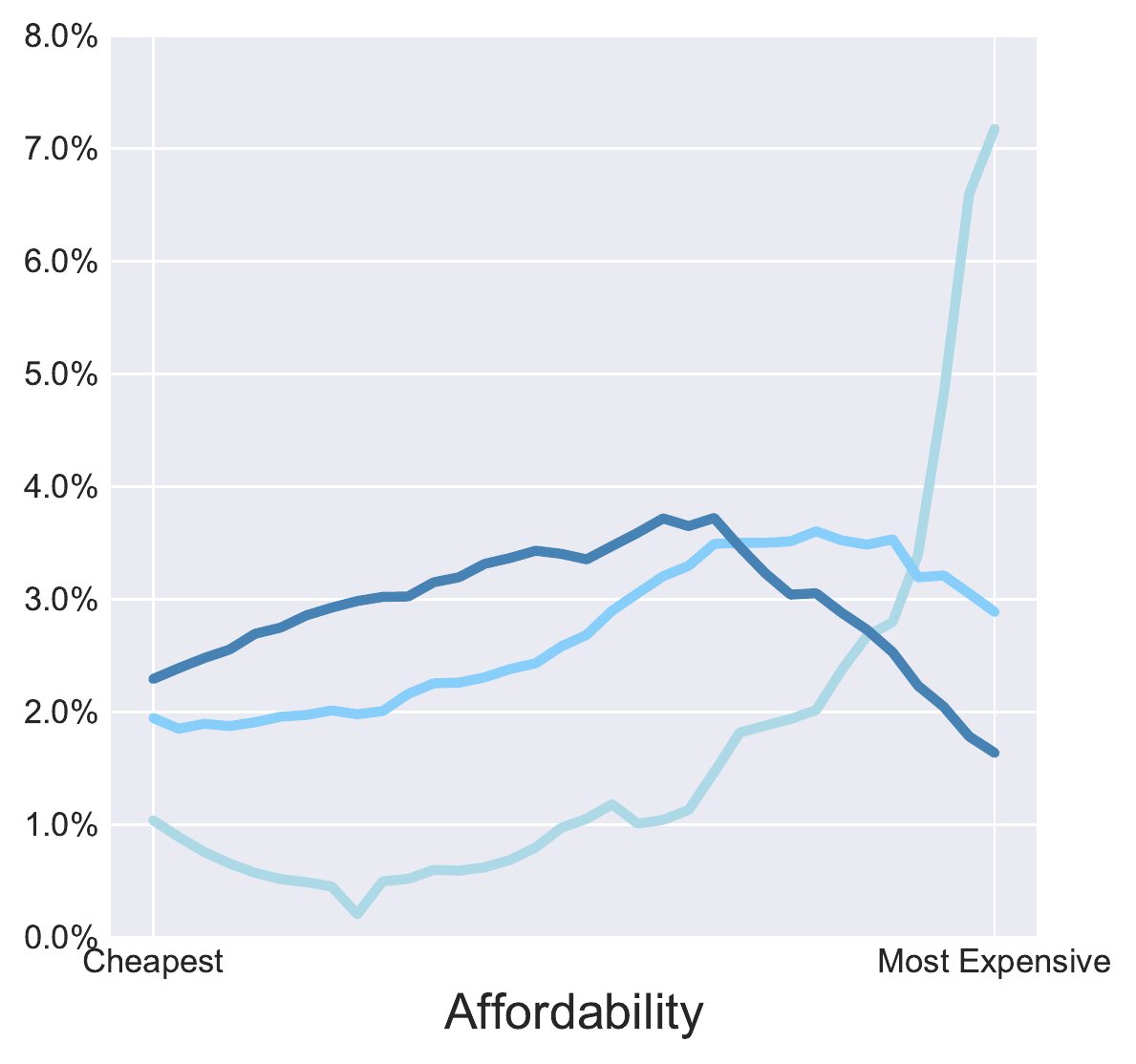}
        \caption{Overseas Investors}\label{figOverseasComp}
    \end{subfigure}
    \hfill
    \begin{subfigure}[b]{0.3\textwidth}  
        \centering 
        \includegraphics[width=\textwidth]{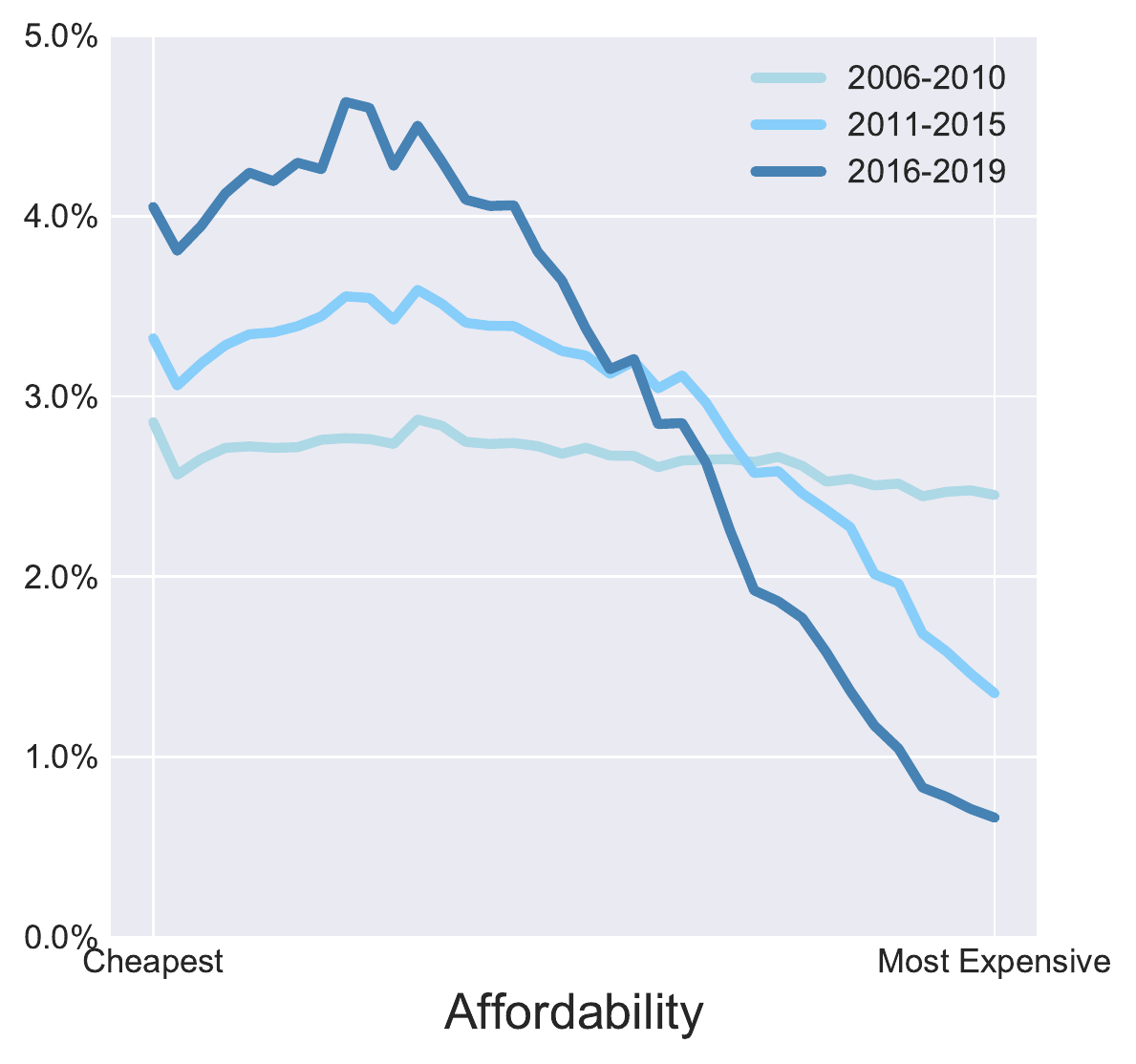}
        \caption{Local Investors}\label{figLocalComp}
    \end{subfigure}
    \hfill
    \caption{Comparison of price influenced mobility between the three periods. The x-axis represents the affordability (most affordable locations on the left). The y-axis represents the (smoothed) percentage of movements to the area. Darker colours represent later years.
    A full breakdown of household mobility is provided in \cref{appendixMovements}.}
    \label{figMovementComparison}
\end{figure}

\subsubsection{New Households/Migrations}\label{secMigrations}

New households are those which are added throughout the simulation based on the projected household growth. New households can result from a variety of sources, such as people moving to Greater Sydney (migration), or households from Greater Sydney splitting, e.g., in the case of divorce or young adults moving out of home. We make no distinction between the two types in the simulation, and for simplicity,  refer to both inter-city and intra-city migration types as migrants.

The most common LGAs into which the new households move are shown in \cref{figMovementMigrate}\footnote{All movements are scaled by the population size to allow a fair comparison, as outlined in \cref{appendixMovements}.}.  The simulation produces a clear trend for migration towards the cheaper areas,  as the price begins to increase throughout the Greater Sydney region over time. This is most apparent for the 2016--2019 period, during which we can detect only a minority of the new households that purchased a dwelling in the expensive areas upon moving to the Greater Sydney. This is markedly different when in comparison to the 2006--2010 period. Furthermore, there is a clear peak in the more affordable LGAs, comprising Western Sydney (such as Campbelltown, Penrith, Blacktown, Fairfield, and Liverpool), and LGAs further away from the metropolitan area (such as Central Coast, Lake Macquarie, and Hawkesbury). A similar trend can be seen with the new renters. This agrees with the discussion in \citep{bangura2019differential}, which names Western Sydney as ``the first port of call for new arrivals, immigrants and refugees''.   This observations also agree with the study of \citep{slavko2020city}, which shows the outward sprawl from the densely populated Sydney metropolitan area.

\subsubsection{First-time home buyers}

Homeownership has long been a goal of many Australians \citep{bessant2013dream}, so simulating the forecasted feasibility in Sydney --- the largest, and most expensive \citep{anz2020}, Australian city --- is essential. The first-time home buyers are defined here as those who have resided in Sydney but have previously been renters, and then purchased their own dwelling. This is in contrast with the analysis in  \cref{secMigrations}, where the new owners are defined as those that had just entered the simulation (by moving to Sydney).  

The first home purchases are visualised in \cref{figMovementFirstHomeowners}. The main diagonal represents an agent purchasing in the same LGA as the one where the household is currently renting.  The area below the main diagonal (which we refer to as lower triangle) shows households purchasing in cheaper LGAs in comparison to those where they are renting, and the area above the main diagonal (the upper triangle) shows the agents purchasing in LGAs more expensive than those where they are currently renting.  In the earlier years (i.e., the 2006--2010 period), we can see that the densities in the heatmaps are relatively evenly distributed. Over time, however, the density of the upper triangle begins to decrease, meaning that the agents are purchasing in the LGAs cheaper than those where they are renting, as the expensive LGAs become increasingly out of reach. This is also reflected in \cref{figFirstHomeownerComp}:  a comparison between the 2006--2010 and the 2016--2019 periods clearly shows that many suburbs are simply becoming out of reach for the first-time home buyers, with a larger percentage of them needing to purchase in the more affordable areas.

This result is in line with \citep{randolph2013first} which shows the distribution of First Home Owner Grants within Sydney statistical districts, over the period 2000--2010, pointing out that such grants were increasingly likely in the lower-income housing markets, such as Western and Southern Sydney. Likewise, in \citep{la2017housing}, it is shown that the average distance to the CBD of dwellings that first-time home buyers can afford has been increasing from 2006 through 2016. Furthermore, \cite{la2017housing} shows that the purchasing capacity of first-time home buyers has been limited to the bottom (i.e., most affordable) 10-30\% of dwellings, where in 2016 the median first-time home buyer could afford only around 10\% of the available dwellings. A similar conclusion is reached in \citep{kupke2011housing} which find key-workers being able to afford fewer dwellings and commute longer distances between 2001-2009, a trend which seems to have been followed ever since, as shown by the mobility patterns here.

\subsubsection{Investors}

Investors are defined as households which own multiple dwellings, or households which live overseas yet own a property in Greater Sydney. We make an explicit distinction between local (residing in Sydney) and overseas investors, since price increases within Sydney are often attributed to the latter category~ \citep{rogers2015politics, rogers2017public, wong2017foreign, guest2017contribution}). This distinction is visualised in \cref{figMovementsInvestors}. The overseas investment approvals are regulated by the Australian government, and details of the approval data are provided in \cref{secOverseasApprovals}.

During the 2006--2010 period, based on the overseas approvals granted, the simulation produces a clear preference for the overseas buyers towards the most expensive regions, purchasing properties almost exclusively in the highest-priced regions. In later periods, we begin to observe a relatively wider-range of preference, although still with a clear trend towards the mid-high range areas. This is reflected in \cref{figOverseasComp}. A driving factor behind this is the higher average government approval for overseas investment given during 2006--2007 years, in comparison to later years, reflected in our simulation (as displayed in \cref{figOverseasApprovals}).  While the foreign investment data in the Greater Sydney housing market is sparse and not fine-grained, this purchasing pattern is in concordance with recent literature. For example, the study of \citep{gauder2014foreign} which mentions that foreign investors tend to prefer inner-city dwellings within Sydney (which tend to correspond to higher-priced LGAs), however recently ``foreign investment has started to broaden out into other areas of Sydney''. 

These findings are in sharp contrast to mobility patterns of the local investors, for which the simulation produces a far wider distribution across areas. For 2011--2015, and 2016--2019 periods, the most expensive LGAs become out of reach for local investors, which is most apparent during 2016--2019 (as shown in \cref{figLocalComp}). Local investors can be seen buying properties in many of the cheaper LGAs, which also falls in line with the simulation showing many renters in these areas. Due to the affordability, these LGAs also exhibit higher population growth rates than other areas. This also agrees with existing studies, for example \citep{pawson2020rental}, which find that many high-income Australian landlords are investing in dwellings in lower socioeconomically developed regions of Sydney.

\section{Conclusions and Future Work}\label{secConclusions}

In this work, we have introduced a spatial element to a model of a large, well-developed housing market (the Greater Sydney region) using an adjacency matrix based on the spatial composition of the city, and introduced several factors which affect agent decision-making within the market. We have shown the model is capable of predicting housing price dynamics as arising from individual buy and sell decisions in the market. The proposed model is capable of capturing a large variety of spatial topologies, for example, monocentric and polycentric cities. Furthermore, the graph-based approach is flexible allowing for any level of granularity, for example, over the differing scales of countries, cities, or suburbs, and extendable to weighted versions which could incorporate transport times between areas.

Using this model we have demonstrated the usefulness of spatial analysis when it is calibrated to the Australian house price data for the Greater Sydney region. The 38 LGAs of Greater Sydney were simulated, and agents (households) were calibrated based on the LGA in which they live. We have shown that the spatial component allows an additional level of fine-tuning that results in better overall fitting of the model to data, as well as producing strong out-of-sample predictions for each individual LGA by optimising only for the overall trend. That is, spatial areas add an additional layer of predictions, while also improving the overall aggregate trend. 

We investigated the agent's spatial awareness of the market, where buyers only have limited knowledge of the market, based on the area in which they reside (imperfect spatial information). We demonstrated through varying $\alpha$ (a parameter that controls how much of the market each agent is able to perceive), that lower values of $\alpha$ capture the true trend better than perfect (whole of market) knowledge, indicating the usefulness of modelling this imperfect spatial information in a housing market and showing that agents often act in a boundedly rational manner.

The spatial component also allows for the analysis of agent preferences in terms of movement patterns, where we have shown differences in mobility and purchase locations between various agent types. For example differences between first-time home buyers and investors, where first-time home buyers are limited to the more affordable locations, with investors being able to purchase higher-priced properties. Likewise between local and overseas investors, where overseas investors are shown to have a strong preference towards mid to high valued areas. We also model new migrations to the city, showing such agents becoming increasingly pushed towards cheaper areas of the city.

We have also introduced a novel fear of missing out component which alters sellers decision-making behaviour. With this parameter, we model how sellers become more likely to sell a listing if many surrounding listings have recently sold, and show a strongly localised fear-of-missing-out occurring throughout the market. This indicates agents' decisions are often motivated by their neighbour's decisions rather than by strict optimisation of their own benefits, i.e., real households are only partially rational in this regard.

While in this work we addressed some key concerns in a housing market, there are still several areas of improvement we would like to focus on in future work. The spatial component opens up a range of additional possibilities, such as overlaying public transport maps on the network, allowing for the distance to key work areas, schools, beaches etc., and further modelling and capturing agent mobility within the simulation. The demographics and household types could be sampled from actual data, which would allow analysis into subgroups of people (i.e., young singles vs retirees vs families), and allow us to model any spatial trends that arise between demographic groups. The internal optimisation functions (for example, what neighbourhood to move to, what kind of dwelling to choose) of agents could also be investigated further, as currently, agents will purchase the most expensive dwelling they can afford based on their knowledge and outreach. There are also three key equations which drive the model that could be further investigated: the bid prices, the listing prices, and the bank approval process. These are currently predefined equations but they could be treated as optimisation problems themselves, finding expressions that match most closely to the training period. %Other key areas we would like to consider in future are the investor renter relationship, and investigation into the self-referential nature and feedback loops of the model and urban housing markets in general.

\appendix

\section{Implementation}

The model is written from scratch in Python3, based on the C++ code from \cite{glavatskiy2020explaining}. 

\section{Baseline Model Parameters}\label{appendixBaseParams}

The work extends the model of \cite{glavatskiy2020explaining} (the baseline). For completeness, all parameters of the baseline method with explanations are given in \cref{tblBaseParams}.

\begin{table}[h]
\centering
\begin{tabular}{lll}
\hline
Parameter                               & Symbol           & Value                                                                                                                                                                 \\ \hline
\textbf{External Paramaters}            &                  &                                                                                                                                                                       \\
                                        &                  &                                                                                                                                                                       \\
Mortgage Rate                           & $\phi_{M}$       & \begin{tabular}[c]{@{}l@{}}2006--2010: 7.3\% -- 9.45\%\\ 2011--2015: 5.53\% -- 7.79\%\\ 2016--2019: 4.95\% -- 5.35\%\end{tabular}                                     \\
                                        &                  &                                                                                                                                                                       \\
Tax Rate                                &                  & \begin{tabular}[c]{@{}l@{}}2006--2010: 0\%, 15\%, 30\%, 40\%, 45\%\\ 2011--2015: 0\%, 15\%, 30\%, 37\%, 45\%\\ 2016--2019: 0\%, 19\%, 32.5\%, 37\%, 45\%\end{tabular} \\
                                        &                  &                                                                                                                                                                       \\
Tax Brackets                                &                  & \begin{tabular}[c]{@{}l@{}}2006--2010: \$0, \$6k, \$25k, \$75k, \$150k\\ 2011--2015: \$0, \$6k, \$37k, \$80k, \$180k\\ 2016--2019: \$0, \$18.2k, \$37k, \$87k, \$180k\end{tabular} \\
                                        &                  &                                                                                                                                                                       \\
Income Expenditure                      & $\phi_I, \phi_b$ & \begin{tabular}[c]{@{}l@{}}2006--2010: [49.37, 0.81]\\ 2011--2015: [76.76, 0.75]\\ 2016--2019: [81.75, 0.80]\end{tabular}                                             \\
                                        &                  &                                                                                                                                                                       \\
Mortgage Duration                       &                  & 30 Years                                                                                                                                                              \\
Annual house tax and fees               &                  & 1.7\%                                                                                                                                                                 \\
Stamp Duty                               &                  & 2.5\%                                                                                                                                                                 \\
House Care                              & $\phi_H$         & $2.5\% \pm 0.5\%$                                                                                                                                                     \\
Purchase fees                           &                  & 2.5\%                                                                                                                                                                 \\
                                        &                  &                                                                                                                                                                       \\
Loan to Value ratio                     &                  & $80\% \pm 10\%$                                                                                                                                                       \\
Income growth rate                      &                  & $1.002 \pm 0.001$                                                                                                                                                     \\
Income Consumption                      &                  & 60\%                                                                                                                                                                  \\
Liquid Consumption                      &                  & 2.5\%                                                                                                                                                                 \\ \hline
\textbf{Internal Paramaters}            &                  &                                                                                                                                                                       \\
                                        &                  &                                                                                                                                                                       \\
Listing probability                     & $p_{b}$          & 1\%                                                                                                                                                                   \\
Amount of houses for reference          & $\overline{Q}_h$ & 10                                                                                                                                                                    \\
List price factor                                   & $b_\ell$         & 1.75                                                                                                                                                                  \\
Sold to list power                      & $b_s$            & 0.22                                                                                                                                                                  \\
Months on market power                  & $b_d$            & -0.01                                                                                                                                                                 \\
                                        &                  &                                                                                                                                                                       \\
Urgency stress                          &                  & 0.2                                                                                                                                                                   \\
Urgency rental                          &                  & 0.02                                                                                                                                                                  \\
Urgency cash                            &                  & 0.2                                                                                                                                                                   \\
                                        &                  &                                                                                                                                                                       \\
Probability of accepting highest-bidder & $p_{m}$          & 80\%                                                                                                                                                                  \\
Expectation downshift ratio             & $p_{d}$          & 60\%                                                                                                                                                                  \\ \hline
\end{tabular}
\caption{Baseline Model Parameters. The external parameters are exogenous, sourced from, the Australian Bureau of Statistics, The Household, Income and Labour Dynamics in Australia Survey \citep{wilkins2015household}. The internal parameters are endogenous, and the default values tend to come from the analysis of \citep{axtell2014agent}. Sensitivity analysis around the internal parameters is given in \cref{figBaseSensitivity}, and a full outline is given in \citep{glavatskiy2020explaining}.}\label{tblBaseParams}
\end{table}

\begin{figure}
    \centering
    \includegraphics[width=\textwidth]{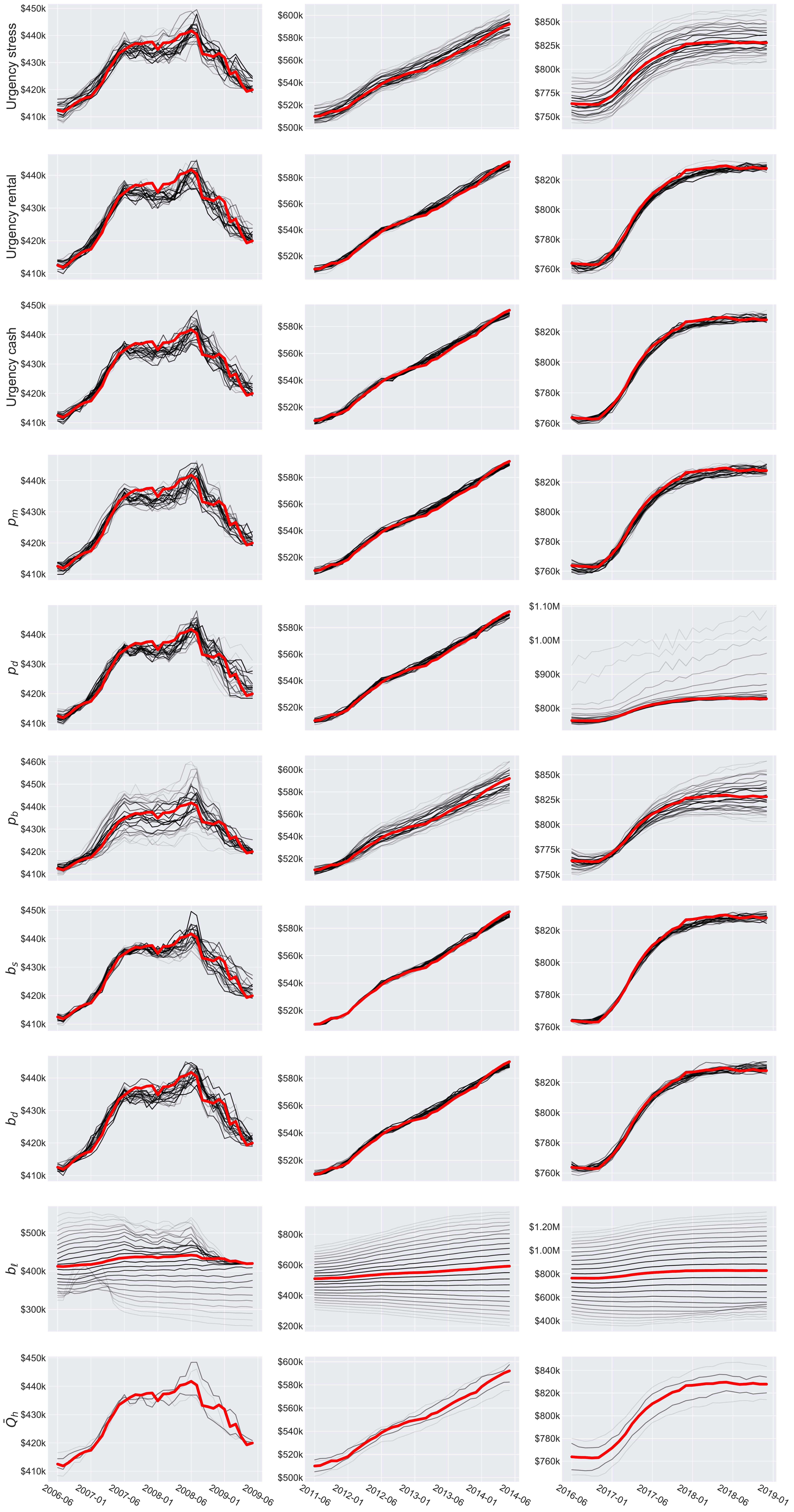}
    \caption{\footnotesize Local sensitivity analysis showing variation in output based on varying the baseline model input parameters one-at-a-time $\pm20\%$ around their default value (with others fixed at their default). The red lines represent the output for the default value, the grey lines represent for the varying values (with shade indicating the distance to the default, black meaning very near the default, light grey being further away). All show relatively small variations in the output based on the input parameters, indicating the robustness of the parameters.}
    \label{figBaseSensitivity}
\end{figure}

Local sensitivity analysis of these parameters is presented in \cref{figBaseSensitivity}, where we vary the internal values around their default ranges (within a range of $\pm 20\%$) one-at-a-time while keeping the other parameters at their default values. The resulting analysis shows the model is robust to small changes in the parameters. The only parameters that stand out from the analysis deserving additional discussion are the listing price factor, the expectation downshift ratio, and the amount quality reference.

The listing price factor $b_\ell$ has relatively large variation in prices, but as shown in \cref{eqListPrice} this is because it acts as a linear scaler on the list prices, so scales the output within the $\pm20\%$ range too, which means the parameter is behaving as expected. The expectation downshift ratio $p_d$ highlights the willingness of agents to downgrade the pre-purchase expectation based on the amount offered by the bank. For higher values, this indicates buyers wanting more from banks, and we see this increases prices overall. It is interesting to note, however, that higher values also increase the volatility of the market, with larger fluctuations seen for example in late 2017 (at the peak of the market). This confirms the discussion in the study of \cite{glavatskiy2020explaining} which highlighted the propensity to borrow as a key explanatory factor of volatility. For the amount quality reference  $\bar{Q}_h$, this is an integer parameter which explains why there are fewer comparison lines than the continuous parameters, but we still observe well-behaved outputs in the $\pm 20\%$ range.

For all other parameters, the outputs only result in small variations from the default value indicating the robustness of the model to variations around the internal parameters.

\section{Market Matching}\label{appendixMarketClearing}
The market matching process where buyers and sellers are matched is relatively simple and given in \cref{algPseudoCode}. We can see the highest bidding buyer gets preference to the listings, and every buyer attempts to purchase the most expensive listing they can afford. In certain cases, deals are rejected due to external influences (modelled by a random 20\% chance of rejection). This matching process is performed once every simulation step, with bids and listings that did not clear persisting into the next step.

\begin{algorithm}
    \caption{Market Matching}
    \label{algPseudoCode}
    \begin{algorithmic}[1] % The number tells where the line numbering should start
        \Procedure{match}{$bids,listings$}
            \State \texttt{sort}  $bids$ \Comment{From highest bid price to lowest}
            \State \texttt{sort} $listings$ \Comment{From highest list price to lowest}
            \For{\texttt{bidder in bids}}
             \State $best\ listing \gets$ \texttt{max listing buyer can afford}
             
             \If {\texttt{no deal breaker}}
                \State \texttt{Make deal between buyer and seller}
                \State $listings - best\ listing$ \ \Comment{Remove this listing from the available listings}
            \EndIf
             
            \EndFor
        \EndProcedure
    \end{algorithmic}
\end{algorithm}

\section{Rental Market}\label{appendixRentalMarket}

In the baseline model, there was no concept of rental matching. Households were randomly assigned a rental, with no regard to the cost of the dwelling or income of the household. Here, we add in an additional matching process based on the idea that households should spend maximum 30\% of their income on housing when possible to avoid housing stress \citep{thomas2016housing, fernald2020}. 

New households are randomly assigned a "local" area (weighted by the population of each area), where they begin and have their characteristics (wealth, income, cash flow etc) assigned. From there, every household attempts to find a vacant rental in their price range (which will likely result in various households moving out of financial requirements). Households with extremely high incomes, where all dwellings are less than 10\% of their income, get the most expensive rental available. Households with extremely low income, where all dwellings are at least 30\% of their income, get the cheapest one they can afford. All other households randomly choose a rental they can afford (in the 10\%-30\% of income range).

Households remain in their rentals for the duration of the simulation. In this work, we do not attempt to capture the rental market in its entirety and leave this for future work where we would like to model the relationship between renters and investors. The changes were made to ensure the cash flow situations of each household match closer to those seen in the real world, where in the previous model many households would be in a poor cash flow situation due to the rental price. Other work such as \cite{mc2010information} looks more in-depth at modelling the rental market.

\section{Bayesian Optimisation}\label{appendixBayesian}

\subsection{Details}

Bayesian optimisation is performed using the Tree of Parzen Estimators approach with hyperopt from \cite{bergstra2013making}. The optimisation process was run for 2000 iterations in all cases. The loss was measured as the average loss over several stochastic runs for each set of parameters, to minimise the effect of randomisation in the model and resulting loss. 

\subsection{Loss Function}\label{appendixLoss}

For measuring the goodness of fit, we use a loss function with two terms - a shape and temporal term to try and capture the nonlinearities overtime when predicting housing price trends. This loss function is a modification of DILATE \citep{vincent2019shape} which was introduced as a loss function for neural networks for time-series predictions, although DILATE has been simplified here (with the removal of smoothing parameters) as Bayesian optimisation does not require the loss function to be differentiable.

The loss function is given in \cref{eqLoss}.

\begin{equation}\label{eqLoss}
    \ell = \lambda * shape  + (1 - \lambda) * temporal
\end{equation}

$\lambda=0.5$ was used throughout since this was the most common in the original paper of \cite{vincent2019shape}. However, as the terms are not normalised, the two do not have an equal contribution, instead, the temporal term serves more like a penalty on the shape (with $\lambda$ controlling the strength of the penalty).

The shape term is based on dynamic time warping (DTW) which has commonly been used in speech recognition tasks \citep{sakoe1978dynamic, myers1980performance}, however, has a wide range of applications in time series data \citep{berndt1994using}. Dynamic time warping can be expressed recursively as a minimisation problem as in \cref{eqDTW}

\begin{equation}\label{eqDTW}
    shape = DTW = d(x,y) + \min \begin{bmatrix} DTW(x-1, y), \\ DTW(x-1, y-1), \\ DTW(x, y-1) \end{bmatrix}
\end{equation}

Which can be read as minimising the cumulative distance (using distance measure $d$, in this case, euclidean distance) on some warped path between $x$ and $y$, by taking the distance between the current elements and the minimum of the cumulative distances of neighbouring points.

Unlike the common applications in speech recognition, where words can be spoken at varying speeds (so the peaks do not necessarily match up), in financial markets, timing such peaks is important. This motivates the introduction of a temporal term, for trying to align such peaks and dips.  The temporal term is based on Time Distortion Index (TDI) \citep{frias2016introducing, frias2017assessing}, which can be thought of as the normalised area between the optimal path and the identity path (where the identify path is ${(1,1), (2,2).., (N,N)}$) \citep{vallance2017towards} and aims to minimise the impact of shifting and distortion in time series forecasting \citep{frias2016introducing}.

\begin{equation}
    P_{l} = \int_{i_{l}}^{i_{l+1}} \left (   x - \frac{(x-i_l) (j_{l+1} - j_{l})}{(i_{l+1}-i_{l})} + j_{l} \right ) dx
\end{equation}

\begin{equation}
    temporal  = TDI = \frac{2 \sum | P_{l}}{N^2}
\end{equation}

\begin{figure}[ht]
    \centering
    \begin{subfigure}[b]{0.45\textwidth}
        \centering
        \includegraphics[width=\textwidth]{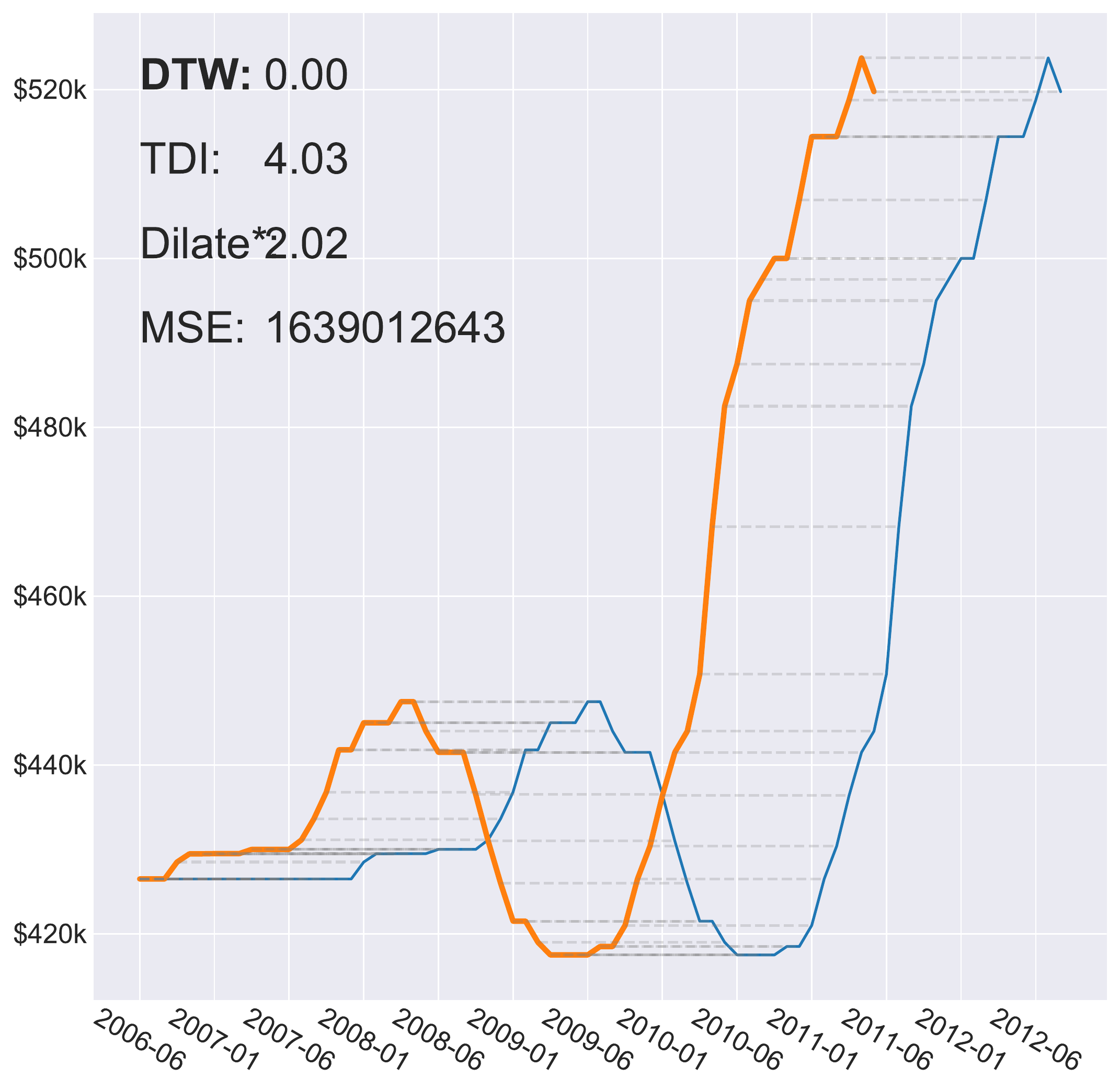}
        \caption{Shifted Predictions}
    \end{subfigure}
    \hfill
    \begin{subfigure}[b]{0.45\textwidth}  
        \centering 
        \includegraphics[width=\textwidth]{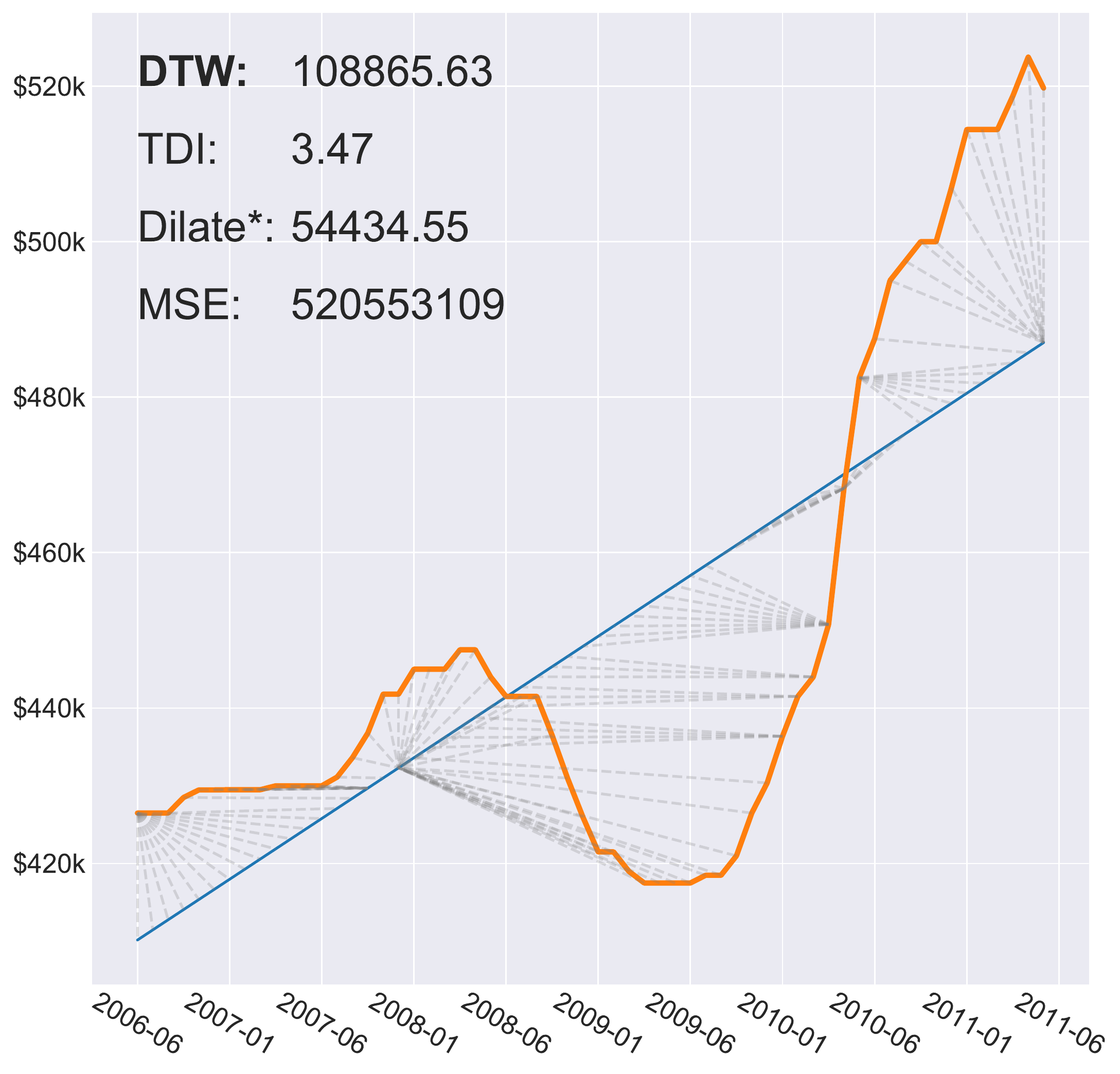}
        \caption{Linear Predictions} 
        \label{figLinearMSE}
    \end{subfigure}
    \hfill
    \caption{Motivation of time series based loss using a constructed example. We can see the line on the right is a very poor predictor of the true trend, failing to capture any of the peaks or dips. However, the MSE is significantly lower than the line on the left. DTW captures the shifts, and incorporating a penalty on time can penalise these shifts. The light grey lines show how DTW matches points together, even if they do not occur at the same time period.}
    \label{figLossExample}
\end{figure}

To see the usefulness over a more standard approach loss function such as MSE for time series, consider the example in \cref{figLossExample}. We can see the MSE can be a problematic approach, and in some cases (as in the example where the linear line \cref{figLinearMSE} has a lower loss) be a misleading measure of goodness of fit. DTW helps to match points in the two time-series, while TDI helps minimise the offset of the predictions (graphically in the example this corresponds to shortening the dotted grey lines). For a full analysis, we refer you to the original DILATE paper of \cite{vincent2019shape}, noting that all smoothing terms have been removed in the modification here. 

\subsection{Global Constraints}\label{appendixConstraints}

\begin{figure}[ht]
    \begin{subfigure}{.3\textwidth}
    \includegraphics[width=\textwidth]{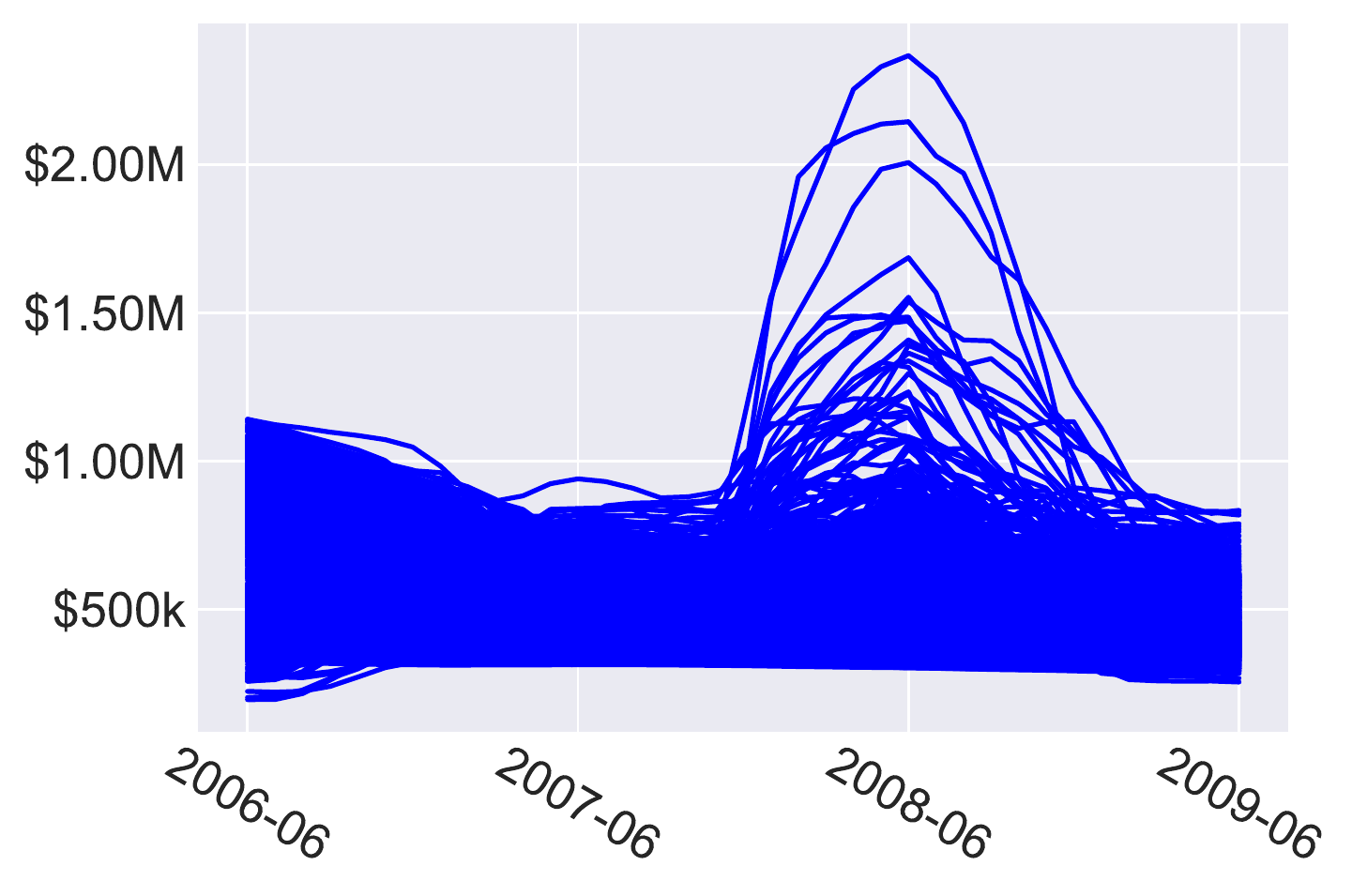}
    \caption{Without Constraints}
    \end{subfigure}
    \begin{subfigure}{.3\textwidth}
      \includegraphics[width=\textwidth]{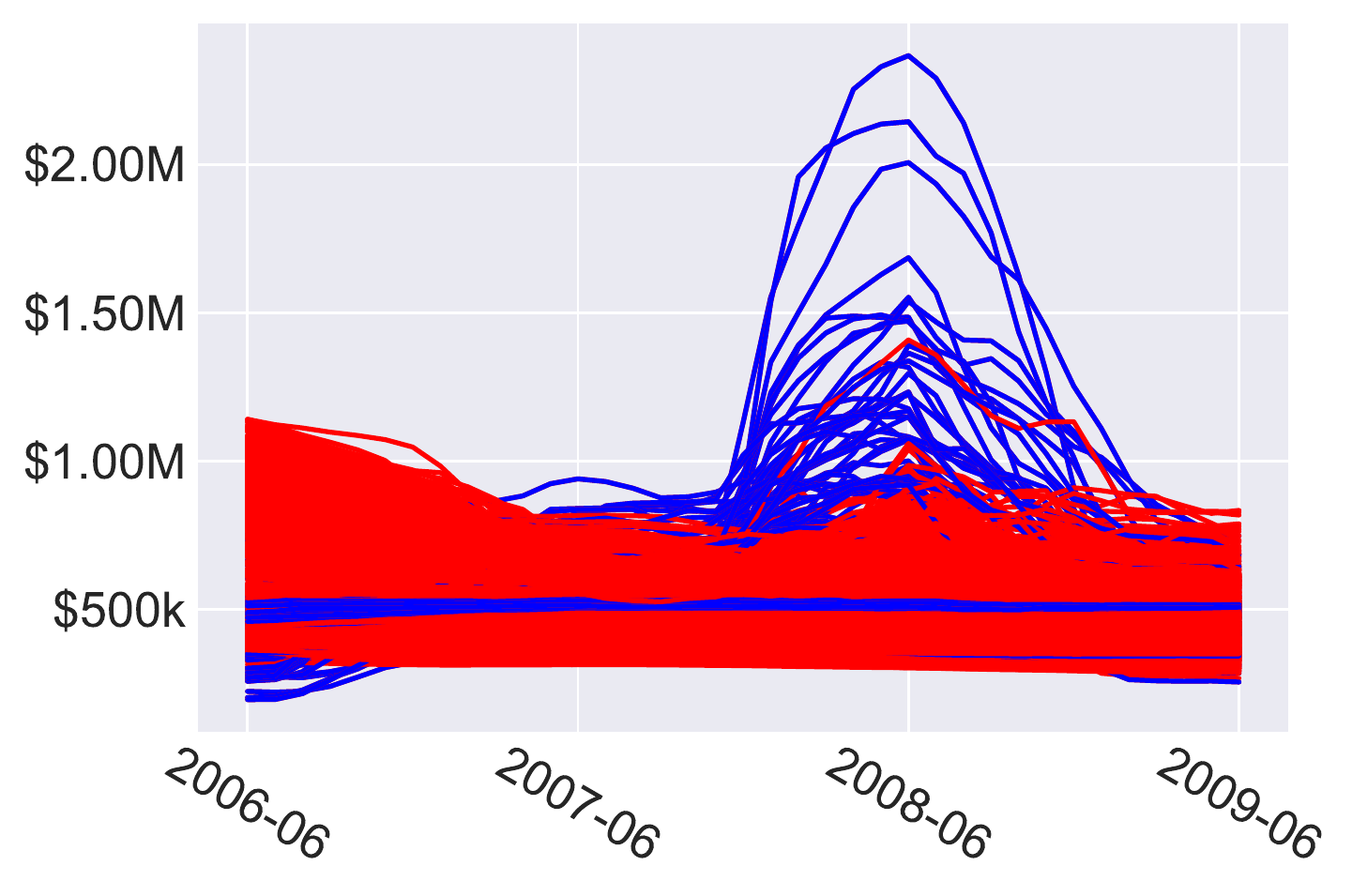}
      \caption{Red shows the paths which will be removed.}
    \end{subfigure}
    \begin{subfigure}{.3\textwidth}
      \includegraphics[width=\textwidth]{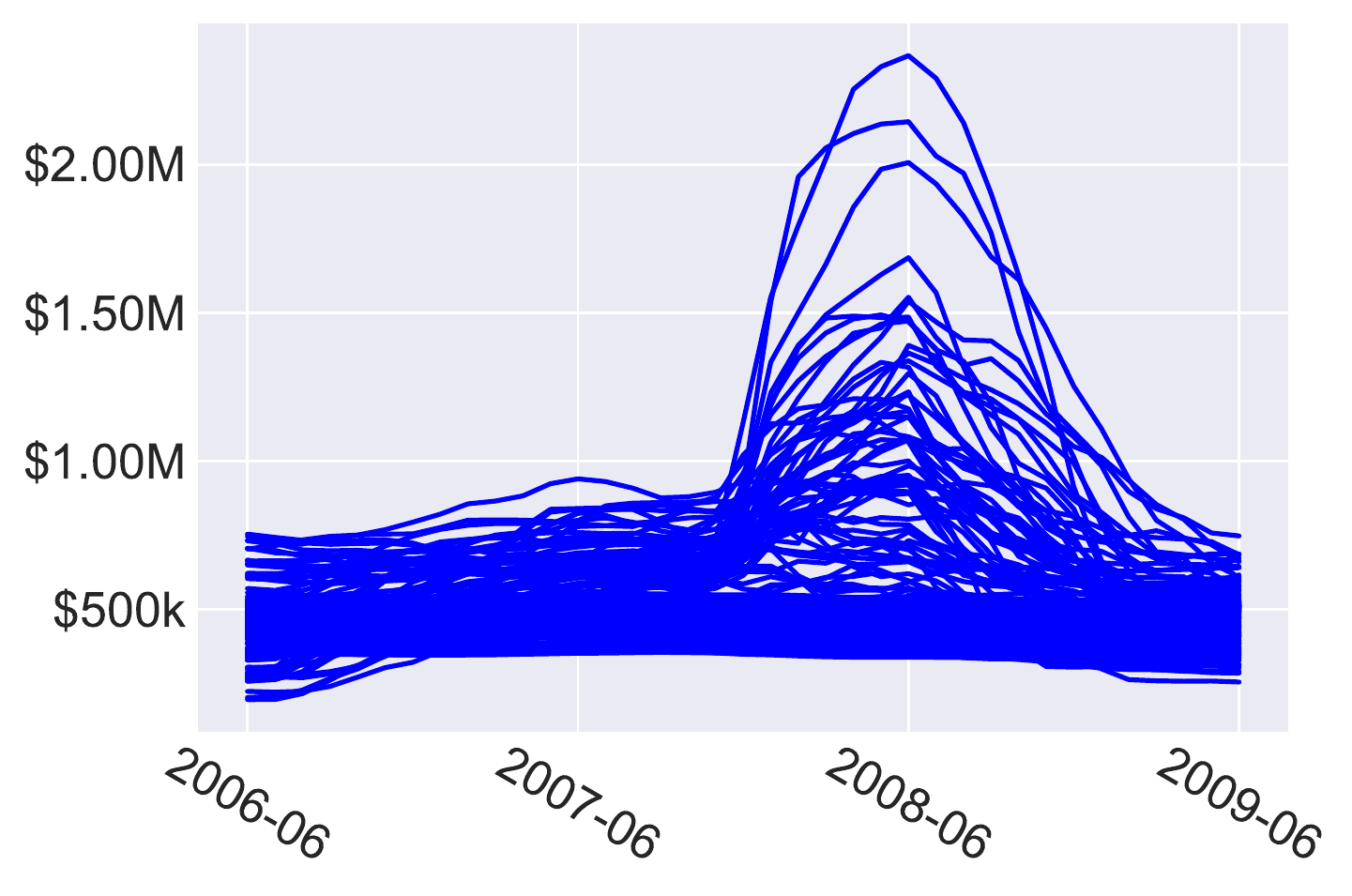}
      \caption{Paths matching constraint}
    \end{subfigure}
    \caption{Global Constraint Process}
    \label{figGlobalConstraints}
\end{figure}

We can see 2011--2015 and 2016--2019 fit the trend very closely, although despite having a low loss, the 2006--2010 simulation path does not follow the dip well, as no distinction is made about being above or below the trend in the loss function. Looking at the individual paths from every run, we can see that a peak and dip is predicted in many of the cases, although the distance is greater than the path with the lowest loss which was perfectly matching across a large portion of the training data but missing the dip. We apply a post optimisation global constraint to 2006--2010, again only using this training period, that the midpoint of the simulation must be higher than the start and ending points (i.e., a peak must occur), and take the parameters with the lowest loss matching this criterion. The process is shown in  \cref{figGlobalConstraints} and the result is shown in \cref{figGlobalConstraintLoss}. We can see for 2006--2010, the $\ell$ is higher than before the constraint, however, clearly, the constraint allows for a closer overall trend following in the training period. The visualisation in \cref{figGlobalConstraints} can also begin to show the wide range of possible market outcomes, for various combinations of the parameters. If a certain section occurs from many parameter outcomes (i.e., with the peak), we can deduce that such dynamics were likely to occur just due to the agent characteristics, regardless of the parameters used. This shows many combinations lead to a peak and dip, perhaps due to mortgage rates and worrying mortgage vs income ratios. This is more in line with suggestions in \cite{edmonds2018using}, which suggest ABMs be used to determine a range of potential future outcomes, which in this case shows a variety of paths leading to a peak and dip. 

\subsection{Parameter Space}

\begin{table}[!ht]
\resizebox{\textwidth}{!}{%
\begin{tabular}{@{}lll@{}}
\toprule
\textbf{}      &  \textbf{Search Space}             & \textbf{Explanation}                                                      \\ \midrule
$\alpha$          &   {[}0,1{]}                  &  Probability of viewing a listing, scaled by the outreach                                                                        \\
$\beta$  &  {[}-10,+10{]}                            & The contribution of surrounding listing sales when considering selling a dwelling \\
$h$      &   {[}-1,+1{]}                         & Trend following aptitude                                                  \\ \bottomrule
\end{tabular}
}
\caption{The three tunable hyperparameters. All parameters are sampled uniformly within these ranges.}\label{tblParams}
\end{table}

The parameter space is defined in \cref{tblParams}.

Even though there are only three parameters to tune, the number of potential combinations exceeds 4 million (this is assuming values are discretized values, so the true number is far greater), making a grid search impractical. 

The three parameters are $h$, $\alpha$, and $\beta$.

\begin{figure}[ht]
    \centering
    \begin{subfigure}{0.325\textwidth}
        \includegraphics[width=\textwidth]{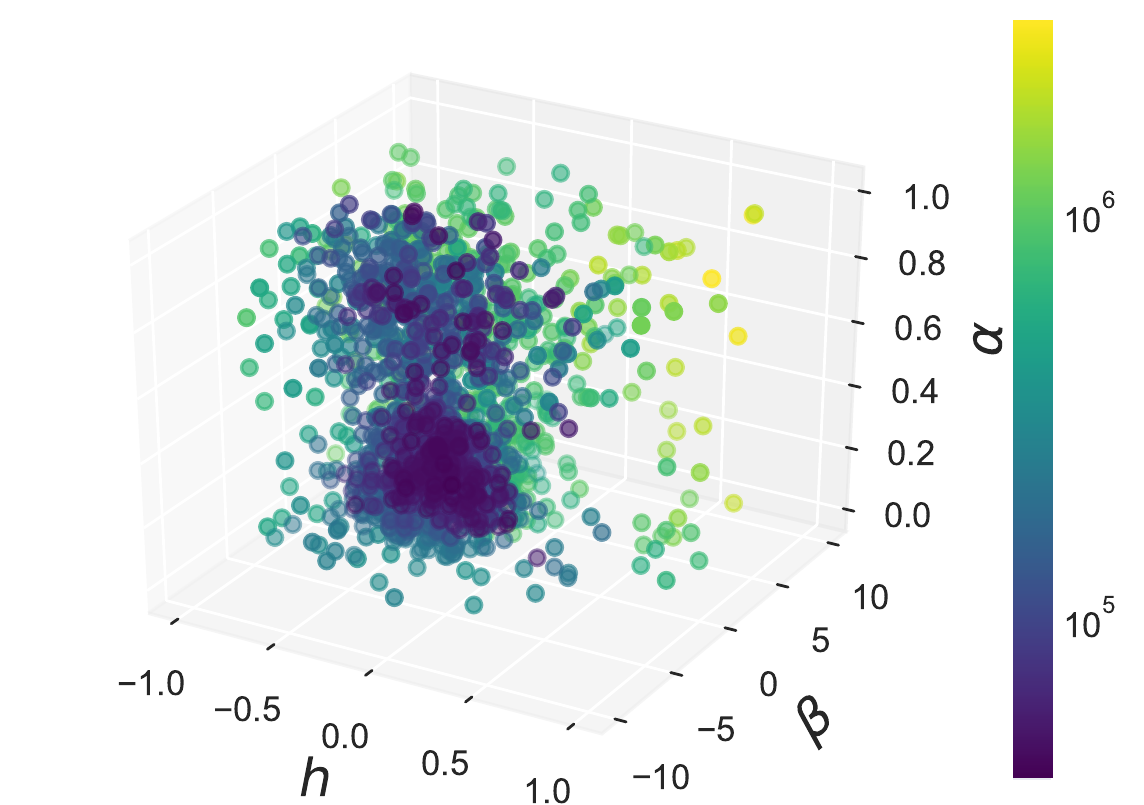}
        \caption{2006--2010}
    \end{subfigure}
    \begin{subfigure}{0.325\textwidth} 
        \includegraphics[width=\textwidth]{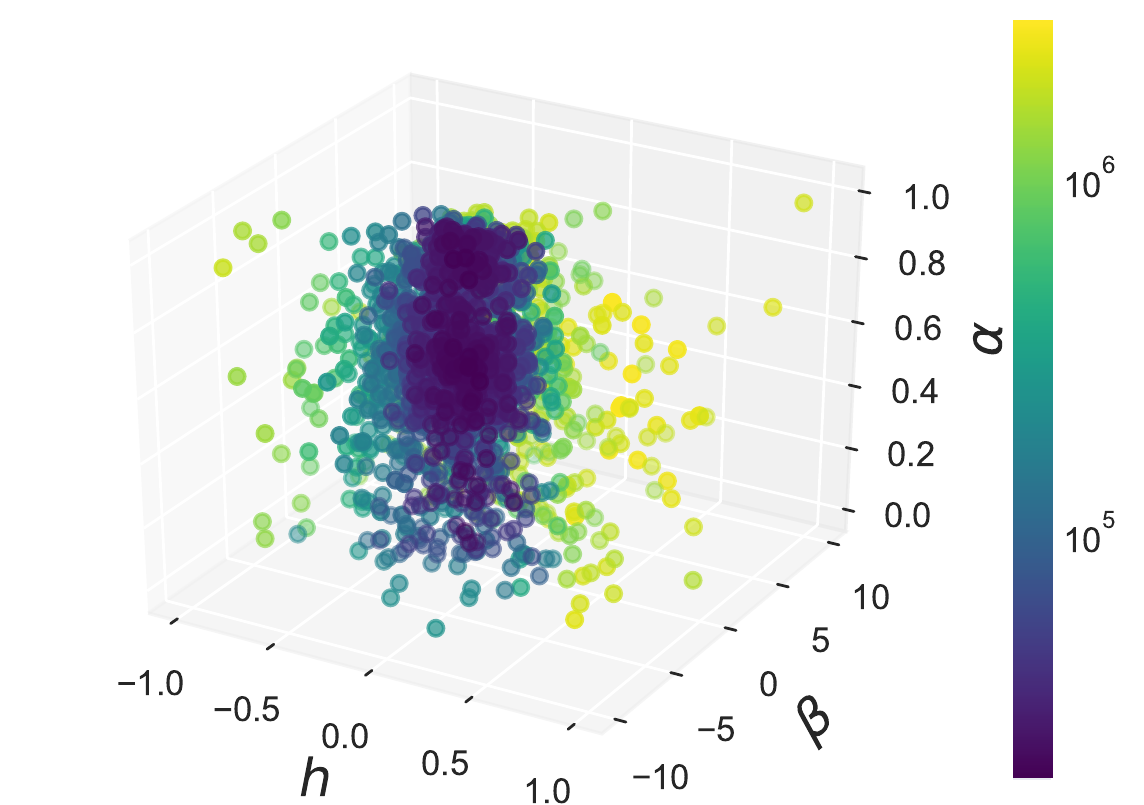}
        \caption{2011--2015} 
    \end{subfigure}
        \begin{subfigure}{0.325\textwidth}
        \includegraphics[width=\textwidth]{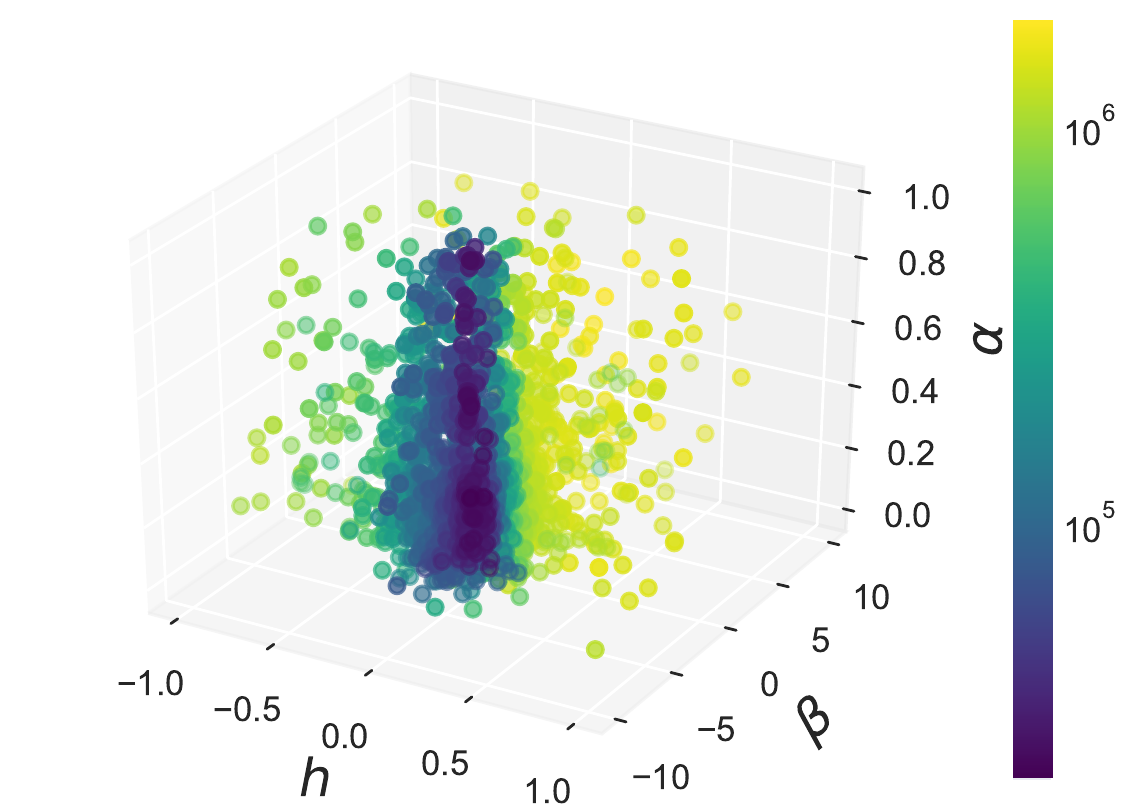}
        \caption{2016--2019}
    \end{subfigure}
    \hfill
    \caption{Search space exploration. Colour indicates the loss.}
    \label{figBayesianExploration}
\end{figure}

\begin{figure}[ht]
    \centering
    \begin{subfigure}[b]{0.3\textwidth}
        \centering
        \includegraphics[width=\textwidth]{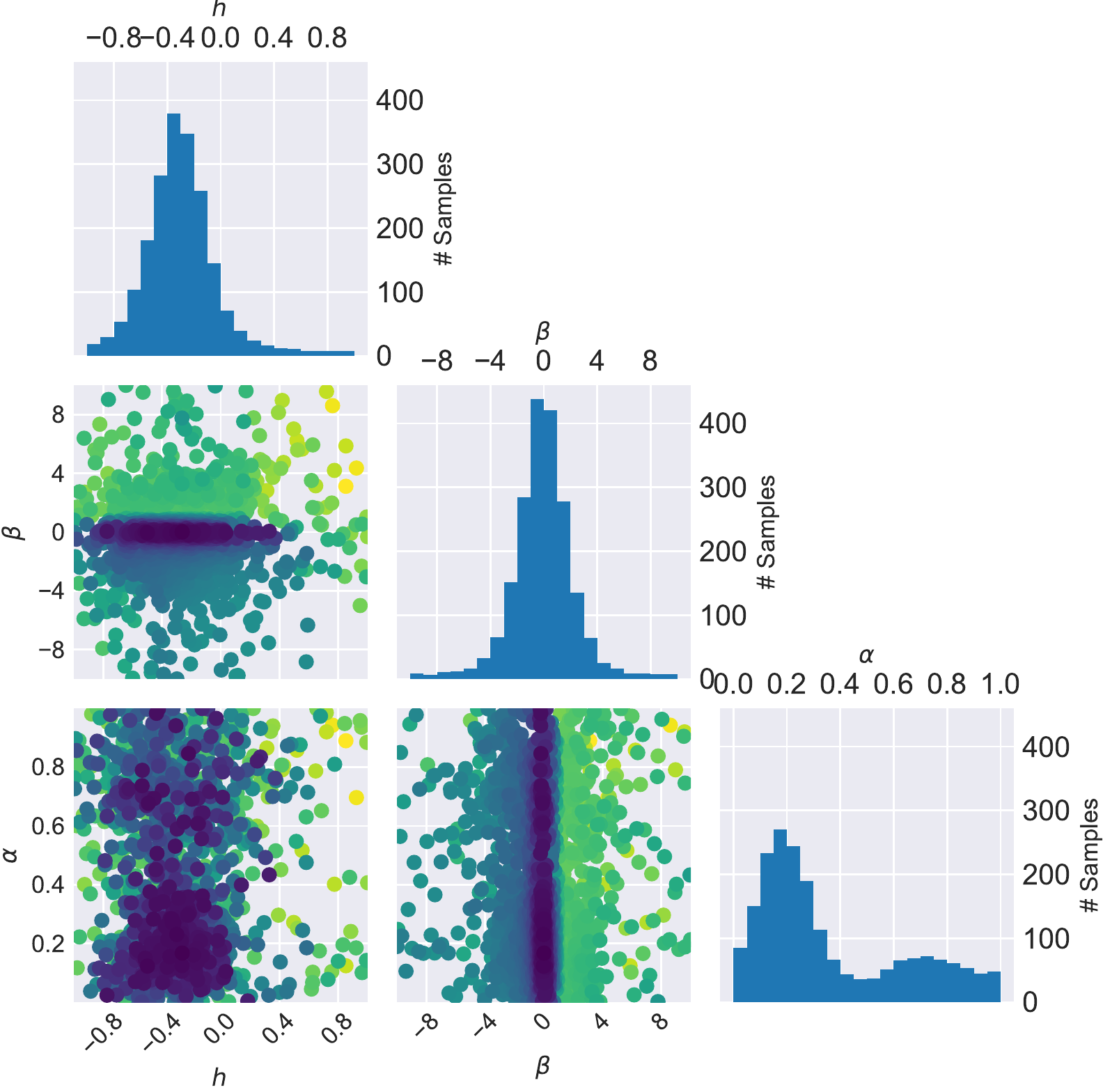}
        \caption{}
    \end{subfigure}
    \begin{subfigure}[b]{0.3\textwidth}  
        \centering 
        \includegraphics[width=\textwidth]{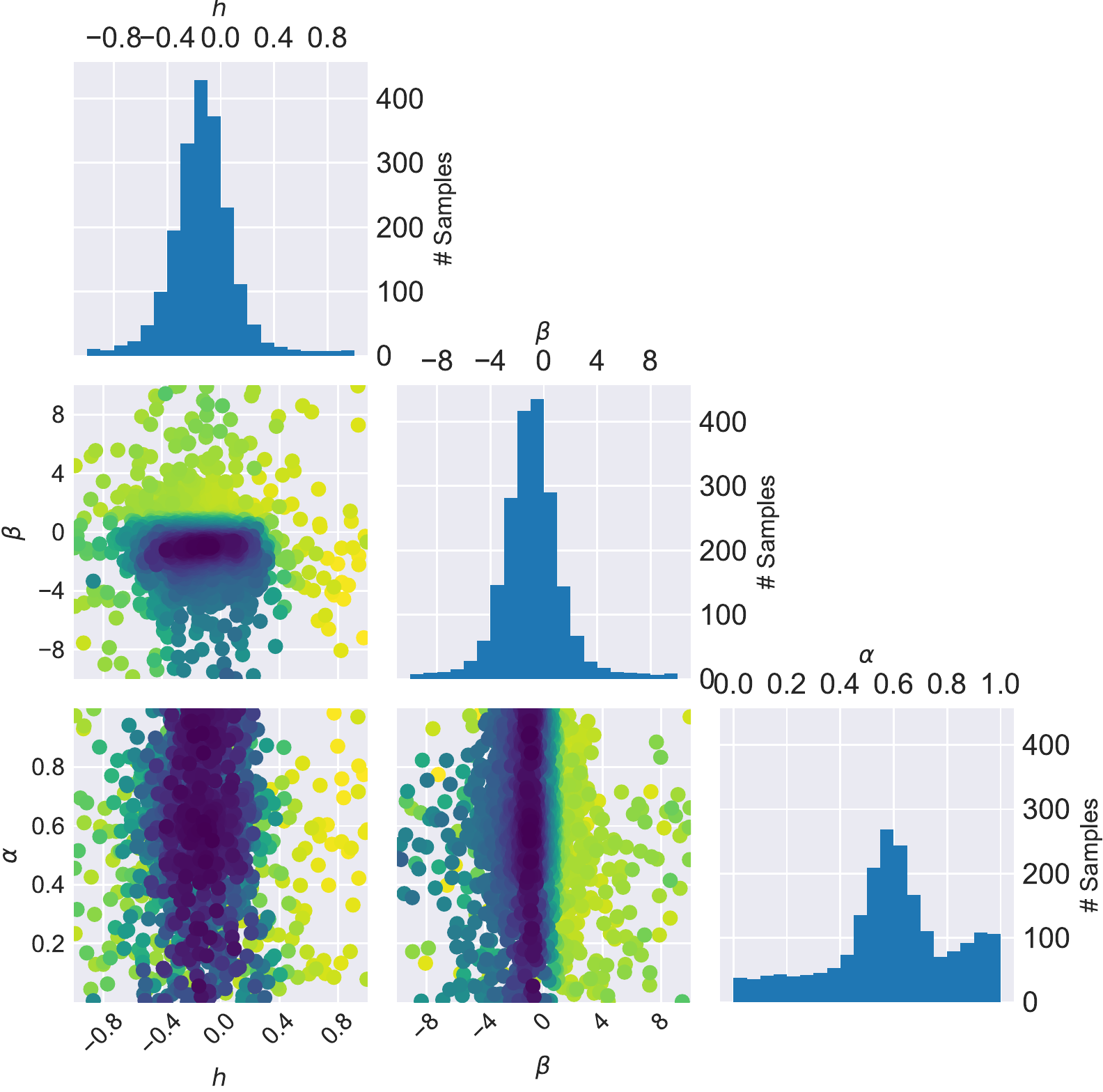}
        \caption{} 
    \end{subfigure}
    \begin{subfigure}[b]{0.3\textwidth}
        \centering
        \includegraphics[width=\textwidth]{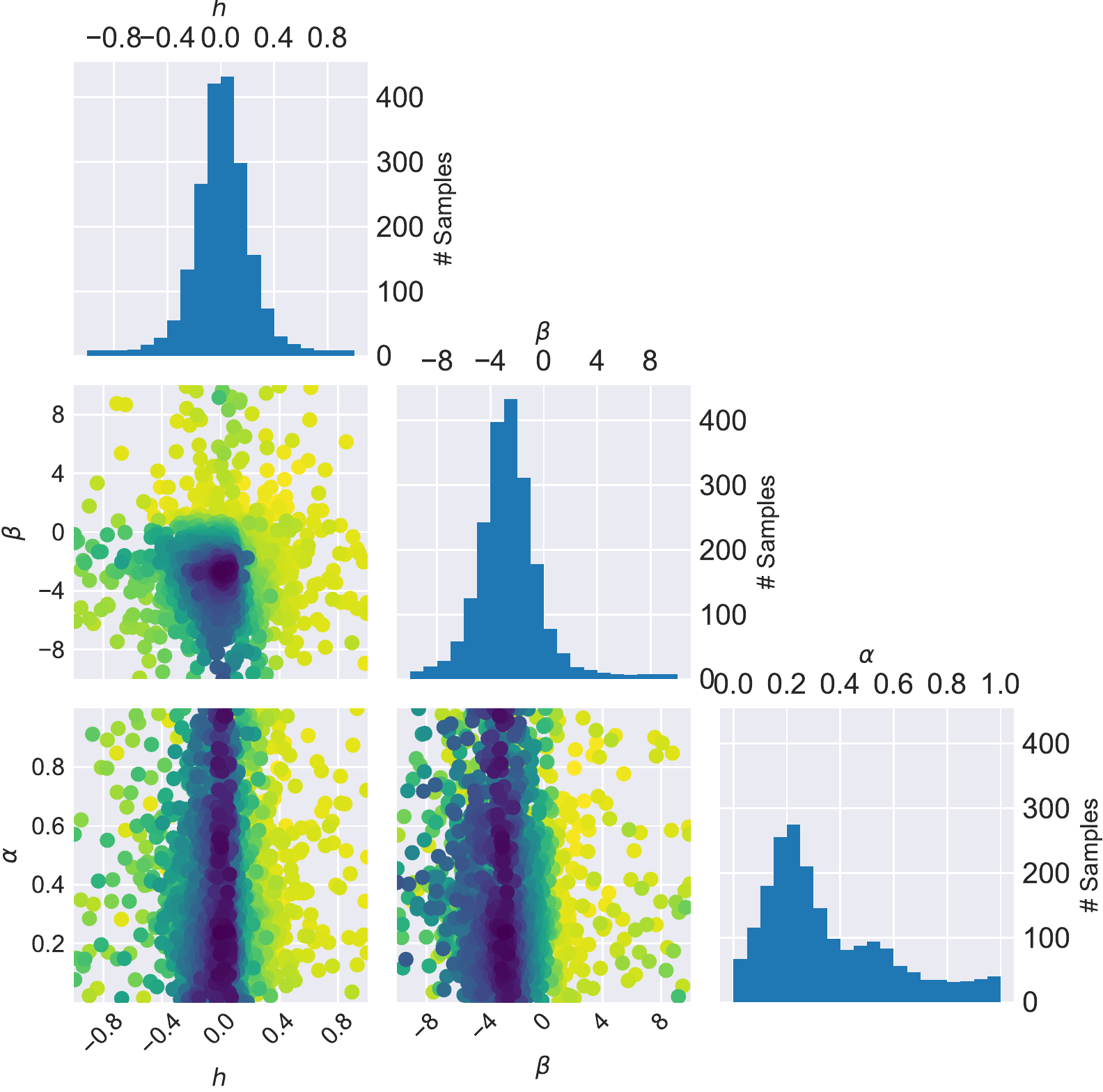}
        \caption{}
    \end{subfigure}
    \caption{Parameter Interactions and Parameter sampling.}
    \label{figBayesianInteractions}
\end{figure}

\section{Sensitivity Analysis}\label{appendixSensitivity}

While in \cref{secOptResults} we analysed the contribution of each new component by comparing the resulting optimised time series after introducing the components one at a time, here we verify and rank the importance of each of the contributions explicitly using global sensitivity analysis (GSA).

Specifically, we analyse the importance of the trend following aptitude ($h$), the social contribution ($\beta$), and the role of $\alpha$ in minimising the loss function.

We use the Morris Method \citep{morris1991factorial} for a GSA, and present the revised $\mu^*$ as suggested in \cite{saltelli2004sensitivity} and $\sigma$. $\mu^*$  represents the mean absolute elementary effect, and can be used to rank the contribution of each parameter, this solves the problem of $\mu$ where elementary effects can cancel out. We also analyse $\sigma$, i.e., the standard deviation of the elementary effects, as a measure of the interactions. 

For parameters for the Morris Method, we use $r=20$ trajectories, $p=10$ levels, and step size $\Delta=p/[2(p-1)]$, i.e., $\Delta \approx 0.52$ with $p=10$. These are within the range of commonly used parameters, e.g. in \cite{campolongo2007effective}.

\begin{figure}[ht]
    \centering
    \begin{subfigure}{0.325\textwidth}
        \includegraphics[width=\textwidth]{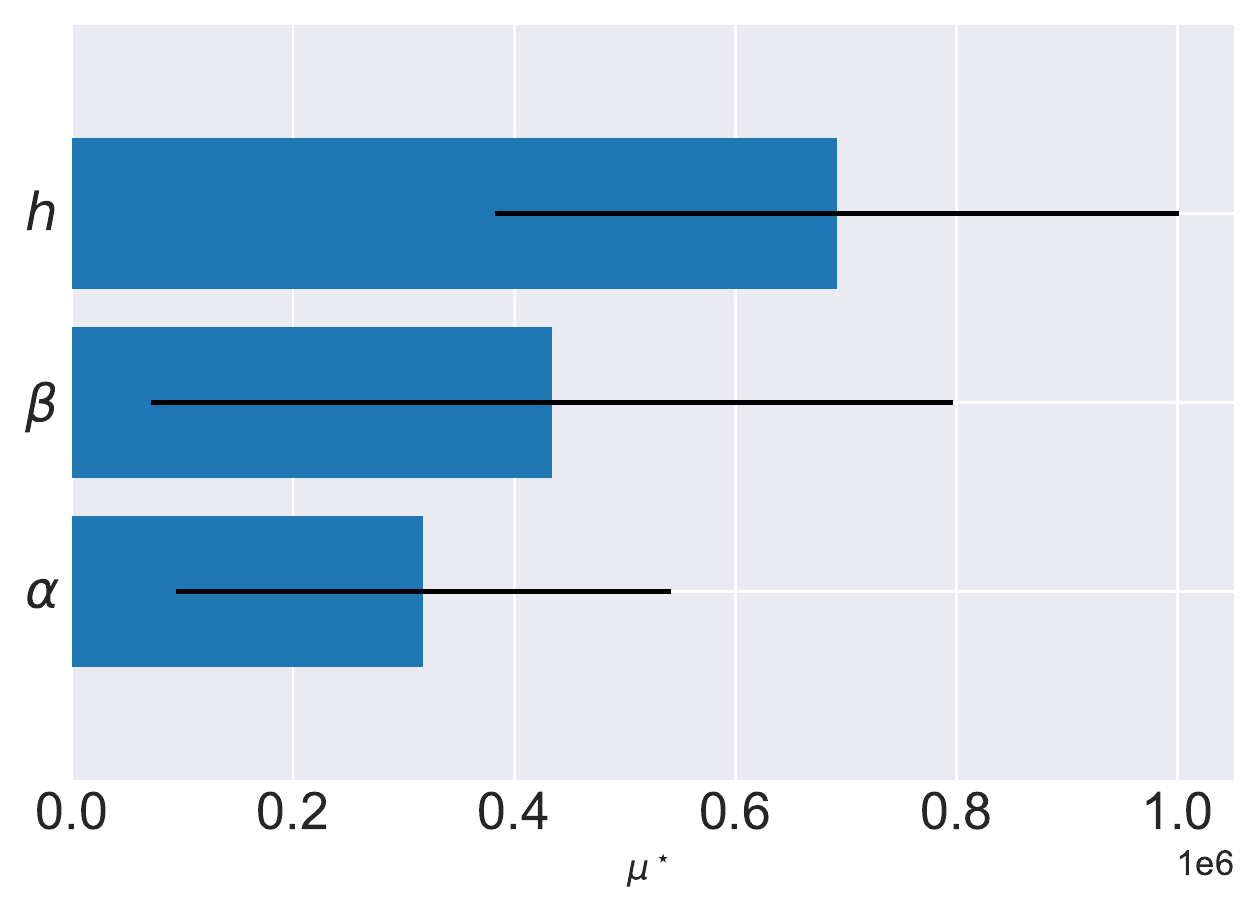}
        \caption{2006--2010}
    \end{subfigure}
    \begin{subfigure}{0.325\textwidth}  
       \includegraphics[width=\textwidth]{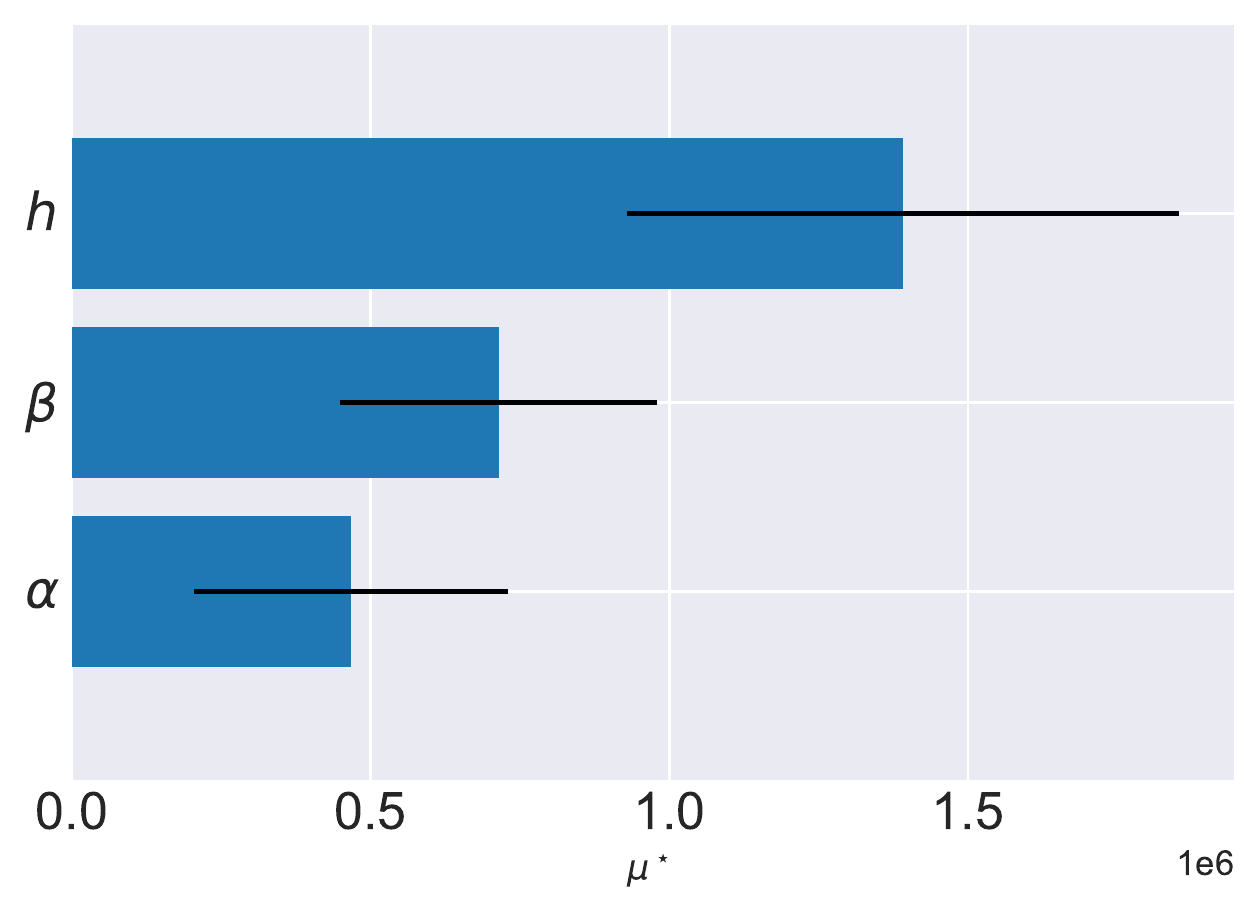}
        \caption{2011--2015} 
    \end{subfigure}
        \begin{subfigure}{0.325\textwidth}
        \includegraphics[width=\textwidth]{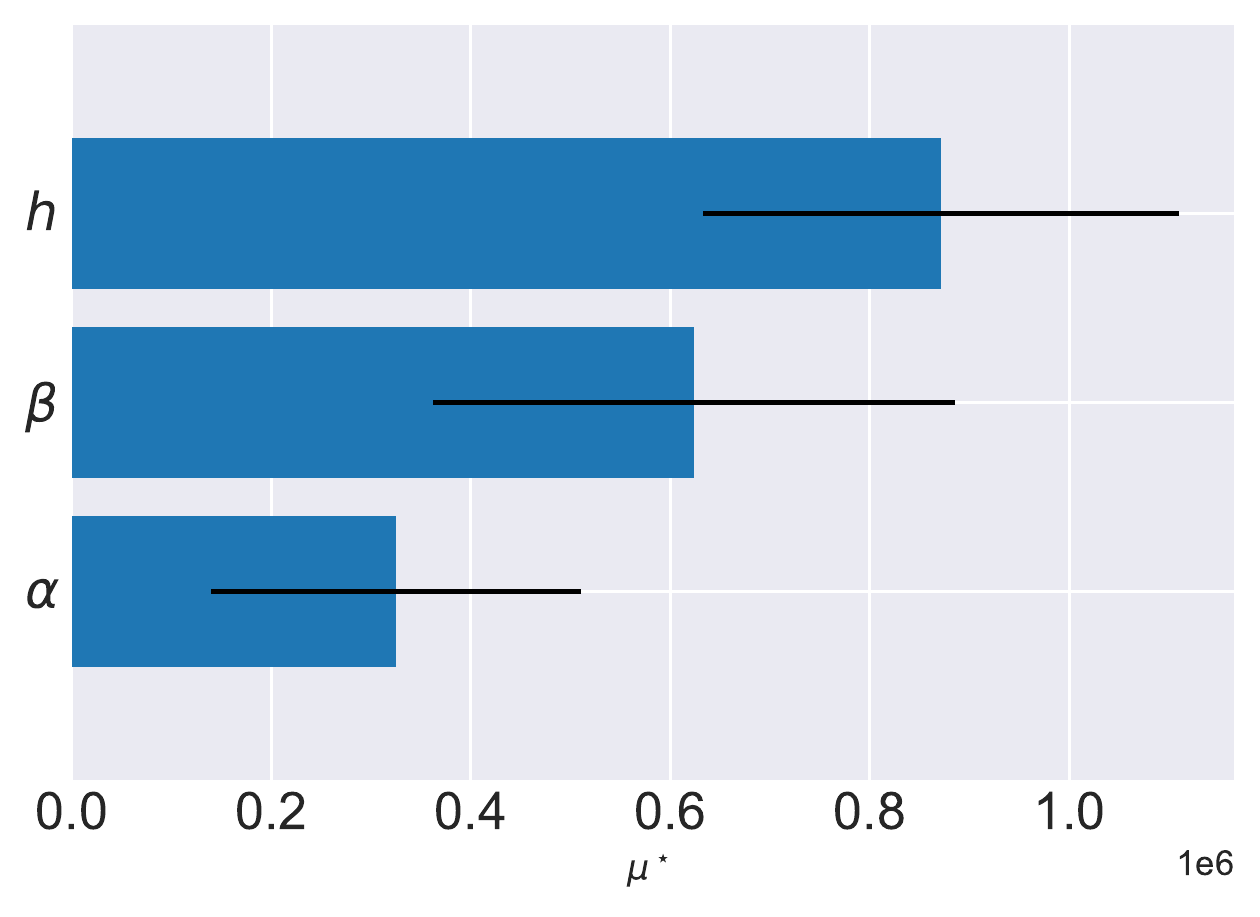}
        \caption{2016--2019}
    \end{subfigure}
    \caption{Importance plot showing $\mu^{\star}$. Error bars are displayed at the 95\% confidence level.}.
    
    \label{figMorrisImportance}
\end{figure}

\begin{figure}[ht]
    \centering
    \begin{subfigure}{0.325\textwidth}
        \includegraphics[width=\textwidth]{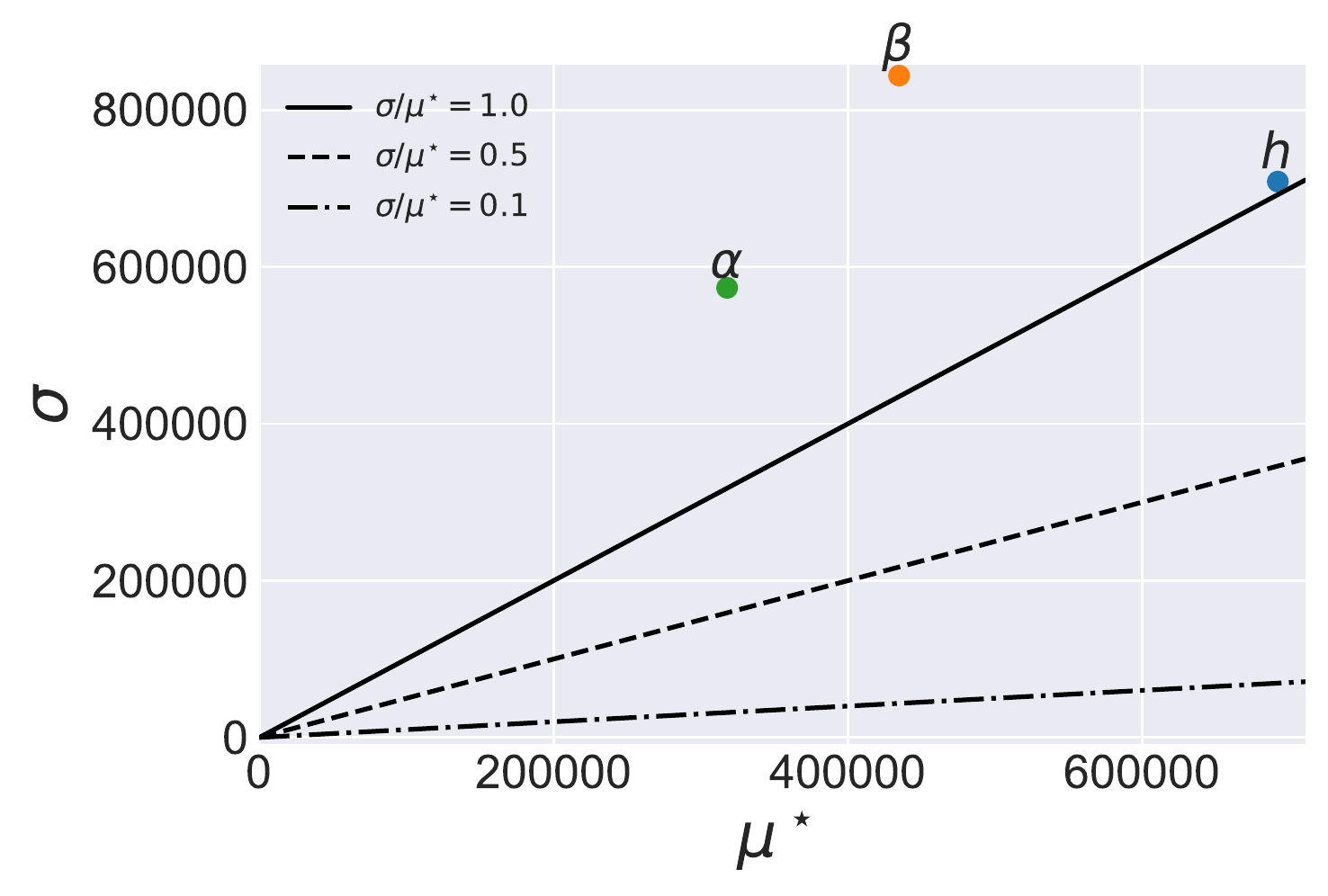}
        \caption{2006--2010}
    \end{subfigure}
    \begin{subfigure}{0.325\textwidth}  
       \includegraphics[width=\textwidth]{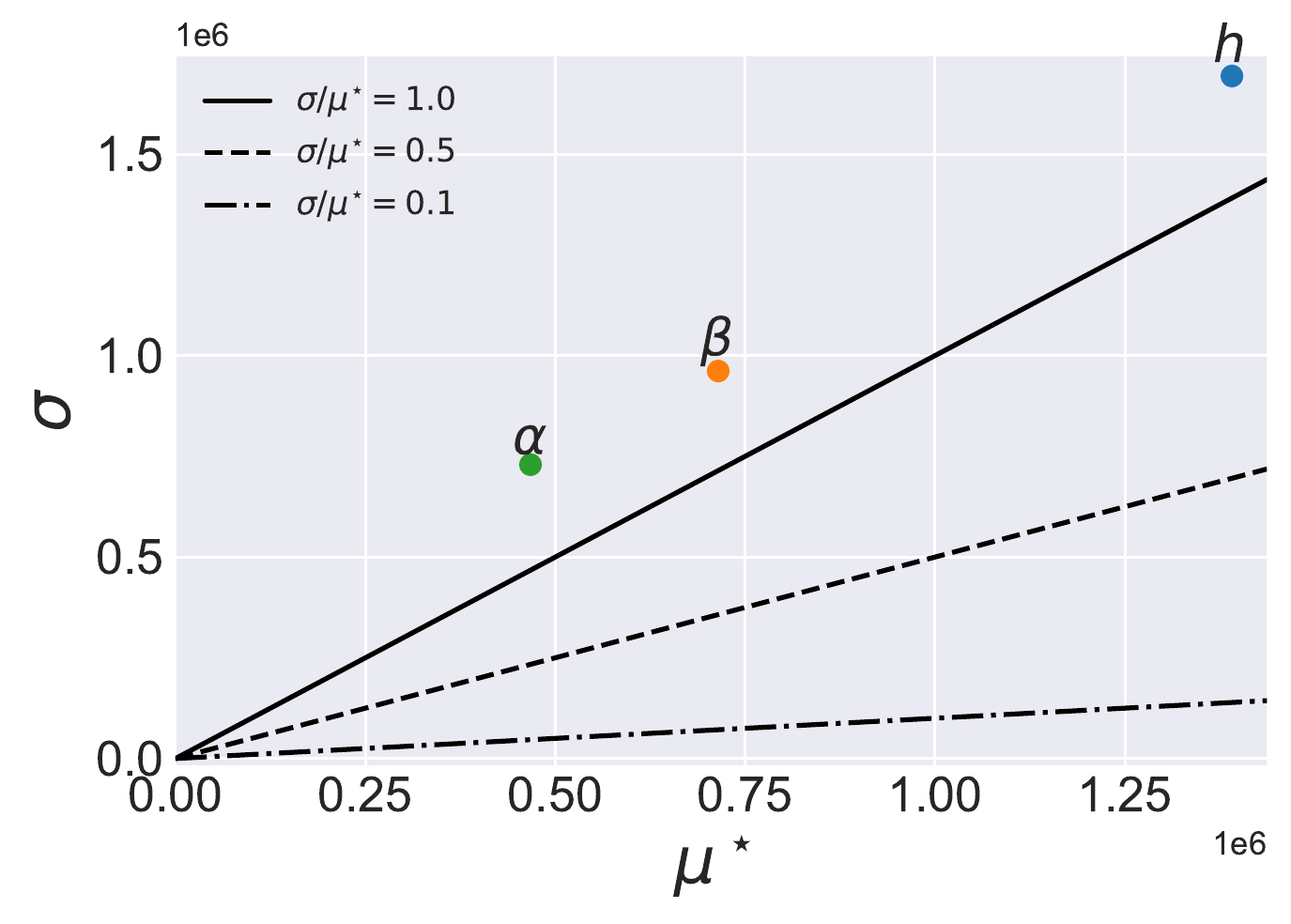}
        \caption{2011--2015} 
    \end{subfigure}
        \begin{subfigure}{0.325\textwidth}
        \includegraphics[width=\textwidth]{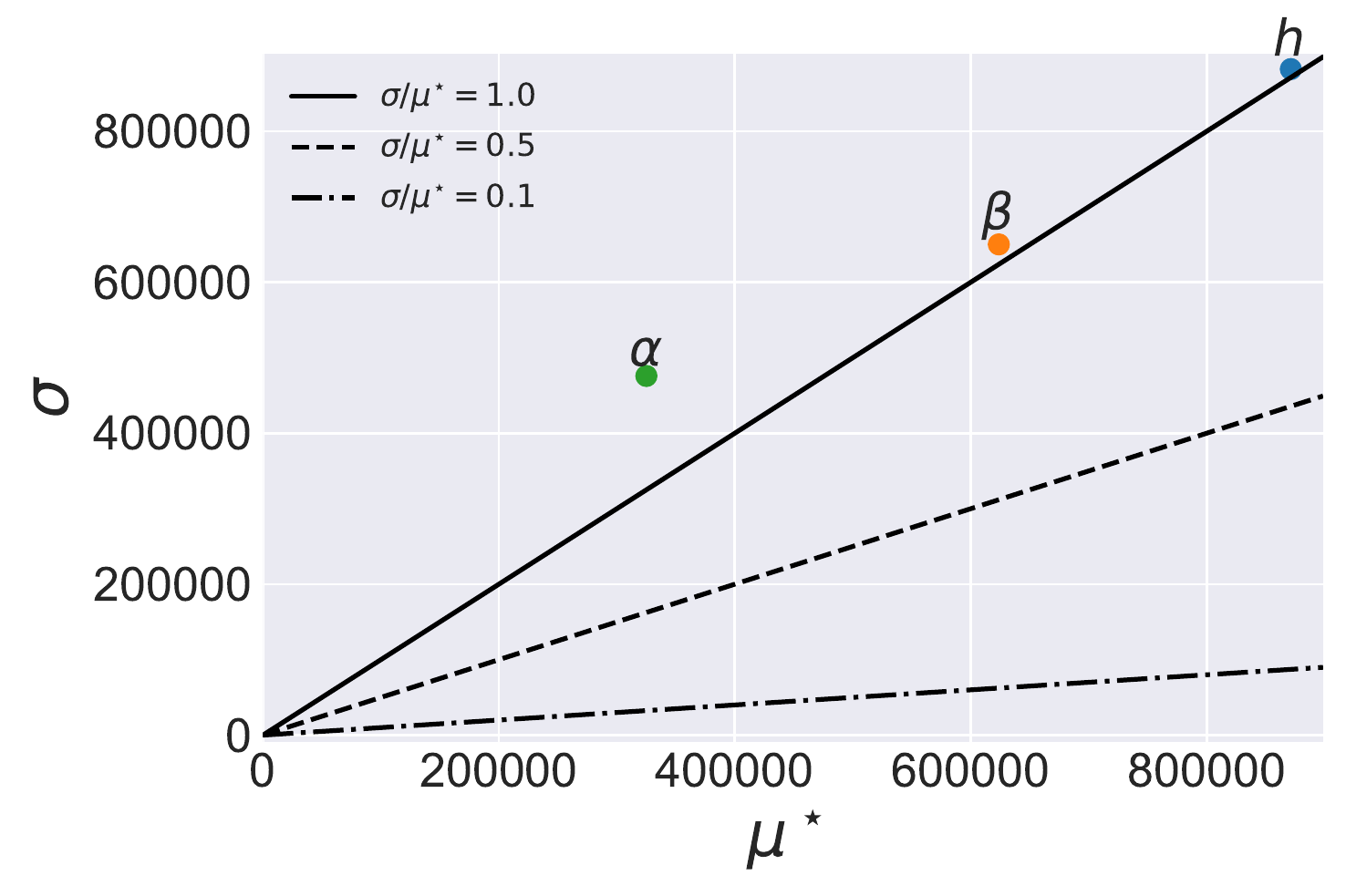}
        \caption{2016--2019}
    \end{subfigure}
    \caption{Global sensitivity analysis with Morris plots. Diagonal lines represent the ranges for $\sigma / \mu^{\star}$. One classification strategy proposed by \cite{sanchez2014application} says factors which are almost linear should be below the 0.1 line, factors which are monotonic between 0.1 and 0.5 lines, or almost monotonic between the 0.5 and 1 line, and factors with non-monotonic non-linearities or interactions with other factors above the 1 line}.
    
    \label{figMorrisPlots}
\end{figure}

The results are presented in \cref{tblMorris}, and visualised in \cref{figMorrisPlots} and \cref{figMorrisImportance}.

\begin{table}[!ht]
\centering
\caption{Morris Method for Sensitivity Analysis}
\label{tblMorris}
\begin{tabular}{lllllll}
\hline
\multicolumn{1}{c}{\textbf{}} & \multicolumn{2}{c}{\textbf{2006--2010}}                                        & \multicolumn{2}{c}{\textbf{2011--2015}}                                        & \multicolumn{2}{c}{\textbf{2016--2019}}                                        \\ \hline
\multicolumn{1}{c}{}          & \multicolumn{1}{c}{$\mu^*$} & \multicolumn{1}{c}{$\sigma$} & \multicolumn{1}{c}{$\mu^*$} & \multicolumn{1}{c}{$\sigma$} & \multicolumn{1}{c}{$\mu^*$} & \multicolumn{1}{c}{$\sigma$} \\
\textbf{$h$}                  & 692019                             & 708678                              & 1390890                            & 1693784                             & 871388                             & 882436                              \\
\textbf{$\beta$}                 & 434626                             & 843830                              & 714507                             & 961618                              & 623839                             & 650213                              \\
\textbf{$\alpha$}            & 317806                             & 573236                              & 467376                             & 729330                              & 325056                             & 475900                              \\ \hline
\end{tabular}
\end{table}

Checking the importance of each parameter, or $\mu^{\star}$, we can see  $h$ consistently ranks the most important, showing its changes have the largest effect on $\ell$. This is followed in importance by $\beta$, and then $\alpha$ each year. However, we see that confidence bars do overlap in \cref{figMorrisImportance}.

Viewing the Morris plots in \cref{figMorrisPlots}, we can see all parameters are deemed important, where unimportant parameters would show up in the bottom leftmost portion of the plot. Using the classification strategy of \cite{sanchez2014application}, all parameters are all considered to be non-monotonic and/or with high levels of interaction, since $\frac{\sigma}{\mu^{\star}} > 1$ in all cases.

This analysis agrees with the preliminary parameter analysis in \cref{bayesianExploration}. 

\begin{figure}
    \centering
    \begin{subfigure}[b]{0.3\textwidth}
        \centering
        \includegraphics[width=\textwidth]{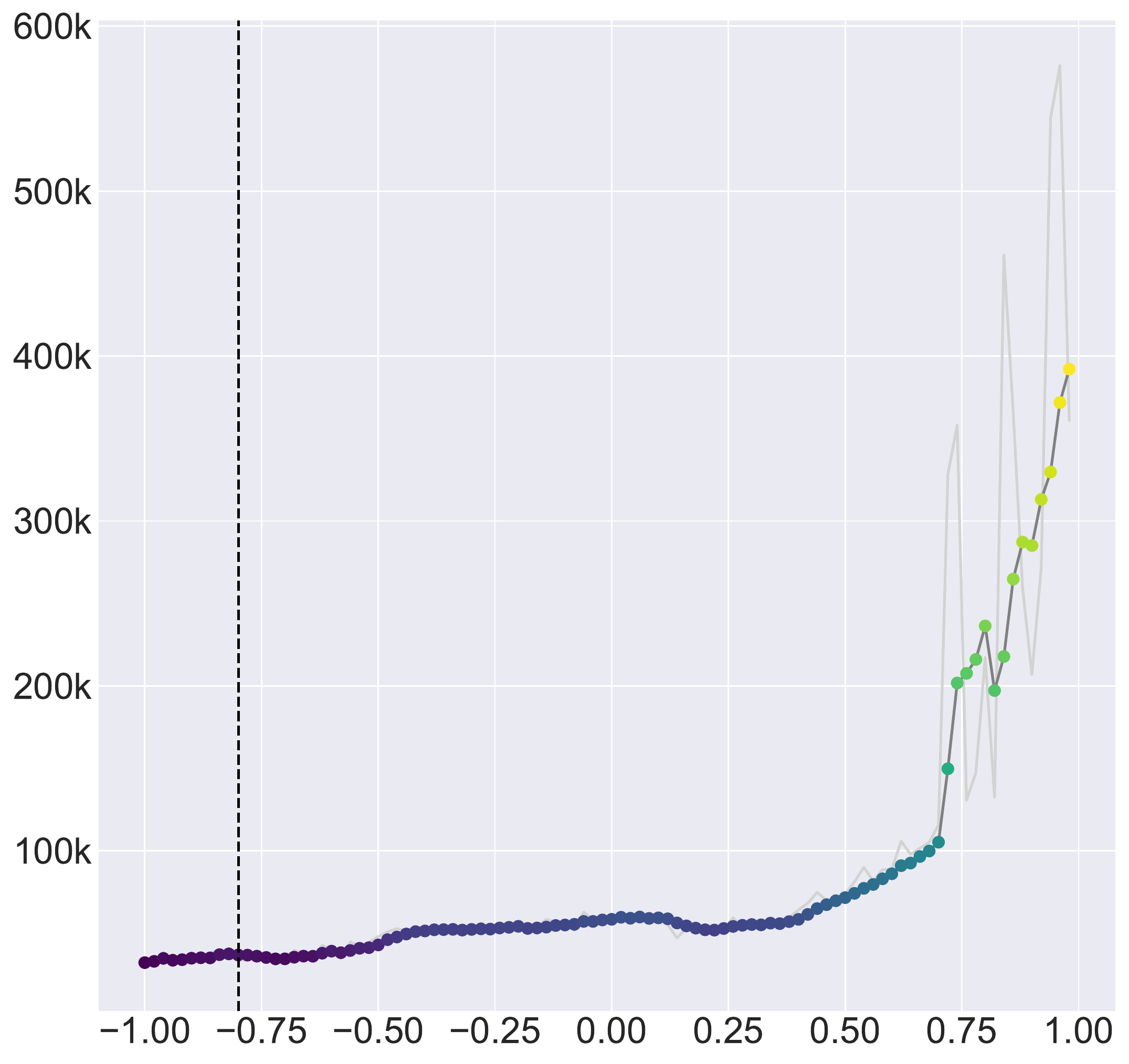}
        \includegraphics[width=\textwidth]{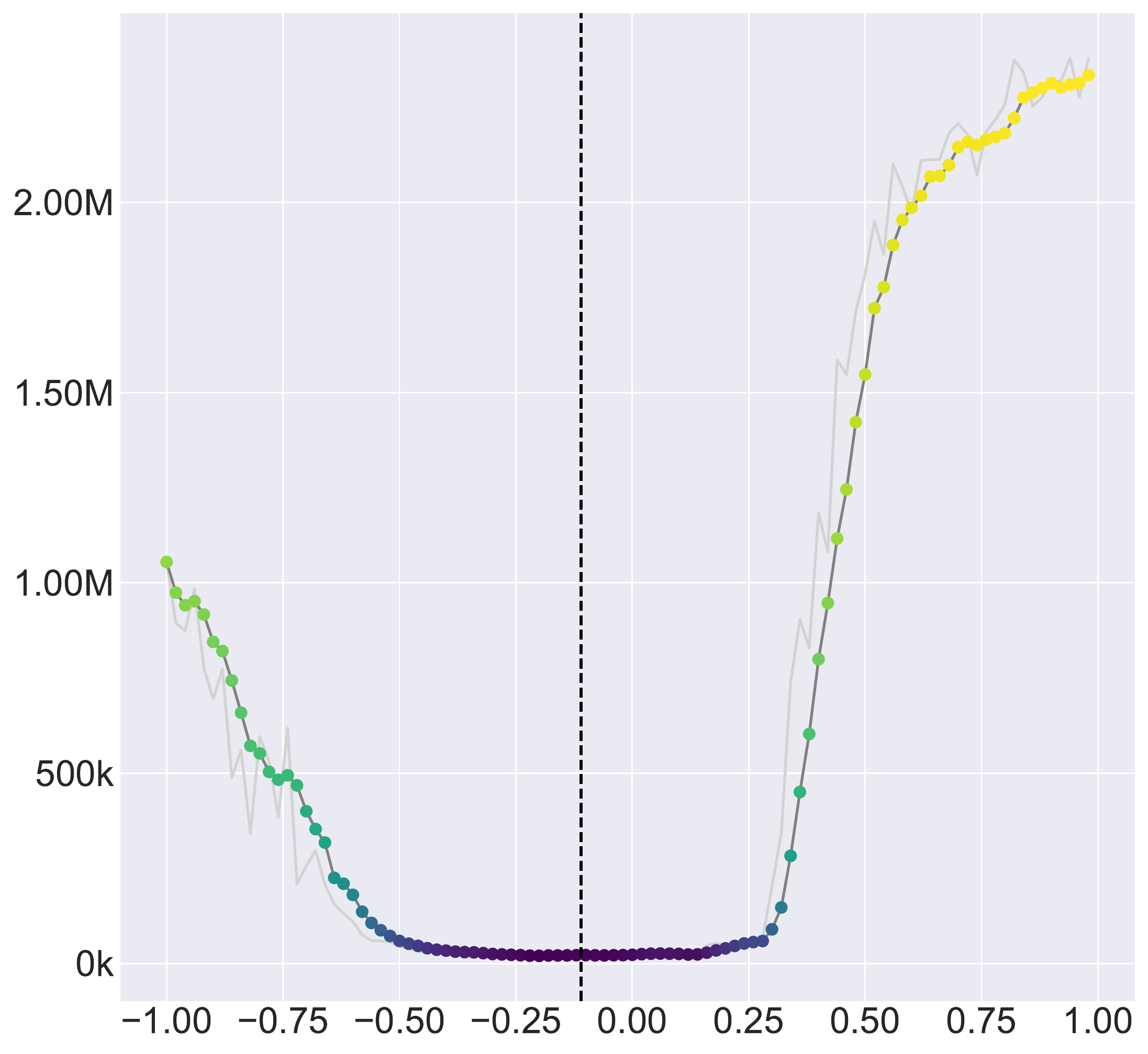}
        \includegraphics[width=\textwidth]{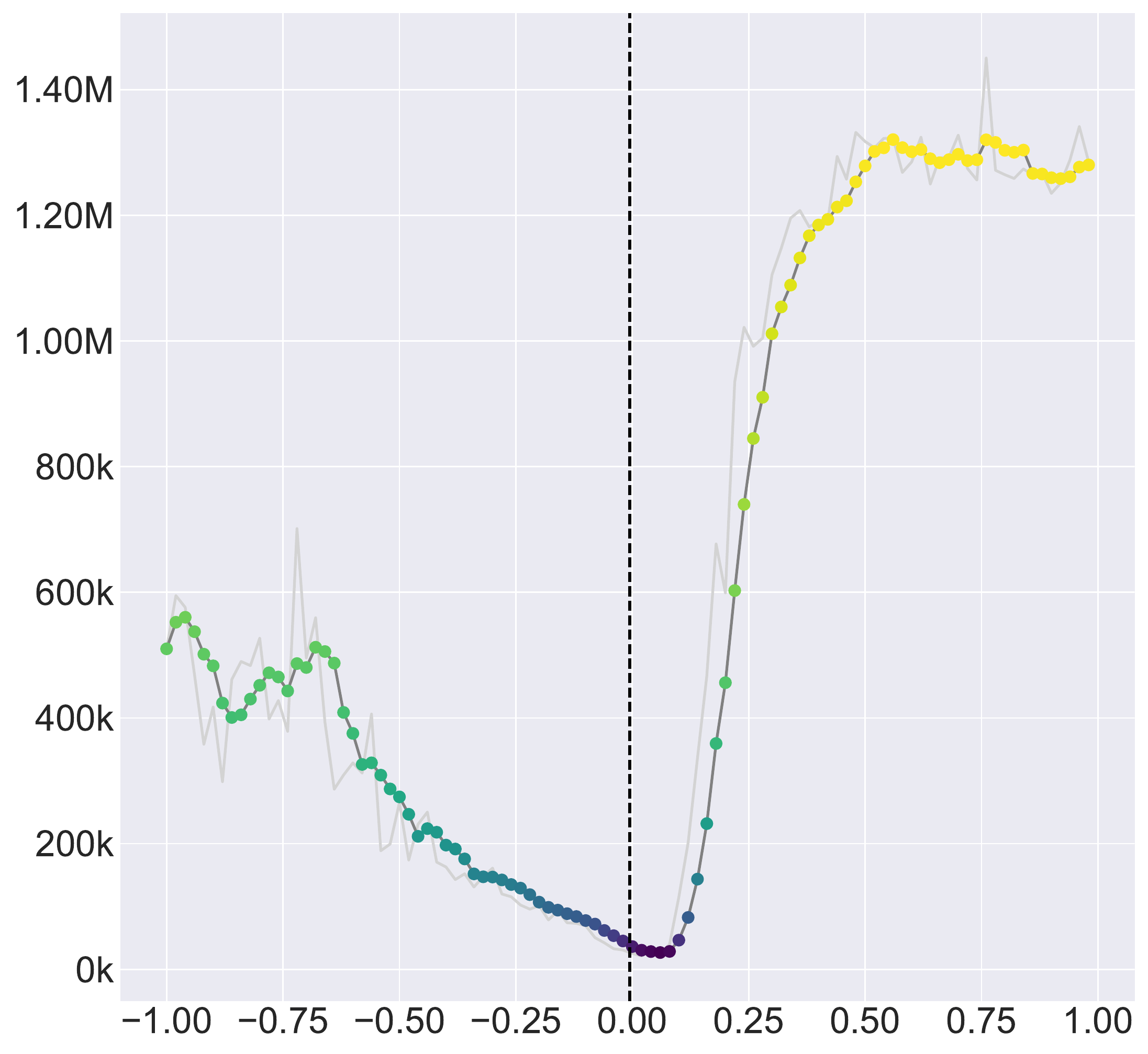}
        \caption{$h$}
    \end{subfigure}
    \hfil
    \begin{subfigure}[b]{0.3\textwidth}  
        \centering 
       \includegraphics[width=\textwidth]{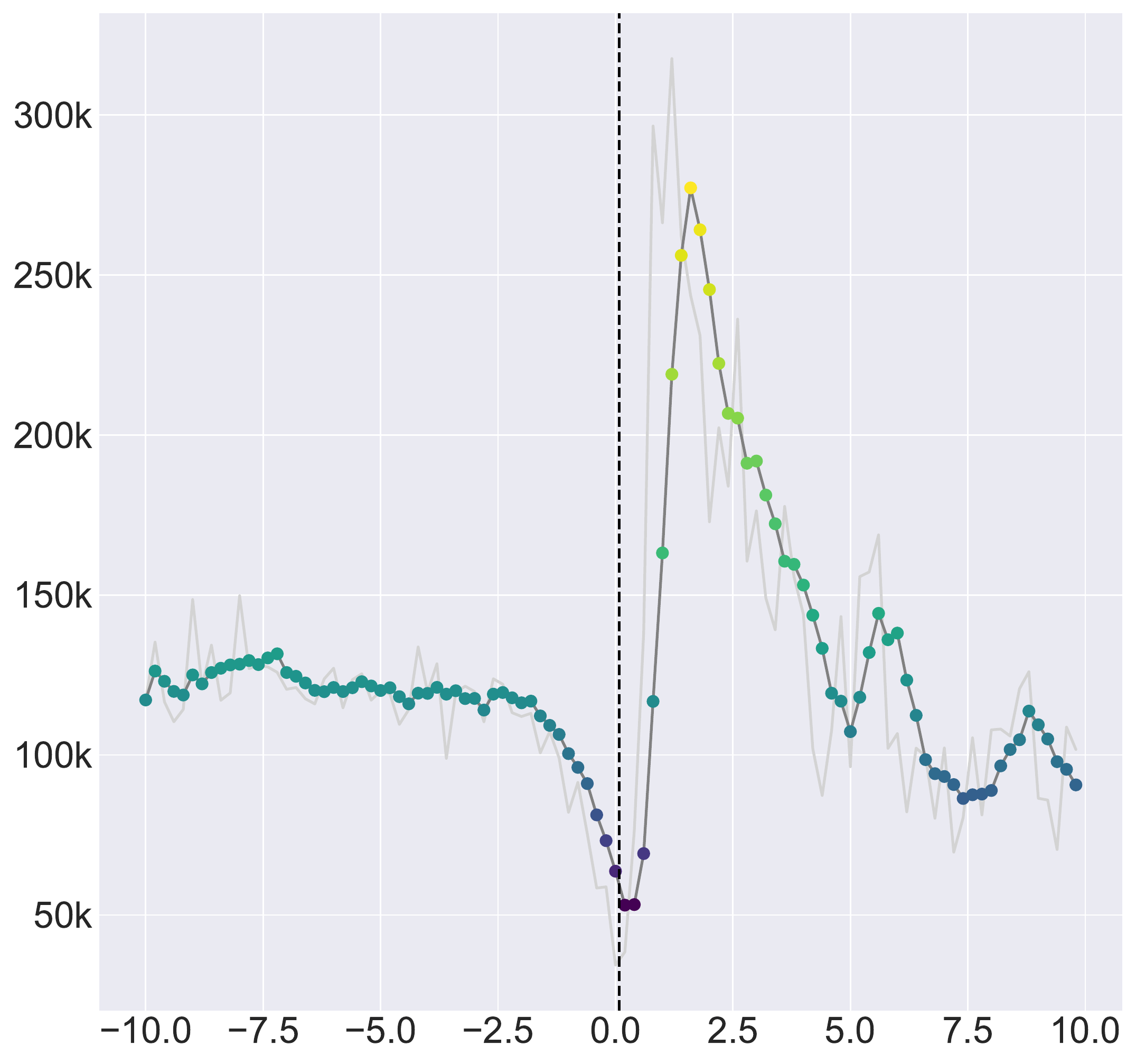}
        \includegraphics[width=\textwidth]{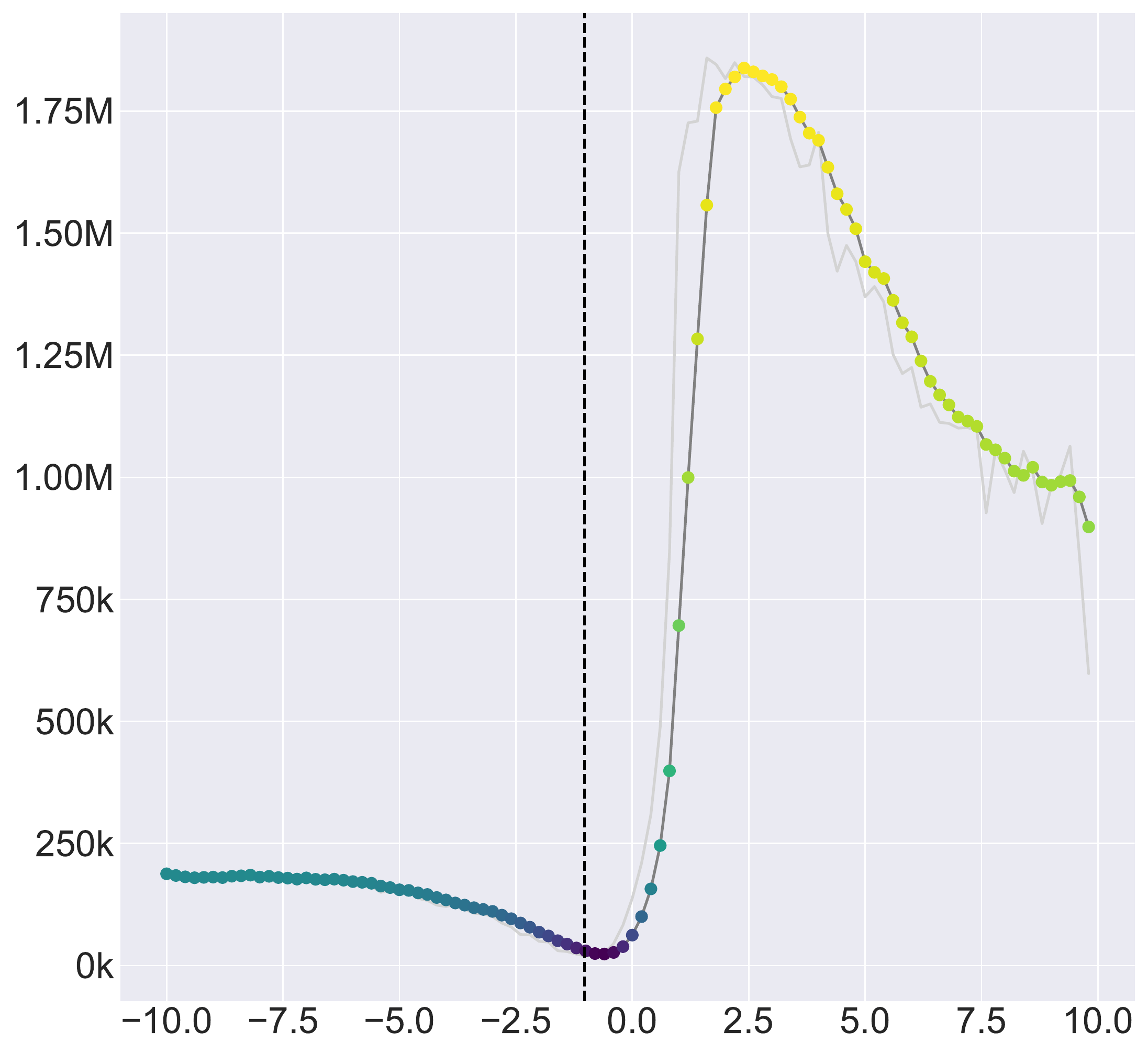}
        \includegraphics[width=\textwidth]{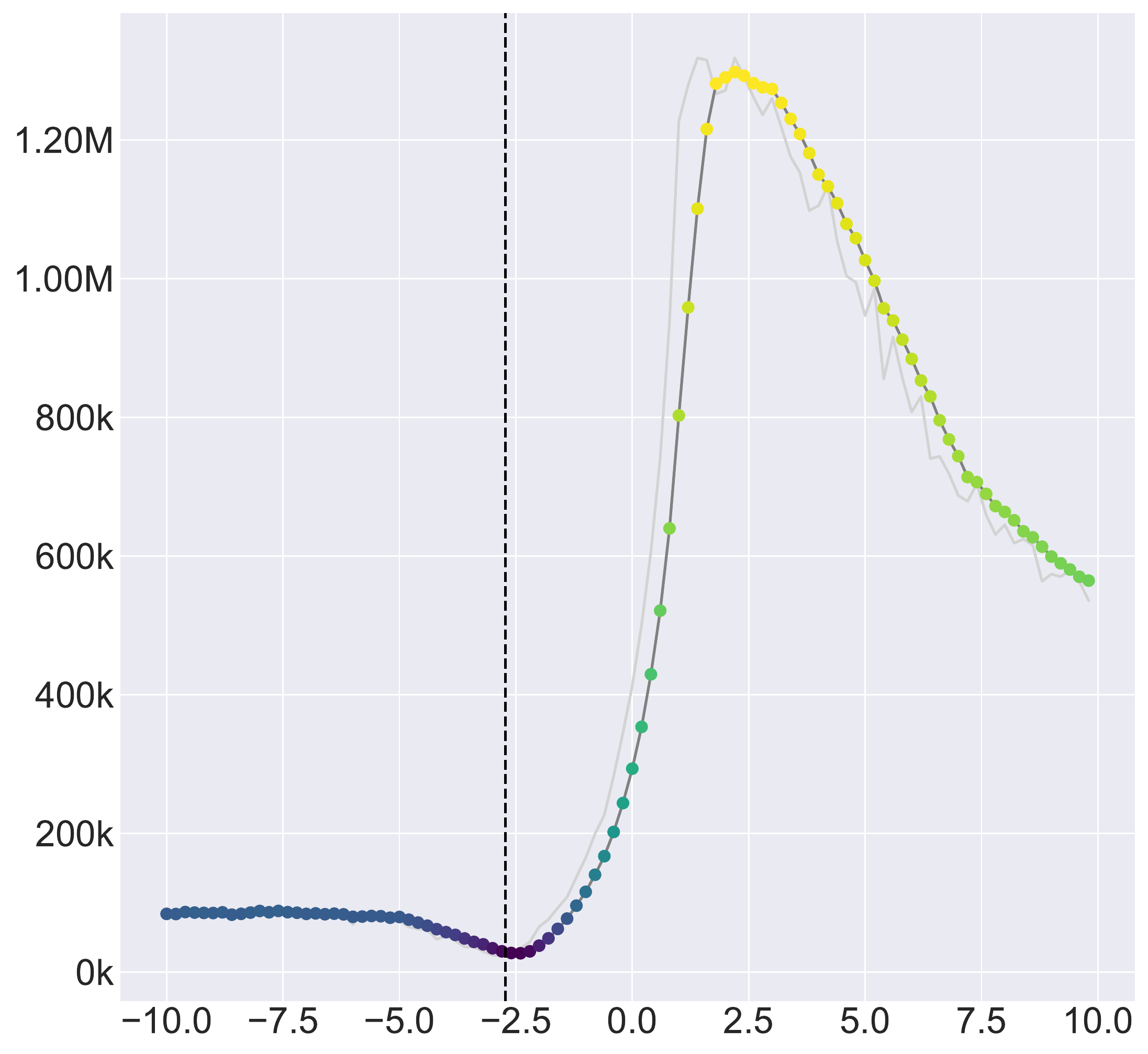}
        \caption{$\beta$} 
    \end{subfigure}
     \hfil
    \begin{subfigure}[b]{0.3\textwidth}
        \centering
        \includegraphics[width=\textwidth]{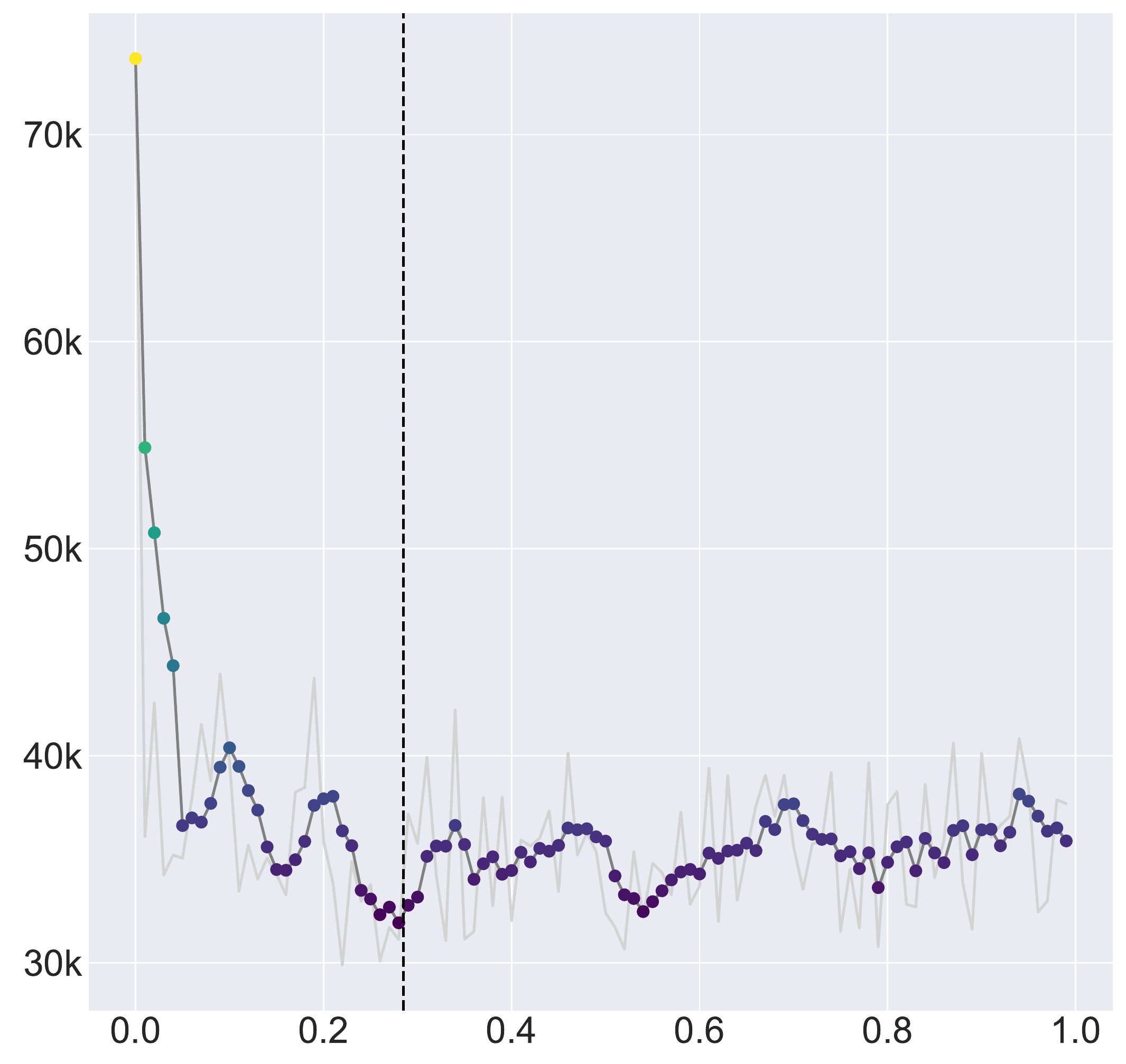}
        \includegraphics[width=\textwidth]{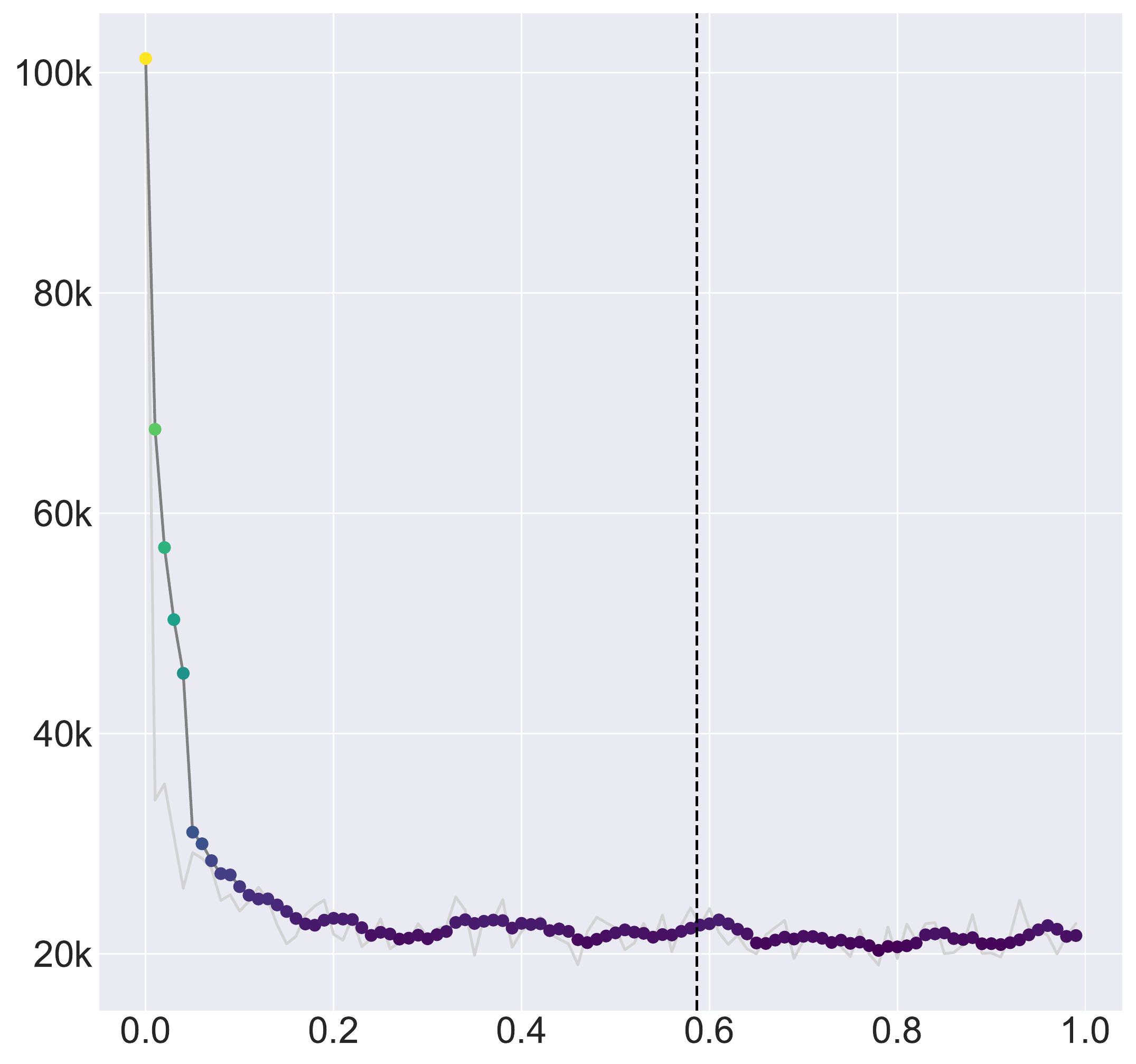}
        \includegraphics[width=\textwidth]{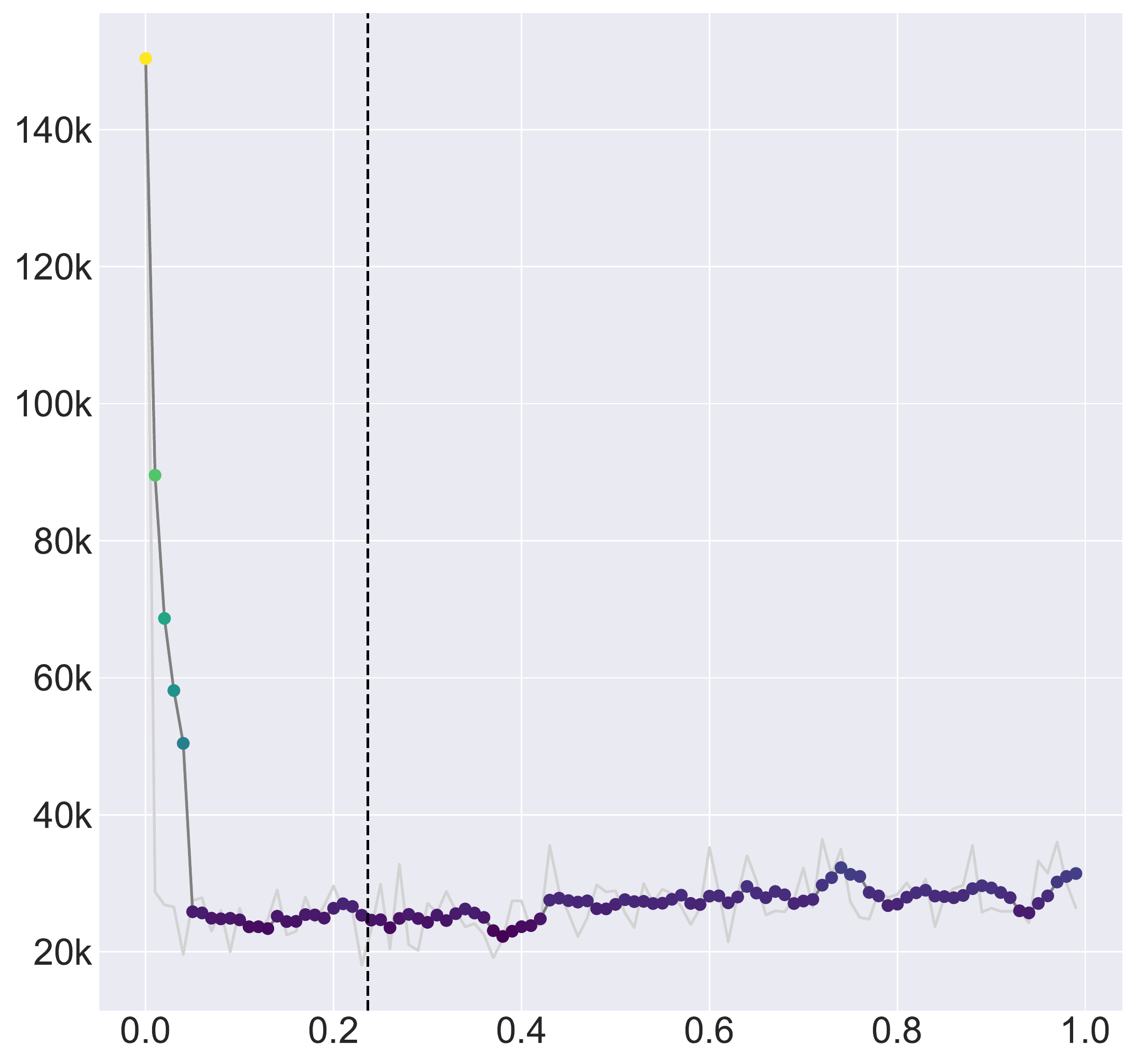}
        
        \caption{$\alpha$}
    \end{subfigure}
     \hfil
     
    \caption{Univariate LSA of default parameters, varying one factor at a time with others at their \textbf{optimised} values. The plots give the change in parameter value (x-axis) vs $\ell$ (y-axis). The dotted vertical black line shows the optimised value.}
    
    \label{figDefaultSensitivity}
\end{figure}

While the Morris method gives us the overall sensitivity across the parameter ranges (in a global way) and allows us to rank the factors in terms of importance, we also provide a fine-grained sensitivity analysis around the default values, i.e., a local sensitivity analysis (LSA).  For this, we use $p=100$ levels, but vary only one parameter at a time while keeping the others fixed at their default values. This is shown in \cref{figDefaultSensitivity}. This analysis shows how robust the resulting default values are to small perturbations, but as this is a local method, the results should be interpreted with caution (and only in conjunction with the GSA method above), since this does not account for any parameter interactions as warned in \cite{saltelli2019so}.

Viewing $h$ (the left column), we can see all values surrounding the default have a similar loss, showing the model is robust to small changes in the aptitude. Looking across the entire search space, we can see choosing from within an appropriate range for the aptitude is important though, but the surrounding parameters are always relatively smooth to the resulting loss.  Viewing $\beta$ (the middle column), we can see the sharp transition above zero. There is a clear optimal range for $\beta$, where the default lies. However, again, the area surrounding the default values is smooth showing robustness to the default parameters (assuming we do not vary past the sharp transition). Looking at $\alpha$ (the final column), the plots initially seem somewhat jagged, although when looking at the scale of the $y-axis$ it becomes clear these are very small shifts in loss (as verified by the plotted time-series with varying $\alpha$ levels). $\alpha$ was deemed the least important of the three parameters by the Morris Method screening but was still important based on the positioning on the Morris plot. We can verify this here, where changes in $\alpha$ do not have a huge impact on $\ell$.

GSA was performed using SALib from \cite{Herman2017}.

\section{Network Topology}\label{secSensitivyArchitecture}

In \cref{secExtension} we introduced a novel graph-based structure for representing the region, at the same time introducing spatial submarkets into the simulation (based on nodes in the graph). In \cref{secOptResults} and \cref{appendixSensitivity}, we have validated the usefulness of the newly introduced parameters and performed a sensitivity analysis of the parameters, whereas in this section, we look to validate the usefulness of the network structure itself.

\begin{figure}
    \centering
    \begin{subfigure}{0.2\textwidth}
    \includegraphics[width=\textwidth]{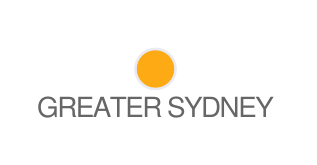}
    \caption{No Spatial Component}
    \end{subfigure}
    \begin{subfigure}{0.5\textwidth}
    \includegraphics[width=\textwidth]{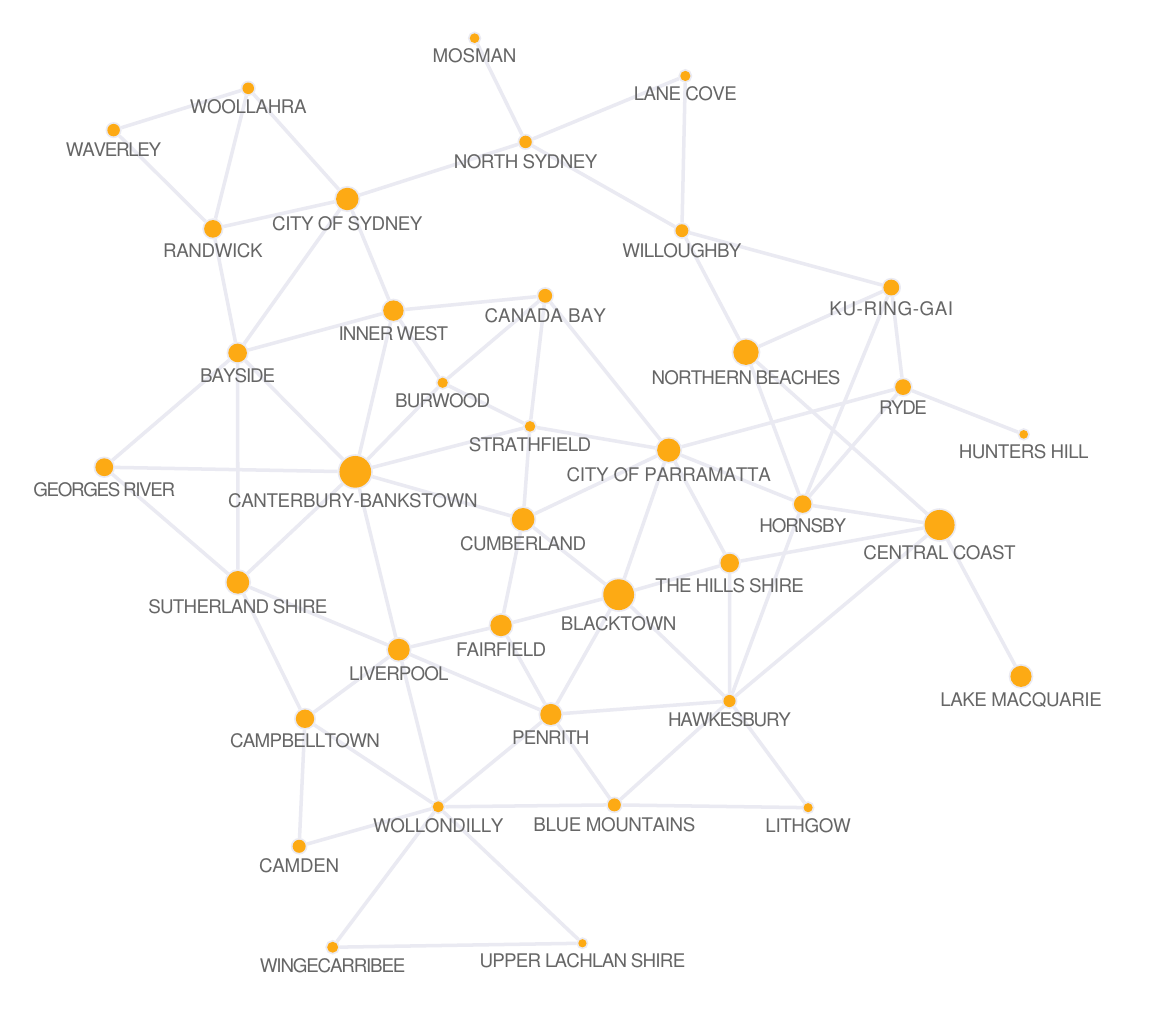}
    \caption{LGAs}
    \end{subfigure}
    \begin{subfigure}{0.2\textwidth}
    \includegraphics[width=\textwidth]{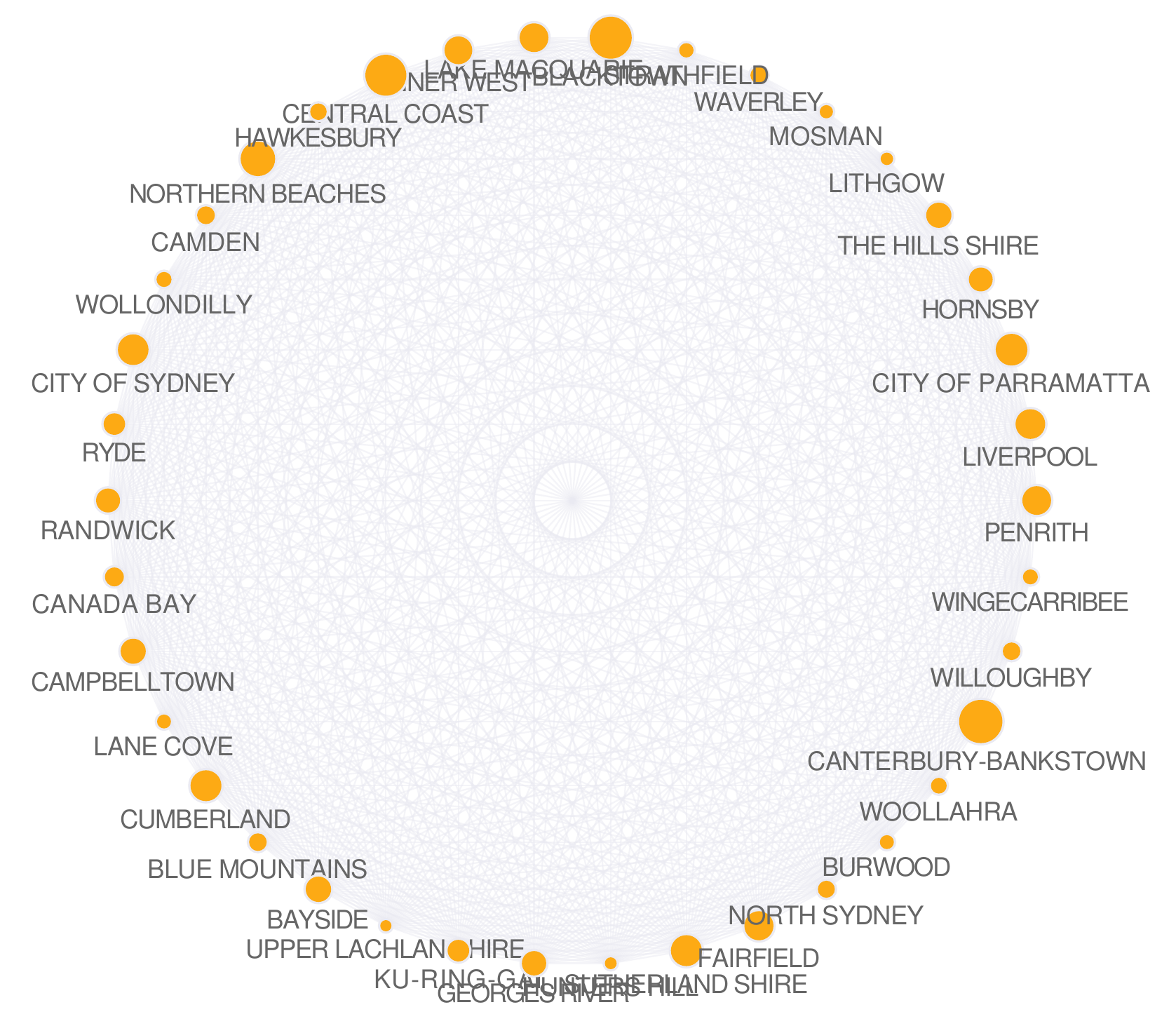}
    \caption{Complete}
    \end{subfigure}
    \caption{Various potential network architectures. The left represents a single node (i.e. no spatial component). The middle is the proposed graph-based approach constructed from the topological layout of the region. The right is a complete graph, with individual areas, but no concept of spatial neighbours (due to the fully-connected nature).}
    \label{figArchitectures}
\end{figure}

To do this, we compare the newly proposed model (with all parameters included), against an identical model with only a single node. We also compare to a fully connected network, i.e., where the spatial element (in terms of neighbourhoods) is not considered directly, but specific areas still exist. These structures are visualised in \cref{figArchitectures}.

\subsection{Topologies}

\subsubsection{No Spatial Component}
To remove submarkets and all spatial components, we use a graph composed of a single node representing the overall Greater Sydney region. That is, agent characteristics and dwelling prices are assigned based on the overall Greater Sydney distributions, rather than specific area distributions. To implement this, rather than $G$ being defined as in \cref{graphs_maps}, instead, $G$ contains a single-node (i.e. a singleton graph) where the node represents the overall Greater Sydney region, i.e., it is the graph $K_{1}$. With this single-node configuration, the spatial outreach from \cref{eqOutreach} is removed, as there is no concept of space. Likewise, $\beta$ is no longer defined, as this is expressed in spatial terms. However, $\alpha$ remains, which controls the boundedness of the agent as discussed in \cref{appendixUtility}, and $h$ keeps the same interpretation.

\subsubsection{Fully-Connected Graph}

To represent the fully connected areas, we use a complete graph where every LGA is connected to every other LGA, i.e. $G=K_{38}$. Again, outreach (from \cref{eqOutreach}) need not be considered, since now every area is directly connected to one another. With this representation, the introduced parameters still remain (i.e $\alpha, \beta, h$). Noting that $\alpha$ again directly corresponds to the boundedness (discussed in \cref{appendixUtility}), and does not encompass outreach. $\beta$ and $h$ have no change to their original meanings introduced in \cref{secExtension}. This introduces individual submarkets (based on LGAs) into the simulation, but does not enforce any spatial-based search costs within the market. As in the proposed approach, agent calibration is also based on the area in which they reside, so agent characteristics match that of their area.

\subsubsection{Analysis}

\begin{figure}
    \centering
    \includegraphics[width=.325\textwidth]{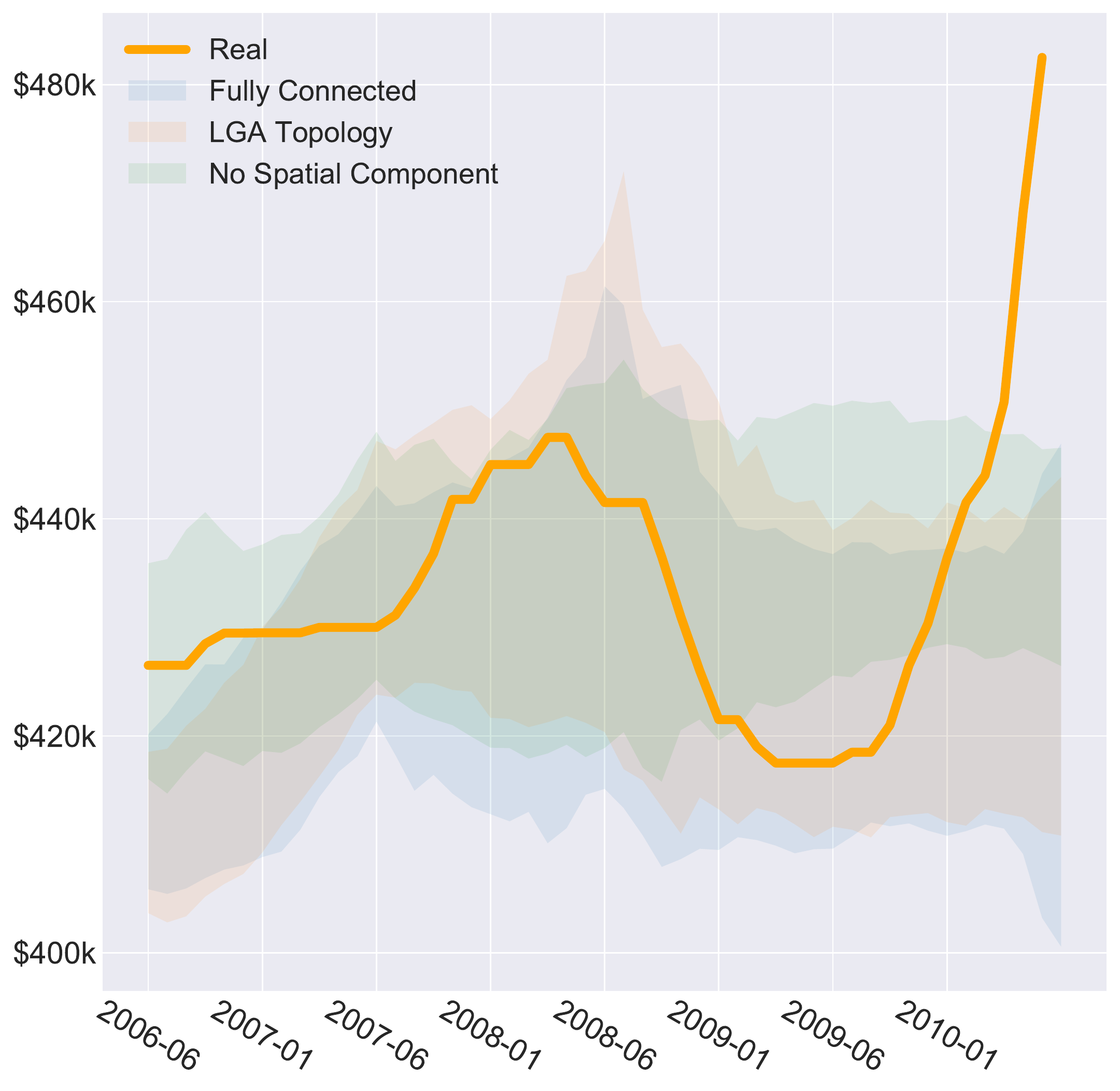}
    \includegraphics[width=.325\textwidth]{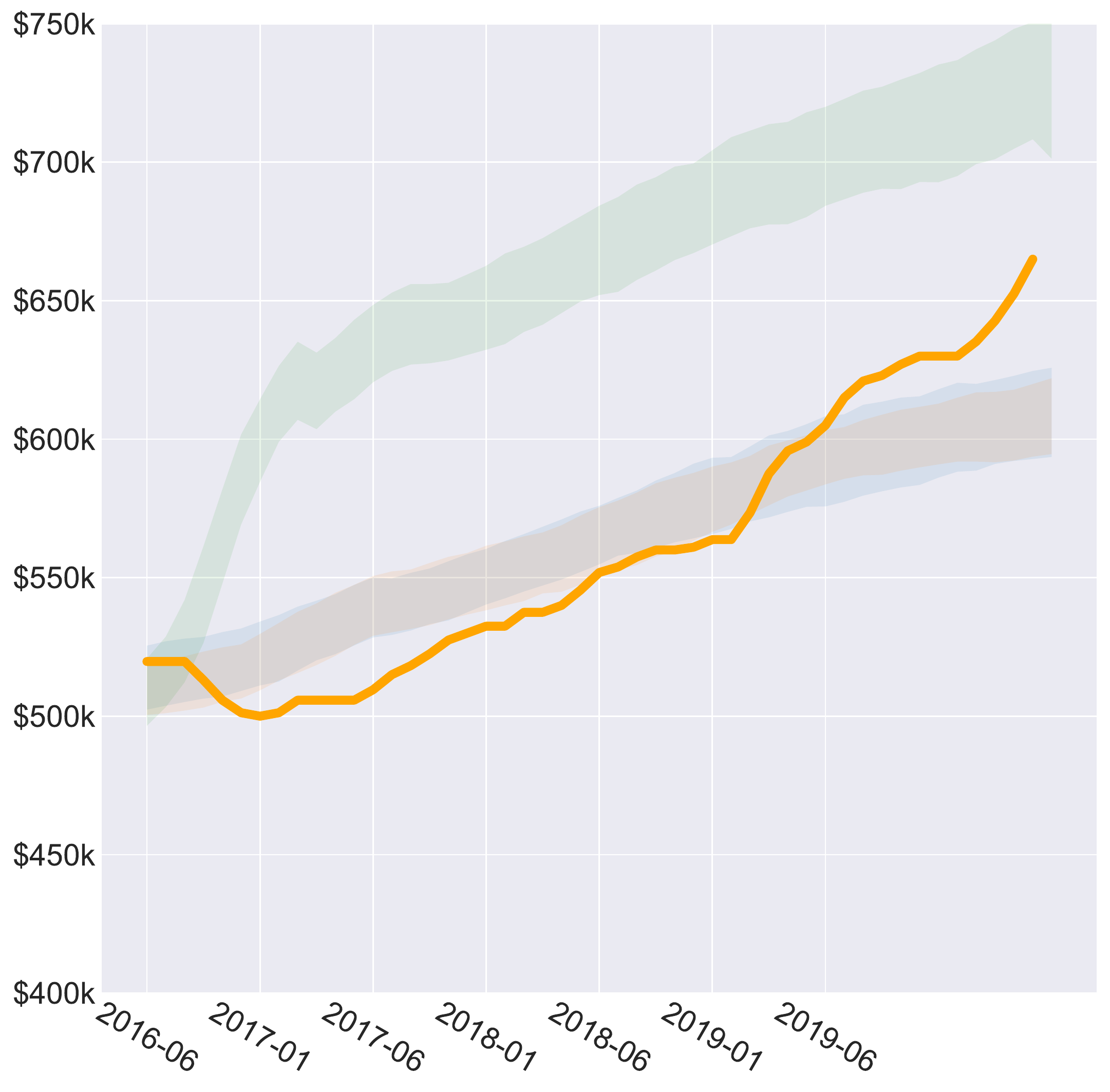}
    \includegraphics[width=.325\textwidth]{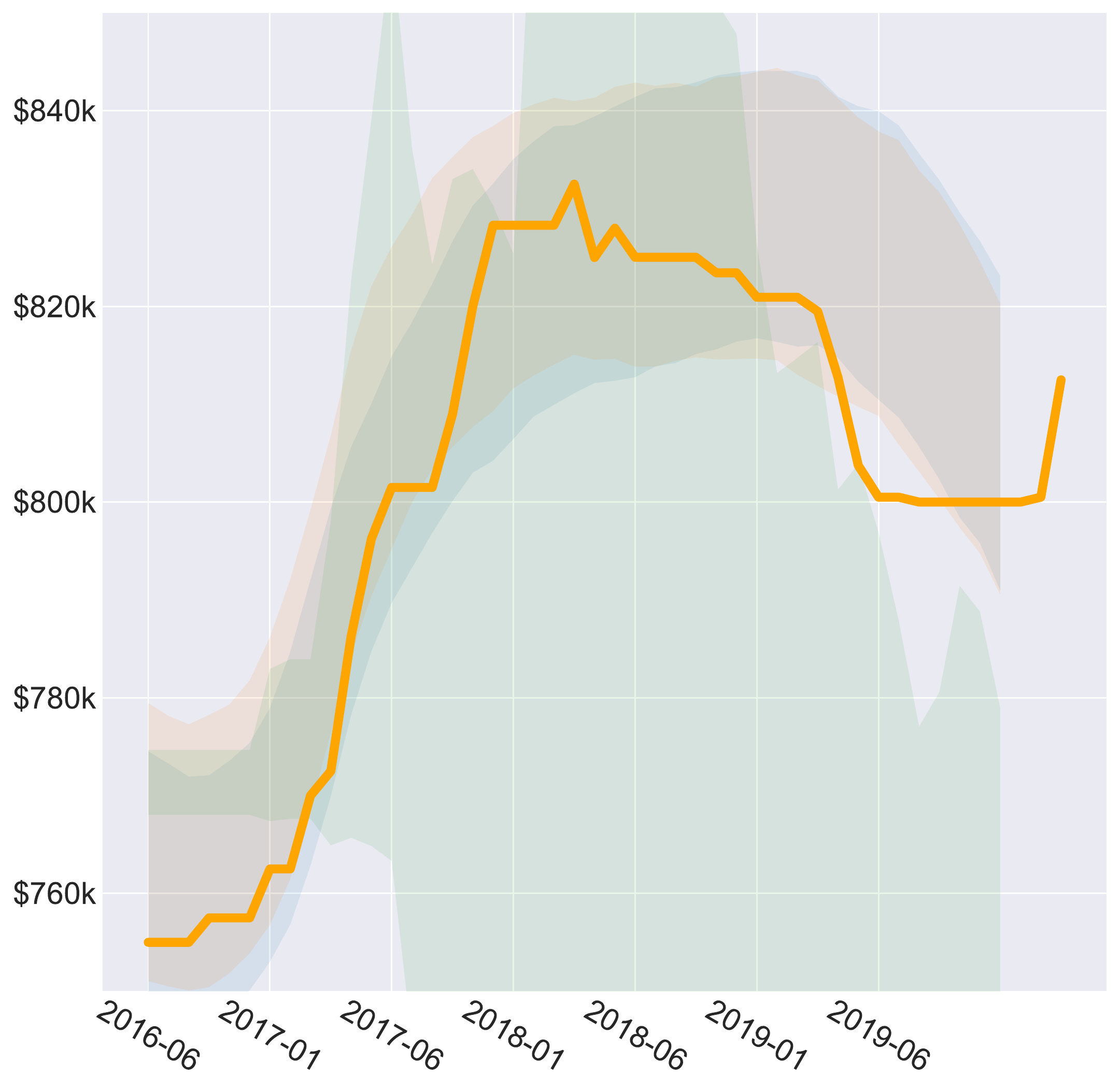}
    \caption{Various network architectures (visualised as the mean $\pm$ standard deviation range), the actual trend is given in orange. We see the individual submarket methods perform well, whereas in this case the single area (meaning no submarkets) method fails to adequately capture the overall trends.}
    \label{figArchitectureLosses}
\end{figure}

The resulting comparisons are visualised in \cref{figArchitectureLosses} which shows the models which include individual areas significantly outperforming the overall Greater Sydney model, indicating the usefulness of area-specific submarkets. Between the two area models, there was little difference in aggregate performance (as shown in \cref{figArchitectureLosses}) which shows the performance improvements come mainly from the introduction of submarkets, not necessarily on the overall spatial structure. This shows that $\beta$, which is based on the individual areas, initialising agent characteristics based on location, and the modification of price setting in terms of $\overline{Q}_h$ based on the area are key for resulting prices (more so than the spatial outreach costs).

\begin{figure}
    \centering
    \begin{subfigure}{0.45\textwidth}
    \includegraphics[width=.95\textwidth]{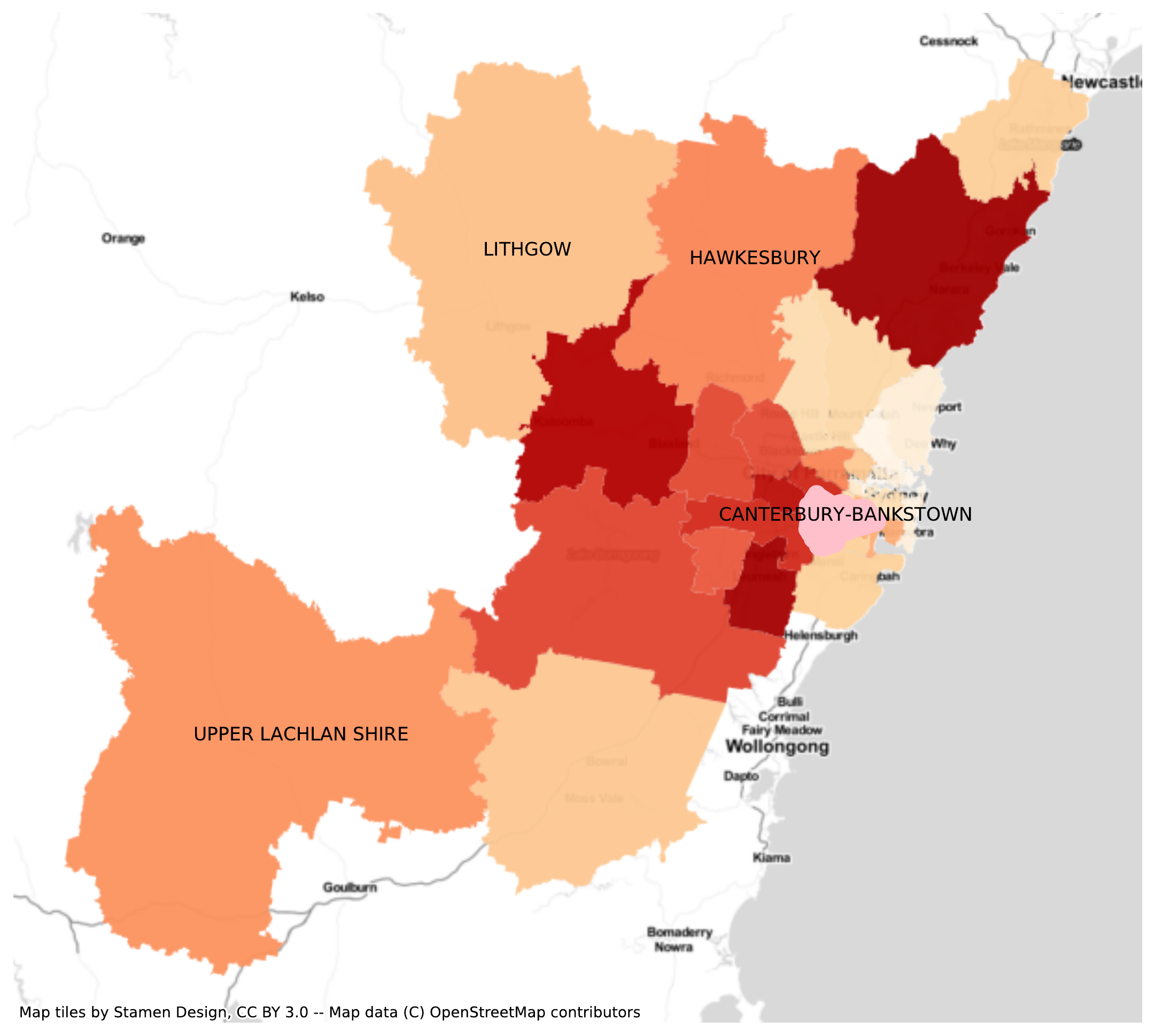}
    \caption{Spatial Structure Enforced}
    \end{subfigure}
    \begin{subfigure}{0.5\textwidth}
    \includegraphics[width=.95\textwidth]{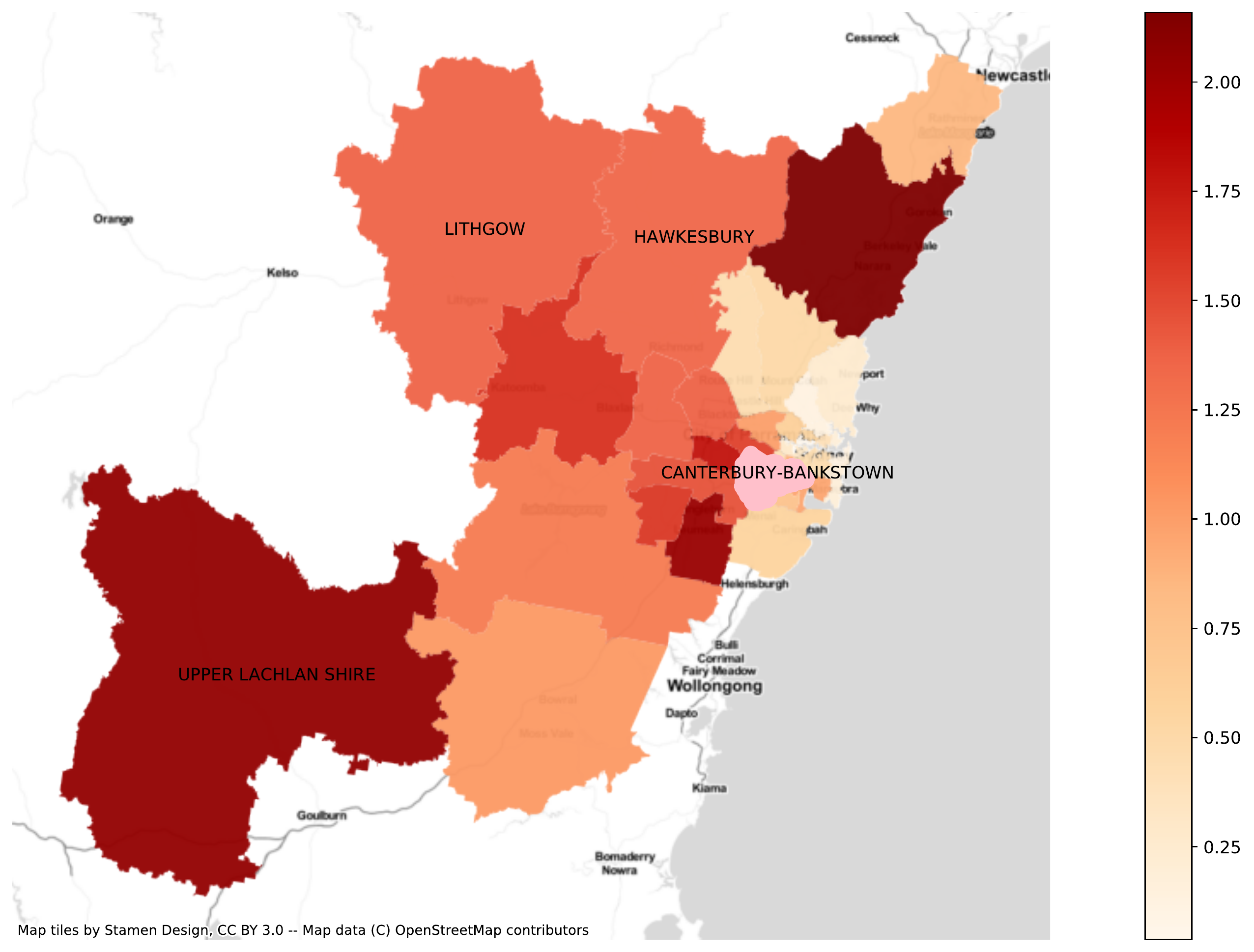}
    \caption{Fully Connected Structure}
    \end{subfigure}
    \caption{Variations in agent preferences with the proposed spatial structure (left) and without a spatial structure enforced (right) from first time home buyers situated in the Canterbury-Bankstown region (visualised in pink and labelled). The colour indicates the percentage of purchases in those areas, with dark indicating a high percentage of relative purchases (controlled for population sizes). Without a preserved spatial structure we see much higher rate of purchases on the outskirts of the Greater Sydney region, such as in Upper Lachlan Shire and Lithgow (labelled). We verify the proposed movements are logical in \cref{secHouseholdMobility}.}
    \label{figArchitectureMovemenets}
\end{figure}

However, when considering the resulting agent preferences (in terms of suburbs to purchase in), the fully connected map had significantly more people moving to more remote regions, whereas the LGA connected spatial topology prevented as drastic movements (with the spatial outreach term), capturing the fact people are often tied to specific areas (i.e. those who work in the CBD and currently reside near there, are unlikely to the outskirts of the Greater Sydney region). This is visualised in \cref{figArchitectureMovemenets}, where we show the proposed movements follow a diffusive-like pattern \citep{slavko2020diffusive} compared to the fully connected topology which had movements to more remote regions of Greater Sydney. We further discuss the resulting movements \cref{secHouseholdMobility}, where we show with the agent movement patterns with the proposed spatial structure are logical and consistent with the actual reported movements in the Greater Sydney region based on the observed trends reported in recent literature.

% Above we showed differences in agent preferences and movement patterns between the fully-connected graph and the proposed LGA spatial topology, showing the LGA topology prevents as drastic movements to remote areas. This could be incorporated into the fully connected network by introducing edge weightings between the areas (as opposed to having a spatial outreach term as in the LGA topology graph) which capture the agent preferences for certain areas. That is, the 38 LGAs in the fully connected graph create a 38x37 matrix (ignoring the self-connection). Each of the 1406 cells in this matrix would now have a weight, and that weight controls the probability of moving (more specifically, viewing a listing and potentially moving) between two LGAS $w(l_i$, $l_j)$. We have not further explored the weighted approach, due to the high number of additional free parameters introduced (from 3 free parameters to 1409 parameters) and prohibitive computational cost of optimisation. The proposed spatial topology can be seen as a special case of the weighted fully-connected topology, where the edge weightings are calculated based on $\alpha$ and the outreach term $O$.

In this section, we have shown the area-specific sub-market extensions significantly outperform an equivalent model which does not include individual areas, highlighting the importance of capturing submarkets. Furthermore, the incorporation of area-specific submarkets allows for additional insights (such as those in \cref{secMigrations} which would not otherwise be possible). We then further validated the choice of the network topology by comparing resulting agent preferences, and showing the proposed architecture (with spatial-based search costs) prevents drastic movements by the agent (in terms of relative distance from an agent's current location), allowing the agents to act based on their location (preferring closer areas) in a manner consistent with the actual observed trends as discussed in \cref{secHouseholdMobility}.

\section{Experiment Settings}
Due to the stochastic and non-deterministic nature of ABMs, we run 100 Monte Carlo simulations per run (unless otherwise stated) and report the aggregate results over all runs to get a robust estimate.

\subsection{Scale}
\begin{figure}
    \centering
    \includegraphics[width=.325\textwidth]{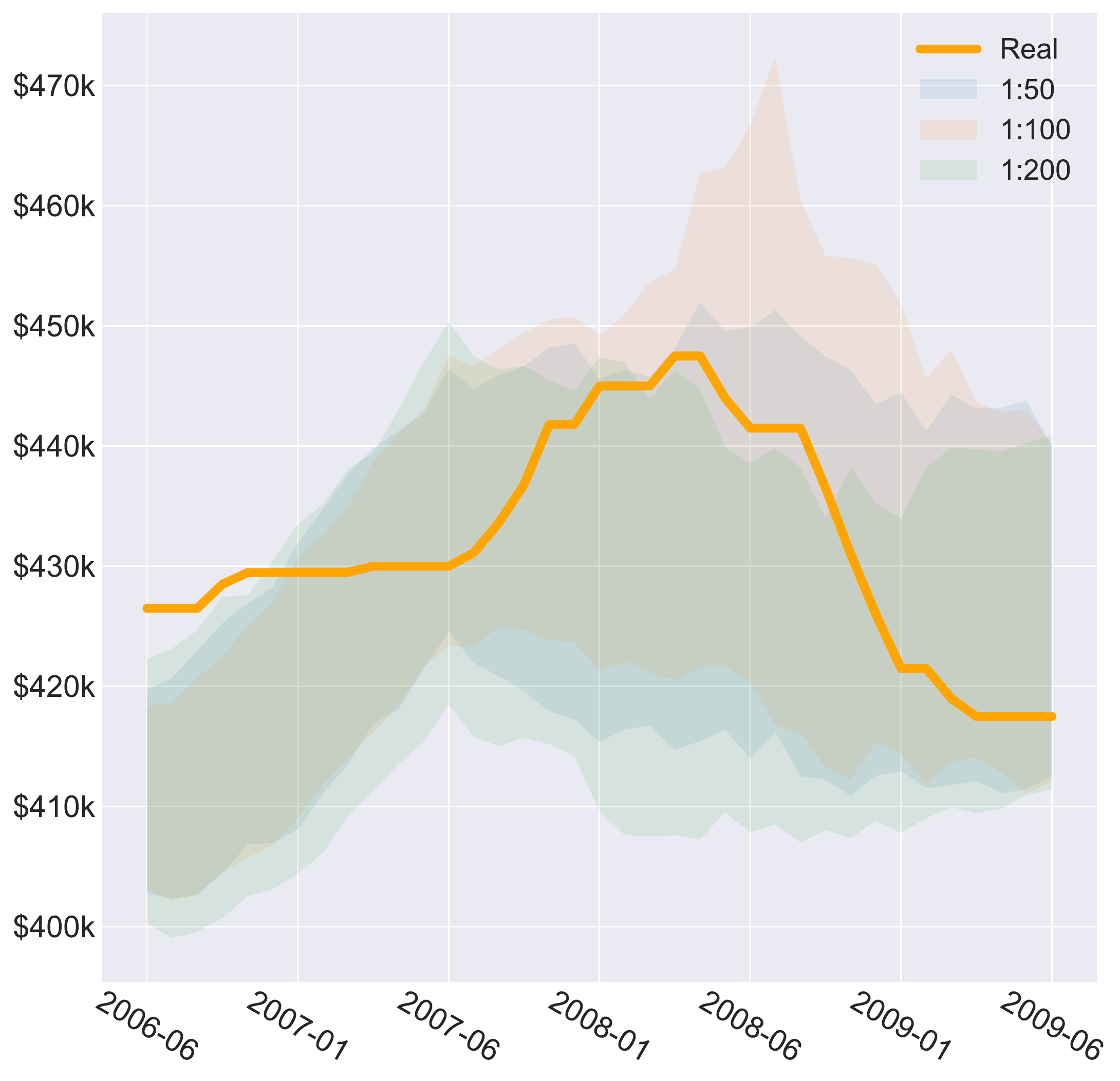}
    \includegraphics[width=.325\textwidth]{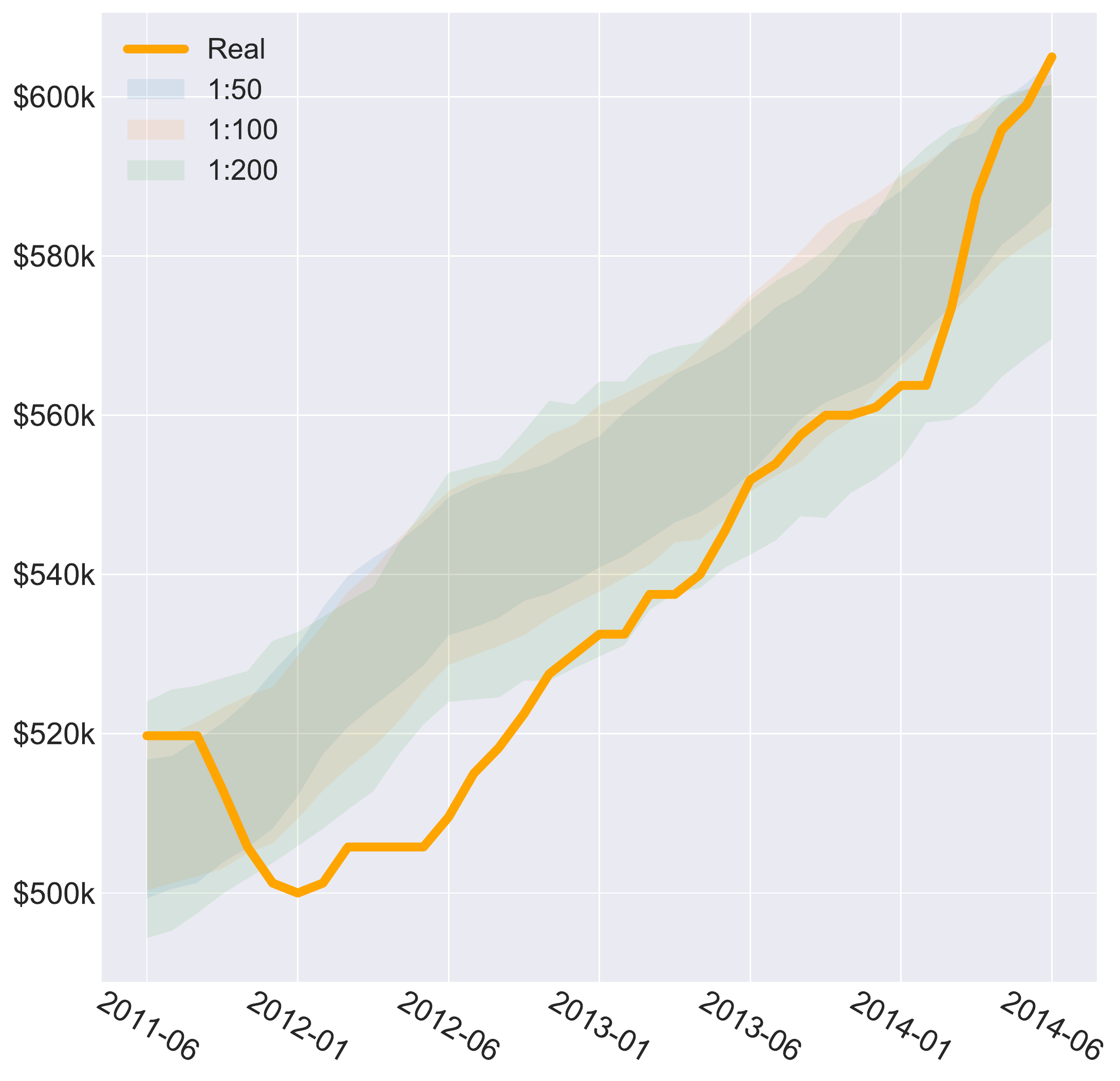}
    \includegraphics[width=.325\textwidth]{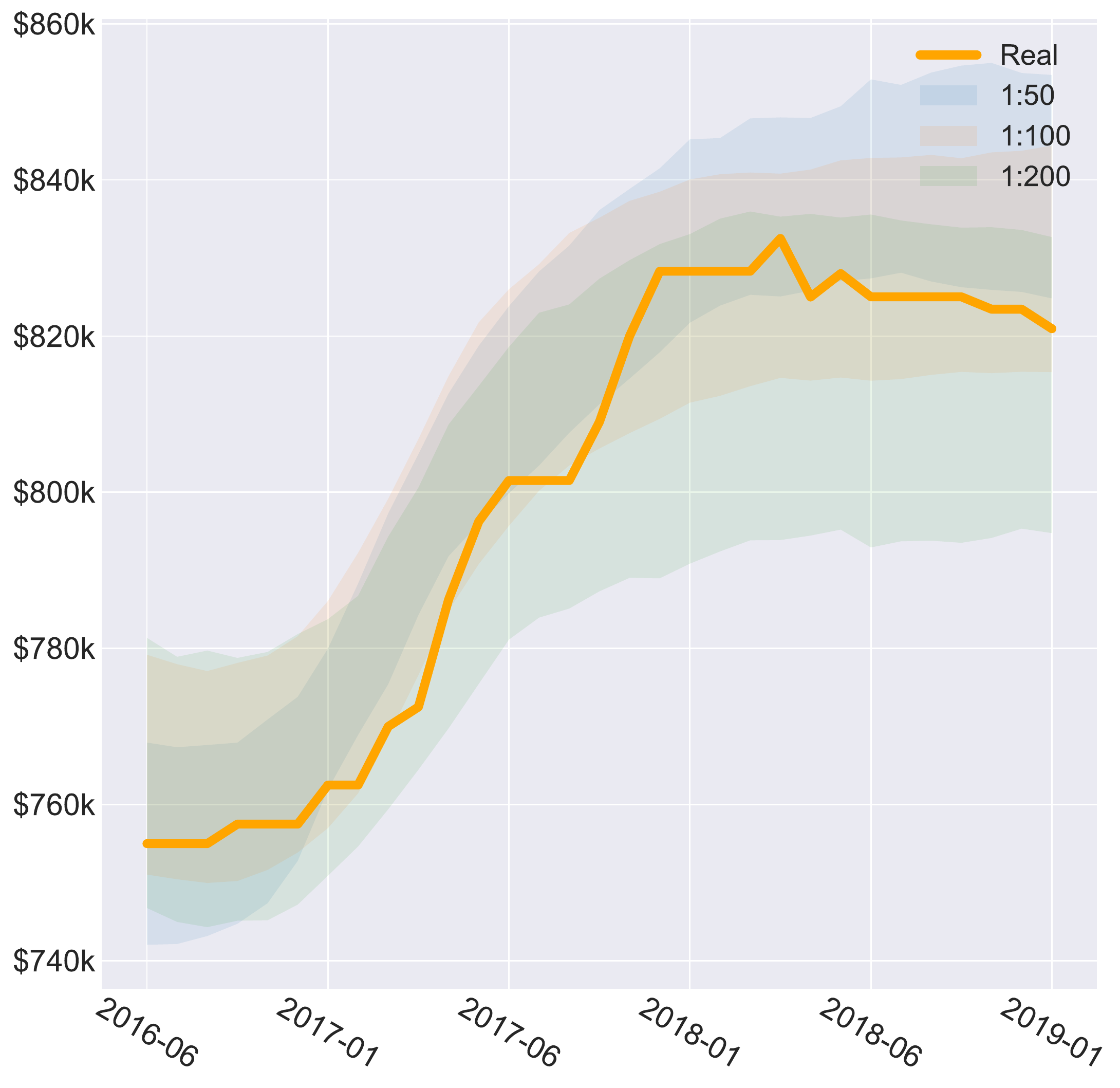}
    \caption{Robustness of scale over the training phases (visualised as the mean $\pm$ standard deviation range for varying scales). We see with varying scales similar results are recovered (with large areas of overlap), indicating in-variance to the scales used (within acceptable bounds).}
    \label{figScales}
\end{figure}

Experiments are run at a 1:100 scale of the true housing market, i.e., every one hundred households in the Greater Sydney region are represented by one household in the model. The 1:100 scale was chosen for efficiency, but results for 1:50, 1:100 and 1:200 are also presented in \cref{figScales} to show robustness to scale. There is an upper limit on the scale where the performance will begin to degrade, for example, the number of overseas investments is given in \cref{tblOverseas}, and by using a scale close to 1:1000, we would lose the contribution of foreign investments (since the values would be $<1$). For lower scales (i.e. 1:1), the results may be more accurate but this comes at the expense of increased computational power, so the 1:100 provided a good trade-off between accuracy and efficiency.

\section{Initialisation Data}\label{appendixInitialisation}

\subsection{Data}
All real estate listings and sales from 2006 to present (2020) were used from SIRCA-CoreLogic, including the sale price, LGA, and sale date. This data is used as the actual price, and to calibrate the ABM.

\subsection{Spatial Initialisation}\label{appendixSpatial}

\subsubsection{Pricing Distributions}
\begin{figure}[ht]
    \centering
    \begin{subfigure}{0.325\textwidth}
        \includegraphics[width=\textwidth]{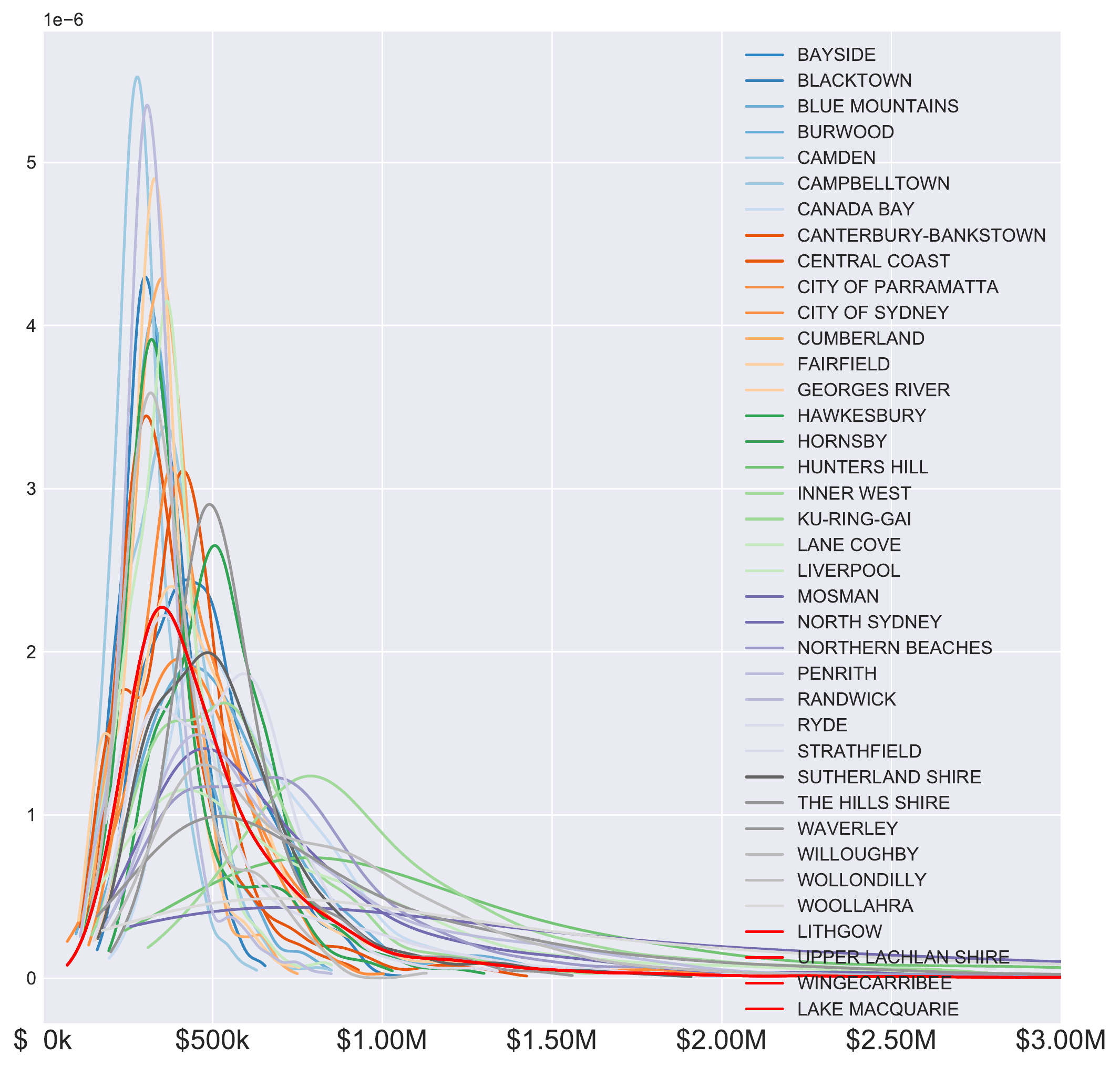}
        \caption{2006}
    \end{subfigure}
    \begin{subfigure}{0.325\textwidth}
        \includegraphics[width=\textwidth]{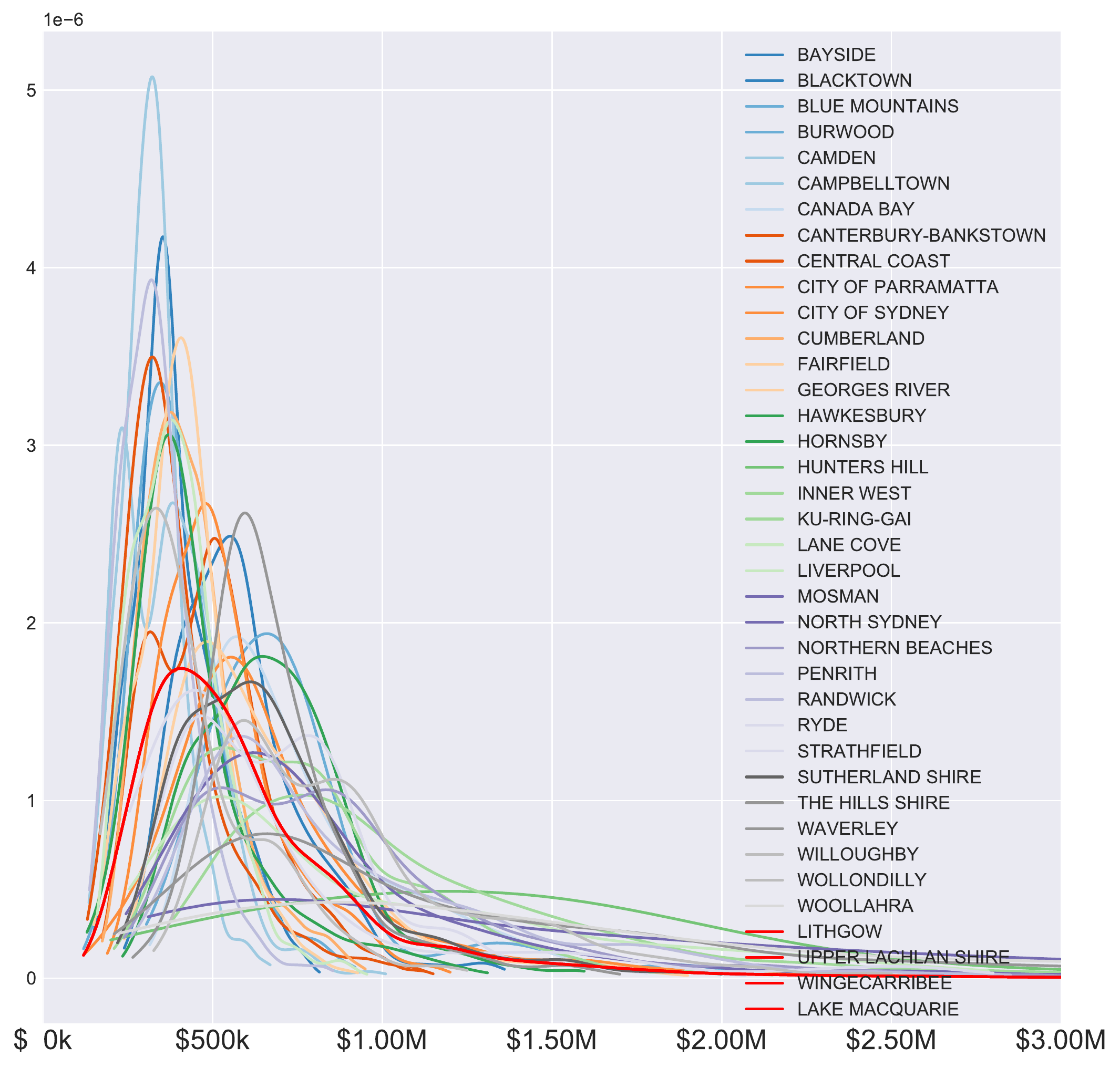}
        \caption{2011} 
    \end{subfigure}
        \begin{subfigure}{0.325\textwidth}
        \includegraphics[width=\textwidth]{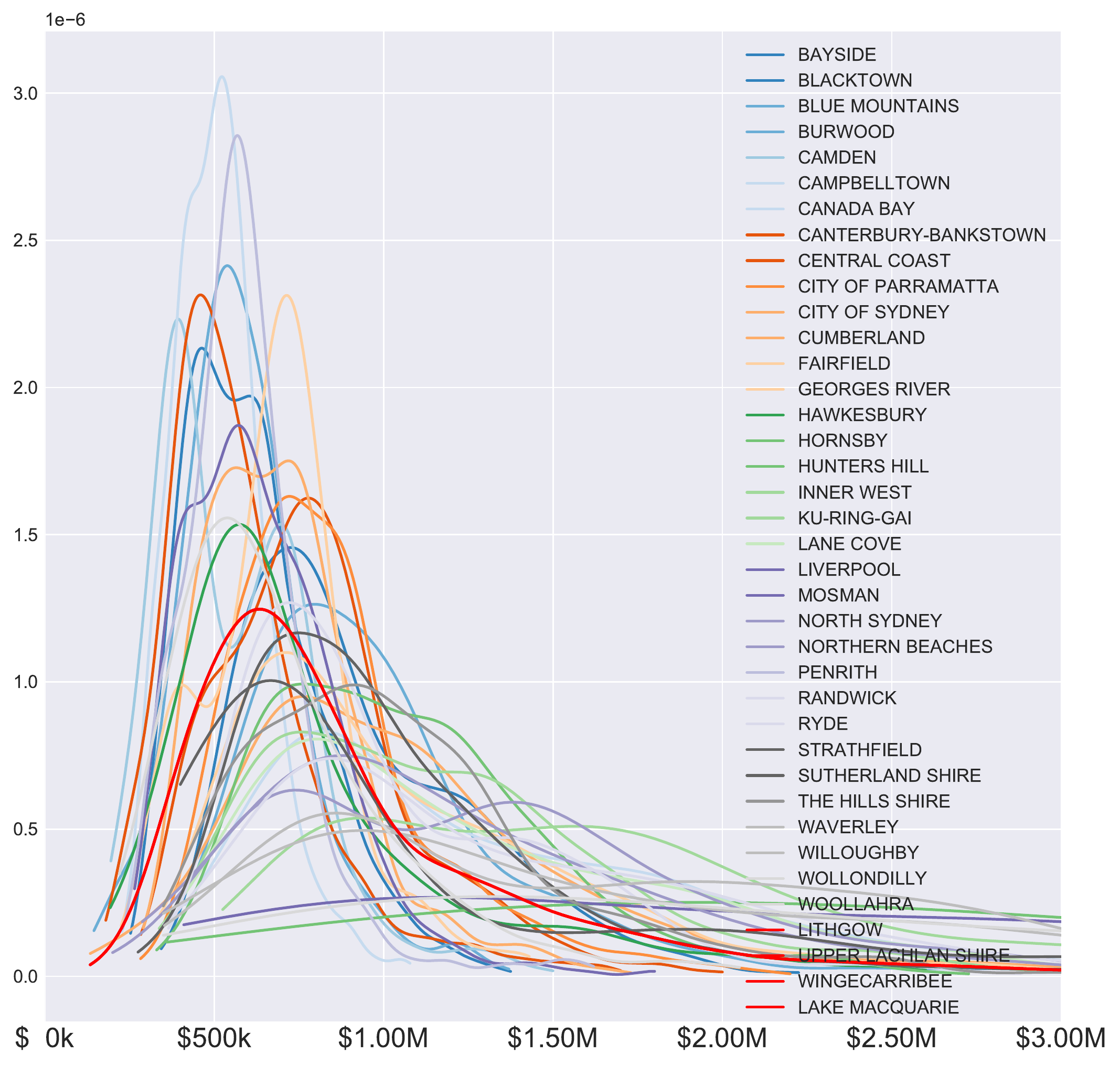}
        \caption{2015}
    \end{subfigure}
    \caption{KDE plots for each LGA based on SIRCA-CoreLogic data. Dark red indicates the Greater Sydney average, and this is assigned to LGAs without enough data to generate their own reliable KDE}.
    \label{figKDE}
\end{figure}

Between LGAs, there is a wide range of dwelling sale prices, and different distributions of prices amongst the LGAs as well. To sample from this effectively, we use kernel density estimation (KDE) to create a probability density function for each LGA for each time period. The previous 3 months of sales from the beginning of the time period are used to generate the density function. Scott’s Rule \citep{scott2015multivariate} is used to assign the bandwidth, which sets the bandwidth to $n ^ {\frac{-1}{d+4}}$, where $n$ is the number of data points (in this case dwelling sales in the LGA at the beginning of the time period), and $d$ is the number of dimensions (in this case $d=1$). When new houses are created for an area, they are set with an initial quality based on this distribution. The resulting KDEs are shown in \cref{figKDE}.

\subsubsection{Positioning}

Households are not assigned to an LGA directly, as households can freely move areas. Instead, the households area is based on the residential dwelling of the household (and thus can change over time). When we reference a households area, we are referring to the LGA of the dwelling where the household currently resides.

At the beginning of the simulation, households which are homeowners are assigned to dwellings to match the population distribution amongst LGAs. The income and liquid wealth for the household are then assigned based on the brackets from the dwellings LGA. Renters are assigned a random LGA to begin with (again weighted by the population of each LGA) and income and wealth based on the distribution of that LGA. Households then try and find a rental they can afford (on with a rental price approximately 10\%-30\% of the household's income) which may mean some have to move LGAs.

\subsection{Time Periods}\label{appendixTimePeriods}

In line with the previous work of \cite{glavatskiy2020explaining}, and following the Australian census timelines (which are performed every 5 years), we choose the three most recent census periods for analysis. These are 2006--2010, 2011--2015, and 2016--2019. The length was chosen such that upon new census information becoming available, a new simulation is run. Meaning, a separate model (and optimisation process) is run for each of the time periods to ensure the model is calibrated to the most recent data available. In doing so, we ensure the agent characteristics of the model most closely match those in the true Greater Sydney market.  
As each period corresponds to the census years, there is a large array of available data for calibration to ensure the models begin in a state as close to possible as the true populations state. Alternate (non-census) dates could be used, however, the model may not begin with as accurate of a reflection on the true underlying agent characteristics (depending on the data availability). While the models are calibrated for the time periods outlined here, such calibrations would also work well for surrounding dates (or alternate run duration's), or the model could be re-calibrated for alternative dates to provide additional forecasting --- for example, after the 2021 census, the agent characteristics could be reassigned and a new optimisation process run to reflect updated agent behaviours, likewise for past market behaviour such as with the 2001 census.

\subsection{Household Characteristics}

\subsubsection{Area}
Agents are initialised into an area based on the Australian census data, meaning the population of each area at the beginning of the simulations corresponds to the proportions from the census data for that time period. For example, if there are three areas "A", "B", "C", and the true proportions in each are 60:25:15, the model will also populate agents into the three areas according to this proportion. Throughout the simulation, agents may move areas. They may be forced to move to a cheaper area if they can not afford their current area, or they may move to a more affluent area if they can afford a dwelling there. So once the simulation begins, the movement dynamics are controlled by the agents' cash-flow position (again from census data, outlined below). Initialising agents into areas based on census data allows for correct agent characteristics (such as income and net worth) that directly line up with those observed throughout the Greater Sydney region.

\subsubsection{Income}

Income is assigned from the distribution based on the households area. This distribution comes from the census data. Income grows throughout the simulation. The income brackets follow those specified in the census data.

\subsubsection{Liquid Wealth}
Again, the liquid wealth (liquidity) of a household is based on the true distributions from census data. However, in this case, liquidity is not available per LGA, only for Greater Sydney as a whole. So to map a household to an appropriate liquidity bracket, the households liquid is based on the income of the household. That is, if a household is in the top X\% of earners in an LGA, the liquidity will be in the top ~X\% as well (approximately, since liquidity is from brackets).

\subsection{Population Distribution}

In this case, there are three measures of interest. The total number of dwellings, the total number of households, and the distribution of these households amongst LGAs. The dwellings and households estimates from the census data are used for each year, and simple linear projections used for forecasting the growth of these. The distribution amongst LGAs is that recorded at the start of the simulation and is assumed to grow linearly with the overall population size. Individual LGA future population projections are available from 2016 onward, but as no projections existed before this date, we used this simplified measure instead of all LGAs growing by a fixed percentage within a given simulation period. As such, higher movements towards one particular LGA throughout simulation could indicate the requirement of additional dwellings being built here to cater for the growth, which is another contribution we consider in later sections of this work.

\section{Movement Pattern Visualisations}\label{appendixMovements}

Over 10 million total movements were tracked across the simulations (approximately 3.3 million per time period). All plots in this section represent the normalised heatmaps of these movements. The total number of movements to a particular LGA is scaled by the population size of this LGA, meaning the results can be interpreted as a preference for certain areas rather than visualising the population size of the LGAs. Therefore, movements are not just reflecting larger populations, instead, reflecting a larger portion of people moving there relative to the size.
All movements are then normalised such that the summation of all cells in the plot is 1, meaning if a particular cell has a value of 0.05, this means 5\% of all matched movements moved to this LGA.

The rows and columns of the plots are always sorted in ascending order based on median price, i.e., the most affordable LGAs first, and the most expensive LGA as the final row or column.

\begin{figure}[ht]
\centering
\begin{subfigure}[b]{0.3\textwidth}
      \includegraphics[height=.4\textheight]{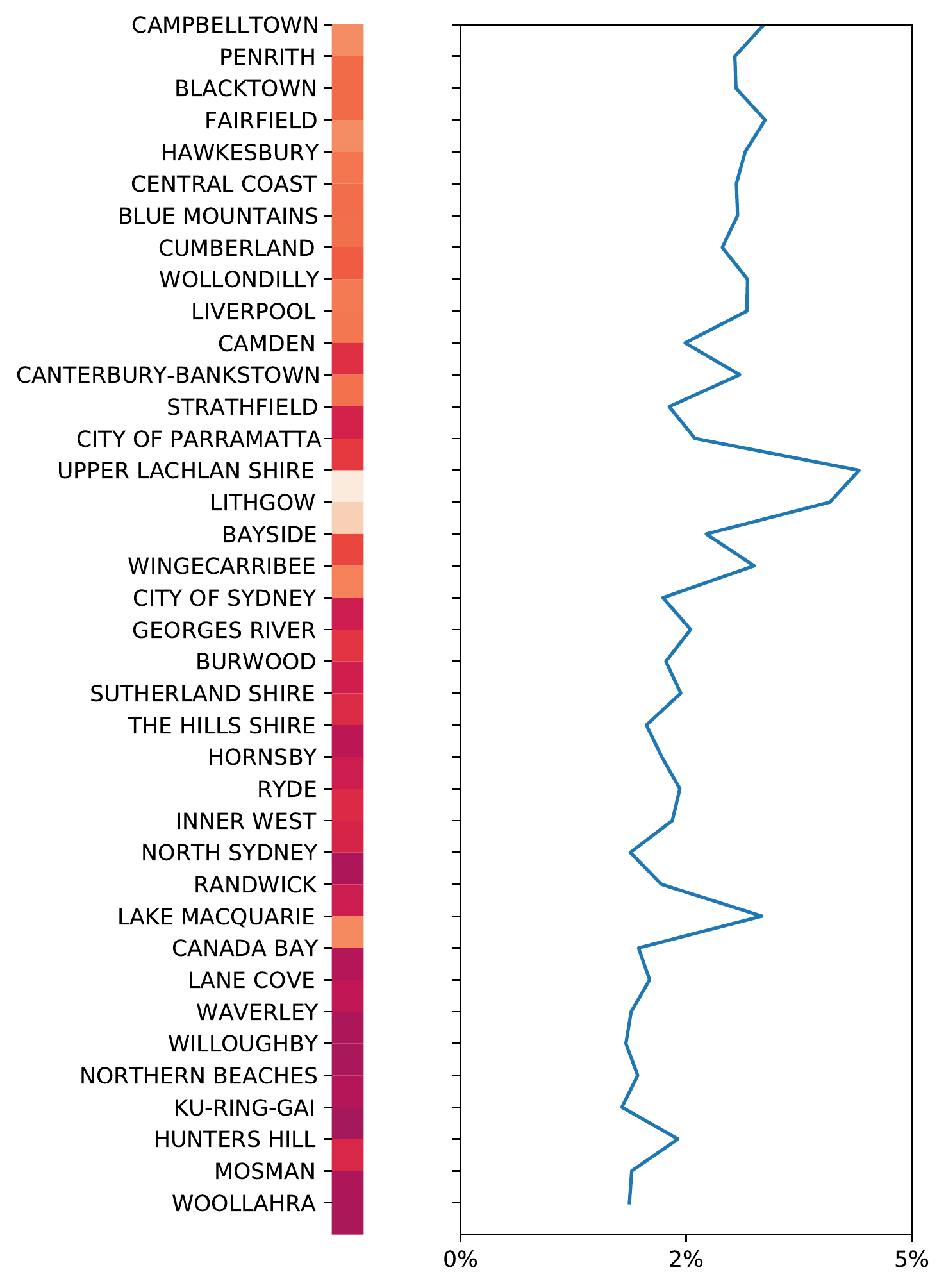}
       \includegraphics[height=.4\textheight]{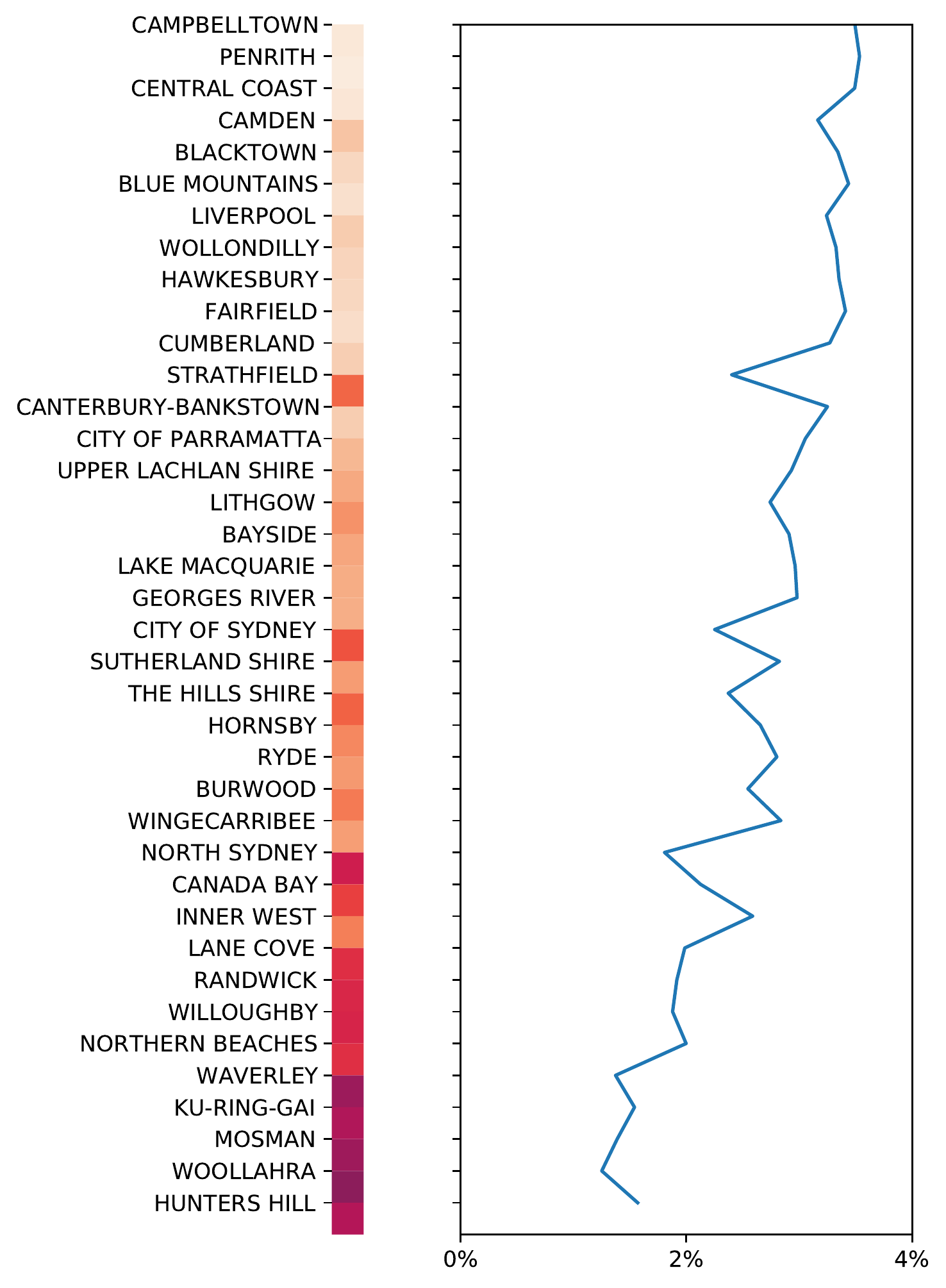}
        \includegraphics[height=.4\textheight]{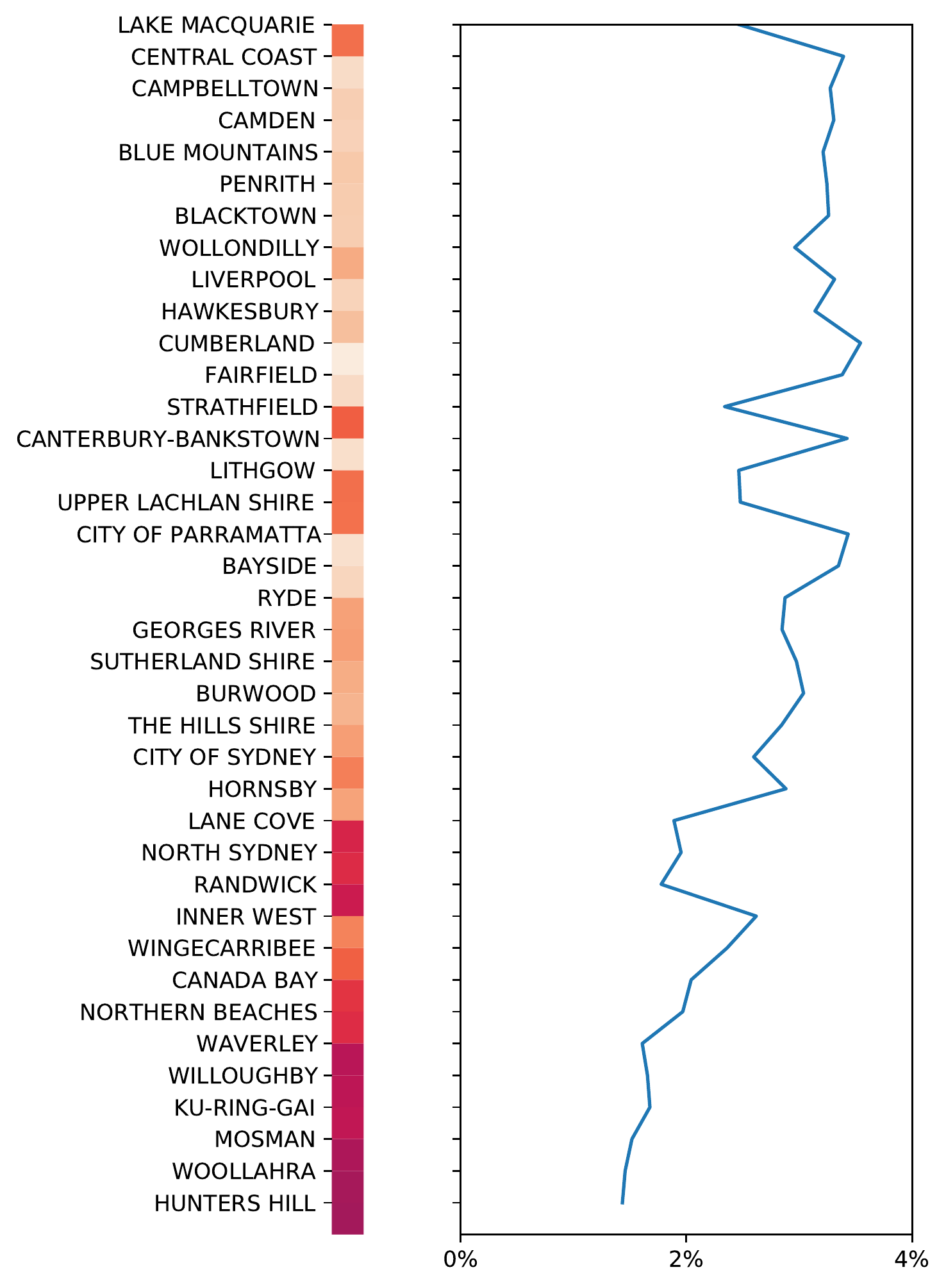}
    \caption{New Renters}
\end{subfigure}
 \hspace{5em}
\begin{subfigure}[b]{0.3\textwidth}
  \includegraphics[height=.4\textheight]{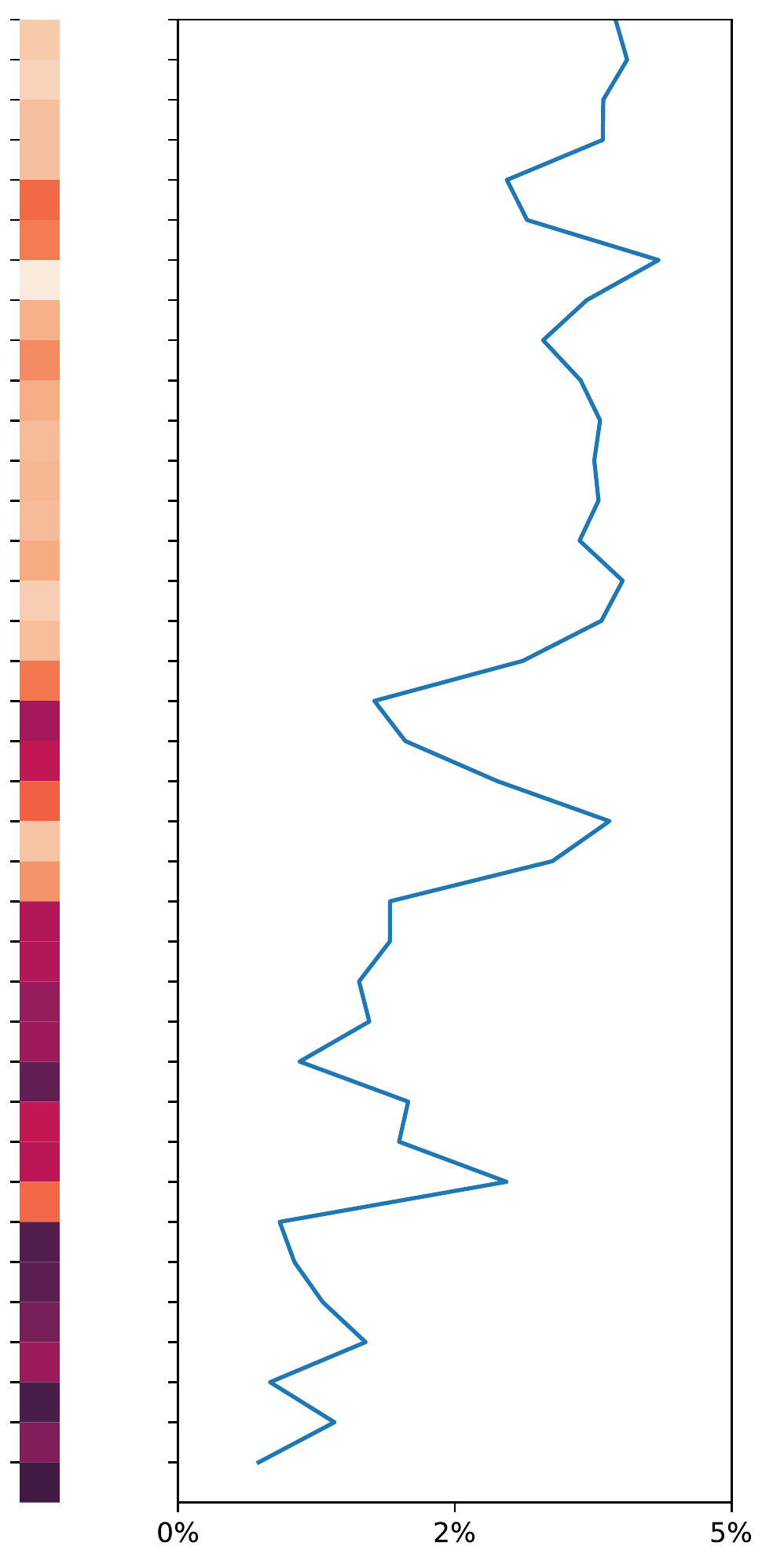}
  \includegraphics[height=.4\textheight]{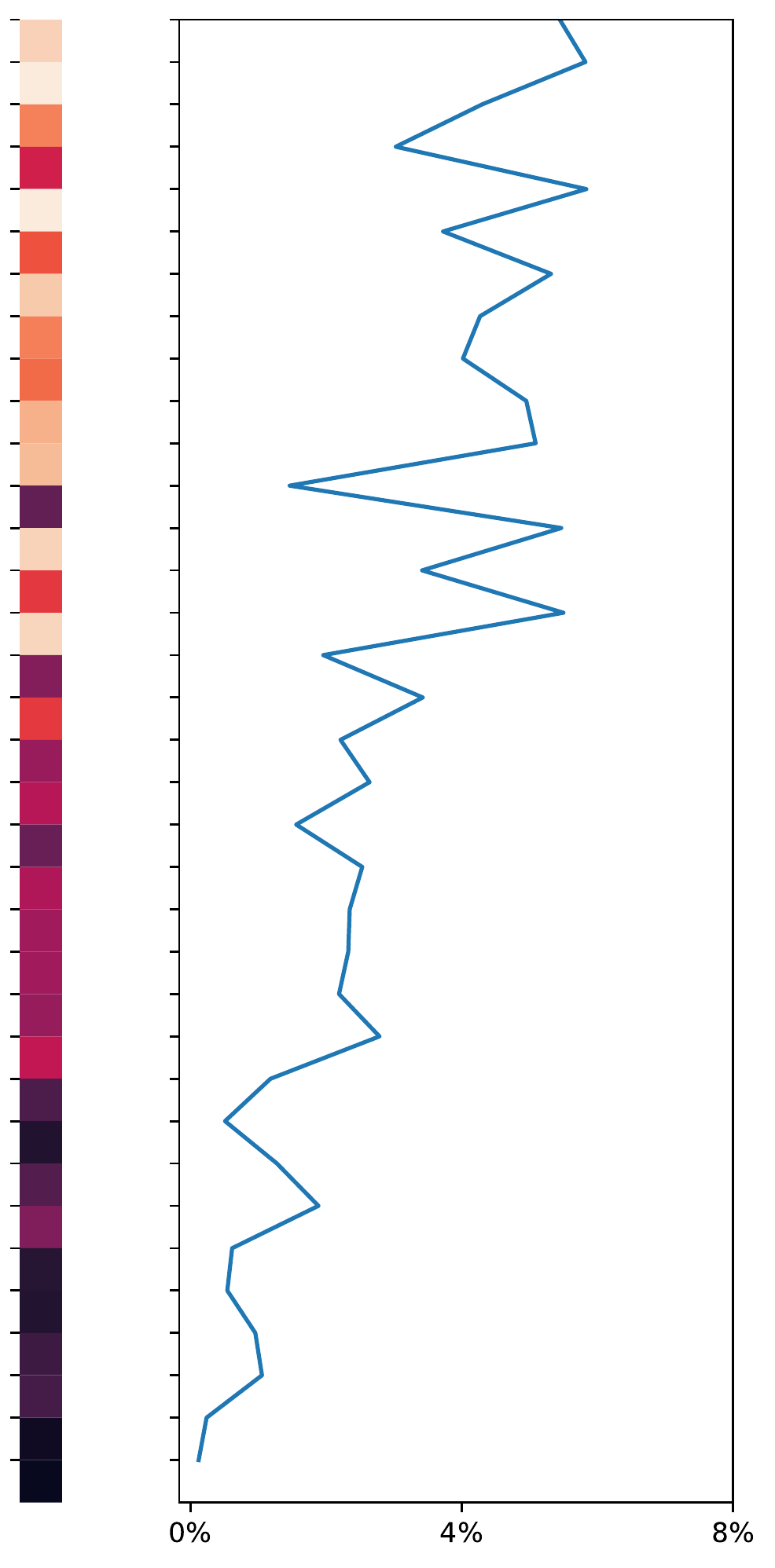}
\includegraphics[height=.4\textheight]{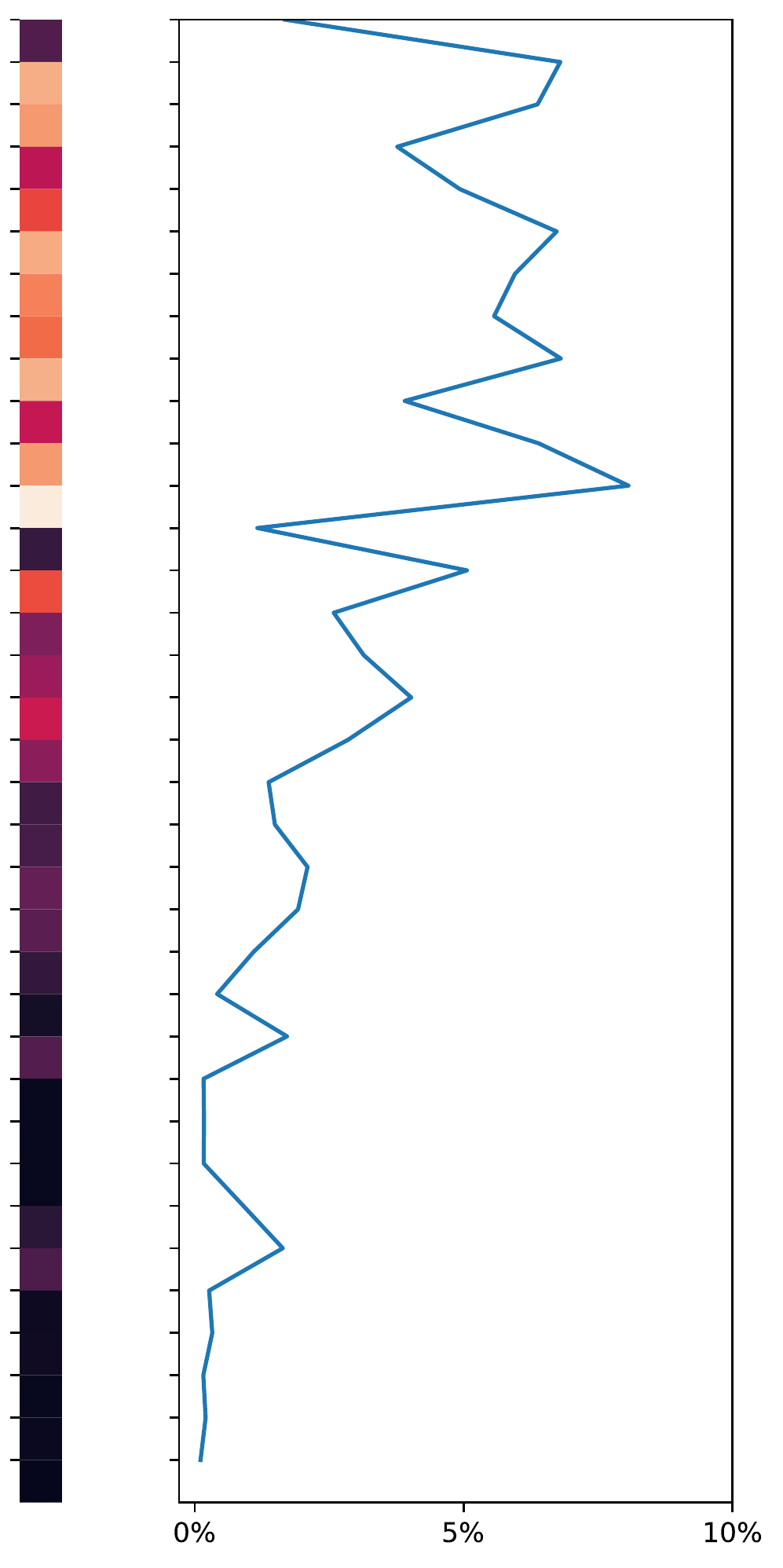}
\caption{New Owners}
\label{figMovementMigrate2006Owners}
\end{subfigure}%

\caption{Migrations. These plots capture new households in Greater Sydney throughout the simulation period, due to either migration or splitting of existing households. The first row is the 2006--2010 period, the middle row the 2011--2015 period, and the final row the 2016--2019 period.}
\label{figMovementMigrate}
\end{figure}

\begin{figure}[ht]
\centering
    \begin{subfigure}[b]{0.3\textwidth}
        \includegraphics[height=.4\textheight]{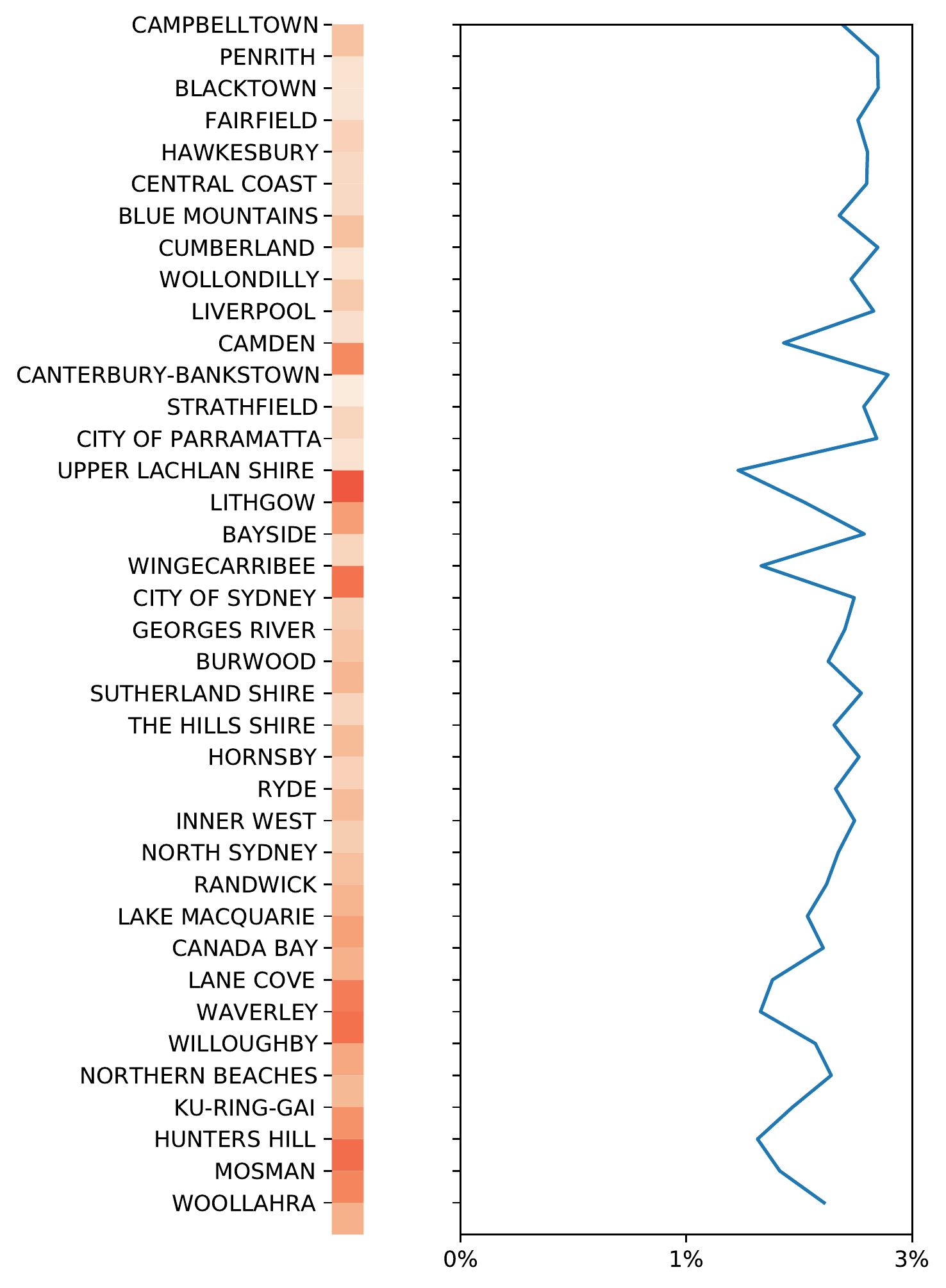}
        \includegraphics[height=.4\textheight]{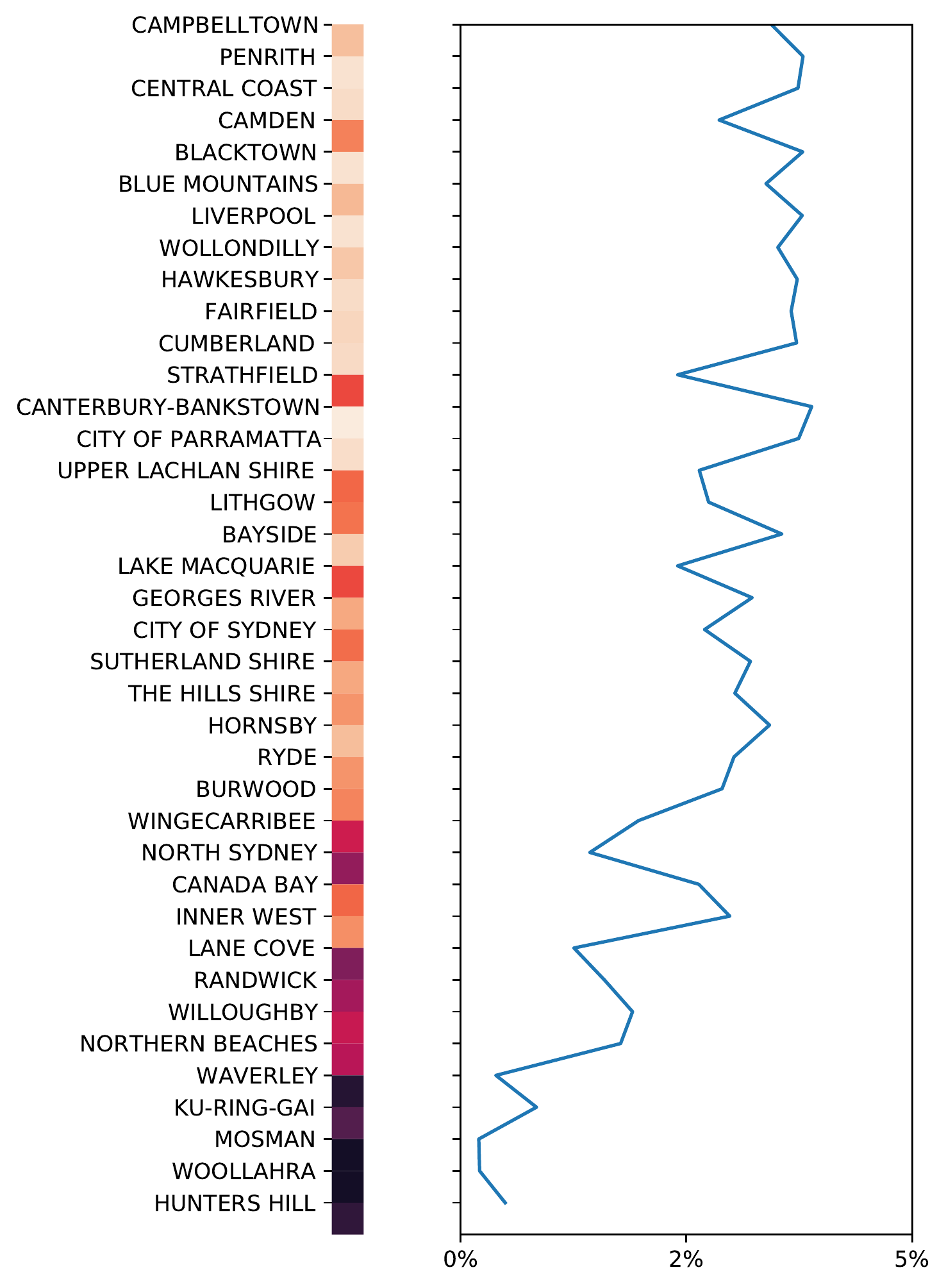}
         \includegraphics[height=.4\textheight]{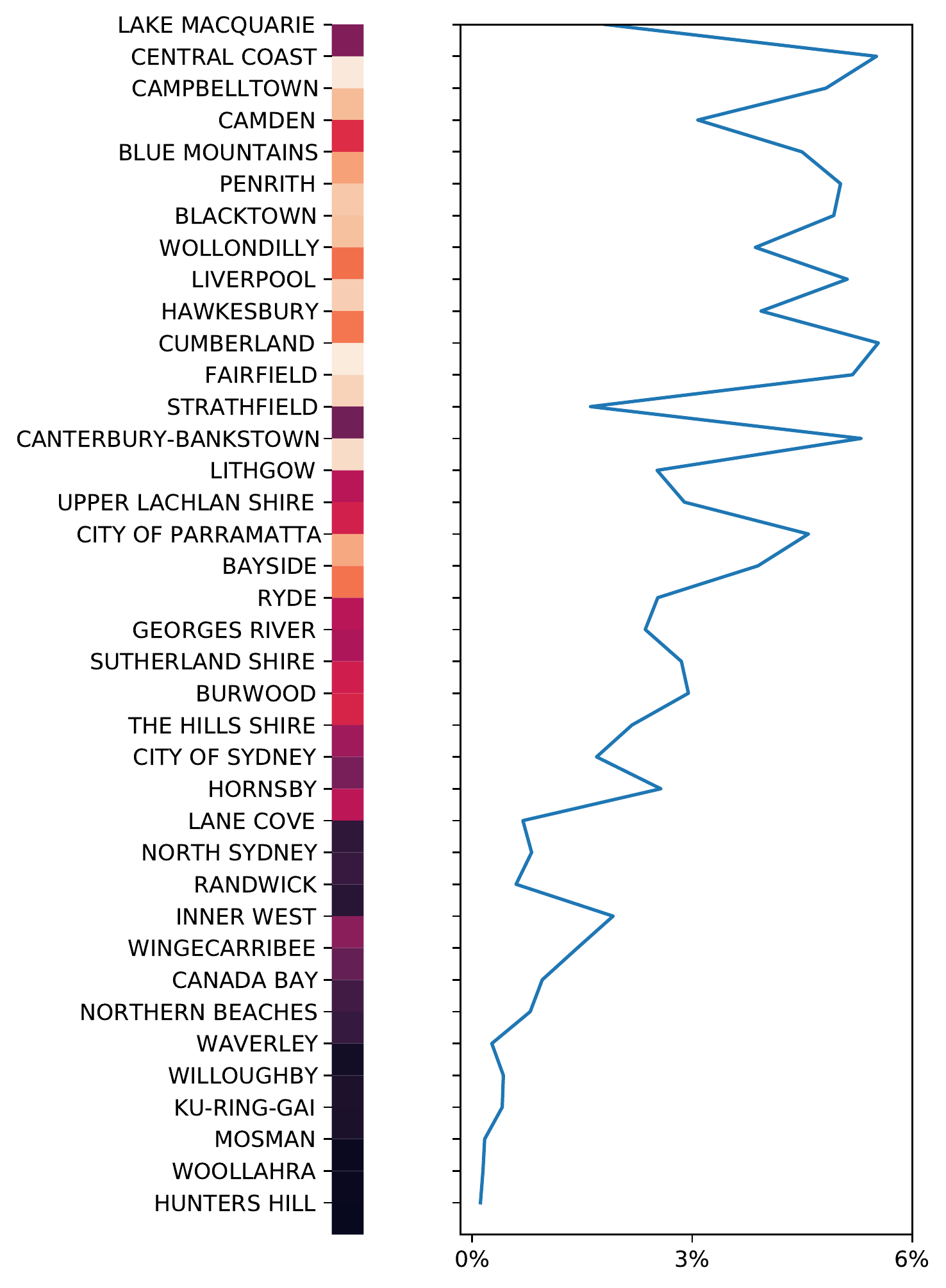}
        \caption{Local}
    \end{subfigure}
    \hspace{5em}
    \begin{subfigure}[b]{0.3\textwidth}
        \includegraphics[height=.4\textheight]{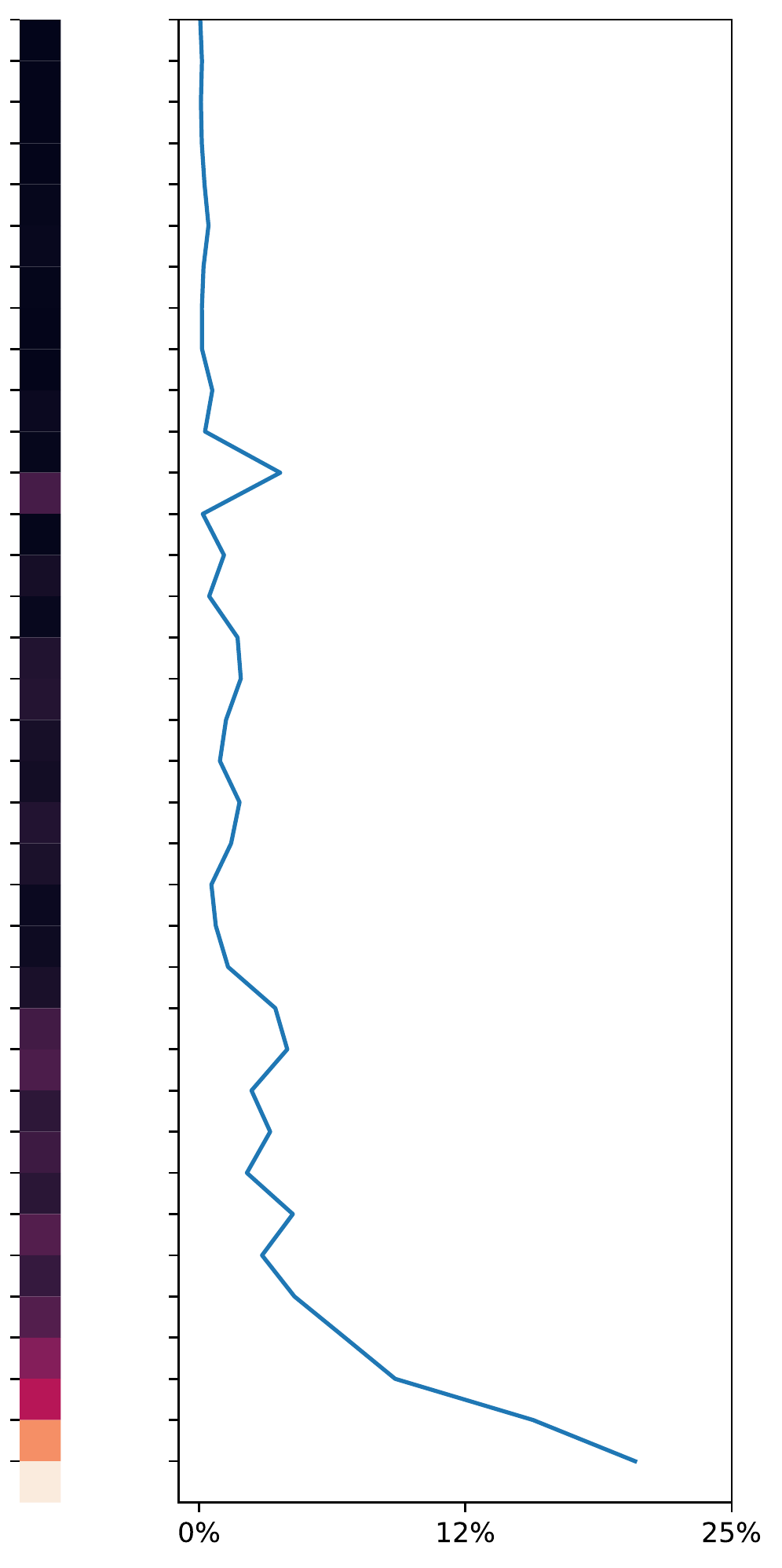}
        \includegraphics[height=.4\textheight]{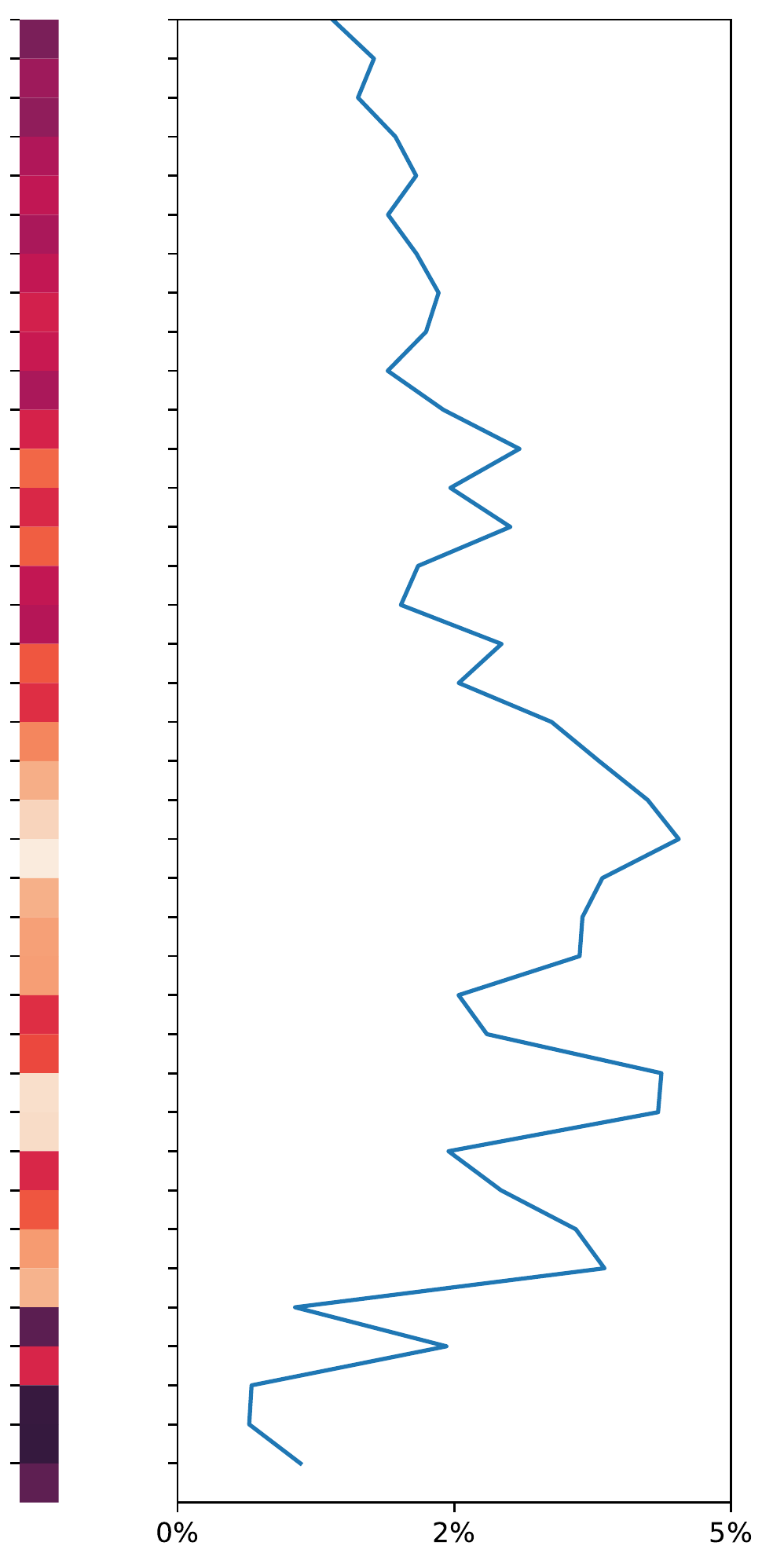}
        \includegraphics[height=.4\textheight]{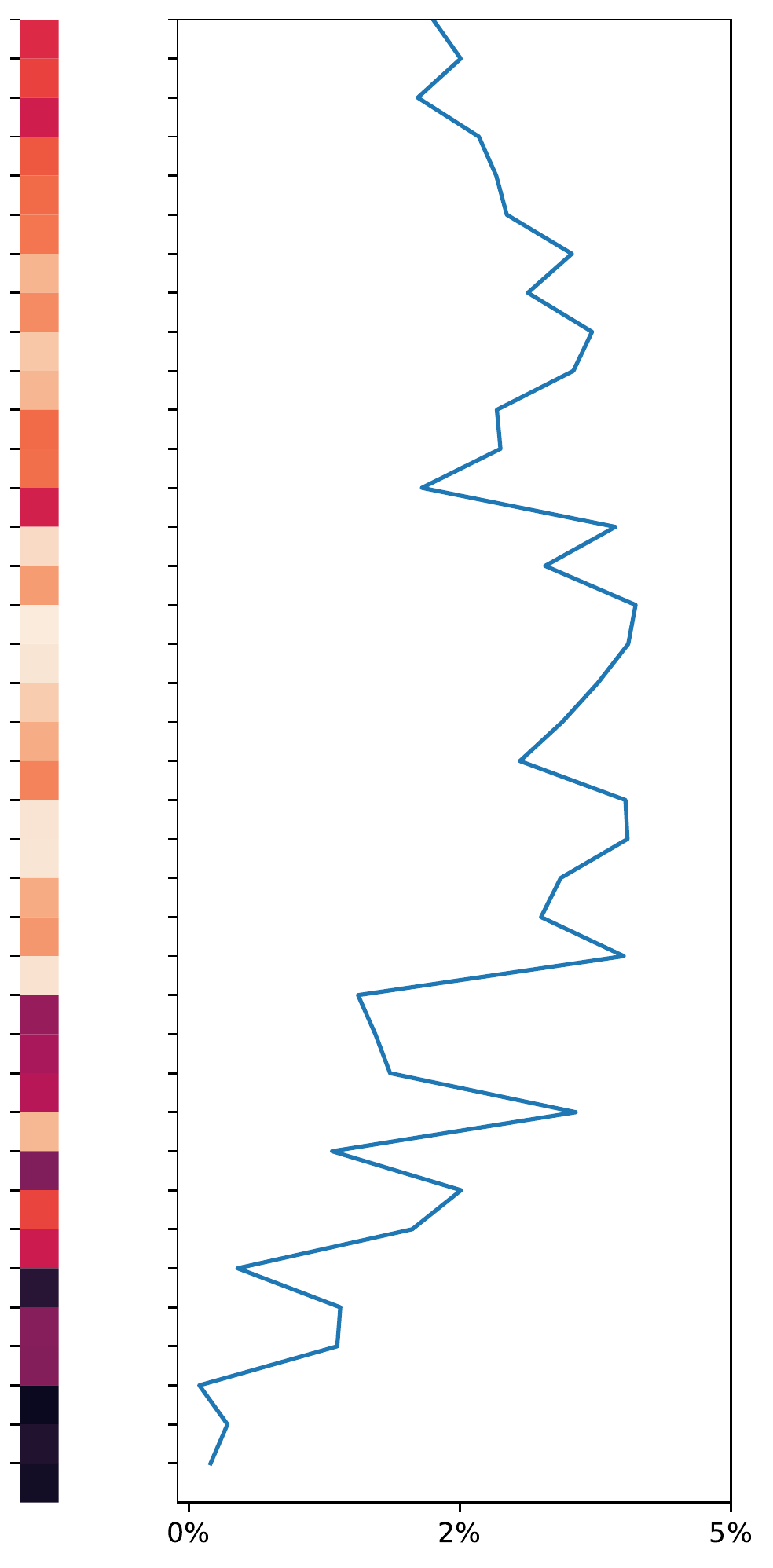}
        \caption{Overseas}
    \end{subfigure}
     
\caption{Investors. These plots show the simulation difference between local and overseas investment patterns.  The first row is the 2006--2010 period, the middle row the 2011--2015 period, and the final row the 2016--2019 period.}
\label{figMovementsInvestors}
\end{figure}

\begin{figure}[ht]
    \centering
    
     \begin{subfigure}[b]{0.7\textwidth}
     \centering
         \includegraphics[height=.38\textheight]{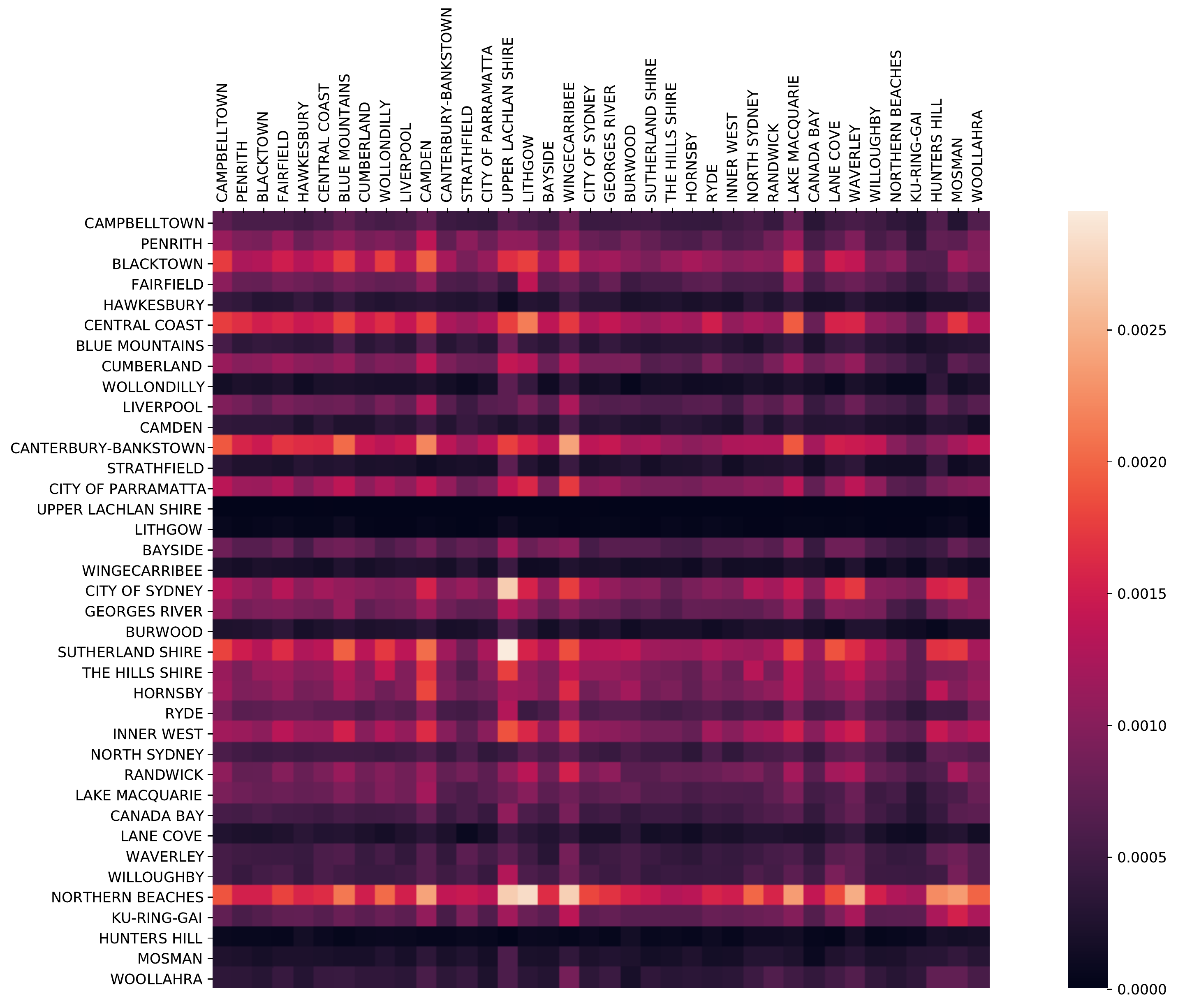}
          \includegraphics[height=.38\textheight]{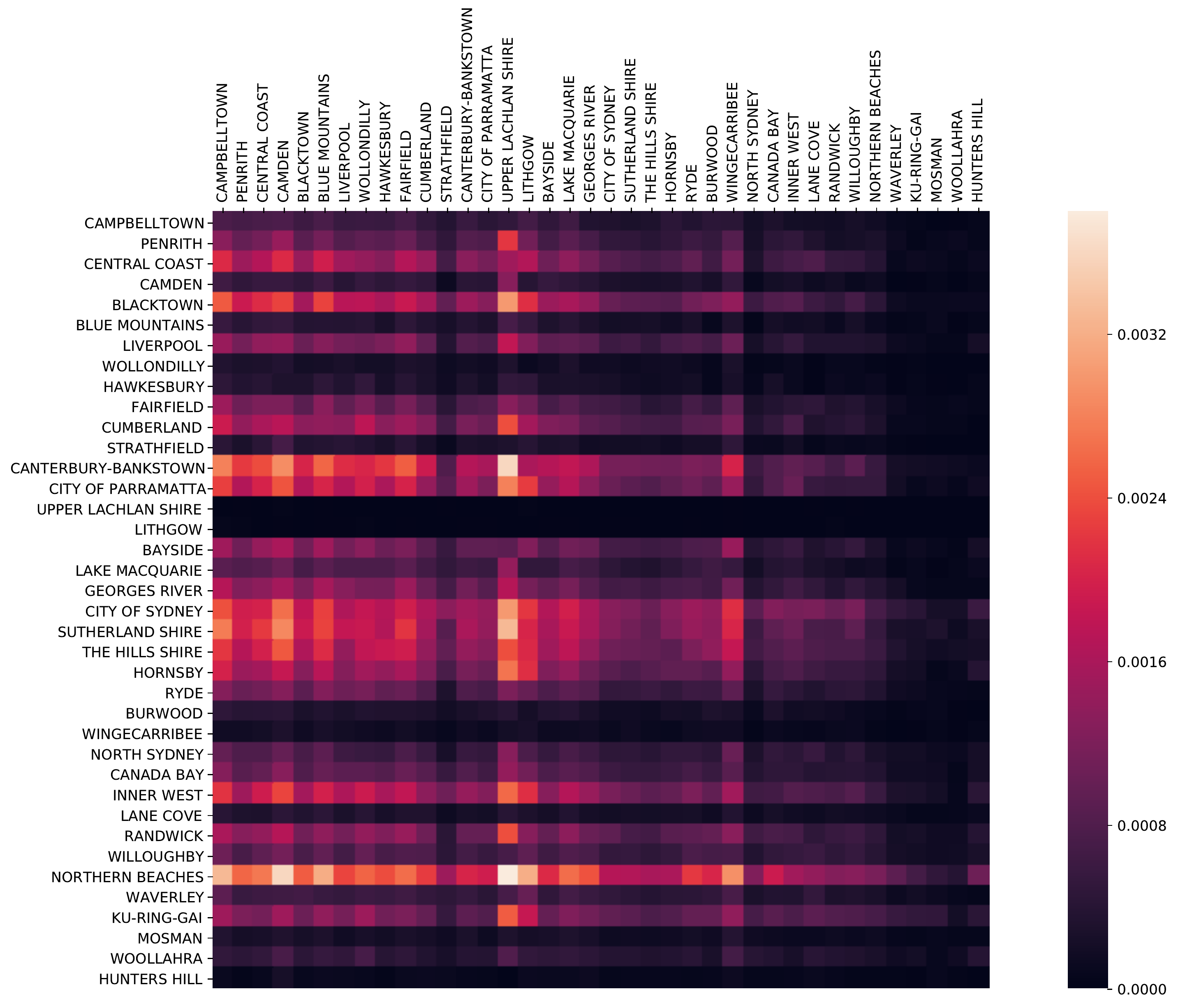}
          \includegraphics[height=.38\textheight]{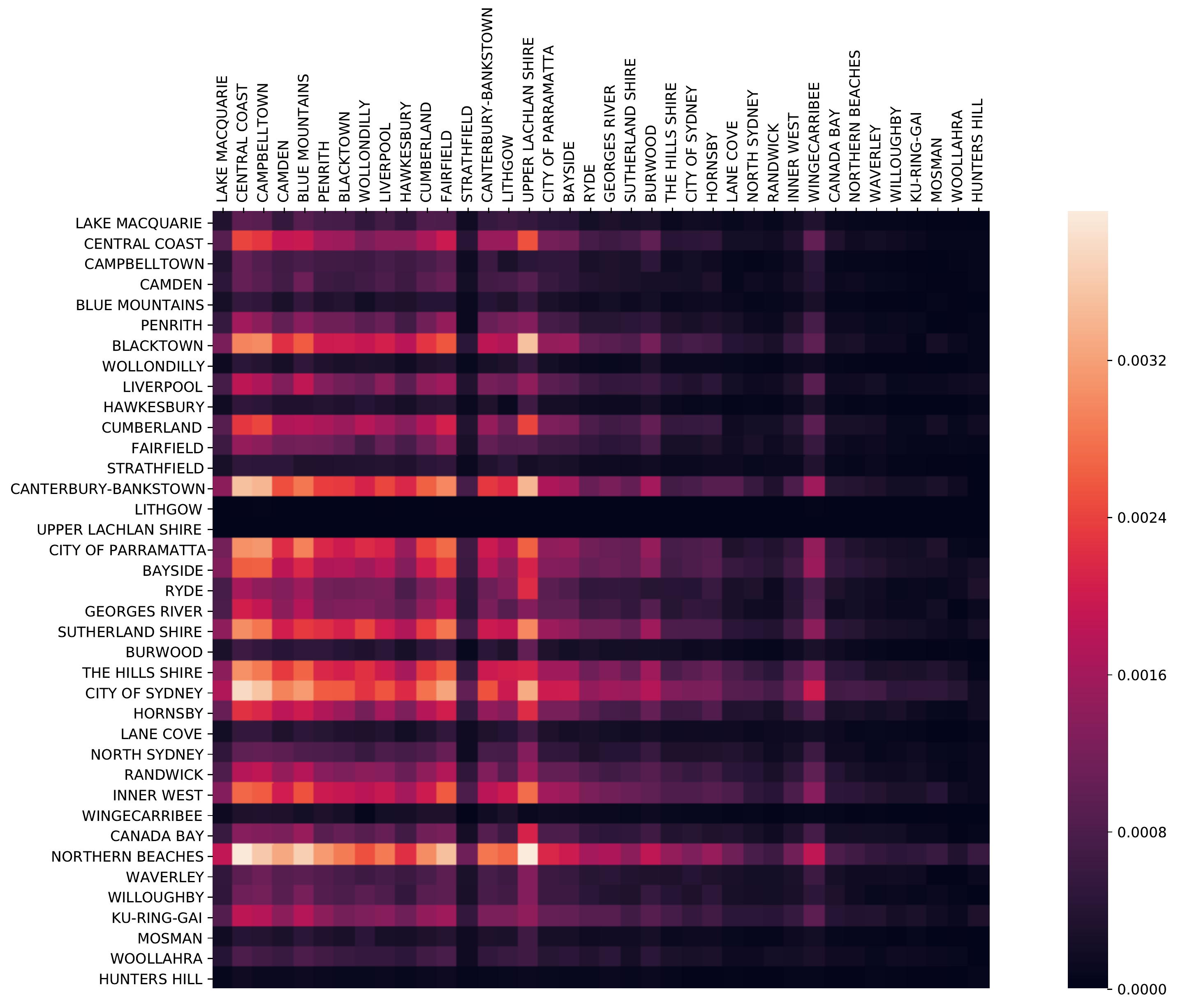}
         \caption{Renting LGA (rows), to Purchase LGA (columns)}
     \end{subfigure}
     \hfill
     \begin{subfigure}[b]{0.25\textwidth}
     \centering
        \includegraphics[height=.38\textheight]{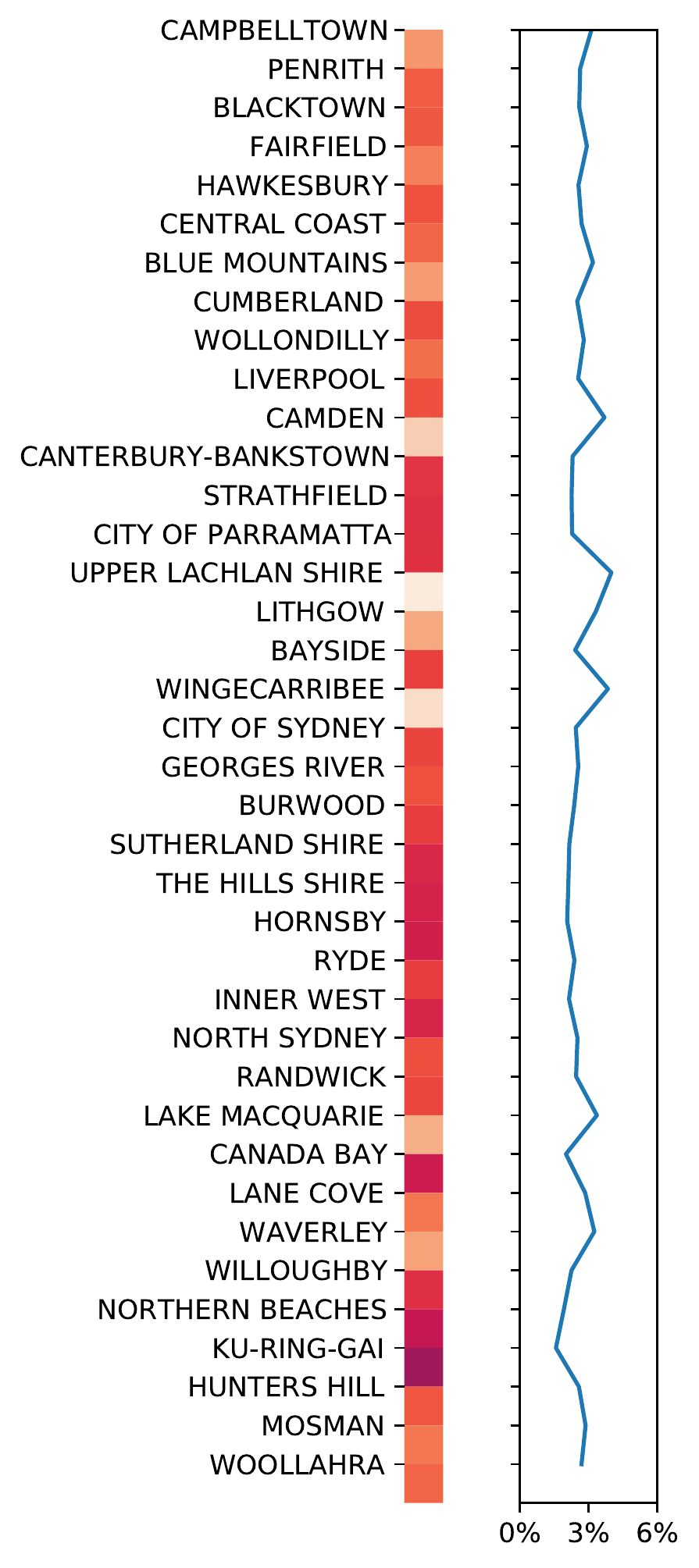}
        \includegraphics[height=.38\textheight]{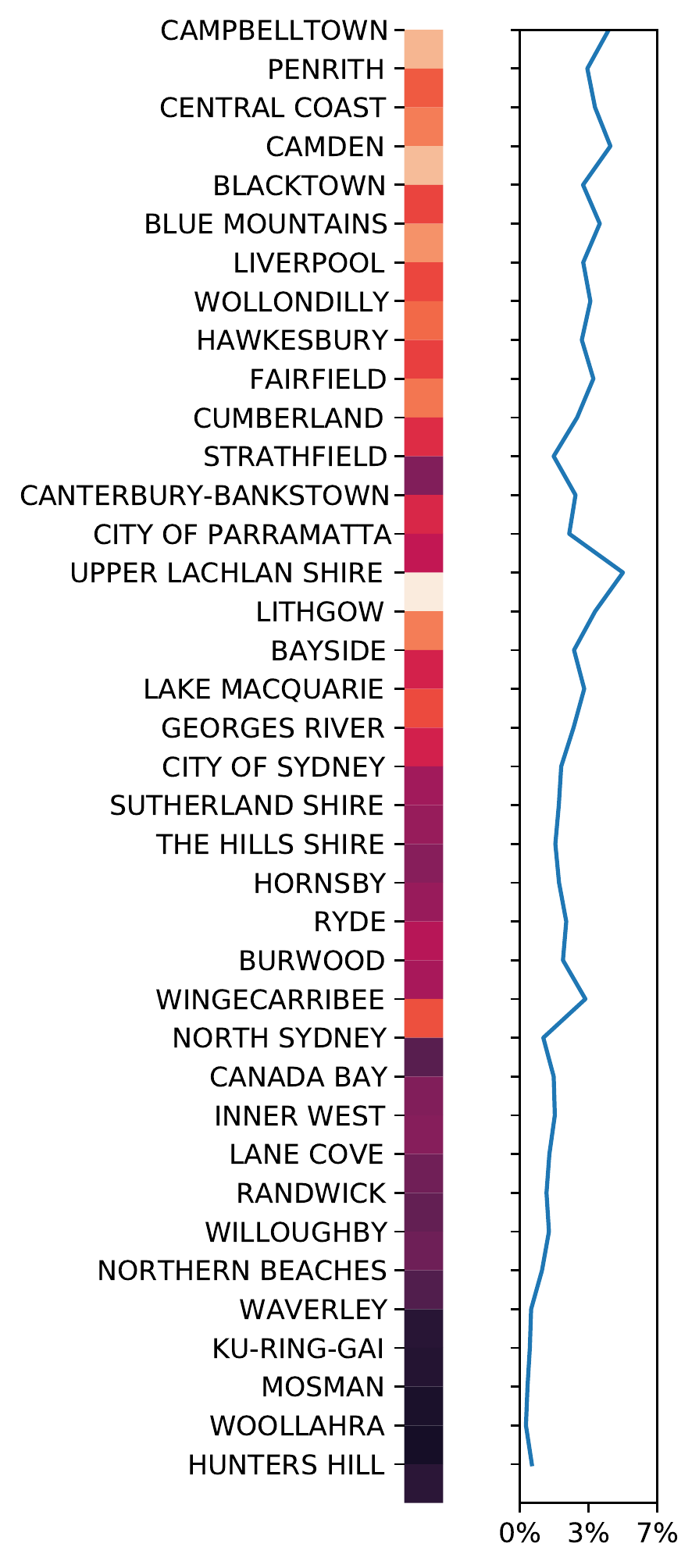}
        \includegraphics[height=.38\textheight]{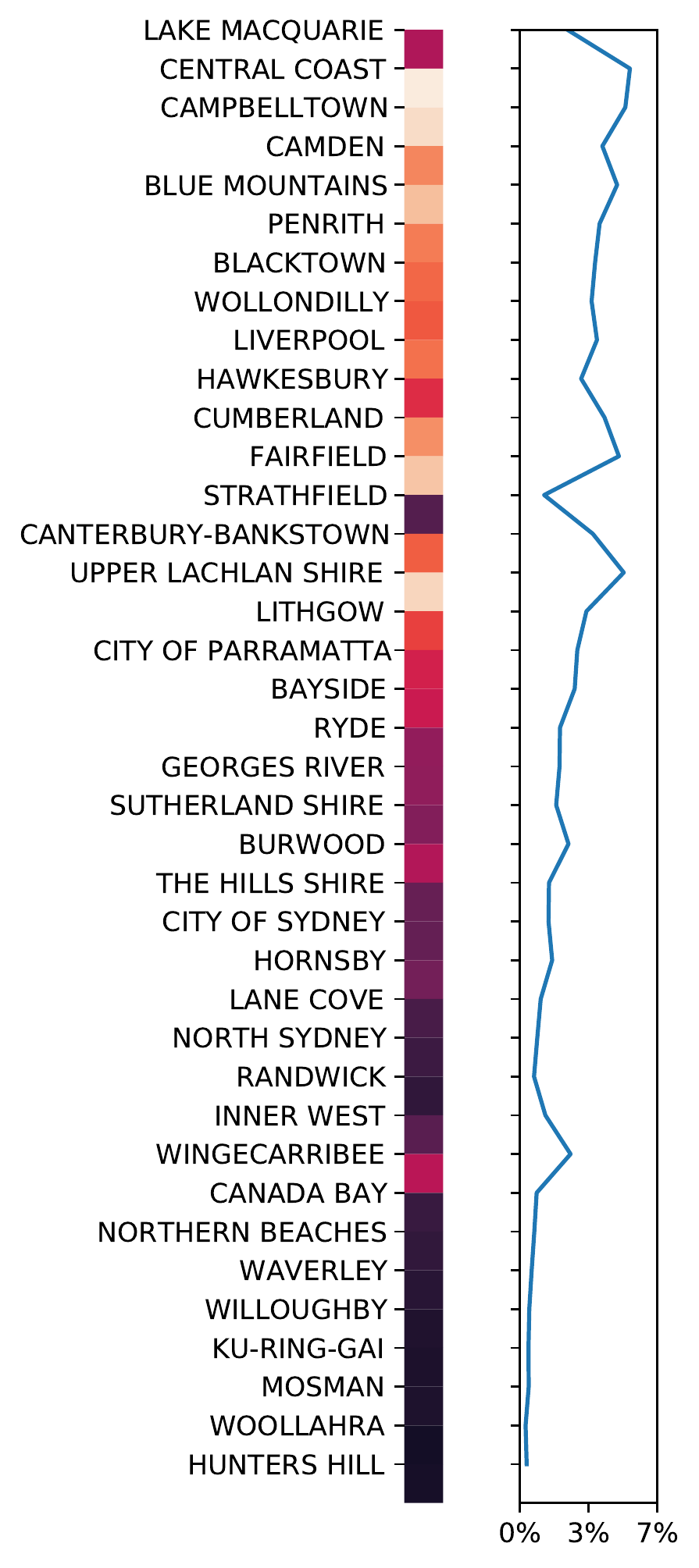}
        \caption{Purchase LGAs}
     \end{subfigure}
    
    \caption{First-time home buyers. The first row is the 2006--2010 period, the middle row the 2011--2015 period, and the final row the 2016--2019 period.}
    \label{figMovementFirstHomeowners}
\end{figure}

\section{Exogenous Variables}

There are two main external influences on the model, which are governed by government approvals (in the case of overseas investments) and the central bank (in the case of mortgage rates).

\subsection{Overseas Investors}\label{secOverseasApprovals}

Overseas investments are often cited as a key driver of price growth in the Australian market \citep{rogers2017public}, and figures show the foreign investment has more than tripled since the mid-1990s \citep{haylen2014house}.  However, actual data on foreign investments is difficult to find. ABS has described their own data on overseas investments to parliament as "hit or miss" \cite{iggulden_2014}.

The purpose of this work is not a full investigation into overseas investments (overviews are given in \cite{gauder2014foreign, australia2014report}), but rather the contribution overseas might have in relation to many other factors with the readily available data (be this complete or not).

\begin{figure}
    \centering
    \includegraphics[width=.9\textwidth]{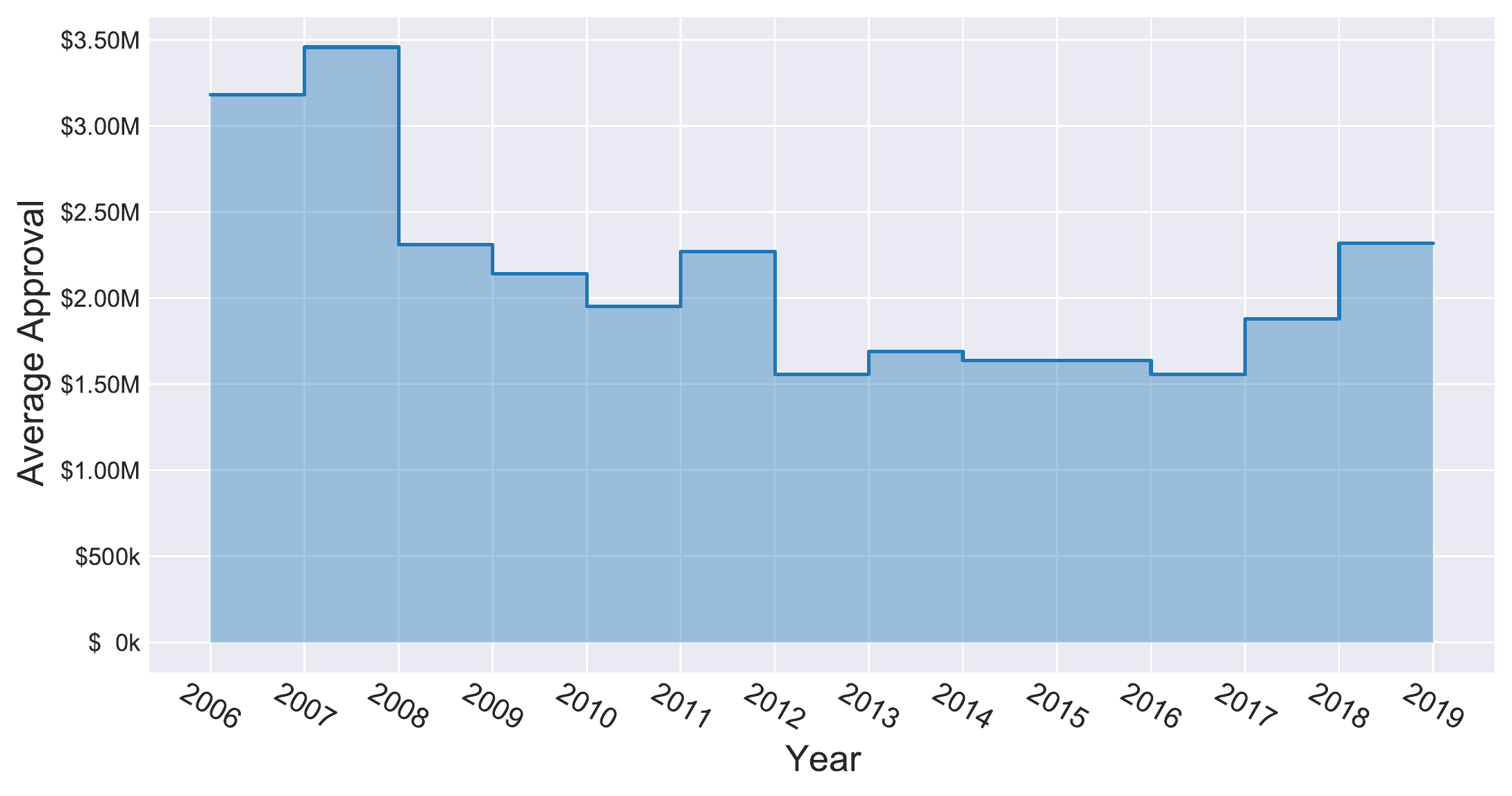}
    \caption{Average Overseas Approval Amount}
    \label{figOverseasApprovals}
\end{figure}

For this, we use the annual reports from the Foreign Investment Review Board (FIRB) from June 2006-June 2019. The June 2019 - June 2020 report was not available at the time of this writing (in 2020), as reports are not made available until the following year. Data is provided yearly at a NSW level, which is converted to monthly (simply dividing by 12). Again, data in this area is sparse, so this is the closest estimate we could derive. This data is provided in \cref{tblOverseas}, and the average approval per year given in \cref{figOverseasApprovals}.

\begin{table}[!ht]
\begin{tabular}{@{}lrrr@{}}
\toprule
\textbf{Period} & \multicolumn{1}{l}{\textbf{Number Approved}} & \multicolumn{1}{l}{\textbf{Value Approved}} & \multicolumn{1}{l}{\textbf{Average Per Approval}} \\ \midrule
2018-2019       & 1337                                         & \$3,100,000,000                             & \$2,318,624                                       \\
2017-2019       & 2340                                         & \$4,400,000,000                             & \$1,880,342                                       \\
2016-2017       & 4224                                         & \$6,580,000,000                             & \$1,557,765                                       \\
2014-2015       & 12349                                        & \$20,230,000,000                            & \$1,638,189                                       \\
2013-2014       & 7814                                         & \$13,220,000,000                            & \$1,691,835                                       \\
2012-2013       & 3580                                         & \$5,580,000,000                             & \$1,558,659                                       \\
2011-2012       & 3048                                         & \$6,920,000,000                             & \$2,270,341                                       \\
2010-2011       & 2598                                         & \$5,070,000,000                             & \$1,951,501                                       \\
2009-2010       & 910                                          & \$1,950,000,000                             & \$2,142,857                                       \\
2008-2009       & 956                                          & \$2,210,000,000                             & \$2,311,715                                       \\
2007-2008       & 1223                                         & \$4,230,000,000                             & \$3,458,708                                       \\
2006-2007       & 908                                          & \$2,890,000,000                             & \$3,182,819                                       \\ \bottomrule
\end{tabular}
\caption{Overseas Investment Approval}\label{tblOverseas}
\end{table}

While the data is provided for the entirety of NSW, it has been shown that foreign investors prefer the inner city over rural areas, and thus the NSW levels have been used for Greater Sydney. This is a fair assumption since the numbers are relatively conservative anyway. For the testing period, the most recent overseas approval value from the training period is used.

\subsection{Mortgage Rates}

Mortgage Rates are those set by the RBA. The final training months mortgage rate is used throughout the testing period since no real value can be read.

\section{Utility Function}\label{appendixUtility}

Following \cite{axtell2014agent}, agents are assumed to choose the most expensive house they can afford, that is, the house price directly corresponds to the utility for the agent.

\begin{figure}
    \centering
    \includegraphics[width=.8\textwidth]{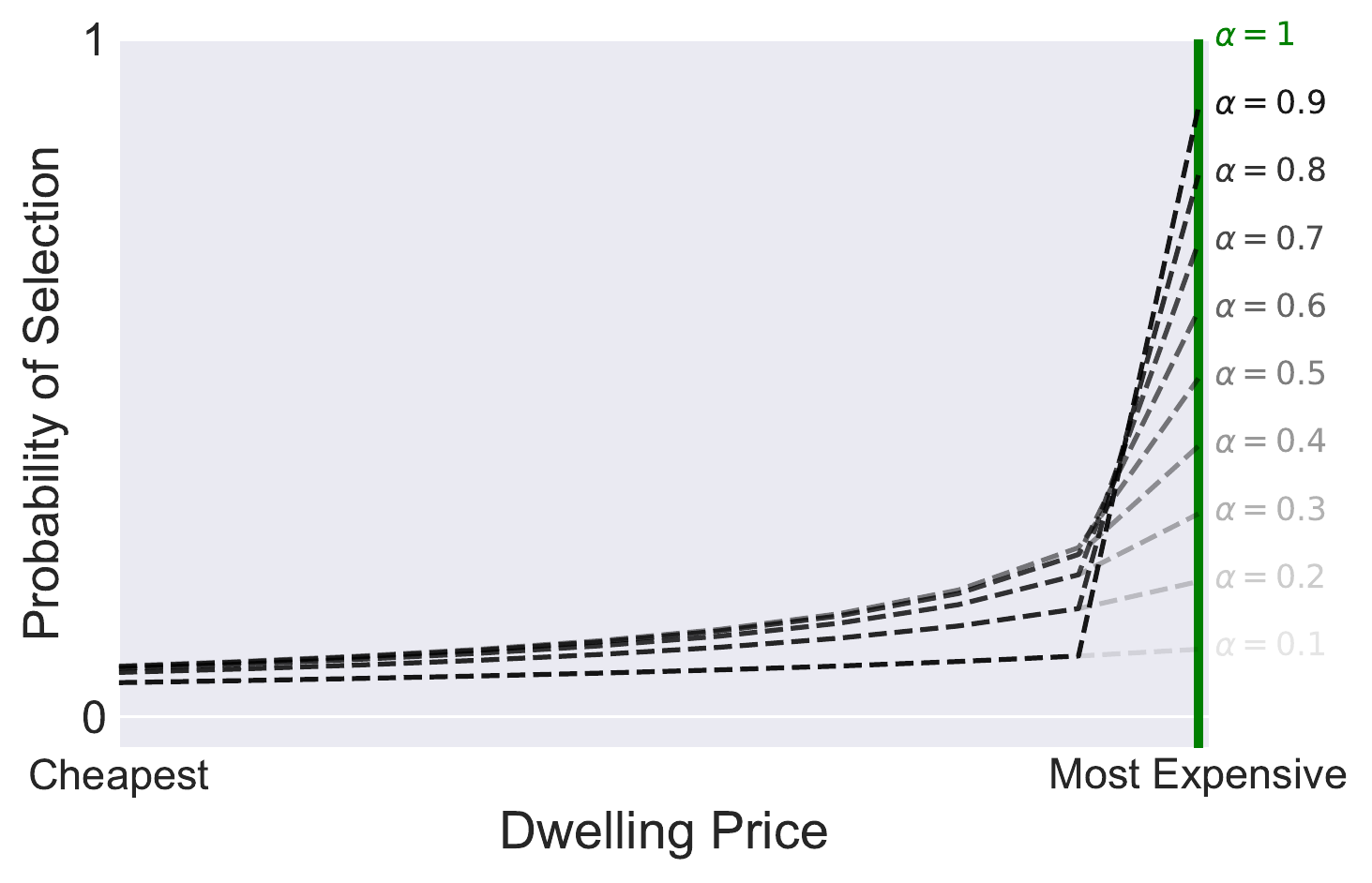}
    \caption{$\alpha$'s effect on utility maximising behaviour.}
    \label{figUtilityAlpha}
\end{figure}

However, the introduction of $\alpha$ alters this, such that there is some uncertainty or error in the agents choice. When $\alpha=1$, the perfect utility maximisation behaviour is recovered where the agent attempts to purchase the most expensive dwelling they can afford. For $\alpha < 1$, the agent buys the most expensive dwelling they can afford with probability $\alpha$, which then decreases for each subsequent listing in turn. This is visualised in \cref{figUtilityAlpha}. For high $\alpha$, we can see the probability mass is contained only in the highest priced dwellings. For lower $\alpha$, this probability mass becomes more distributed, meaning less focus on utility, and potential for cheaper houses to be purchased. For $\alpha=0$, the utility is not considered at all and a random house within the agents budget is chosen (i.e. the probability mass is uniform across options). $\alpha$, therefore, corresponds to the boundedness of the agent. 

\begin{figure}
    \centering
    \includegraphics[width=.8\textwidth]{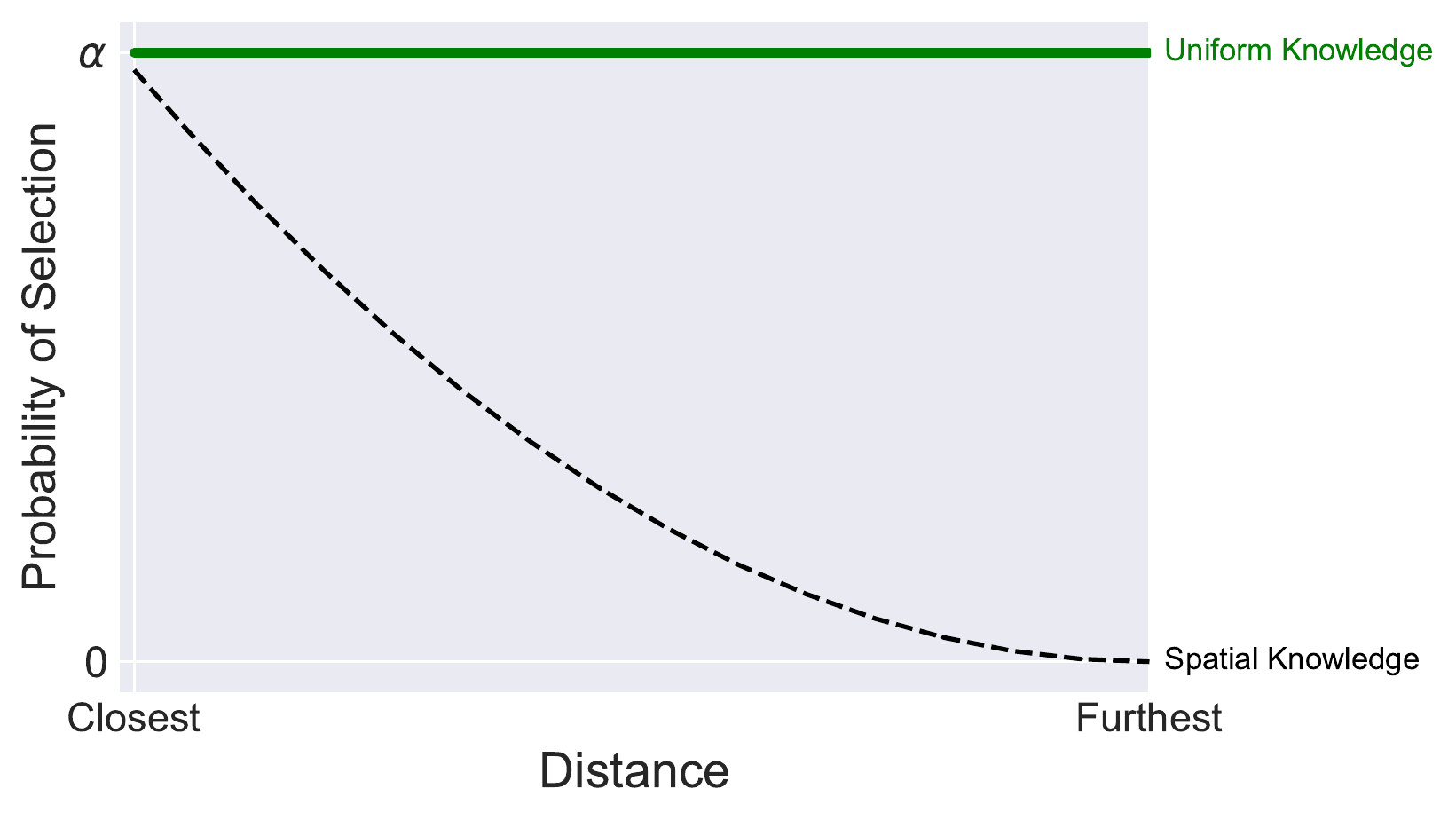}
    \caption{Probability of viewing based on distance to agents location.}
    \label{figUtilityDistance}
\end{figure}

The above description considers the case of uniform knowledge, i.e., for investors where they are assumed to be invariant to the areas available. However, for first time home buyers, we propose a spatial based knowledge where buyers are more likely to consider listings close to where they are renting. The probability associated with the distance to the agents' location is visualised in \cref{figUtilityDistance}. The uniform knowledge of investors is given in green, and the spatial knowledge of first-time home buyers is given as the dotted black line.

\begin{figure}
    \centering
    \includegraphics[width=\textwidth]{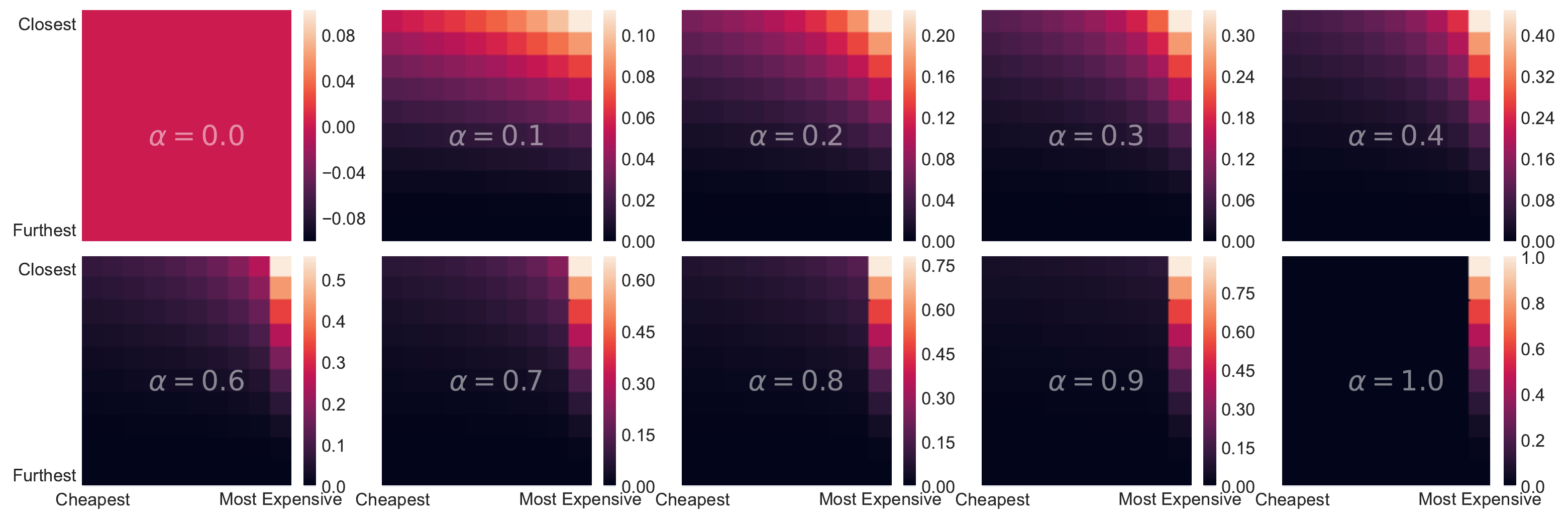}
    \caption{First time home buyers probability of viewing a listing for various $\alpha$'s. Showing the relationship between dwelling price (x-axis) and distance to dwelling (y-axis), and how $\alpha$ adjusts this distribution. Low $\alpha$'s correspond to higher dispersion, and less focus on utility maximising behaviour. High $
\alpha$'s focus the agent on dwellings which maximise utility.}
    \label{figUtility2D}
\end{figure}

For first time home buyers, the probability of viewing a listing is therefore controlled by both the proximity of the listing to the agents current (rental) location, and the price of the listing. This is visualised in \cref{figUtility2D}. For $\alpha=0$, the agent preference is uniform across all choices, placing no emphasis on utility (from either price or difference). As $\alpha$ increases, the focus shifts to the more expensive dwellings, and does so based on the distance to the listing. This can be seen in \cref{figUtility2D}, where with increasing $\alpha$ the emphasis focuses on the top right corner, which is the optimal value for both distance (closest) and price (most expensive in the agents budget). We can see that price remains the most important term in the agents' utility though, with close listings with low prices having a low resulting probability, indicating the agent likely wants to move to a more affluent area if they can afford to do so. However, given an equal price, agents will prefer the closer listing. 

 % Read from external appendix.tex file

\bibliographystyle{spbasic}
\bibliography{bib}

\end{document}